# Air-sea interaction in tropical atmosphere: influence of ocean mixing on atmospheric processes

Dariusz Bartłomiej Baranowski

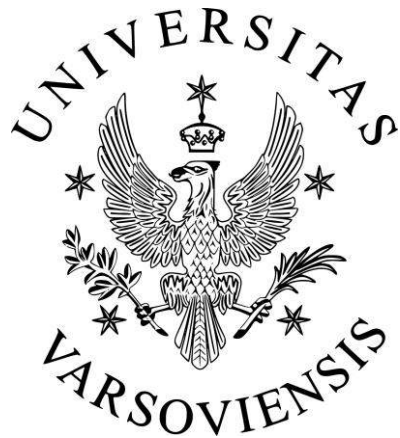

Doctoral dissertation
prepared at the Institute of Geophysics
Faculty of Physics, University of Warsaw

under the supervision of
Piotr J. Flatau and Szymon P. Malinowski
Warsaw, December 2014

Submitted to the Faculty of Physics of the University of Warsaw
in fulfillment of the requirements for the degree of Doctor of Physics

# Acknowledgements

Completing this dissertation would not have been possible without the support of many people to whom I am deeply grateful. First and foremost, I would like to thank my PhD advisors: Dr Piotr Flatau for his guidance, encouragement, mentoring and patience, and Professor Szymon Malinowski for his constructive suggestions and help throughout my doctoral program at the Institute of Geophysics of Warsaw University. I am especially thankful to Dr Maria Flatau for discussions, inspiration, support and her hospitality during my several visits at the Naval Research Laboratory in Monterey. The collaboration with her has been my privilege and great pleasure. I would like to extend my gratitude to Dr Adrian Matthews from University of East Anglia for inspiring and very fruitful collaboration as well as very useful discussions about tropical meteorology. The experimental part of this thesis is based mostly on data collected during the DYNAMO, an international field campaign in which I had an opportunity to participate. I would like to thank Dr Scott Harper of the Office of Naval Research who supported my efforts through the Office of Naval Research DRI project and Dr William Brown of the National Center for Atmospheric Research for letting me participate in the DYNAMO cruise on R/V Roger Revelle. I would like to thank Dr Dean Roemmich of the Scripps Institution of Oceanography, Dr Jim Price and Dr Iam-Fei Pun of the Woods Hole Oceanographic Institution, Professor Kerry Emanuel of the Massachusetts Institute of Technology, Dr Ben Webber, Dr Bastien Queste, Dr Rob Hall and Professor Karen Heywood of the University of East Anglia, Dr Sue Chen, Dr Jerome Schmidt, Dr Jim Ridout and Dr Carolyn Reynolds of the Naval Research Laboratory for their help and support during this research.

This thesis is based on the research supported by Office of Naval Research award number N00014-10-1-0898, Polish National Science Center award number UMO-2012/07/N/ST10/03303, Office of Naval Research Global NICOP award number N62909-11-1-7061 and POGO-SCOR Fellowship award.

# Streszczenie

Głęboka konwekcja i jej zmienność w skali od jednego dnia do kilku miesięcy to ważne zjawiska odpowiedzialne za przebieg przepływów atmosferycznych w tropikach. Jednym z głównych czynników determinujących powstawanie i ewolucję głębokich chmur konwekcyjnych jest temperatura powierzchni morza. Temperatura ta jest zależna z kolei od stratyfikacji górnych warstw oceanu, na którą wpływa dopływ energii słonecznej, słodkiej wody z opadów, a także procesy mieszania zależne od wiatru nad powierzchnią morza. W rozprawie przedstawiam wyniki badań wzajemnych oddziaływań pomiędzy procesami głębokiej konwekcji w atmosferze i własnościami górnych warstw oceanu. W badaniach wykorzystuję dane z pomiarów in-situ, dane z radarów i obrazów satelitarnych a także wyniki symulacji numerycznych.

W Rozdziale 2 przedstawiam analizy, które pokazują, że procesy głębokiej konwekcji w atmosferze wpływają na rozkład energii termicznej górnych warstw oceanu. Podczas spokojnych i bezchmurnych warunków atmosferycznych, przy niewielkim wietrze o średniej dobowej prędkości poniżej 6 ms$^{-1}$ i wysokim usłonecznieniu o średnim dobowym strumieniu energii przekraczającym 80 Wm$^{2}$, przy powierzchni oceanu formuje się głęboka na kilka metrów warstwa ciepłej wody. Ta ciepła warstwa może być interpretowana jako dobowa anomalia (fluktuacja) temperatury powierzchni morza osiągająca amplitudę do 0.8 °C, co powoduje fluktuacje strumienia ciepła oceanu do atmosfery o amplitudzie 4 Wm$^{-2}$. Wykorzystując te wyniki wprowadzam model prognostyczny fluktuacji temperatury powierzchni morza w funkcji nasłonecznienia i prędkości wiatru przy powierzchni morza. Obliczone modelem anomalie temperatury powierzchni morza wykorzystuję w analizie rozwoju i ewolucji konwekcji atmosferycznej, zorganizowanej w postaci równikowych fal Kelvina związanych konwekcyjne (ang. Convectively Coupled Kelvin Waves) oraz ich zmienności w skali kilku tygodni powodowanych oscylacjami Maddena-Juliana.

W celu zbadania wpływu anomalii temperatury powierzchni morza na powstawanie i rozwój fal Kelvina, wykorzystując dane satelitarne utworzyłem nową nową bazę danych o trajektoriach fal Kelvina wędrujących wokół globu wzdłuż równika.

W Rozdziale 3 przedstawiam analizy inicjalizacji fal Kelvina. Wyniki badań strumieni energii z powierzchni oceanu, wzbogacone modelowaniem numerycznym, pokazują że znaczący ułamek obserwowanych fal Kelvina powstaje dzięki oddziaływaniu z inną istniejącą już falą Kelvina. Zdefiniowałem dwie rozłączne kategorie takich dwu- i wielokrotnych inicjalizacji, które zostały przeanalizowane niezależnie. Do pierwszej kategorii zaliczam przypadki inicjalizacji zachodzącej przy wysokiej dobowej anomalii temperatury powierzchni morza. Druga kategoria to inicjalizacja typu „spin off", czyli wskutek wzrostu prędkości wiatru i strumienia ciepła utajonego na powierzchni oceanu wywołanego wcześniejszą falą Kelvina.

W Rozdziale 4 przedstawiam wyniki badań prowadzonych wzdłuż trajektorii rozwiniętych fal na Oceanie Indyjskim dotyczących oddziaływania pomiędzy falami Kelvina a górnymi warstwami oceanu. Wykazuję w nim, że szybkie fale Kelvina wpływają na zmienność dobową temperatury powierzchni morza oraz na prędkość wiatru i strumień ciepła utajonego na powierzchni oceanu. Tworząc przez uśrednienie wielu pojedynczych fal Kelvina tzw. „falę kompozytową" pokazuję, że zmiany w prędkości wiatru, strumieniu ciepła utajonego i anomalii temperatury powierzchni morza mają pewne cechy charakterystyczne. Podczas przejścia fali prędkość wiatru i strumień ciepła utajonego rosną, a anomalia temperatury powierzchni morza maleje. Efekty wzrostu i spadku zależą od fazy oscylacji Maddena-Juliana, w której występuje fala.

Dalej, w Rozdziale 5, badam własności fal Kelvina propagujących się z Oceanu Indyjskiego nad Azję Południowo-Wschodnią. Ten obszar ma duże znaczenie dla dynamiki przepływów tropikalnych, a wiele modeli prognoz pogody ma trudności w prognozowaniu przepływu w tym regionie. Pokazuję, że fale Kelvina na określonej długości geograficznej - nad Afryką, Oceanem Indyjskim i Azją Południowo-Wschodnią, charakteryzują się tą samą fazą cyklu dobowego (ang. phase locking). Wynika z tego, że fale Kelvina zazwyczaj przechodzą nad danym obszarem o konkretnej porze dnia. Nad Azją Południowo-Wschodnią ta stała faza jest w zgodzie ze średnim, lokalnym cyklem dobowym konwekcji w atmosferze.

Fale, które nie są zablokowane przez Azję Południowo-Wschodnią, czyli propagują się dalej na Zachodni Pacyfik, różnią się od takich które są zablokowane, czyli kończą propagację nad Azją Południowo-Wschodnią. Kombinacja prędkości fazowej fali Kelvina i pory dnia, o której fala nadchodzi nad Azję Południowo-Wschodnią wpływa na prawdopodobieństwo jej skutecznej propagacji na Zachodni Pacyfik. Fale Kelvina o prędkości fazowej od 10 do 11 stopni na dzień nad centralnym i wschodnim Oceanem Indyjskim, które przecinają południk 90E pomiędzy 9UTC a 18UTC mają największą szansę przejścia na Zachodni Pacyfik. Dla takich fal odległość pomiędzy Sumatrą a Borneo – dwoma głównymi ośrodkami konwekcji nad lądem w tym regionie – jest zgodna z dystansem jaki taka fala Kelvina przemierza w ciągu jednego dnia. Sugeruje to, że wyspy Azji Południowo-Wschodniej mogą działać jako „filtr" dla fal Kelvina, sprzyjając skutecznej propagacji fal, które poruszają się zgodnie z lokalnym cyklem dobowym konwekcji. Dla uproszczonej analizy tego zjawiska wprowadzam prostą metrykę lokalnego cyklu dobowego biegnących fal Kelvina - indeks AmPm. Pokazuję, że jest on użyteczny i przedstawia kluczowe cechy konwekcji związanej z propagującymi się przez Azję południowo-Wschodnią falami Kelvina.

Głównym wynikiem pracy jest wykazanie, że oddziaływanie pomiędzy konwekcją w atmosferze a górnymi warstwami oceanu ma charakter dwukierunkowego sprzężenia zwrotnego. Rozwinięta konwekcja w atmosferze wpływa na rozkład energii termicznej w górnych warstwach oceanu, a w szczególności na temperaturę powierzchni morza. Zmiany temperatury powierzchni morza wpływają z kolei na rozwój konwekcji w atmosferze, a w szczególności konwekcji zorganizowanej w postaci równikowych fal Kelvina. Szczegółowe wyniki pracy pokazują konsekwencje tego sprzężenia dla własności fal Kelvina.

# Abstract

Changes in activity of deep convective clouds, developing over warm waters of tropical oceans, are dominant mode of diurnal to intraseasonal variability of the atmospheric circulation in the tropics. One the major factors determining the development and evolution of atmospheric convection are the sea surface temperature. Thus, the variability of the upper ocean stratification, which impacts the sea surface temperature, is an important factor in the variability of the atmospheric convection in the tropics. In this thesis, the two way interactions between atmospheric convection and upper ocean are investigated on the basis of the in-situ measurements, satellite data and numerical simulations.

The results show that state of atmospheric convection impacts the diurnal distribution of thermal energy in the upper ocean. Under calm and clear sky conditions, with daily mean wind speed below 6 ms$^{-1}$ and daily mean solar radiation flux above 80 Wm$^{2}$, a shallow warm layer of several meters depth develops on the surface of the ocean. This warm layer may be interpreted as a diurnal sea surface temperature anomaly which often reaches amplitude of 0.8 °C and drives an anomalous flux of 4 Wm$^{2}$ from the ocean to the atmosphere. Based on these results a predictive model of sea surface temperature anomaly, as a function of surface insolation and wind speed, is developed. The derived sea surface temperature anomaly and surface fluxes are used in analysis of the development and evolution of atmospheric convection organized in equatorial convectively coupled Kelvin waves.

A novel Kelvin wave trajectory database based on satellite data is introduced in this study. The investigation of surface fluxes and remote sensing data, augmented by the numerical modeling, shows that substantial fraction of Kelvin waves is initiated as a result of interaction with another Kelvin wave. Two distinct categories are defined and analyzed independently. The first one accounts for two- and multiple Kelvin wave initiations which occur when diurnal sea surface anomaly is high. The second category is a "spin off" initiation which occurs when a Kelvin wave initiates over the area through which another Kelvin wave passed within a few days. Results show that primary forcing of such waves are increased wind speed and latent heat flux at the ocean surface.

In the following Chapter investigation of interactions between Kelvin wave and upper ocean is presented. Variability of the ocean surface and subsurface along Kelvin wave trajectories over Indian Ocean is investigated. It is shown that fast propagating Kelvin waves impact diurnal variability of the sea surface temperature, surface wind speed and latent heat flux. Composites of all the Kelvin waves show that changes in wind speed, latent heat, and sea surface temperature anomaly have similar signature. Wind speed and latent heat flux increase and a sea surface temperature anomaly decreases during Kelvin wave passage. Such changes depend on the phase of the Madden-Julian Oscillation in which Kelvin wave propagates.

In the next Chapter the properties of convectively coupled Kelvin waves in the Indian Ocean and their propagation over the Maritime Continent are studied. It is shown that Kelvin waves are longitude-diurnal cycle phase locked over the Africa, Indian Ocean and Maritime Continent. This means that they tend to propagate over definite areas during specific times of the day. Over the Maritime Continent, longitude-diurnal cycle phase locking is such that it agrees with mean, local diurnal cycle of convection in the atmosphere. The strength of the longitude-diurnal cycle phase locking differs between non-blocked Kelvin waves, which make successful transition over the Maritime Continent, and blocked waves that terminate within it. It is shown that a specific combination of Kelvin wave phase speeds and time of the day at which a wave approaches the Maritime Continent influences the chance of a successful transition into the Western Pacific. Kelvin wave that maintains phase speed of 10 to 11 degrees per day over the central-eastern Indian Ocean and arrive at 90E between 9UTC and 18UTC has the highest chance of being non-blocked by the Maritime Continent. The distance between the islands of Sumatra and Borneo agrees with the distance travelled by an average convectively coupled Kelvin wave in one day. This suggests that the Maritime Continent may act as a "filter" for Kelvin waves, favoring successful propagation of those waves which are in phase with the local diurnal cycle of convection. The AmPm index, a simple measure of local diurnal cycle for propagating systems, is introduced and shown to be a useful metric depicting key characteristics of the convection associated with propagating Kelvin waves.

Thus, the main message of this thesis is that interaction between atmospheric convection and upper ocean are characterized by two-way feedbacks. Mature atmospheric convection influences the distribution of the energy in the upper ocean affecting the sea surface temperature. On the other hand, changes in the sea surface temperature impact the organization of the atmospheric convection in equatorial convectively coupled Kelvin waves.

# Table of contents

# Chapter 1. Introduction

Organization and evolution of the atmospheric convection in the tropical regions is one of the key problems in meteorology but improvements in understanding and representation of physical processes responsible for organization of local, mesoscale and global circulations in the tropics are constant challenge [*Gray*, 1968; *Madden and Julian*, 1971; *Riehl and Malkus*, 1958]. Over the years many international observational campaigns in the tropical regions were conducted which lead to advancements in understanding of convection organization in that region. The three main projects covered all major tropical ocean basins: the Global Atmospheric Research Program (GARP) Atlantic Tropical Experiment (GATE) in 1974 in the Atlantic Ocean [*Kuettner*, 1974], the Tropical Ocean Global Atmosphere Coupled Ocean Atmosphere Response Experiment (TOGA COARE) in 1992-1993 over the tropical Western Pacific [*Webster and Lukas*, 1992] and the Dynamics of Madden Julian Oscillations/Cooperative Indian Ocean experiment on the intraseasonal variability in the Year 2011 (DYNAMO/CINDY2011) in 2011-2012 over the tropical Indian Ocean [*Zhang et al.*, 2013]. Such progress of understanding dynamics of the atmospheric circulation has been augmented by development of measurement techniques that include ground and satellite based systems in particular observations of precipitation in tropical regions [*Houze*, 2003]. Figure 1.1 illustrates the importance of the tropics for the global distribution of precipitation. Figure 1.1a presents global sea surface temperature (SST) from the World Ocean Atlas 2009 [*Locarnini et al.*, 2010] and Figure 1.1b presents the average monthly rain rate in the tropics. It can be seen that areas of high precipitation coincide with areas of the highest SST, especially over eastern Indian Ocean, Maritime Continent and Western Pacific. This area is often referred to as "warm pool" and its importance stems from the fact that it provides most of the energy for the global circulation.

It was the GATE campaign during which the quantitative measurements of heating profiles in the tropics were intensively conducted and they have shown that convection is organized in mesoscale systems [*Houze*, 1982; *Houze et al.*, 1980] in addition to large scale organization. The TOGA COARE campaign was centered in the Western Pacific warm pool and focused on the coupled (atmosphere-ocean) processes responsible for the organization of the convection in that region. Results from TOGA COARE showed that the atmospheric convection has trimodal characteristics [*Johnson et al.*, 1999] and indicated that multiscale interactions between convection and atmospheric waves are important [*Kikuchi and Takayabu*, 2004; *Kiladis and Wheeler*, 1995; *Kiladis et al.*, 1994; *Yanai et al.*, 2000]. It was established that the intraseasonal (30-90 days) variability of convection is organized in Madden-Julian Oscillations (MJO) [*Madden and Julian*, 1971; 1972] which are leading to variability in oceanic conditions, primarily surface fluxes and the sea surface temperature [*Lau and Sui*, 1997; *Shinoda and Hendon*, 2001; *Shinoda et al.*, 1998; *Weller and Anderson*, 1996]. Since then, the variability of the ocean surface and subsurface on intraseasonal time scales has become of interest to the tropical meteorology.

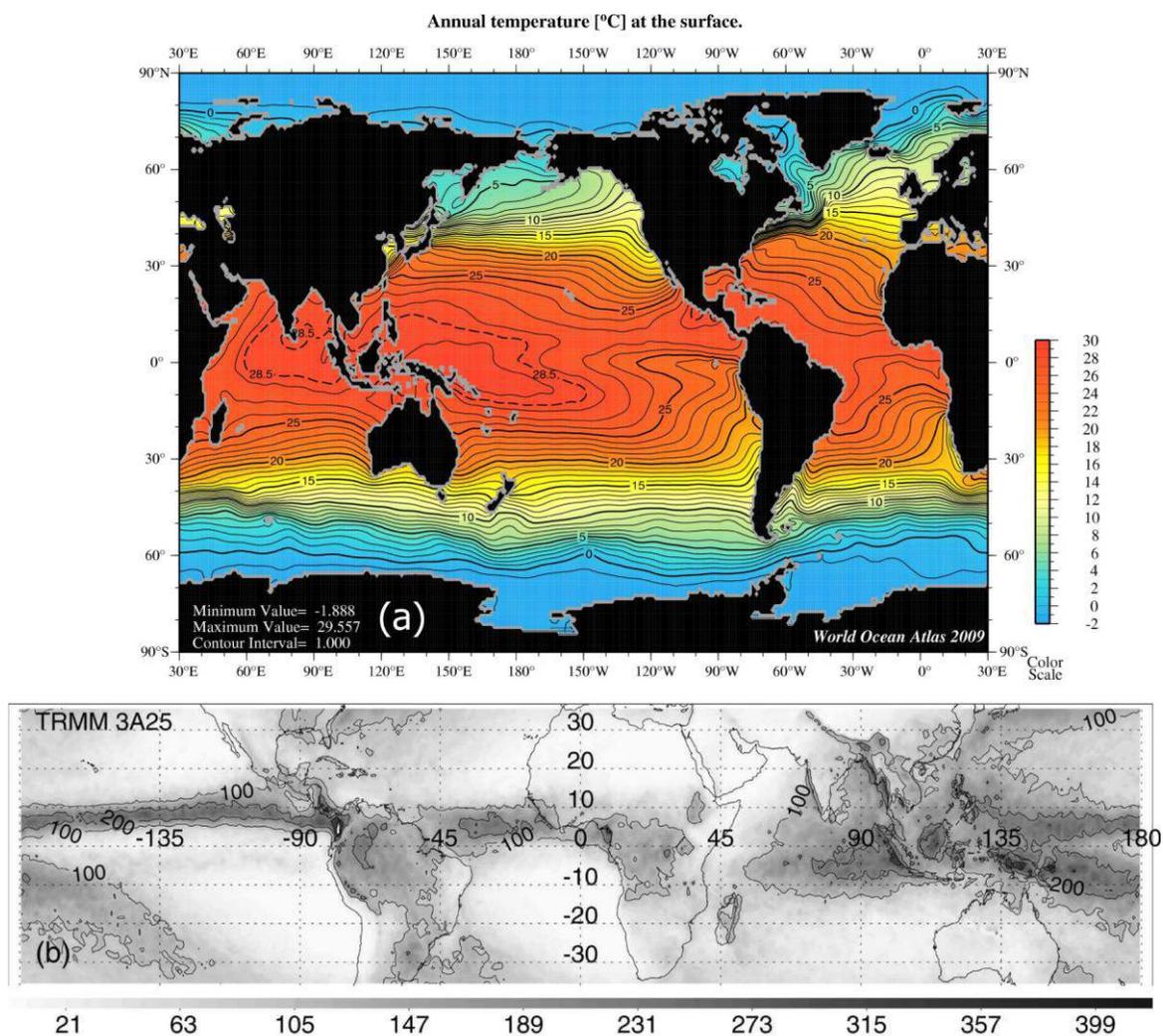

**Figure 1.1 (a) World Ocean Atlas (WOA) 2009 Sea Surface Temperature (SST). Color scale indicate temperature in [°C]. (b) The 6-yr (1998–2004) average of monthly rain rate [mm per month] from TRMM 3A25 product. Shades of grey indicate level of monthly rain rate(this is a reprint of Figure 1b from Liu *et al.* [2007])**

In spite of progress in understanding of the MJO dynamics the problem of initiation of MJO constituted a great challenge after TOGA-COARE. Zhang [2005] provided a review of understanding of the intraseasonal variability of the atmospheric convection and associated oceanic conditions in the tropical Indo-Pacific region in that time. Subsequently, the international DYNAMO/CINDY2011 project has been designed to resolve gaps in our knowledge of MJO initiation. The field phase of the experiment took place between October 2011 and January 2012. The observational campaign was supported by multiple numerical modeling efforts [*Fu et al.*, 2013]. The research activities of DYNAMO/CINDY2011 focused on the MJO initiation over the tropical Indian Ocean. One of the working hypotheses related to MJO initiation included a crucial role of air-sea interaction mechanisms on its development [*Yoneyama et al.*, 2013].

This thesis focuses on the processes responsible for the upper ocean variability relevant to the overall DYNAMO/CINDY2011 objectives and the role of the upper ocean in MJO initiation. The main goal is to improve our understanding of the processes governing interactions between atmospheric convection and upper ocean and their two-way feedbacks. In particular, the short term ocean surface fluxes and SST variability are investigated on the basis of in-situ measurements. The SST variability can be driven by atmospheric and oceanic processes; therefore its short term variability is forced mostly by changes in solar radiation, precipitation and wind stress at the ocean surface. Absorption of the solar radiation influences temperature of the ocean surface, while evaporation and precipitation influence salinity. Wind effects involve mixing of warm and fresh mixed layer waters (~50m deep in the tropics) with colder and saltier waters from beneath the thermocline. Therefore, a complete assessment of vertical profiles in the upper ocean is important for proper representation of the atmosphere induced SST variability.

Ocean–atmosphere interactions are important for tropical weather and climate on various temporal and spatial scales. For example, moisture flux from the ocean to the atmosphere increases approximately exponentially with SST through the Clausius–Clapeyron and bulk flux relationships [*Fairall et al.*, 1996a]. This is one of the factors important for evolution of the El Niño–Southern Oscillation (ENSO) [*Fairall et al.*, 1996a] on interannual time scales. On shorter, intraseasonal time scales, ocean-atmosphere interaction has a significant role in the development and eastward propagation of MJO [*Drushka et al.*, 2011; *Flatau et al.*, 1997; *Matthews*, 2004; *Shinoda et al.*, 1998; *Woolnough et al.*, 2000]. Also, initiation, evolution and dynamics of the tropical cyclones strongly depend on the two way interaction between atmosphere and ocean [*Baranowski et al.*, 2014; *Black et al.*, 2007; *Cione and Uhlhorn*, 2003; *Emanuel et al.*, 2004; *Fisher*, 1958; *Mrvaljevic et al.*, 2013].

Recently, the ocean–atmosphere interaction on even shorter, diurnal time scales has been recognized as an important factor in forcing of atmospheric convection. A strong diurnal cycle of SST was observed during the TOGA–COARE [*Sui et al.*, 1997]. This was particularly prevalent during conditions of high surface solar radiation flux causing surface heating and low surface wind speeds resulting in weak vertical mixing. Such diurnal warm layer can have the significant effect on atmospheric convection. Typically, the diurnal cycle of precipitation over the tropical open ocean is weak, with a peak before sunrise [*Bowman et al.*, 2005; *Gray and Jacobson*, 1977]. However, during periods of high solar radiation and weak winds, and a strong diurnal SST cycle, atmospheric convection over the ocean can behave similarly to that over land, with successive development of deeper clouds and a precipitation maximum in the late evening [*Johnson et al.*, 1999]. Therefore, a proper assessment and representation of the variability of the upper ocean diurnal cycle is important for representation of the atmospheric convection in the tropics. In particular, for coupled ocean–atmosphere general circulation models (GCM) this rectification of daily mean SST by the diurnal variability of SST can increase the long term mean SST by 0.2–0.3 °C which also improves the mean precipitation fields [*Bernie et al.*, 2008]. These processes are not resolved in most of current ocean–atmosphere GCMs (with their current vertical grid spacing of the order of 10 meters) and the atmosphere-ocean coupling which is performed only once daily. This leads to degradation in the simulation and forecasting of the MJOs [*Woolnough et al.*, 2007] in GCMs and errors in the derived mean climate.

Another aspect of the circulation in the tropical Indo-Pacific region is related to the complex topography and shallow seas within the Maritime Continent. The Maritime Continent is located within the tropical warm pool and separates Indian Ocean Basin from the Western Pacific. It includes the archipelagos of Indonesia, New Guinea and Malaysia with surrounding shallow seas with the warmest ocean temperatures in the world. The importance of the Maritime Continent for the global climate was identified by Ramage [1968]. The strong atmospheric moist convection in this region provides the heat source for circulation anomalies and through teleconnections influences the global circulation [*Neale and Slingo*, 2003].

In MJO modeling the Maritime Continent is recognized as a "predictability barrier" due to inability of dynamic models to properly represent the passage of MJO through combination of shallow seas, islands and large topographic features [*Seo et al.*, 2009]. This problem always exists in dynamic models but is not evident in the statistical forecasts. It was hypothesized previously [*Inness and Slingo*, 2006] that propagation of MJO through the Maritime Continent depends on ability of atmospheric equatorial convectively coupled Kelvin waves (CCKW) to cross the Maritime Continent barrier. CCKW are important part of the atmospheric equatorial dynamics because they are the leading modes of eastward moving convection on time scales between several days to three weeks [*Kiladis et al.*, 2009]. Convectively coupled Kelvin waves, together with other equatorial eastward and westward propagating perturbations form "building blocks" of the active phase of the MJO [*Majda and Khouider*, 2004; *Mapes et al.*, 2006]. Their importance was first recognized by Nakazawa [1988] who noticed the eastward moving cloud "superclusters" embedded in the MJO envelope. Kelvin waves can be separated from the MJO by the phase speed that is more than twice that of MJO and can reach 17 ms$^{-1}$ [*Straub and Kiladis*, 2002; *Wheeler and Kiladis*, 1999]. While the phase speed of MJO observed in the Indian Ocean is about 4-5 ms$^{-1}$ the phase speeds of convectively coupled Kelvin waves in this region ranges from about 13.9 ms$^{-1}$ for suppressed MJO condition to 11.5 ms$^{-1}$ during the convective phase of MJO [*Roundy*, 2008]. Thus, CCKW may have limited effect on SST due to fast propagation but still influence short term variability such as a diurnal cycle of the SST.

Figure 1.2 (which is a reprint of Figure 2.1 from Masunaga *et al.* [2006]) presents Hovmöller diagrams of convective activity for 3 different time periods, one period in each row. It illustrates multiscale interactions between convective disturbances, including MJO, Kelvin and Rossby waves, over Indo-Pacific basin. In each diagram time goes down along y-axis. In column a, the total unfiltered precipitation is plotted with thin black contours. Superimposed color shading indicate anomalies of the filtered MJO component of precipitation. Positive anomalies of the MJO component are additionally marked with thick black line. Column b shows the total unfiltered precipitation (thin black contours), positive anomalies of the filtered MJO component (thick black contours) and the filtered Kelvin wave component indicated by the color shading. In column c the total unfiltered precipitation is plotted with colors, anomalies of the filtered MJO component are plotted with thick black lines, anomalies of the filtered Kelvin component wave are plotted with thin, black dashed line and anomalies of the filtered Rossby wave are plotted with thin, black solid line. The anomalies of the filtered MJO and filtered Kelvin wave propagate eastward. Anomalies of the filtered Rossby wave propagate westward. Column d presents schematic plots of interaction between MJO, Rossby and Kelvin waves during these three periods. This figure illustrates how various types of tropical disturbances interact with each other. Although Kelvin wave maintain larger phase speed than MJO, it can be seen that they can interact with MJO envelope in various stages of its evolution. Although temporal and spatial scales of Kelvin and Rossby waves are different from the typical MJO scales, their activity is high during MJO propagation.

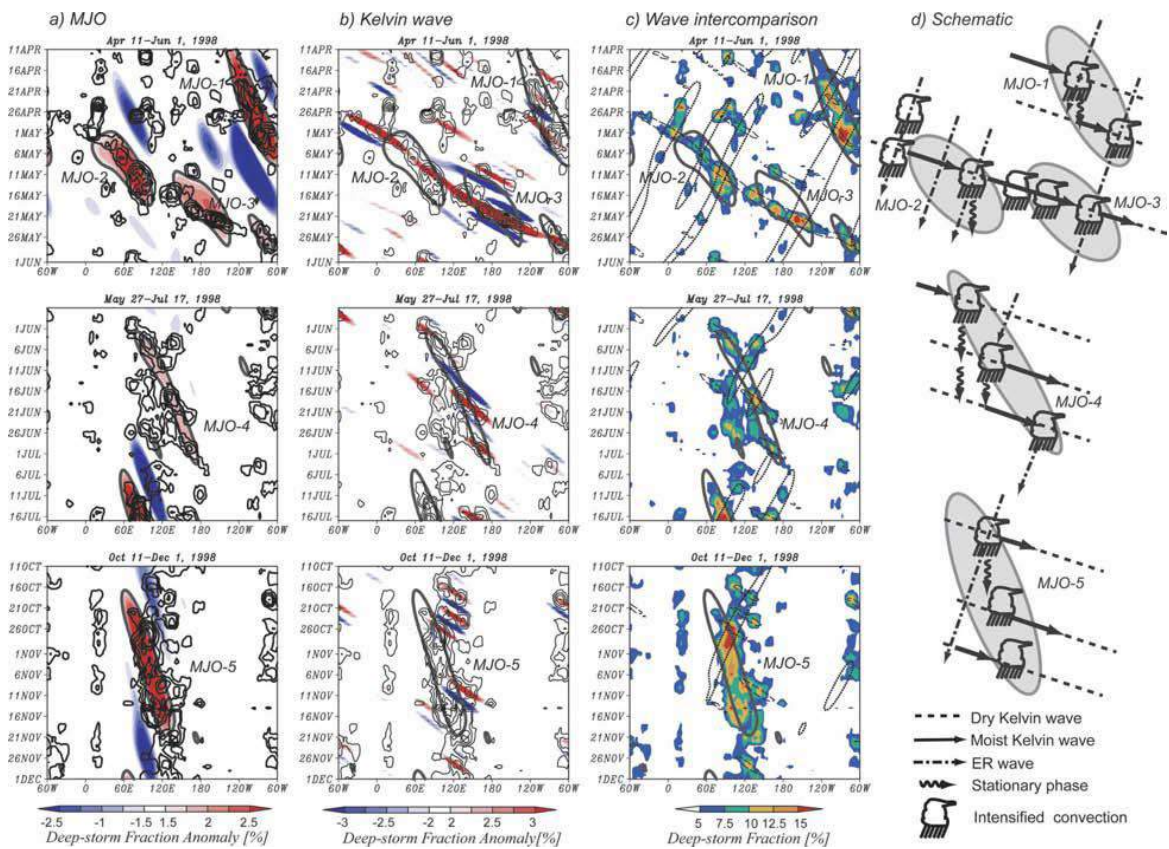

**Figure 1.2** Time–longitude diagram of deep-storm fraction for different time periods. (a) The MJO-filtered anomaly (color) and unfiltered field contoured at an interval of 2.5% starting with 5%. (b) The Kelvin-filtered anomaly (color) as well as the unfiltered field. The MJO-filtered anomaly at the +1.5% level is superimposed for reference. (c) The unfiltered field in colors superimposed with the anomalies of the filtered Kelvin wave (at +2% in dashed contour) and Rossby wave (at +1% in dotted contour). (d) A schematic representation of wave interactions and a resultant stationary convective phase. The dry Kelvin wave is simply extrapolated from the moist counterpart in the schematic, where difference in propagation speed between the dry and moist waves is ignored for brevity. Reprint of Figure 2.1 from Masunaga *et al.* [2006]

In this thesis, the interactions between short term variability in the upper ocean temperature structure and ocean surface fluxes and atmospheric convection are investigated. First it is shown on the basis of in-situ SeaGlider measurements how changes in atmospheric convection on time scales from diurnal to intraseasonal impact development and evolution of the warm layer at the ocean surface. Next we present how the existence and strength of this warm layer and associated variability in surface fluxes feed back into development of the atmospheric convection organized in CCKW. Please note that any future reference to the Kelvin waves is meant in a sense of atmospheric convectively coupled Kelvin waves, unless otherwise clearly stated. We use 15 years of precipitation data from satellite Tropical Rainfall Measurement Mission (TRMM) to study characteristics of the atmospheric convection.

The structure of this thesis is as follows. Chapter 2 describes the development and evolution of upper ocean diurnal cycle based on in-situ measurements. It shows how the diurnal SST variability responds to the state of the atmospheric convection providing prescriptive model of the oceanic warm layer. This model is used in subsequent chapters. The contents of Chapter 2 is based on paper published recently in the Journal of Climate by Matthews *et al.* [2014] to which the thesis's author is a major contributor.

Chapter 3 describes initiation of CCKW in the Indian Ocean basin. It presents the results from numerical modeling and TRMM trajectory data analysis. Two regimes of Kelvin wave initiations are described. Environmental conditions, including surface fluxes and upper ocean temperature diurnal variability, for each regime are analyzed.

In Chapter 4 eastward propagation of CCKW through Indian Ocean and typical variability of surface fluxes and upper ocean temperature is of primary interest. It is shown how atmospheric convection, associated with mature, propagating Kelvin waves impacts the ocean surface and subsurface layers. Furthermore, zonal variability across the Indian Ocean and temporal, intraseasonal variability of the typical response to a Kelvin wave passage are investigated.

In Chapter 5 we focus on propagation of Kelvin waves initiated over Indian Ocean over the Maritime Continent. In particular, diurnal cycle of precipitation along a CCKW trajectory is investigated. The analysis is complemented by statistics of Kelvin wave characteristics calculated for waves "blocked" and "non-blocked" by the Maritime Continent. Characteristics include area of origin, associated precipitation, phase speed and time of the day at which a Kelvin wave approaches the Maritime Continent.

Chapter 6 summarizes all the results. The elements of methodology used in Chapters 2-5 are included in Appendices. In particular, they contain detailed description of the development of Kelvin wave trajectories database and handling of the flux data are included.

# Chapter 2. The surface diurnal warm layer in the Indian Ocean during CINDY/DYNAMO

## 2.1. Introduction

Ocean–atmosphere interaction is a key process in tropical weather and climate. The moisture flux from the ocean to atmosphere increases approximately exponentially with sea surface temperature (SST) through the Clausius–Clapeyron and bulk flux relationships [*Fairall et al.*, 1996a]. These processes are core to the evolution of the El Niño–Southern Oscillation (ENSO) [*Fairall et al.*, 1996a] on interannual time scales. On shorter, intraseasonal time scales, ocean-atmosphere interaction has a significant role in the development and eastward propagation of the MJO [*Drushka et al.*, 2011; *Flatau et al.*, 1997; *Matthews*, 2004; *Shinoda et al.*, 1998; *Woolnough et al.*, 2000], and also in the northward-propagating intraseasonal oscillations observed in the Indian Ocean during the boreal summer (BSISO) [*Fu and Wang*, 2004; *Seo et al.*, 2007].

Recently, attention has focused on ocean–atmosphere interaction on even shorter, diurnal time scales. A strong diurnal cycle of SST (up to 2 $^{o}$C magnitude) in the western Pacific was observed during the TOGA–COARE experiment [*Sui et al.*, 1997]. This was particularly prevalent during the inactive stage of the MJO, with in-situ conditions of high surface solar radiation flux and surface heating, and low surface wind speeds and weak vertical mixing. This diurnal cycle in SST is part of a diurnal warm layer that grows during the day through absorption of solar radiation in the top few meters of the ocean [*Fairall et al.*, 1996b; *Gentemann et al.*, 2009; *Price et al.*, 1986; *Prytherch et al.*, 2013]. Nocturnal surface cooling and destabilization then mix up the diurnal warm layer overnight into the deeper mixed layer.

Observations of the diurnal warm layer have characterized its temperature profile as being either approximately isothermal [*Delnore*, 1972; *Soloviev and Lukas*, 1997], decreasing approximately linearly with depth [*Delnore*, 1972; *Prytherch et al.*, 2013; *Webster et al.*, 1996], or decreasing approximately exponentially with depth [*Halpern and Reed*, 1976; *Soloviev and Lukas*, 1997; *Webster et al.*, 1996], depending on ambient wind conditions.

However, due to the shallow nature of the diurnal warm layer, it is difficult to measure. Observations of the diurnal warm layer have been mainly made using shipboard

conductivity–temperature–depth (CTD) instruments or from moorings. However, both of these platforms have disadvantages. A vertical resolution of 1 m or finer is required near the surface. This is difficult to achieve using a conventional shipboard CTD, due to the wake around the CTD package disturbing the fine-scale structure of the diurnal warm layer, and motion from the ship. Additionally, long ship deployments, such as would be necessary to capture the diurnal cycle in different stages of the MJO, and are very expensive. For moorings, the buoy structure can distort the near-surface flow. In the presence of a diurnal warm layer, this has been found to bring warm water from near the surface down to the level of the sensor, leading to an overestimation of temperature of up to 1 °C [*Prytherch et al.*, 2013]. Standard Argo floats [*Gould et al.*, 2004] do not have the vertical or temporal resolution required to resolve the diurnal warm layer, typically only making one temperature measurement in the top 10 m, and one profile every 5–10 days. However, ongoing deployment of SOLO-II Argo floats, with higher vertical resolution and profile frequency, may partially address this in the future.

This diurnal warm layer and diurnal SST cycle can have a significant effect on atmospheric convection. Typically, the diurnal cycle of precipitation over the tropical open ocean is weak, with a peak before sunrise [*Bowman et al.*, 2005; *Gray and Jacobson*, 1977]. However, during periods of high solar radiation and weak winds, and a strong diurnal SST cycle, atmospheric convection over the ocean can behave similarly to that over land, with a strong diurnal cycle, growth of cumulus congestus, and a precipitation maximum in the late evening [*Johnson et al.*, 1999].

This diurnal variability can then impact onto longer time scales. For example, precipitation over the seas and islands of the maritime continent is predominantly accounted for by the diurnal cycle. Over this region, 80% of the MJO precipitation signal is directly accounted for by changes in the amplitude of the diurnal cycle of precipitation [*Peatman et al.*, 2014]. There is also a strong indirect effect that also operates over the open ocean. The diurnal warm layer is the result of solar heating during the day being concentrated in a shallow layer only a few meters thick. This layer has a lower heat capacity, compared to a situation where the entire deeper mixed layer is heated. Hence, even though the diurnal warm layer is mixed back into the underlying mixed layer overnight, the daily mean SST is higher than it would have been in its absence [*Mujumdar et al.*, 2011; *Shinoda*, 2005].

In a coupled ocean–atmosphere GCM, this rectification of daily mean SST by the diurnal variability of SST can increase the long term mean SST by 0.2–0.3 °C [*Bernie et al.*, 2008], with a subsequent improvement in the mean precipitation. In their simulations, the MJO is also improved, due to the increase in intraseasonal SST variability that stems from the improved diurnal SST variability [*Bernie et al.*, 2007]. These and other studies (e.g.,[*Klingaman et al.*, 2011]) concluded that a very fine vertical grid spacing, of approximately 1 m, is required in the upper layers of the ocean component to resolve the diurnal warm layer processes there. As most current ocean–atmosphere GCMs have a grid spacing of order 10 m and are often only coupled daily, these processes are not resolved. This leads to a degradation in the simulation and forecasting of the MJO [*Woolnough et al.*, 2007], and also errors in the mean climate.

Ocean gliders are a relatively new technology for observing the ocean [*Eriksen et al.*, 2001]. They can provide very high resolution data right to the surface, without the drawbacks of a shipboard CTD, and are relatively inexpensive to operate.

In this Chapter, the diurnal warm layer is analyzed using measurements from an ocean glider deployed as part of the Central Indian Ocean Experiment on Intraseasonal Variability / Dynamics of the Madden–Julian Oscillation (CINDY/DYNAMO) international field experiment. CINDY/DYNAMO was designed to investigate ocean–atmosphere interactions and the initiation of the MJO in the Indian Ocean [*Gottschalck et al.*, 2013]. The high-quality, high-resolution data measured by the glider over a long deployment (approximately 100 samples of the diurnal cycle) allows the detailed structure of the diurnal warm layer to be analyzed in unprecedented detail. A further focus is how this varies under different environmental forcing conditions, particularly those associated with active and inactive phases of the MJO? Simple models of the diurnal warm layer under different environmental conditions are then developed, with the aim of informing (climate) model development.

## 2.2. Data processing

### 2.2.1. External data sources

Sea surface temperature data were extracted from the NOAA Optimum Interpolation (OI v2) data set [*Reynolds et al.*, 2002]. These were obtained on a 1° × 1° grid as weekly means, which were then interpolated to daily means for ease of analysis. Precipitation was diagnosed using the Tropical Rainfall Measuring Mission (TRMM) merged precipitation (3B42) data set [*Kummerow et al.*, 2000]. The data were on a 0.25° × 0.25° grid at 3 h resolution. Daily mean surface wind speeds were analyzed using the OAFlux data set [*Yu et al.*, 2008]. A self-consistent set of surface wind speed and shortwave radiative flux data was extracted from the TropFlux archive [*Praveen Kumar et al.*, 2013]. These daily mean data were obtained on a 1° × 1° grid.

### 2.2.2. CINDY/DYNAMO and Seaglider deployment

The DYNAMO field experiment concentrated on the 10° × 10° box from 70–80°E and 0°S–0°N in the equatorial Indian Ocean (Figure 2.1a). It is at the heart of the high SST and maximum precipitation zone during this time of year (line contours and color shading, respectively, in Figure 2.1a). The global maximum in MJO precipitation is also in this region [*Hendon and Salby*, 1994].

During CINDY/DYNAMO, three distinct MJO events passed through the study region. These can be seen as eastward-propagating bands of enhanced precipitation (above 10mmd$^{-1}$) that cross the CINDY/DYNAMO 70–80 °E sector, in late October, late November and late December 2011 (Figure 2.1b). In between the active MJO events, precipitation is suppressed (below 5 mmd$^{-1}$). Westerly wind bursts were associated with each MJO event [*Moum et al.*, 2013], which forced a strong thermodynamical and dynamical ocean response [*Shinoda et al.*, 2013b]. These MJO events were generally well forecast; inclusion of ocean–atmosphere coupling extended the skill of the MJO forecasts over this period by approximately one week, emphasising the important role that such interactions played in these MJO events [*Fu et al.*, 2013].

The MJO is conveniently diagnosed by the real-time multivariate MJO index (RMMI) [*Wheeler and Hendon*, 2004]. Time series of the amplitude and phase of the RMMI index during CINDY/DYNAMO are shown in Figure 2.2a by the thin and thick black lines, respectively. The RMMI phase is expressed as an integer between 1 and 8. Phases 1–4 indicate enhanced precipitation over the Indian Ocean; these are grouped into the "active" stage here. The RMMI phases 5–8 indicate suppressed precipitation over the Indian Ocean: the "inactive" stage.

The Seaglider is a 1.8 m, 50 kg unmanned buoyancy-driven autonomous underwater vehicle (AUV), instrumented for oceanographic research to measure pressure, temperature, conductivity, dissolved oxygen, chlorophyll fluorescence and turbidity [*Eriksen et al.*, 2001]. A typical glider dive cycle has a sawtooth profile, with a dive phase from the surface to a specified depth (maximum 1000 m) and a climb phase back to the surface.

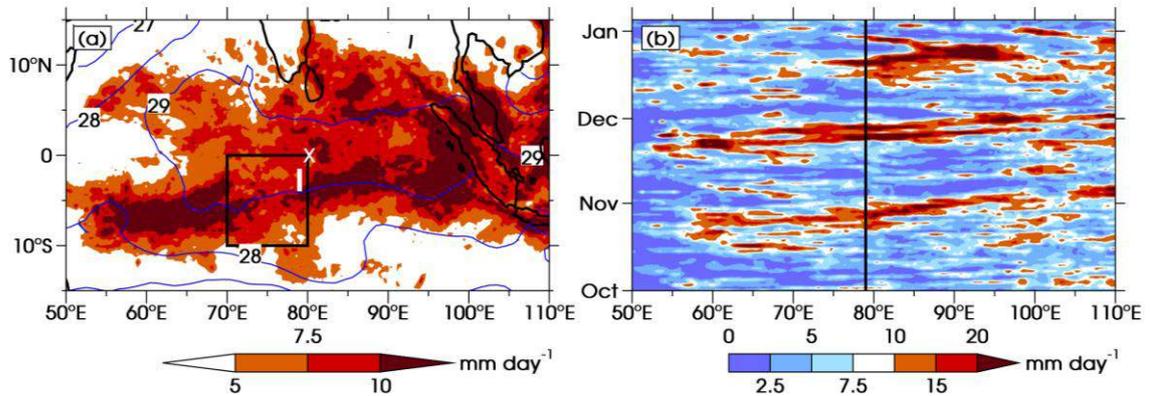

**Figure 2.1 (a)** Time-mean TRMM 3B42 precipitation rate (colour shading; mm d$^{-1}$) and SST (blue line contours; interval 1 °C) over the study period of glider deployment during CINDY/DYNAMO (1 October 2011 to 5 January 2012). The box shows the approximate location of the CINDY/DYNAMO study area. The thick white line along 78°50'E, between 1°30'S, and 4°S, shows the glider track. The white cross at 0°N, 80°E shows the location of the R/V Roger Revelle. **(b)** Time-longitude diagram of TRMM 3B42 precipitation rate (mm d$^{-1}$), averaged from 15°S to 15°N. The thick black line shows the envelope of the glider tracks.

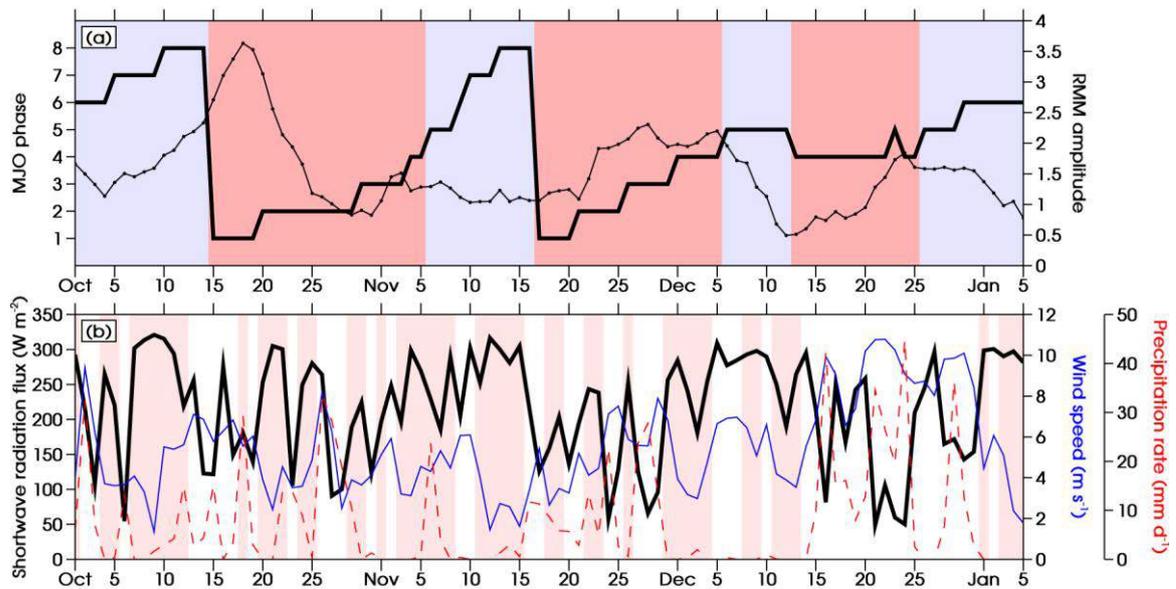

**Figure 2.2** Daily time series from 1 October 2011 to 5 January 2012 of (a) MJO RMMI phase (thick line, left axis), MJO RMMI amplitude (thin line, right axis). MJO active phases (1–4) are shaded pink, and inactive phases (5–8) are shaded light purple. (b) Meteosat7-derived shortwave radiation flux (black thick line, left axis), OAFlux surface wind speed (thin blue line, right axis), TRMM precipitation rate (dashed red line, far right axis), at glider location. Days when a surface diurnal warm layer developed are shaded pink.

The Seaglider SG537 "Fin" was deployed in the equatorial Indian Ocean on 14 September 2011 at 79°50'E, 1°30'S. It was then piloted southwards along the 79°50'E meridian to 3°S (30 September 2011). The glider then made 10 transects between 3°S and 4°S, arriving at 3°S for the final time on 5 January 2012. It then continued northward to the retrieval location (same as deployment, at 79°50'E, 1°30'S) on 23 January 2012, with total mission duration of 131 days (Figure 2.1a).

### 2.2.3. Seaglider measurements and processing

During the mission, the glider carried out 738 dive cycles, an average of 5.6 dive cycles per day or a time interval between the start of successive dives of 4.3 hours. Of these, 564 dives (76%) were to 1000 m depth; the remaining dives were to 300 or 500 m. The glider vertical velocity is in the range 0.15–0.25 ms$^{-1}$. Temperature and salinity were sampled every 5 s, hence the effective vertical resolution is approximately 1 m.

The glider data were corrected for thermal lags, sensor response time, and non-synchronicity of sensor readings. Temperature and salinity data are accurate to 0.01 °C and 0.01 psu, respectively. A further correction was made to the pressure measurements, to remove long term drift of the pressure sensor over the mission duration, and also to account for hysteresis within each dive. Further details are provided in [*Webber et al.*, 2014]. The analysis was then carried out on the data gathered in the north–south sections from 3–4°S, between 1 October 2011 and 5 January 2012.

Before the dive phase of a dive cycle, the glider floats at the surface, engaged in satellite communications. The temperature sensor can be subjected to solar radiation and exposure to the atmosphere. When the dive phase starts, the temperature sensor can take several seconds to re-equilibrate to the water temperature. Hence, temperature measurements in the top few meters in the dive phase are unreliable. There are no such problems in the climb phase. Therefore, only data from the climb phases were used in this analysis.

The temperature time series was optimally interpolated [*Bretherton et al.*, 1976] onto a two-dimensional grid, with regularly spaced pressure $p_j$ points (from 0.5 m to 1000 m) and time $t_j$ points (every hour from the start to the end of the mission). Both the observed temperature $T(t_i)$ and pressure $p(t_i)$ time series were used to create the gridded temperature field $T(p_j,t_j)$. First, an initial background temperature field was created at each grid point, using a weighted average of temperature observations from nearby in the

pressure–time space. A Gaussian weighting function $w_{ij}$ was used to calculate the weighting that each temperature observation $T(p_i,t_i)$ contributes to the gridded value $T(p_j,t_j)$:

$$w_{ij} = \exp\left\{-\left[\left(\frac{p_i - p_j}{p_r}\right)^2 + \left(\frac{t_i - t_j}{t_r}\right)^2\right]\right\} \quad (2.1)$$

The "radii of influence" $p_r$ and $t_r$ in the pressure and time dimensions are chosen to reflect the scales of variability of interest, and also the inherent resolution of the observational data. For the diurnal cycle studied here, values of $p_r$=1 dbar (or about 1 m) and $t_r$=3 h were chosen. The background field was then used as the input to the optimal interpolation scheme. The covariances of the data were parameterised using the same Gaussian function (Eq.(2.1)) to calculate the analysis increment, which was then added to the background field to create the final optimally interpolated field.

The grid spacing for the optimally interpolated field were chosen to be Δp = 0.5 dbar (or about 0.5 m) and Δt = 1 h. Note that it is the radii of influence $p_r$ and $t_r$ that govern the form of the gridded field. The choice of the output grid spacing is just a matter of presentational convenience.

An example of the optimal interpolation, and of the development of a diurnal warm layer, is shown in Figure 2.3. The optimally interpolated temperature (color shading in Figure 2.3a) on a sample day (3 December 2011) shows a well mixed isothermal layer over at least the top 8 m at midnight (0000 local solar time, LST). There is then further cooling through the night until a minimum temperature of 28.87 °C is reached at sunrise at 0600 LST.

Note that there is no gridded level at 0 m, as all contributing glider measurements would be below this level, introducing a bias. The first gridded level is at 0.5 m, which allows for contributing measurements above and below. Hence, for the purposes of this analysis, the temperature at 0.5 m is referred to as the surface temperature. Also, as a 1 dbar pressure change is very close to a 1 m depth change, pressure and depth are used interchangeably in the following discussions.

During the day the surface warms, until it reaches a maximum of 29.21 °C at 1500 LST. This diurnal warming is confined to the upper few meters. The temperature at 5 m at 1500 LST is 28.95 °C, 0.26 °C colder than the surface. This is the stably stratified diurnal warm layer. As the solar radiation flux decreases through the afternoon, it is eventually overwhelmed by the cooling fluxes of latent heat, infrared radiation, and sensible heat, and the temperature of the diurnal warm layer decreases. After sunset (1800 LST) there is rapid cooling and mixing, and a return to isothermal conditions at 0000 LST the next day.

The colored vertical lines in Figure 2.1a show the times of the seven glider profiles during that day (Figure 2.3b), that the optimally interpolated temperature field in Figure 2.3a was effectively constructed from. Visual inspection reveals a very close agreement of the optimally interpolated temperature with the input temperature profiles.

## 2.2.4. Meteosat7-derived surface solar radiation flux

The development of the diurnal warm layer depends on the surface flux of solar radiation. Direct measurements of this flux are not available at the glider location. However, variations in surface solar radiation flux are mainly driven by changes in convective clouds, which also drive changes in outgoing longwave radiation (OLR). Using a novel method described in Appendix A, a proxy time series of solar radiation flux at the glider location can be constructed from satellite measurements of OLR at the glider location, using a conversion derived from surface solar radiation flux measurements made at the R/V Roger Revelle, stationed nearby (Figure 2.1a). The details of the methodology are given in Appendix A. The proxy time series for the surface solar radiation flux at the glider location (black line in Figure 2.2b) shows clear variability on day-to-day and intraseasonal time scales.

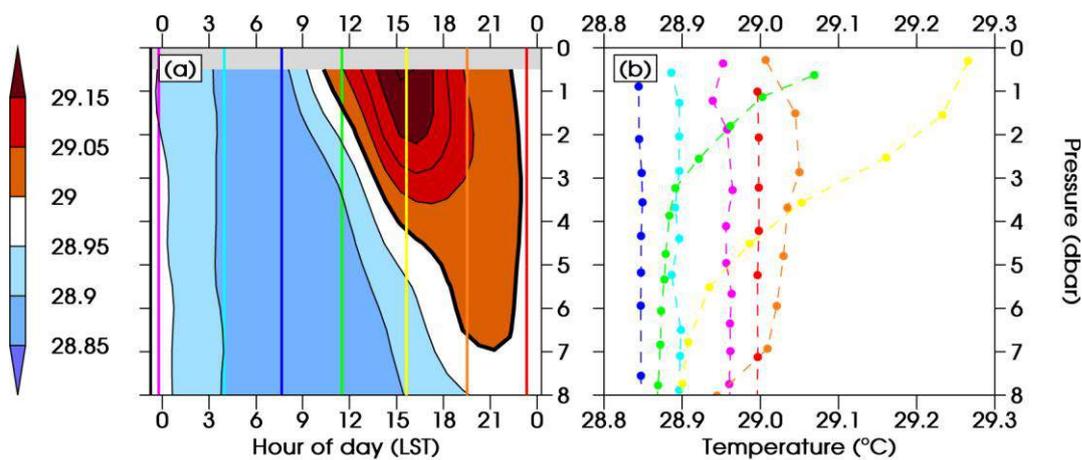

**Figure 2.3** (a) Optimally interpolated temperature [°C] on 3 December 2011. Contour interval for line contours and shading is 0.05 °C. The thick line indicates the isotherm T* that is used to define the base of the diurnal warm layer. Temperatures above T* are shaded red. For this day, T*=29 °C. (b) Individual glider profiles on the same day. The vertical coloured lines in (a) show the times of the glider profiles of the same colour in (b).

## 2.3. Overview

### 2.3.1. Mean profile

The diurnal and intraseasonal variability of the upper ocean are perturbations to, and strongly influenced by, the background thermodynamic structure. The mean temperature profile over the Seaglider deployment (1 October 2011 to 5 January 2012) has a surface mixed layer at 29.0 °C over the top 30 m depth (Figure 2.4, red solid line). Below this, the temperature decreases with depth through the thermocline, with, for example, the depth of the 20 °C isotherm at 105 m. Between 150 and 200 m, the temperature gradient decreases; below that is the deep ocean. Temperature decreases monotonically with depth, hence the temperature stratification is stable everywhere.

The mean salinity profile (Figure 2.4a, blue dashed line) exhibits a surface fresh layer (34.32 psu). At 25 m depth, the salinity increases rapidly, to a maximum of 35.23 psu at 75 m depth. This "subtropical underwater" [*O'Connor et al.*, 2005] originated at the surface in the subtropics, where evaporation exceeds precipitation, leading to the high salinity values. Below the salinity maximum, the mean profile becomes fresher with depth. Hence, the salinity stratification is unstable here. However, as the (potential) density is mainly controlled by the temperature variation over these tropical temperature and salinity ranges, this unstable salinity stratification does not lead to unstable density stratification. Density increases monotonically with depth (Figure 2.4a, thick black line), and the mean profile is statically stable at all levels.

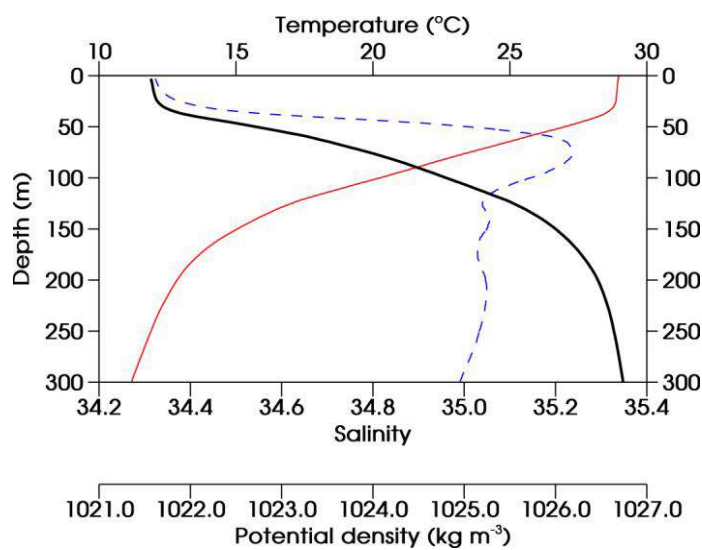

**Figure 2.4 Mean profiles of temperature (red solid line), salinity (blue dashed line), and potential density (black thick line), from the optimally interpolated glider data.**

## 2.3.2. Generic diurnal variability

The day-to-day evolution of the temperature structure of the ocean surface boundary layer is shown for the sample month of November 2011 (Figure 2.5). At the beginning of the month, the seasonal mixed layer is relatively cool (less than 29.2 °C). This coincides with the end of the active stage of the first MJO event on 5 November (Figure 2.2a). By mixed layer here, we mean the bulk mixed layer of depth order 20 m, upon which a diurnal warm layer is superimposed at the surface. The MJO then moves into its inactive stage over the Indian Ocean. The mixed layer warms considerably, to above 29.6 °C by 15 November. On 17 November the MJO enters the active stage of the second MJO event, and the mixed layer rapidly cools to below 29.0 °C by 25 November. This local evolution of the mixed layer temperature (as measured by the Seaglider) is remarkably consistent with the expected sea surface temperature cycle in a canonical MJO cycle (e.g., Shinoda *et al.* [1998]), both in magnitude and phase.

Superimposed on this intraseasonal variability is a clear diurnal cycle in temperature, especially in the upper 5 m. The tick marks on the time axis in Figure 2.5 are at 0000 (midnight) LST. A diurnal warming in the upper few meters by over 1 °C is observed on many days, peaking in the early afternoon. This diurnal warm layer occurs throughout the cycle of the MJO during November 2011. Similar behavior was observed in the other months (not shown). However, the diurnal variability appeared to be particularly strong during the inactive stage of the MJO (6–16 November 2011).

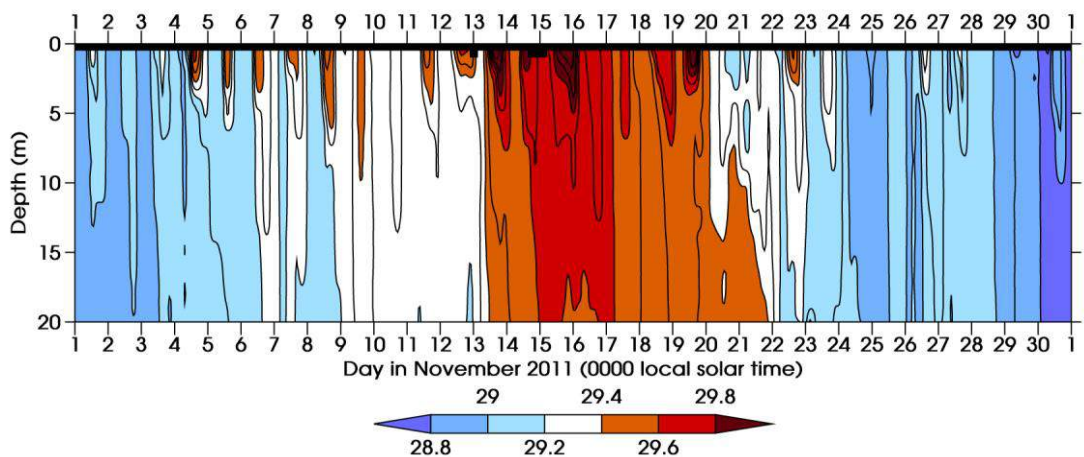

**Figure 2.5 Depth-time section for optimally interpolated glider temperature during November 2011. Contour interval is 0.2 °C. See legend for shading levels. Tick marks on the horizontal axis correspond to 0000 LST for each day.**

## 2.4. Diurnal warm layer occurence

The sample day of 3 December 2011, that was used to illustrate the optimal interpolation technique, is also a clear example of the development of a diurnal warm layer (Figure 2.3a). The surface (0.5 m) temperature increases from a minimum at sunrise (06 LST) of $T_{06}$ = 28.85 $^{\circ}$C, to a maximum value of $T_{max}$ = 29.20 $^{\circ}$C, at 1500 LST. In the following analysis, a technique is derived to objectively determine the existence of a surface diurnal warm layer, and to quantify the effect the diurnal warm layer has on the SST.

A convenient measure of the depth of the warm layer is the depth $d_{WL}$ of a fixed isotherm ($T*$).

$$d_{WL} = -z(T = T_*) \tag{2.2}$$

Here, we use z as the vertical coordinate (positive upward, with z = 0 at the surface), and d for depth (which is always positive). The value of the isotherm is determined for each day individually. It should lie between the minimum and maximum values, and is defined as

$$T_* = max\left[T_{06} + \alpha\left(T_{max} - T_{06}\right), T_{06} + 0.1^{\circ}C\right]. \tag{2.3}$$

The parameter $\alpha$ can be chosen from the range 0< $\alpha$ <1. It should be as small as possible to allow for as much of the diurnal warming to be included in the definition of the diurnal warm layer. However, if $\alpha$ is too small, the analysis is sensitive to noise. A value of $\alpha$=0.3 was found to be suitable. To account for days when a diurnal warm layer did not develop or was not well defined, a minimum value of $T_*=T_{06}+0.1$ $^{\circ}$C is specified, which also reduces sensitivity to noise.

For the sample day in Figure 2.3a, $T_*$ = 29.0 $^{\circ}$C. The thick contour at 29.0 $^{\circ}$C, and the red shading for temperatures above 29.0 $^{\circ}$C in Figure 2.3a illustrate the depth and extent of the diurnal warm layer on 3 December 2011.

The development of the (optimally interpolated) temperature profile during 3 December 2011 is shown from the surface to below 35 m in Figure 2.6a. In an idealized framework, the vertical structure can be characterized by a three-layer model, with a temporally developing diurnal warm layer above a deeper mixed layer, which itself lies above a thermocline layer. For example, at 1700 LST on 3 December 2011, there is a clear diurnal warm layer with a surface (0.5 m) temperature of 29.2 $^{\circ}$C that extends downward by a few meters (Figure 2.6a). Below this, there is a seasonal mixed layer at a temperature of approximately 28.7 $^{\circ}$C.

An idealized representation of the diurnal warm layer, and underlying mixed layer and thermocline layer, will now be constructed, with the ultimate purpose of calculating the effect of the warm layer on the mean SST. At any given time, an idealized representation of the warm layer is an isothermal layer of temperature $T_{WL}$ and depth $d_{WL}$ (Figure 2.6b). The warm layer temperature $T_{WL}$ is calculated such that the idealized warm layer has the same heat content as the actual water column between the surface and the base of the warm layer at $d_{WL}$.

$$T_{WL} d_{WL} = \int_{-d_{WL}}^{0} T(z) dz. \tag{2.4}$$

A useful definition of the mixed layer is the depth $d_{MLsfc}$ at which the temperature is 0.2 °C less than the surface temperature $T_{sfc}$ [*de Boyer Montégut et al.*, 2004], and is typically several tens of metres.

$$d_{MLsfc} = -z\left(T = T_{sfc} - 0.2°C\right). \tag{2.5}$$

Note that the surface temperature defined here is actually the temperature at the highest available level, which is at 0.5 m depth. However, with a well developed diurnal warm layer in the afternoon, this definition would give a mixed layer depth of only 4 m (Figure 2.6b). This is not a useful definition in this case. The diurnal warm layer does not typically extend below a few meters. Hence, a reference level at 10 m is chosen, and the depth of the seasonal mixed layer $d_{ML}$ is defined here as the depth at which the temperature is 0.5 °C less than the temperature at 10 m.

$$d_{ML} = -z\left(T = T_{10m} - 0.5°C\right) \tag{2.6}$$

These two definitions of the mixed layer depth can be used to define whether a diurnal warm layer develops on any given day. If a diurnal warm layer does develop, then the minimum value during that day of the mixed layer depth defined relative to the surface [min($d_{MLsfc}$)] will be small, compared with the (daily mean) value of the mixed layer depth defined relative to the 10 m reference level [mean($d_{ML}$)]. If no diurnal warm layer develops, then the two definitions of the mixed layer depth will return similar values. A scatter plot of [min($d_{MLsfc}$)] against [mean($d_{ML}$)] (Figure 2.7) shows two clusters, conveniently divided by the line min($d_{MLsfc}$) = 14.5 m. Hence, a diurnal warm layer is defined to exist if

$$min\left(d_{MLsfc}\right) < 14.5 \ m. \tag{2.7}$$

An idealized isothermal mixed layer is then defined, with a mixed layer temperature $T_{ML}$ such that the heat content of the idealized mixed layer is the same as the heat content of the water column between the base of the diurnal warm layer and the base of the mixed layer

$$T_{ML}(d_{ML} - d_{WL}) = \int_{-d_{ML}}^{-d_{WL}} T(z)dz. \qquad (2.8)$$

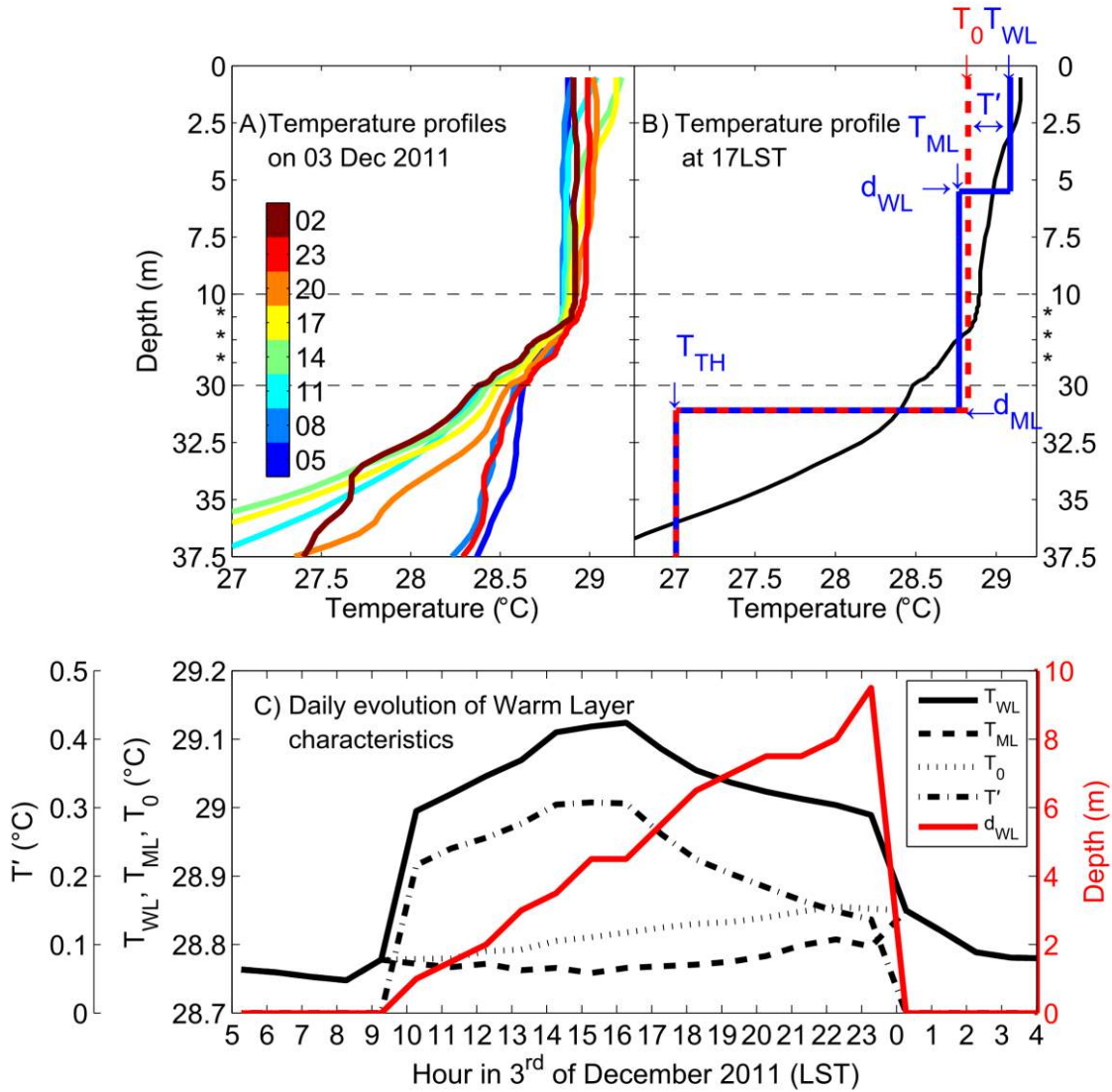

**Figure 2.6 (a) Temperature profiles of optimally interpolated glider data every 3 hours from 0500 LST on 3 December 2011 to 0200 LST on 4 December 2011. The colours of the individual profiles correspond to the times (hours local solar time) in the legend. (b) Temperature profile of optimally interpolated glider data at 1700 LST on 3 December 2011 (black solid line). The idealised three-layer (two-layer) model fitted to this profile is shown by the thick blue (dashed red) line. Annotations show the values of the layer temperatures and depths of interfaces between the layers. (c) Time series of warm layer temperature TWL (thick black solid line), mixed layer temperature TML (black dashed line), mixed layer temperature in two-layer model T0 (thin black solid line), temperature anomaly due to existence of warm layer T', and depth of warm layer dWL (red solid line), during 3 December 2011.**

To complete the three-layer model, the thermocline layer is defined to be the layer at depths greater than $d_{ML}$. It has a constant temperature $T_{TH}$, which is taken to be the temperature at the level 5 m below the mixed layer depth:

$$T_{TH} = T(z = -d_{ML} - 5) \qquad (2.9)$$

The thermocline layer is not used in the subsequent analysis, but is included here for completeness, as it will be used in future modeling studies.

To quantify the effect of the diurnal warm layer, an alternative two-layer model is constructed, where the diurnal warm layer and the mixed layer are combined, and the entire water column is mixed from the surface to the base of the mixed layer. This "super" mixed layer has temperature $T_0$, where

$$T_0 d_{ML} = \int_{-d_{ML}}^{0} T(z) dz. \qquad (2.10)$$

The second layer of this two-layer model is the thermocline layer, as before. This two-layer model would be appropriate for a numerical model that does not resolve the details of the diurnal warm layer development. It can be seen that

$$T_{ML} \leq T_0 \leq T_{WL}. \qquad (2.11)$$

The increase in SST due to the existence of the warm layer is then

$$T' = T_{WL} - T_0. \qquad (2.12)$$

Time series of these quantities are shown in Figure 2.6c for the sample day, 3 December 2011. Before 0930 LST, the surface temperature is below the value of $T_*$ for that day. Hence, there is no warm layer, and $d_{WL}=0$, $T_{WL}=T_{ML}$, and $T'=0$. At 0930 LST, the surface temperature is equal to $T_*$, and the warm layer begins to develop. Its depth, $d_{WL}$, increases approximately linearly, from 0 m at 0930 LST, to 9.5 m at 2330 LST, i.e., at a rate of approximately 0.7 mhr$^{-1}$. At 2330 LST there is then rapid mixing of the water column, presumably due to nocturnal destabilization by cooling surface fluxes of infrared radiation, and sensible and latent heat flux. The surface temperature rapidly decreases to below $T_*$, the warm layer vanishes, and $d_{WL}=0$ again.

The temperature of the warm layer $T_{WL}$ is set to $T_{ML}$ before 0930 LST when the warm layer does not exist. It then increases rapidly to 29.0 °C by 1000 LST, then steadily to 29.12 °C at 1600 LST. It then steadily decreases throughout the rest of the day. By contrast, the mixed layer temperature $T_{ML}$ from the three-layer model, and the "super"-mixed layer temperature $T_0$ from the two-layer model both remain fairly constant throughout the day. Hence, T', the increase in SST due to the existence of the warm layer, increases from 0 at 0930 LST to a maximum of 0.3 °C at 1400–1600 LST, then decreases to zero at 0030 LST when the warm layer vanishes.

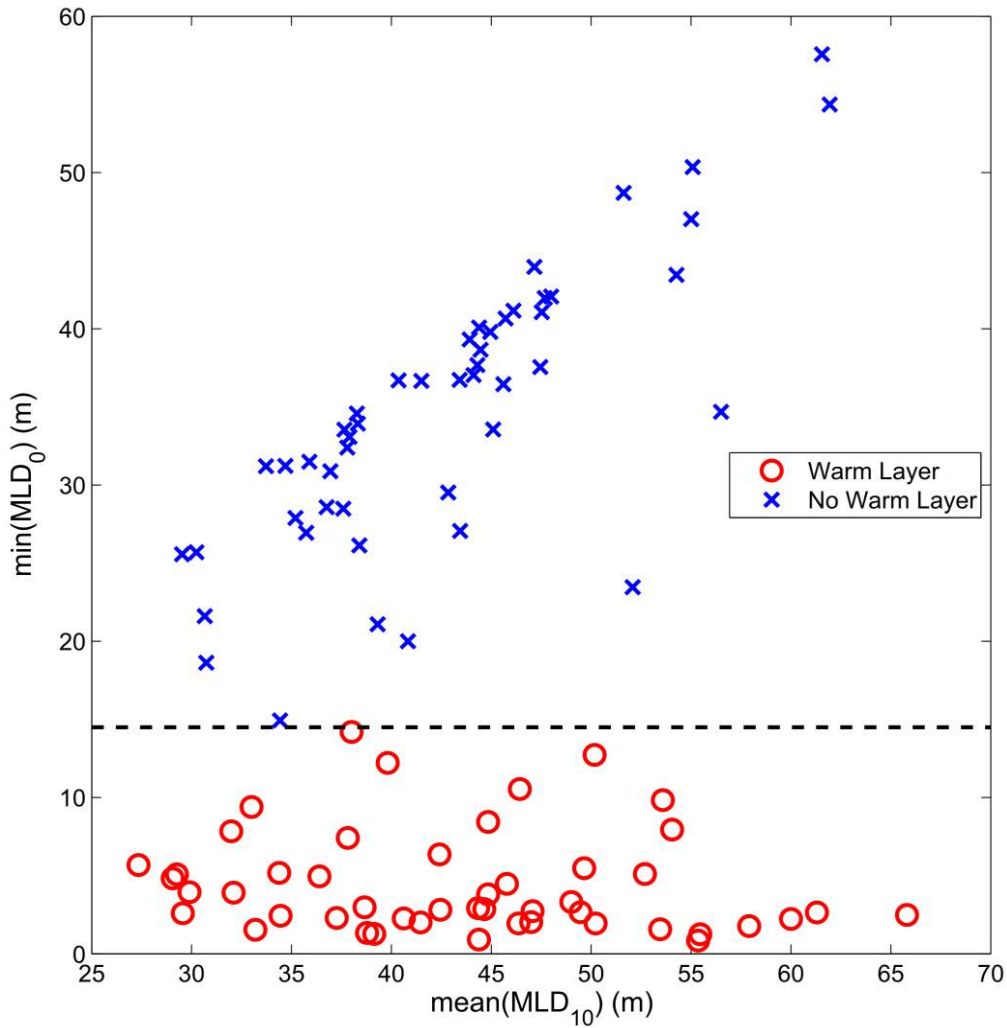

**Figure 2.7 Scatter plot of daily minimum value of the mixed layer depth defined relative to the surface [min($d_{MLsfc}$)], against the daily mean value of mixed layer depth defined relative to the 10 m reference level [mean($d_{ML}$)]. Days with a diurnal warm layer are shown by the red circles; days without a diurnal warm layer are shown by the blue crosses. The horizontal line at min($d_{MLsfc}$)=14.5 m divides the two regimes.**

Although the diurnal warm layer is, by definition, a temperature structure, it may also have an effect on the near-surface salinity. At the glider location, there was a large freshwater input at the surface from precipitation (Figure 2.2b). In the presence of mixing with the high-salinity water below, a steady state salinity profile would result (Figure 2.4). However, as the temperature stratification increases in the afternoon with the development of the diurnal warm layer, the mixing would be weakened. A transient fresh anomaly might be expected to develop at the surface at this time. Such a signal was sought for, but data quality problems prevented any firm conclusions being made. This was because calculation of salinity from conductivity has temperature dependence. The long response time of the conductivity sensor compared with the temperature sensor, the glider crossing a high temperature gradient in the diurnal warm layer, and only the climb phase data being able to be used here, all contributed to prevent salinity data of sufficient accuracy being produced in the upper few meters.

## 2.5. Variability of diurnal warm layer formation

The surface diurnal warm layer framework of section 2.4 is now extended from the sample day of 3 December 2011 (Figure 2.6) to the whole study period, through a series of "stacked" diurnal cycles in Figure 2.8. In these figures, each row represents the time series of a variable for a particular day, from 0000 UTC (approximately 0500 LST) to 2400 UTC (approximately 0500 LST on the next day). The rows are stacked above each other, so that the vertical axis represents time in (integer) days. There are 97 days in the study period (1 October 2011 to 5 January 2012), corresponding to 97 rows in the stacked diurnal cycles. Hence, the row corresponding to 3 December 2011 in the stacked diurnal cycle of T' (Figure 2.8a) contains the same data as the time series of T' for this day in Figure 2.6c.

Over the 97-day study period, there are clusters of days (e.g., 4–13 October 2011) where the diurnal warm layer develops, and T' regularly exceeds 0.4 °C in the afternoon. There are other periods where there is no diurnal warm layer development, and T'=0 throughout the day (e.g., 15–31 December 2011). The periods of surface diurnal warm layer development show a tendency to occur preferentially in the inactive stage of the MJO (light purple shading in vertical bar to the right of the stacked diurnal cycle in Figure 2.8a). Conversely, periods with no diurnal warm layer development occur preferentially in the active stage of the MJO (pink shading in vertical bar).

However, this relationship between the occurrence of the diurnal warm layer and the state of the MJO is certainly not a perfect one. Convection and related conditions at a single geographical point (the glider location) will be subject to large variability, as the planetary scale MJO envelope is made up of contributions from multiple scales. This can clearly be seen in the time series of precipitation, wind speed, and shortwave radiation at the glider location Figure 2.2b). The tendency for wet, windy and cloudy conditions in the active MJO stage, and dry, calm and clear conditions in the inactive MJO stage, can be seen. However, there is much day-to-day local variability. Hence, the analysis of the diurnal warm layer development in this limited data set is made based on local conditions, rather than on the planetary scale MJO state.

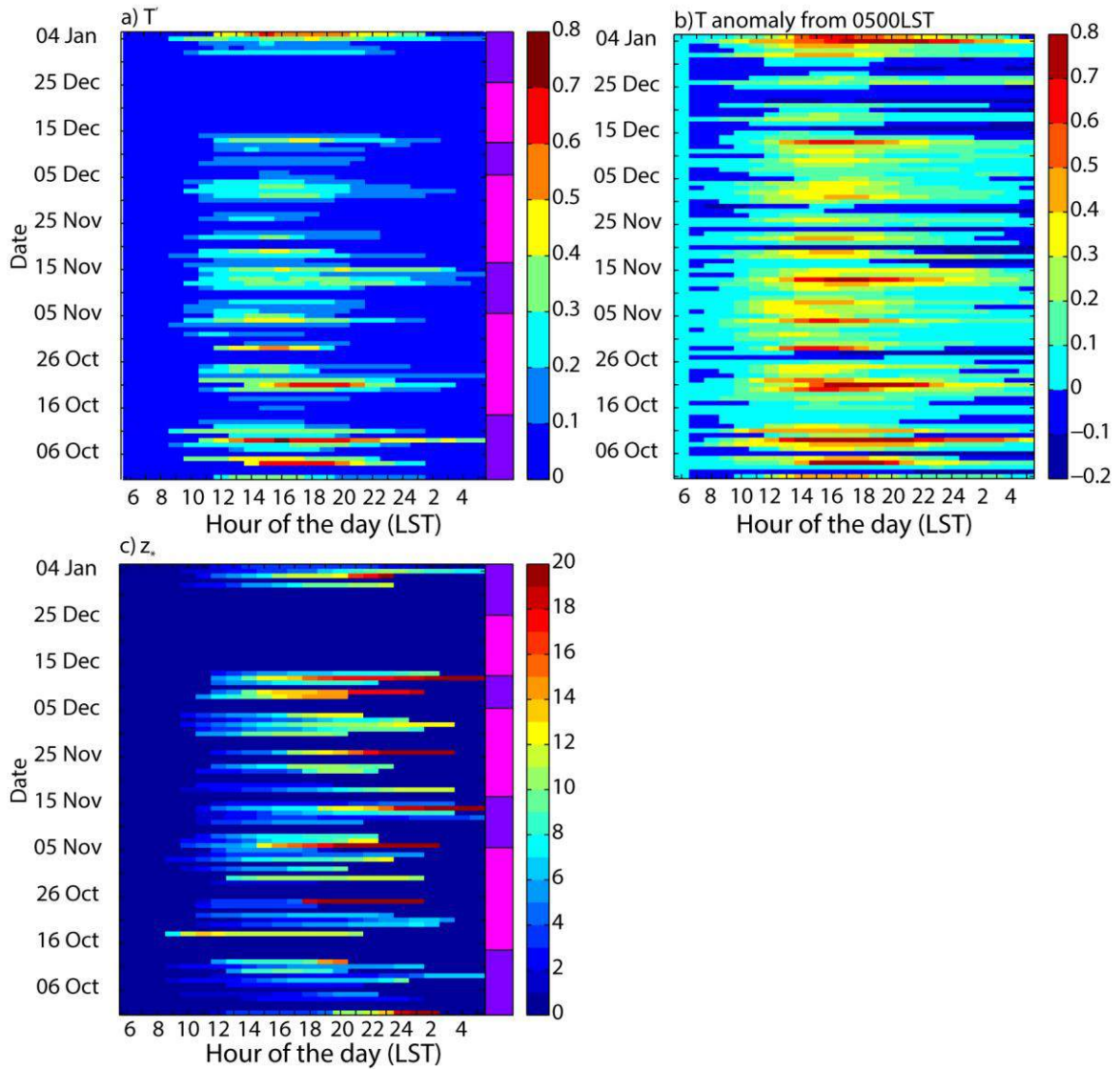

**Figure 2.8** Stacked diurnal cycles of (a) temperature anomaly due to the existence of the diurnal warm layer (T'), (b) temperature anomaly (from 0600 LST) at 0.5 m (T†), (c) depth of the diurnal warm layer ($z_*$). The vertical bars, with pink and light purple shading, mirror the background in Figure 2.2a and indicate the stages of the MJO (pink for active, light purple for inactive).

The scatter plot in Figure 2.7 and the stacked diurnal cycles in Figure 2.8 show two clearly distinct regimes: diurnal warm layer days, and non-warm layer days. Using the criterion of Eq. (2.7), the population of 97 days in the study period is split in half, with 48 (49%) warm layer days and 49 (51%) non-warm layer days.

The apparent relationship between diurnal warm layer formation and the state of the MJO can be tested statistically using a contingency table (Table 2.1). The background or unconditional probability, over all N=97 days, of warm layer formation on a particular given day is $p_0$=48/97=0.495. The conditional observed probability of warm layer formation, given that the MJO is in its inactive stage, increases to $\hat{p}$=25/43=0.581. A test statistic

$$Z = \frac{\hat{p} - p_0}{s_{\hat{p}}}, \quad (2.13)$$

where

$$s_{\hat{p}} = \sqrt{\frac{p_0(1-p_0)}{N}}, \quad (2.14)$$

follows a Gaussian distribution with zero mean and unit standard deviation (e.g., Hall et al., 2001). Here, Z=1.69, which is above the critical value of Z=1.64 at the 90% significance level. Hence, the null hypothesis, that the probability of a diurnal warm layer forming is independent of the state of the MJO, is rejected at the 90% level.

|  | Warm layer | No warm layer | All |
| --- | --- | --- | --- |
| Active MJO | 23 | 31 | 54 |
| Inactive MJO | 25 | 18 | 43 |
|  | 48 | 49 | 97 |

**Table 2.1 Contingency table of number of days when a diurnal warm layer formed or did not form, against the state of the MJO (active or inactive).**

The calculation of T' (Eq. (2.3) - (2.12)) is rather involved. To verify that the existence of the surface diurnal warm layer is not an artifact of this processing, a stacked diurnal cycle is constructed for a simpler measure of the diurnal temperature variation

$$T^{\dagger}(t) = T(t) - T_{06}. \quad (2.15)$$

Here, the surface (0.5 m) temperature at any particular time is expressed as an anomaly $T^{\dagger}$ from its value at sunrise (0600 LST) on that same day. Hence, all the anomalies in the first column of the stacked diurnal cycle of $T^{\dagger}$ (Figure 2.8b), at 0600 LST, are zero by definition. The broad pattern and variability of the $T^{\dagger}$ anomalies are reproduced, confirming that the analysis is not sensitive to the detailed processing.

The stacked diurnal cycle of the depth ($z_*$) of the diurnal warm layer (Figure 2.8c) shows coherent growth in the layer depth throughout the days when a warm layer is found. Typical maximum values lie between 5 and 20 m. There is also considerable variability in the duration of the warm layer, with the terminal mixing and disappearance of the layer generally occuring between sunset (1800 LST) and the early hours of the next morning.

## 2.6. Structure of diurnal warm layer

### 2.6.1. Exponential profile

The mean diurnal cycle for the warm layer days (Figure 2.9a) shows a clear development of a diurnal warm layer, as expected. For this mean diurnal cycle, T06 = 28.94 °C, and $T_{max}$ = 29.36 °C. Hence, the isotherm that defines the warm layer is T* = 29.05 °C (equation 3). Temperatures above T* are shaded red in Figure 2.9a. This mean diurnal warm layer starts at 0900 LST, reaches a maximum depth of $d_{WL}$ = 8 m at 1900 LST, and finally disappears at 0100 LST the following morning.

Note that the surface temperature minimum at sunrise ($T_{06}$) is actually lower than the temperature immediately below. Hence, there is a temperature inversion at the surface. However, because salinity is also a minimum at the surface (Figure 2.4), the density increases monotonically with depth, and the profile remains statically stable.

At all depths within the diurnal warm layer, the temperature minimum and maximum are separated by approximately 12 hours. Hence, the diurnal harmonic (i.e., a shifted cosine wave with a period of exactly 24 hours) will be a good approximation to the full diurnal cycle. At each level, the diurnal harmonic of temperature is calculated. The semidiurnal and higher harmonics were also calculated, but found to be very weak. The amplitude of the diurnal harmonic of temperature decreases monotonically with depth d down to 20 m (thin black line with dots in Figure 2.9a). The phase of the diurnal harmonic (expressed as time of maximum temperature) is shown by the thick line in Figure 2.5a. It increases slowly with depth, from 1600 LST at the surface, to 1800 LST at 13 m, then decreases slowly down to 20 m. Below this level, sampling variability due to gravity waves and their vertical movement of the thermocline masks the structure of the diurnal cycle.

The observed amplitude $T_{dh}$ of the diurnal harmonic is very well modeled by an exponential function of the form

$$T_{dh}(d) = T_{dh0} e^{-d/H} + \varepsilon, \qquad (2.16)$$

where $T_{dh0}$ = 0.22 °C is the (extrapolated) amplitude of the diurnal harmonic at the surface (z= 0 m, i.e., the SST), the scale depth is H = 4.2 m, and ε = 0.004 °C is a residual constant term. The values of these parameters were determined by least squares regression of Eq (2.16) onto the observations, and the modeled curve is shown by the dotted red line in Figure 2.9a. The root mean square error is 0.0017 °C. Hence, the diurnal harmonic of temperature can be modeled as

$$T(d,t) = T_{ref}(d) + T_{dh0} e^{-d/H} \cos\left[\frac{2\pi}{24}\left(t - 16 - \frac{2d}{13}\right)\right], \qquad (2.17)$$

where t is time (hour of day local solar time), and $T_{ref}(d)$ is the background or reference temperature profile.

This exceptionally close agreement between the observed profile of the temperature diurnal harmonic and an exponential function is consistent with the theoretical and observed profiles of solar radiation in seawater. The downward solar radiation flux in seawater, which is responsible for the diurnal warming, also follows an exponential profile, through the Beer–Lambert law for absorption.

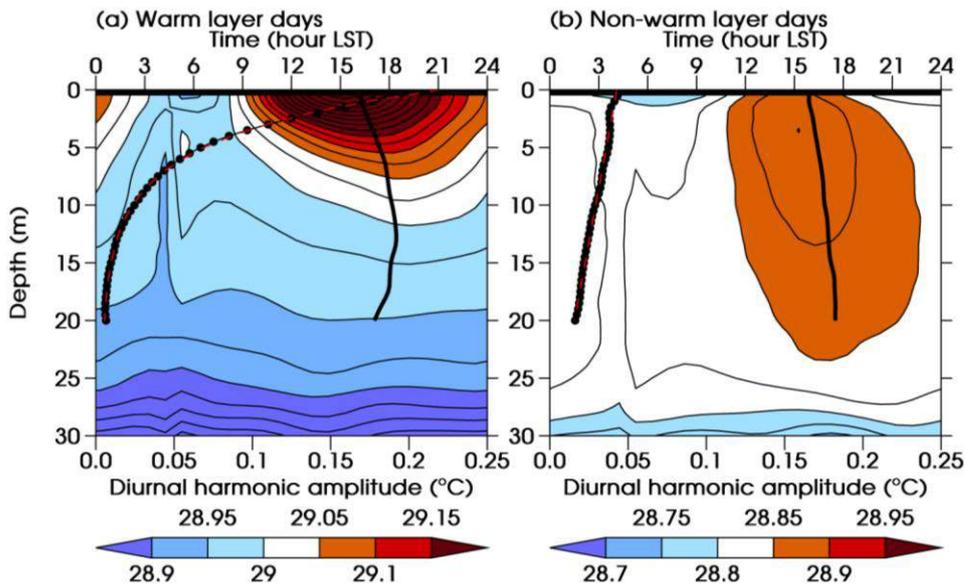

**Figure 2.9** Mean diurnal cycle of temperature during (a) warm layer days, (b) non-warm layer days. Shading interval is 0.05 °C in both panels; see legends for shading intervals. Line contour interval is 0.025 °C. The profile of the diurnal harmonic amplitude is shown by the black dots in each panel; see lower axis for scaling. The dashed red line shows the best-fit exponential function in (a), and linear function in (b). The profile of the diurnal harmonic phase (time of maximum temperature) is shown by the thick black line in both panels.

In fact, the absorption of solar radiation might be expected to follow a double exponential function with two scale depths. One corresponds to the absorption by the red part of the solar spectrum and has a scale depth in the range 0.35–1.4 m [*Jerlov*, 1968; *Mobley*, 1994; *Ohlmann et al.*, 1998; *Paulson and Simpson*, 1977]. The other corresponds to absorption of blue–green light, with a scale depth in the range 7.9–23 m. The single scale depth of H=4.2 m found here is between these values, and corresponds to a mean value for absorption across the whole solar spectrum. However, when a double exponential function was fitted to the observational data, no improvement was found over the single exponential function of Eq (2.16). The two scale depths produced were both very close to 4.2 m.

## 2.6.2. Effect of chlorophyll on diurnal warm layer

The range of scale depths from previous studies (e.g., Paulson and Simpson [1977]) is attributable to different water "types". These have different concentrations of inorganic and organic particles that absorb solar radiation in addition to the absorption by water molecules. Limited measurements of chlorophyll concentration were made by the Seaglider, using a Wetlabs EcoPuck measuring chlorophyll fluorescence. High chlorophyll concentrations are an indication of high phytoplankton abundance, and therefore potentially stronger solar absorption of solar radiation, and a lower scale depth in Eq (2.16).

There were 14 warm layer days when chlorophyll data were available. These separated into two clear regimes. There were 10 days in a low chlorophyll regime, where the mean 0–10 m chlorophyll concentration was in the range 1.1–5.8 mg m$^{-3}$, with a mean of 3.3 mg m$^{-3}$. The remaining 4 days were in a high chlorophyll regime, in the range 14.4–21.7 mg m$^{-3}$, with a mean of 18.8 mg m$^{-3}$. The mean diurnal cycle was calculated separately for each of these two regimes. Despite the low sample sizes, a well defined diurnal warm layer with an exponential temperature profile developed in each regime.

Recall that the scale depth was H = 4.2 m when calculated over all 48 warm layer days. The low chlorophyll regime had a higher scale depth of H = 4.7 m, corresponding to weaker absorption. The high chlorophyll regime had a lower scale depth of H = 4.0 m, corresponding to stronger absorption. Hence, even with the limited measurements available, chlorophyll variability has a clear and significant effect on the depth of the diurnal warm layer, and then potentially on the rectification of SST by the diurnal cycle. The variability in chlorophyll, at the Seaglider location during CINDY/DYNAMO, was controlled by Ekman pumping by the local surface winds and propagation of oceanic equatorial Rossby waves, as part of the dynamical ocean-atmosphere MJO system [*Webber et al.*, 2010; *Webber et al.*, 2014].

## 2.7. A predictive model for diurnal warm layer formation

The diurnal cycle of temperature at the ocean surface will be controlled by two competing processes. The diurnal cycle of surface shortwave radiation flux will heat the surface, stabilize the water column, and force a strong diurnal cycle. Wind-driven mixing will mix up cold water from below and weaken the diurnal cycle. In this section, these two effects are quantified, and a predictive model for the strength of the diurnal cycle is developed, based on the daily mean surface shortwave radiation flux and wind speed.

Daily mean values of the diurnal temperature change are calculated, along with daily mean values of the Meteosat7-derived surface shortwave radiation flux (SWR) at the glider location, and surface wind speed V from the TropFlux data set [*Praveen Kumar et al.*, 2013].

First, the dependence of the simple measure of diurnal temperature anomaly from sunrise $T^{\dagger}$, on SWR and V, is examined. A scatter plot of $T^{\dagger}$ against SWR and V, colored by the value of $T^{\dagger}$ (Figure 2.10a), shows a clear dependence. High values of $T^{\dagger}$ occur on days with high SWR and low V. Days on which a diurnal warm layer formed, according to the criterion of Eq (2.7), are indicated by an additional cross. These occur exclusively on days with high shortwave flux and low wind speed. A predictive model is calculated by non-linear regression, such that

$$T^{\dagger}_{predicted} = \alpha_1 SWRV + \alpha_2 SWR + \alpha_3 V + \alpha_4, \qquad (2.18)$$

where $\alpha_1$ = -2.16×10$^{-4}$, $\alpha_2$ = 0.00208, $\alpha_3$ = 0.0152, and $\alpha_4$ = -0.182, with $T^{\dagger}_{predicted}$ expressed in °C, SWR in Wm$^{-2}$, and V in ms$^{-1}$. Contours of $T^{\dagger}_{predicted}$ are overlain in Figure 2.10a, and fit the data well. A more physically useful measure of the diurnal temperature change is T', the change in SST due to the existence of the diurnal warm layer (Eq (2.12)). Values of T' show a similar relationship with SWR and V (Figure 2.10b). As T' is only meaningful on days when a diurnal warm layer forms, the predictive model for T' is calculated only using data from these days (crosses in Figure 2.10b). The model is

$$T'_{predicted} = \beta_1 SWRV + \beta_2 SWR + \beta_3 V + \beta_4, \qquad (2.19)$$

with $\beta_1$ = -0.000208, $\beta_2$ = 0.00130, $\beta_3$ = 0.0159, and $\beta_4$ = -0.0556. Contours of T'$_{predicted}$ are overlain in Figure 2.10b. The predictive Eq (2.19) for T' asymptotes to a constant value of

$T'_{predicted} = \beta_4 - \beta_2\beta_3/\beta_1 = 0.044$ °C ≈ 0, at SWR = $-\beta_3/\beta_1$ = 76 Wm$^{-2}$, and also at V = $-\beta_2/\beta_1$=6.2 ms$^{-1}$. These are non-linear cutoffs, giving the minimum value of SWR and maximum value of V under which a diurnal warm layer can form. In practice, it can be seen that diurnal warm layer days (crosses in Figure 2.10b) occur almost exclusively on days when

$$T'_{predicted} > 0.1°C. \qquad (2.20)$$

Assuming that this locally-derived relationship holds approximately over the tropics, the importance of diurnal warm layer development can now be assessed across the tropical ocean basins. Daily maps of predicted warm layer occurrence, using Eq (2.20), are calculated from daily maps of TropFlux surface shortwave radiation flux and wind speed, from 1 January 1979 to 31 December 2011. A diurnal warm layer is predicted to occur on over 25% of days over the tropical warm pool region of the Indian Ocean and western Pacific (Figure 2.11a). Other regions of importance are the far eastern Pacific and the eastern Atlantic. The predicted mean excess of SST (T') over nearly all the tropical oceans, when a warm layer develops, is in the range 0.12–0.18 °C (Figure 2.11b).

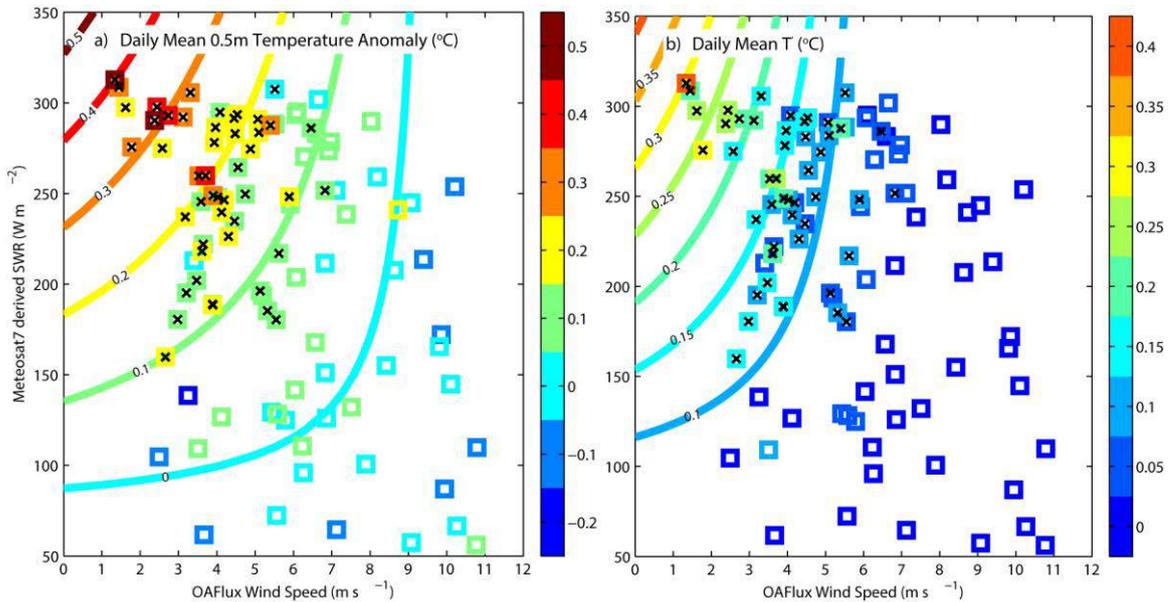

**Figure 2.10 (a) Scatter plot of daily mean OAFlux wind speed versus Meteosat7-derived SWR, coloured by the daily mean temperature anomaly T† at 0.5 m depth. Points marked with an "×" are days when a warm layer developed. Colour contours show the best fit regression lines of T†predicted. (b) As (a) but for T'. The regression lines here are calculated only using data from warm layer days.**

In previous sections, it was shown that warm layer days at the glider location occurred preferentially in the inactive stage of the MJO (RMMI phases 5–8). This deduction was based on data measured at a single location over a relatively short time period. The generality of this observation can now be tested using the global maps of predicted T'. The mean predicted T', averaged over all, not just warm layer, days in MJO phase 4 is approximately 0.08 °C across the equatorial Indian Ocean (Figure 2.11c). Hence, even in the active stage of the MJO, when the effects of the diurnal cycle are at their weakest, a model that cannot resolve these processes would underestimate the SST by approximately 0.08 °C. During the opposite part of the MJO cycle (phase 8, in the inactive stage), this error would be larger, with mean T' values over 0.12 °C over the equatorial Indian Ocean (Figure 2.11d).

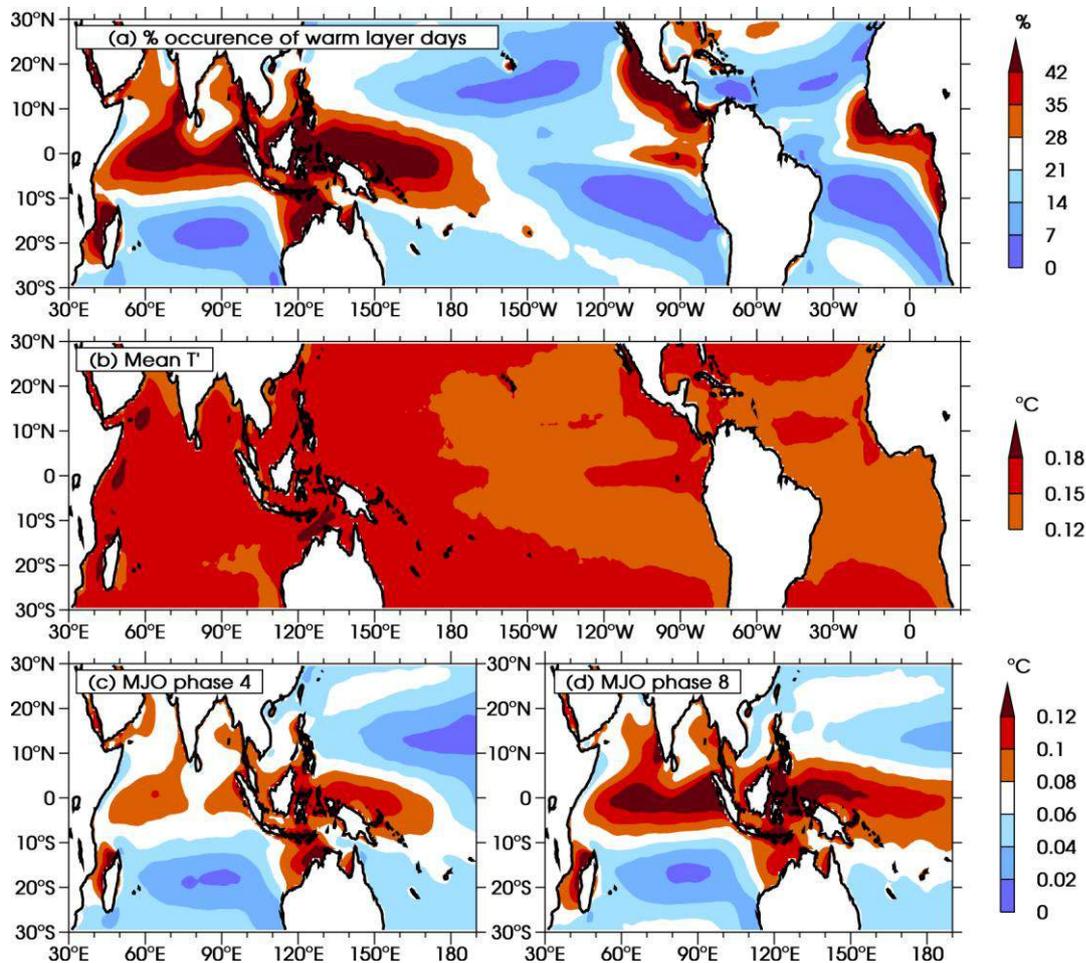

**Figure 2.11 (a)** Percentage of days when a diurnal warm layer is predicted to occur. **(b)** Mean predicted T' on those warm layer days. **(c)** Mean predicted T' for all days in MJO phase 4. **(d)** As (c) but for MJO phase 8.

## 2.8. Diurnal cycle rectification: implied anomalous surface fluxes

The development of the diurnal warm layer increases the SST by an amount T', compared with the situation if the warm layer did not exist. Alternatively, the negative of T' is the SST error that a model would incur if it could not resolve these processes. Instantaneous peak values of T' in the afternoon can reach 1 °C (Figure 2.8a), and daily mean values are in the range 0.1–0.3 °C (Figure 2.10b, Figure 2.11b,c,d). The existence and strength of the diurnal warm layer depends on the surface fluxes of shortwave radiation and momentum (through the wind speed). However, the changes in SST by the increment T' will have a feedback onto the surface fluxes, and a subsequent effect on ocean-atmosphere interaction. In this section, the effect of T' on the surface fluxes is quantified. A (daily mean) value of T'=0.2 $^oC$ will be used to produce numerical values of anomalous surface fluxes.

The surface fluxes are linearised about the mean state to calculate the perturbation fluxes due to the surface temperature anomaly T'. The details of the individual fluxes of longwave radiation, latent and sensible heat are described in Appendix B. The total perturbation flux is

$$Q' = Q_{LW}' + Q_{L'} + Q_{S'}$$
$$= \left[ 4\sigma T_b^3 + \left( \frac{L^2 \varepsilon e_{s_0}}{RR_w T_b^3} + \rho_a c_p \right) c_E V \right] T' \quad (2.21)$$
$$\approx 23T' \approx 4 W m^{-2}.$$

This is the extra cooling surface flux that occurs due to the existence of the diurnal warm layer. Models that do not simulate these processes would have an SST that is too cool, and therefore an erroneous downward surface flux of approximately 4 $Wm^{-2}$. This represents a physical process by which heating on longer time scales is rectified by the diurnal cycle of solar heating.

## 2.9. Residual diurnal cycle in absence of diurnal warm layer

A surface diurnal warm layer developed on 48 of the 97 days in the study period. These days were characterised by high solar radiation flux and low wind speed. However, although a surface diurnal warm layer did not develop on the remaining 49 days, there was still a residual diurnal cycle, which exhibits qualitatively very different behavior to the surface diurnal warm layer (Figure 2.9b). By design, the residual diurnal cycle is much weaker with a range of only 0.08 $^oC$ in the surface temperature, compared with a range of 0.44 $^oC$ for the warm layer days. On the non-warm layer days, the water column is also much cooler overall, with a mean surface temperature of 28.82 $^oC$, compared with 29.14 $^oC$ for the warm layer days. This is consistent with the diurnal warm layer preferentially occurring in the inactive stage of the MJO, when the SST is increasing to its maximum value.

However, in addition to these expected differences, there is also a clear qualitative difference in the spatial structure of the diurnal cycle of temperature. The much weaker surface warming in the non-warm layer days is spread out over a much larger depth than in the warm layer days. A linear function provides a more appropriate fit to the observed structure (black line with dots in Figure 2.9b), where

$$T_{dh} = T_{dh0} - \gamma d, \qquad (2.22)$$

with the surface amplitude $T_{dh0}$=0.04 °C, and the gradient γ=0.0014 °Cm$^{-1}$. By extrapolation, the amplitude of the diurnal harmonic becomes zero at depth D, where

$$D = \frac{T_{dh0}}{\gamma} = 32\ m. \qquad (2.23)$$

The linear profile might be expected. Surface heating is accomplished through the absorption of solar radiation with an exponential profile in the upper few meters. In low wind conditions, with weak mixing, this leads to formation of a diurnal warm layer, with a similar exponential temperature profile. However, in high wind conditions, this warm surface layer will be mixed down into the cooler water below. The linear profile is then a steady-state solution of the (eddy) diffusion equation, with boundary conditions of constant temperature at the surface and the base of the diurnal layer (at D≈30 m here). Note that the temperature gradient in this linear diurnal layer is very weak, and is only detectable due to the high-quality, high-resolution measurements available from the glider.

As wind speed increases, this already very weak temperature gradient would be expected to decrease further, tending toward isothermal conditions, with a weaker vertical temperature gradient γ and lower surface amplitude $T_{dh0}$. The stronger mixing would also extend over a deeper layer, hence the depth of the linear diurnal layer might be expected to increase with wind speed. To test this hypothesis, the 49 non-warm layer days were ordered by the strength of the OAFlux daily mean wind speeds at the glider location. They were then stratified into three wind regimes of approximately equal sample size (16, 16, and 17 days respectively). In the low wind speed regime, the wind speed was in the range 2.5–6.1 ms$^{-1}$, with a mean of $\overline{V}$=4.9 ms$^{-1}$. The medium regime had winds of 6.1–7.9 ms$^{-1}$, with a mean of $\overline{V}$ = 6.8 ms$^{-1}$. The high regime consisted of all days with winds above 7.9 ms$^{-1}$, with a mean of $\overline{V}$ = 9.4 ms$^{-1}$.

A mean diurnal cycle was constructed separately for each of the three regimes. A clear linear temperature profile was reproduced in each regime. As hypothesized the linear temperature gradient did monotonically decrease with wind speed (Figure 2.8; black solid line), by a factor of four from γ = 0.0020 °Cm$^{-1}$ at $\overline{V}$=4.9 ms$^{-1}$ to γ=0.0005 °Cm$^{-1}$ at $\overline{V}$=9.4 ms$^{-1}$.

The surface amplitude was approximately constant at $T_{dh0} \approx 0.05$ °C in the low and medium wind speed regimes, but then decreased to $T_{dh0} \approx 0.027$ °C in the high wind regime (Figure 2.12; blue dotted line). The depth of the linear diurnal layer increased consistently with wind speed, from D=25 m at $\bar{V}$=4.9 ms$^{-1}$ to D=55 m at $\bar{V}$=9.4 ms$^{-1}$.

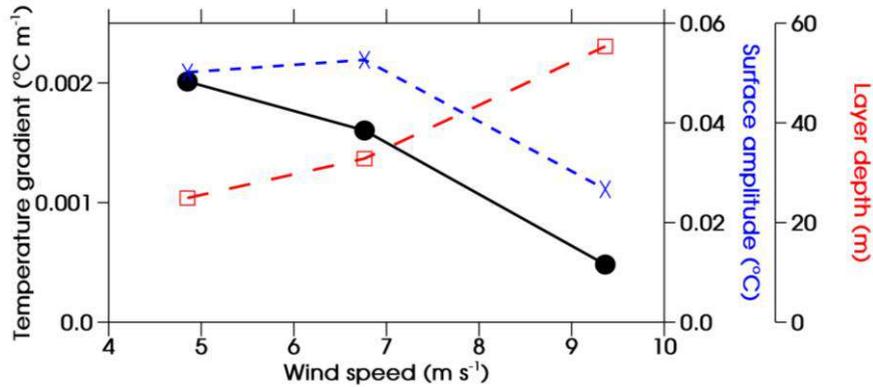

**Figure 2.12 Dependence of linear temperature profile parameters on wind speed for the residual diurnal cycle on non-warm layer days: temperature gradient γ (black solid line); surface amplitude $T_{dh0}$ (blue dotted line); layer depth D (red dashed line).**

## 2.10. Summary

The time-developing structure of the diurnal warm layer in the equatorial Indian Ocean was diagnosed, using high-resolution measurements from a Seaglider during the CINDY/DYNAMO field experiment. Two distinct regimes were found, summarized in the schematic in Figure 2.13.

On half of the days in the study period, a diurnal warm layer developed (Figure 2.13a). It was characterized by a temperature structure with a surface maximum that peaked in the mid-afternoon at 1600 LST. The temperature anomaly decayed exponentially with depth over a scale depth of 4–5m, depending on chlorophyll concentration. This is consistent with heating by absorption of solar radiation. In the late afternoon and after sunset, the cooling fluxes of longwave radiation, and latent and sensible heat fluxes, take over, and the diurnal warm layer decays. Eventually, just before sunrise, the surface cooling is enough to destabilize the water column, which rapidly overturns, creating a deeper isothermal mixed layer. Because the solar heating is effectively trapped in the shallow diurnal warm layer during the day, the daily mean SST is higher than it would be in the absence of the diurnal warm layer. Hence, the SST on longer time scales, including the climatological mean, should be rectified by the diurnal cycle. This rectification is quantified by a simple model. The mean effective SST anomaly due to the existence of the diurnal warm layer is T'≈0.2 °C.

On the remaining half of the days, a diurnal warm layer did not develop. Instead, a much weaker residual diurnal cycle was observed, whose amplitude showed a linear decrease with depth down to approximately 25–50 m (Figure 2.13b), dependent on wind speed. This is consistent with control by vertical mixing.

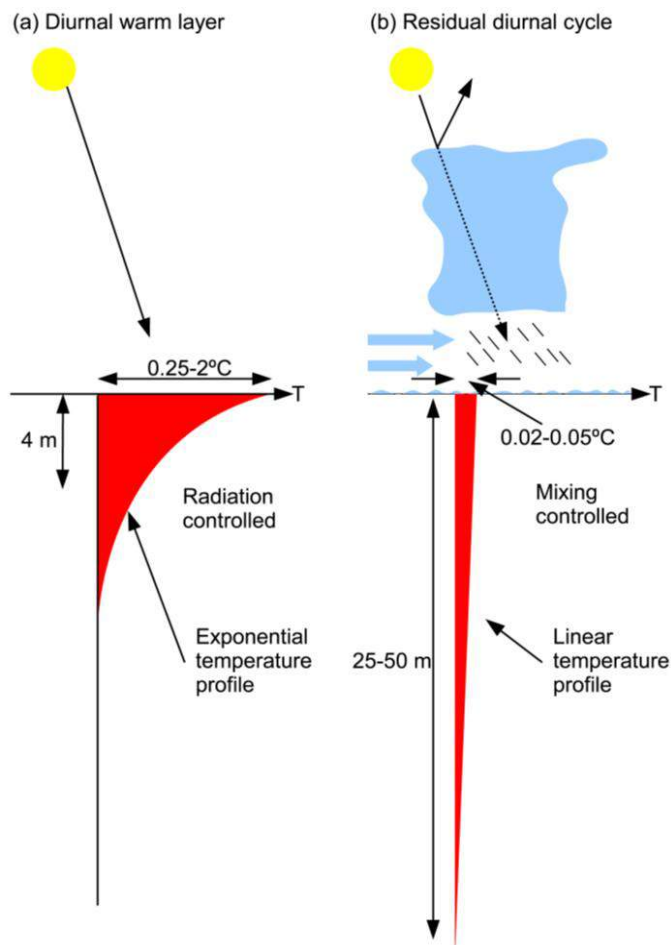

**Figure 2.13 Schematic diagram of processes leading to the (a) diurnal warm layer, (b) residual diurnal cycle.**

The days when a diurnal warm layer did develop were characterized by high values of solar radiation flux (above approximately 80 Wm$^{-2}$) and low wind speeds (below approximately 6 ms$^{-1}$). Conversely, the days with no diurnal warm layer had low solar radiation and high wind speed.

A predictive model for the existence and strength of the diurnal warm layer was developed, using the TropFlux gridded solar radiation flux and wind speed products. Surface diurnal warm layers are predicted to occur on over 30% of days across the warm pool region, and over the tropical eastern Pacific and eastern Atlantic. At first, this might seem surprising, as these are regions of maximum mean precipitation and cloudiness, and minimum mean solar radiation flux, compared to other locations at the same latitude. However, there is much temporal variability in precipitation and cloud cover due to the multi-scale nature of tropical convection. On days when cloud cover is absent, the wind speed is low enough to allow a surface diurnal warm layer to form. This variability is linked with the MJO, with a higher proportion of diurnal warm layer days within the convectively suppressed phase of the MJO, compared to the convectively active phase.

Conversely, regions of high mean solar radiation flux, such as the relatively cloud-free central equatorial, and subtropical, Pacific, had a much lower proportion of diurnal warm layer days. This is due to the consistently high wind speeds in these trade wind regimes.

The daily mean SST anomaly due to the existence of the diurnal warm layer will drive anomalous fluxes of longwave radiation and latent and sensible heat. The flux equations were linearised about the mean state to quantify this. The anomalous flux is of the order of 4 $Wm^{-2}$ upwards, cooling the ocean and warming the atmosphere. A coupled ocean–atmosphere model, that does not resolve the diurnal warm layer, would incur the negative of this flux anomaly as an error, i.e., an erroneous flux of 4 $Wm^{-2}$ downwards, warming the ocean and cooling the atmosphere.

A flux anomaly, or error, of this magnitude is significant. For example, it is comparable to the differences between surface flux estimates from various reanalysis-based climatologies for the Indian Ocean at these latitudes [*Schott et al.*, 2009]. On intraseasonal time scales, surface flux anomalies drive the ocean–atmosphere interactions with the MJO. During the inactive stage of the MJO, these anomalies are order 20 $Wm^{-2}$ and are downward, warming the ocean ahead of the MJO convection [*Shinoda et al.*, 1998; *Woolnough et al.*, 2000]. The flux anomaly due to the diurnal warm layer is upward, cooling the ocean, and will act as a weak negative feedback on the coupled ocean–atmosphere component of the MJO.

Future modeling studies will investigate the SST and surface flux contributions from the diurnal warm layer, and the errors incurred when the processes are not resolved. These modeling studies, and the observational analysis here, can then feed into the development of a parameterization scheme for the effect of the diurnal warm layer in coarse resolution coupled ocean–atmosphere models.

# Chapter 3. Influence of heat source variability on atmospheric Kelvin wave initiation

## 3.1. Introduction

In this Chapter will investigate how the air sea interaction and diurnal variability in the upper ocean could contribute to generation of Kelvin waves and, in particular, multiple Kelvin waves. We use shallow water model because it offers simplest tool to investigate behavior of the atmospheric waves and the results can be used for interpretation of observations and more complicated dynamical simulations. We also use observational data to derive case study analysis and statistics of a Kelvin wave initiation.

In the past a nonlinear shallow water model on the sphere was used to simulate barotropic aspects of Intertropical Convergence Zone (ITCZ) breakdown [*Ferreira and Schubert*, 1997]. In their work the ITCZ was simulated by a prescribed zonally elongated mass sink near the equator. The mass sink produced a cyclonic potential vortices and the shallow water model simulated ITCZ break down into several cyclones that could be observed in the satellite images.

A nonlinear shallow-water model on the sphere was also used to study barotropic aspects of the formation of twin tropical disturbances by MJO convection [*Ferreira et al.*, 1996]. In that work the effect of MJO convection upon the lower-tropospheric tropical circulation was simulated by an eastward moving, meridionally elongated mass sink straddling the equator. Again it explained patterns shown by Nakazawa [1988] of western moving tropical cluster emanating from easterly propagating MJO envelope. Recently, a shallow-water model was used to simulate the MJO [*Yang and Ingersoll*, 2013]. In this model setup, convection is parameterized as a short-duration, localized mass source and is triggered when the layer thickness falls below a critical value. Radiation is parameterized as a steady uniform mass sink. Many aspects of MJO were simulated in such a simple setup. In particular those authors propose that the simulated MJO signal is an interference pattern of westward and eastward inertia–gravity (WIG and EIG) waves.

In this Chapter we use a shallow water model to investigate the influence of diurnal variability of the heat source on the zonal structure of tropical waves. We will show that while the Rossby waves are not sensitive to the temporal variability of the heat source, the Kelvin wave response is highly sensitive to the heat source frequency. The shallow water we use was developed by Giraldo [2000] and our heat sources are based on those used by Ferreira *et al.* [1996].

These results suggest that in real atmosphere, the high frequency variability of forcing may result in change of the zonal and temporal characteristics of the Kelvin waves. We hypothesize that short term variability of the forcing may be influenced by the upper ocean diurnal cycle. Therefore, such high frequency upper ocean variability may contribute to initiation of the Kelvin waves. To this end, we will examine the upper ocean and surface variability during Kelvin waves initiations.

A recent study [*Ruppert and Johnson*] shows that diurnal SST variability is reflected in the diurnal cycle of the heating profiles in the atmosphere. Those authors show that the low-to-midlevel moistening (or preconditioning) during the suppressed phase of the MJO is accomplished by a population of shallow cumulus and congestus clouds that exhibits a pronounced diurnal cycle in response to oceanic diurnal warm layers. The analysis of the suppressed phase revealed that the moistening characteristic of this period is accomplished by the diurnal cycle of cumulus clouds. This diurnal cycle in atmospheric convection is driven by the cycle in sea surface temperature and air-sea fluxes linked to shallow oceanic diurnal warm layer development. This coupled diurnal cycle, and the associated afternoon peak in convective cloud depth, cloud areal coverage, and cumulus moistening, likely drives more vigorous overall moistening than would occur without this diurnal cycle.

Figure 3.1 compares diurnal cycles of precipitation for the periods prior to (A) October 2011 and (B) November 2011 MJO events. While rainfall in the October period largely persists through night before tapering off in the morning, rainfall in the November period exhibits a more distinct early-morning maximum offset by ~12 h from the afternoon peak. Both cases occur in the suppressed phase of the MJO and the main result is that that both show afternoon peak in convection suggesting the link to ocean diurnal variability. Figure 3.2 provides diurnal composite specific humidity (q), vertical motion (ω), and vertical heating profiles (apparent heat source $Q_1$ and apparent moist sink $Q_2$) derived from the DYNAMO northern array (defined by the location of the islands of Gan, Male and Ceylon, and location of R/V Roger Revell at the equator and longitude 80E) gridded sounding analysis for the suppressed phases for: (A) October 2011 and (B) November 2011. Specific humidity q is shown with composite mean removed (q'), though composite means are retained in the other fields. There is marked consistency in the diurnal cycle between the October and November suppressed phases. Both periods exhibit a morning-evening q' swing of 0.4~0.6 gkg$^{-1}$ in the layer 900~550 hPa, with moister conditions in the evening. Since temperature varies negligibly, this diurnal variation in q' directly correlates with relative humidity. The evolution of $Q_1$ largely reflects the diurnal cycle of shortwave heating, though enhanced warming near and within the boundary layer relates to the large eddy flux of sensible heat from the surface as the mixed layer develops and deepens.

This analysis suggests that using diurnally varying heat source in the shallow water model to study initiation of Kelvin waves can justified by the DYNAMO observations. This is especially true when considering the suppressed phase of the MJO.

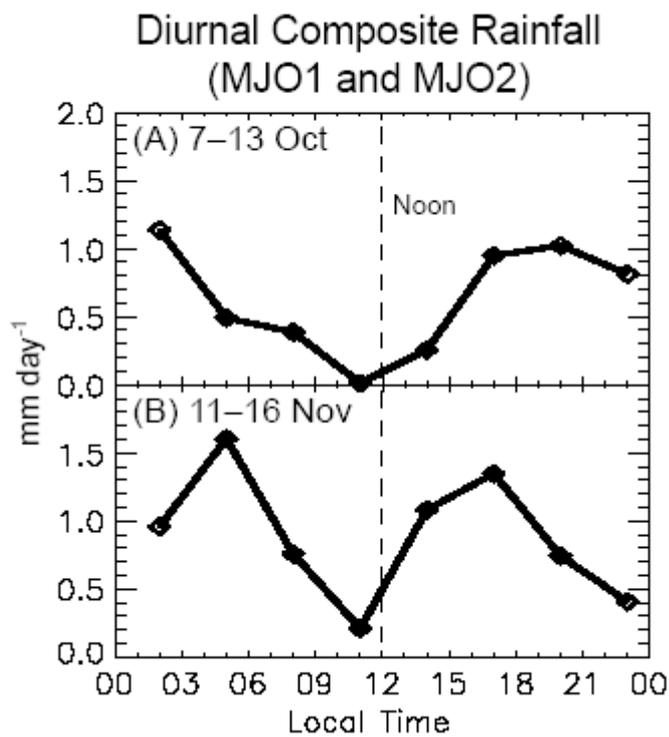

**Figure 3.1** Diurnal composites of TRMM 3B42 rainfall [mmday$^{-1}$] averaged over the northern sounding quadrilateral for the (a) October 2011 and (b) November 2011 suppressed phases. Composite date ranges are (a) 7-13 October and (b) 11-16 November. Local time is indicated along the abscissa. This is reprint of Figure 16 from Ruppert and Johnson [2014]

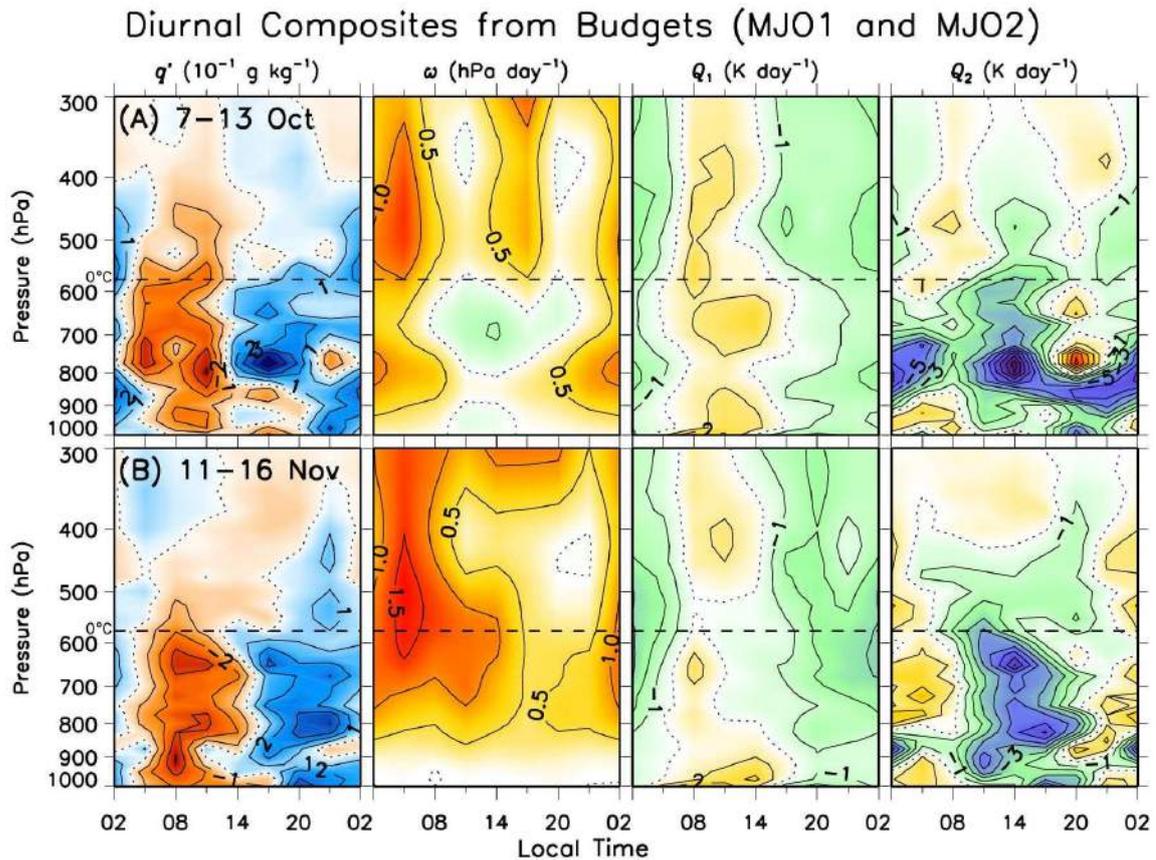

**Figure 3.2** Diurnal composite variables from the northern-array gridded analysis for the (a) October and (b) November suppressed phases, with (from left-right) water vapor mixing ratio with composite mean removed q' [$10^{-1}$ g kg$^{-1}$], ω [hPa day$^{-1}$], $Q_1$, and $Q_2$ [K day$^{-1}$]. Composite date ranges are (a) 7-13 October and (b) 11-16 November (end-dates are inclusive). Horizontal dashed lines indicate the 0ºC level. Vertical smoothing has been applied to all fields using a three-point running mean. This is reprint of Figure 17 from Ruppert and Johnson [2014]

## 3.2. Shallow water model description and simulation set up

To test the hypothesis that diurnal cycle and its characteristics may impact structure of the equatorial Kelvin Wave is tested using global, 2 dimensional spectral element shallow water [*Giraldo*, 2000] model. This model solves shallow water equations (3.1) on spherical geodesic grid. It is suitable for testing dynamic response of the global atmospheric circulation to the prescribed and idealized forcing. The equations are solved for the four variables φ, φu, φv and φw, where φ is geopotential, u, v and w are velocity components. The right hand site of Eq. (3.1),(3.2) represents forcing which consists of the force due to pressure ($F_P$), the force due to the rotation of the earth ($F_R$), the force required to constrain the fluid particles to remain on the surface of the sphere ($F_C$), and additional mass source/sink (Eq. (3.2)) which represents external forcing ($F_F$).

$$\frac{\partial \boldsymbol{\varphi}}{\partial t} + \nabla \cdot (\varphi \mathbf{u}) = \mathbf{S}(\varphi) \tag{3.1}$$

$$\boldsymbol{\varphi} = \begin{bmatrix} \varphi \\ \varphi u \\ \varphi v \\ \varphi w \end{bmatrix}, \quad \mathbf{u} = \begin{bmatrix} u \\ v \\ w \end{bmatrix}, \quad \mathbf{S}(\varphi) = \mathbf{F}_P + \mathbf{F}_R + \mathbf{F}_C + \mathbf{F}_F \tag{3.2}$$

$$\mathbf{F}_P = -\varphi \nabla \varphi, \quad \mathbf{F}_R = -f(\mathbf{r} \times \varphi \mathbf{u}), \quad \mathbf{F}_C = \mu \mathbf{r}$$

where f is Coriolis parameter and µ is the Lagrange multiplier used to constrain the fluid particles to remain on the surface of the sphere condition. Numerical experiments were conducted using different temporal characteristics of the external forcing term **F**$_F$.

Hexahedral model grid was used in all simulations; its horizontal resolution is roughly equivalent to the grid point resolution of 2.2 degrees.

All runs were conducted using mass source/sink of circular shape, centered at the equator at [0,0] [*Ferreira et al.*, 1996]. Source had constant spatial distribution with radius equivalent to 10 degrees along the equator. Elsewhere **F**$_F$ value was set to 0. Eq (3.3) defines spatial distribution of external forcing, where f(t) refers to temporal variability of mass sink described by Eq. (3.4) and r is distance from the source center, time is in days, T is period of oscillation set to 24 hours, and τ is the source amplitude. Length of the simulation was always 10 days. External forcing had non zero value during first 5 days of the simulation. Eq. (3.4) describes temporal characteristics of the mass sink for the first 5 days; "s1" refers to the constant forcing experiment, "s2" refers to the case of forcing that oscillates with period T equal 1 day and constant magnitude τ and "s3" refers to one with forcing oscillating with period T and magnitude changing with time. Integrated external forcing is the same for all simulations.

$$F_F = \begin{cases} f(t), & r \leq 10 \text{ degrees} \\ 0, & r > 10 \text{ degrees} \end{cases} \tag{3.3}$$

$$f(t \leq 5) = \begin{cases} s1: \tau \\ s2: \tau \left[ 1 + \sin\left(\frac{2\pi}{T} t\right) \right] \\ s3: \tau \left\{ 1 + \sin\left(\frac{2\pi}{T} t\right) - 0.5 \cdot \frac{2}{\pi} \cdot \arctan\left[ \left(\frac{t}{T} - 2.5\right) \cdot 10 \right] \right\} \end{cases} \tag{3.4}$$

## 3.3. Shallow water model results

Figure 3.3 presents control simulation "s1" at 7.75 days, when the edge of the Kelvin wave propagated around the equator and approached the area where it had been initiated. Color shading shows geopotential height anomaly, arrows show wind anomaly. This set up is the same in all figures presented in this section. In this simulation forced Kelvin wave, Rossby wave and their interaction are apparent. Kelvin wave response is manifested by negative geopotential height anomaly west of longitude -100E. After 7.75 days Kelvin wave travelled eastward along the equator and its edge is visible at longitude -100E. The Rossby wave response is manifested as a pair of symmetric cyclonic gyres at [15N,-50E] and [-15N,-100E], and is associated with the pair of positive geopotential height anomalies at [20N,-20E] and [-20N,-20E]. The equatorial area between the two Rossby gyres and between longitudes -87E and 3E is associated with westerly wind anomaly. The eastern edge of the Kelvin wave is also associated with westerly flow, and along the Kelvin wave, the easterly flow dominates. After 7.75 days the edge of the Kelvin wave interacts with the Rossby wave. The area between -100E and -87E is characterized by shift in zonal flow from easterly to the west of that location to westerly to the east of that location. Control simulation shows differences in phase speed between Kelvin and Rossby wave responses. In 7.75 days Kelvin wave almost travelled around the entire equator and Rossby wave "only" travelled distance of 50 degrees. Hence, Kelvin waves propagate much faster than Rossby waves. This is typical Gill's response to the equatorial heat source [*Gill*, 1980].

Figure 3.4 and Figure 3.5 show results for experiments "s2" and "s3". It can be seen in both figures that Rossby wave response does not change much in comparison with the simulation "s1", but the structure of the Kelvin wave response is different. In experiment "s2" the Rossby wave response is manifested by the pair of symmetric cyclonic gyres at [15N,-50E] and [-15N,-50], and is associated with them a pair of positive geopotential height anomaly at [20N,-20E] and [-20N,-20E]. Although the magnitude of these anomalies is a little different from control simulation, the overall structure is similar. In simulation "s2" flow and geopotential height structure is little different from control simulation at longitude -60E, where geopotential height minimum is present at the equator. This minimum is first of the sequence of five geopotential height minima associated with Kelvin wave response, which exhibits highly disturbed structure in comparison with control simulation. A sequence of negative geopotential height anomalies along the equator with minima at longitudes -70E, -130E, -170E, 155E and 120E represent Kelvin waves in the presence of varying heat source. The first of the consecutive Kelvin waves is interacting with Rossby wave and therefore is associated with westerly wind burst, which extends between longitudes -87E and 3E. Each of the following Kelvin waves is associated with easterlies at the equator along the geopotential height minima and weak westerlies at each wave's edge. Therefore, the dynamic structure of the Kelvin wave sequence results in shift from westerlies to easterlies at the beginning of the wave and change from easterlies to westerlies at its end. Thus, the edge of each wave is associated with flow divergence and the end of the wave is characterized by flow convergence.

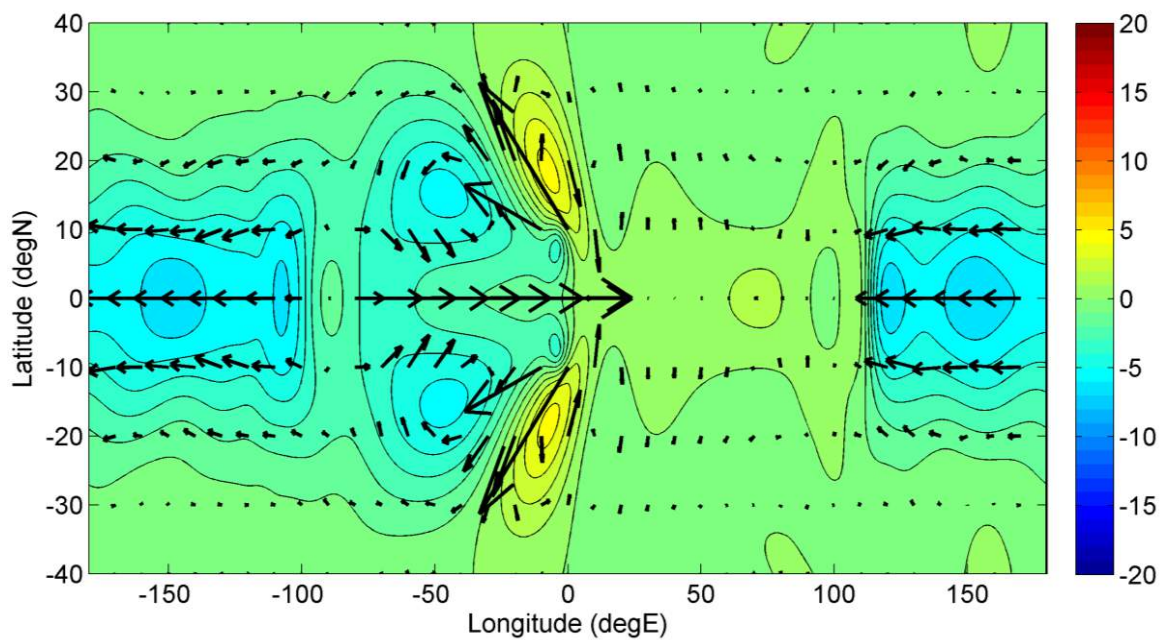

**Figure 3.3 Constant heat source experiment "s1" (control) at 7.75 days. The x-axis is longitude; the y-axis is latitude. Color shading indicates geopotential anomaly in [m] and arrows indicate wind anomaly in [ms$^{-1}$]. The heat source is located at [0N,0E] and has 10 degrees spatial extent. At 20N and 20S are positive geopotential anomalies associated with Rossby wave. Negative geopotential anomalies west of longitude -25E are related to Kelvin wave response which were initiated at the heat source location and circulated around the globe. Westerly wind burst can be seen between -50E and 0E close to equator and it is related to Kelvin wave and Rossby wave interaction. The (dry) Kelvin wave extends between 160E and -30E.**

The limited westerlies are associated with each of the consecutive Kelvin wave's edge, within the sequence, but the first one. They are visible at longitudes -115E, -140E and -175E and 145E. The first Kelvin wave within the sequence is already interacting with Rossby wave. Thus, westerlies associated with the first of the Kelvin waves are part of a broader area of the westerly burst. Between longitudes 50E and 100E, the symmetric structure consisting of two negative and two positive geopotential height anomalies is visible. It is the only area where the meridional flow at the equator is visible. Such geopotential height and flow characteristics are consistent with Easterly Inertial Gravity (EIG) wave.

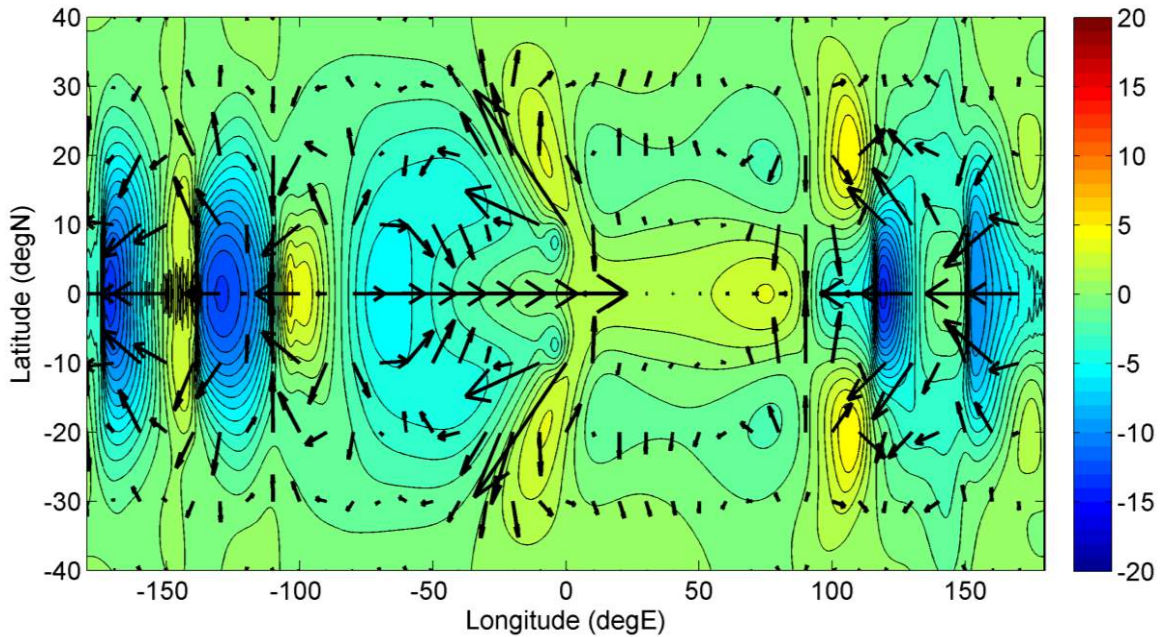

**Figure 3.4 Varying heat source with constant amplitude experiment "s2" at 7.75 days. The x-axis is longitude; the y-axis is latitude. Color shading indicates geopotential anomaly in [m] and arrows indicate wind anomaly in [ms$^{-1}$]. The heat source is located at [0N,0E] and has 10 degrees spatial extent. At 20N and 20S are positive geopotential anomalies associated with Rossby wave. Sequence of negative geopotential anomalies west of longitude -25E are related to Kelvin waves which were initiated at the heat source location and circulated around the globe. Strong westerly wind burst can be seen between -50E and 0E close to equator and it is related to Kelvin wave and Rossby wave interaction. Westerly bursts can also be seen at longitudes -110E, -150E and 150E, and are related to sequential Kelvin waves interaction. Between longitudes 100E and 150E the structure consisting of two negative and two positive geopotential anomalies is associated with Westerly Inertial Gravity (WIG) wave.**

It can be seen that the differences between "s2" (Figure 3.4) and "s3" (Figure 3.5) are relatively small if compared with the difference between these experiments and the control run. Experiment "s3" shows the structure of the Kelvin wave similar to one simulated in "s2" experiment. In Figure 3.5 the sequence of waves can be seen with the geopotential height minima located at longitudes -60E, -130E,-170E, 155E and 120E. All but the first wave are at the same locations as in the simulation "s2". Each of the consecutive waves is associated with divergence at its leading edge and convergence at its tail due to the shift in zonal wind. This is similar to behavior simulated in "s2" experiment. Between longitudes 80E and 100E the structure consisting of two negative and two positive geopotential height anomalies is associated with EIG wave, though these magnitudes are smaller than in "s2" simulation. The relative difference is attributed to relative difference in heat source magnitude between the two simulations. The first of the consecutive Kelvin waves is interacting with the Rossby wave, as in simulation "s2".This wave is located in broad area of strong westerly burst between longitudes -85E and 5E. Therefore, the observed structure at the location of the first of the consecutive Kelvin waves is in fact a superposition of Rossby and Kelvin waves. Although the locations of the Kelvin waves within the sequence are similar in both simulations, the differences in their magnitudes between the "s2" and "s3" experiments can be seen. In simulation "s3" the magnitude of diurnal heat source variability is higher for the days one and two, and lower for days four and five than in experiment "s2". Hence, the relative differences between geopotential height anomalies associated with individual Kelvin waves are likely due to differences in heat source magnitude. Figure 3.6 shows comparison of (a) geopotential height and (b) wind speed at the equator between "s1", "s2" and "s3" simulations. This analysis confirms that the negative magnitude of negative geopotential height anomaly associated with the second (-130E) and the third (-170E) Kelvin wave is larger in "s3" simulation than in "s2" simulation. On the other hand, the forth (155E) and fifth (120E) waves in Kelvin wave sequence have smaller negative geopotential height anomaly in simulation "s3" than in experiment "s2". Comparison of all 3 simulations shows that the differences in temporal distribution of the heat source do not affect Rossby wave response. The characteristics in the area between -90E and 5E are very similar in all simulations. The fact that westerlies begin a little further to the east in simulations "s2" and "s3", if compared with control run, may be attributed to the interaction with the first of the consecutive Kelvin waves. Because Kelvin waves are associated with easterlies, they can compensate westerlies associated with Rossby response at that location. On the other hand, the Kelvin wave response to the temporally disturbed heat source shows significant differences. Oscillating structure in both geopotential height and zonal wind speed is clearly associated with daily "pulses" of increased heat source.

Based on these 3 idealized simulations it has been shown that characteristics of the heat source may affect Kelvin waves characteristics and their spatial distribution, but have very limited impact on Rossby waves. In the case of variable source, the geopotential height anomaly, the wind anomaly and the wind convergence are stronger than in control simulation. The symmetric structure in the tail of the group of Kelvin waves is due to the existence of intertial-gravity waves in simulations with diurnally varying source; these waves were not present in simulation "s1" with constant forcing. This is consistent with the analysis presented in Figure 3b in [*Wheeler and Kiladis*, 1999] which shows that in symmetric OLR power spectrum for these frequencies range EIGs and Kelvin waves coexist.

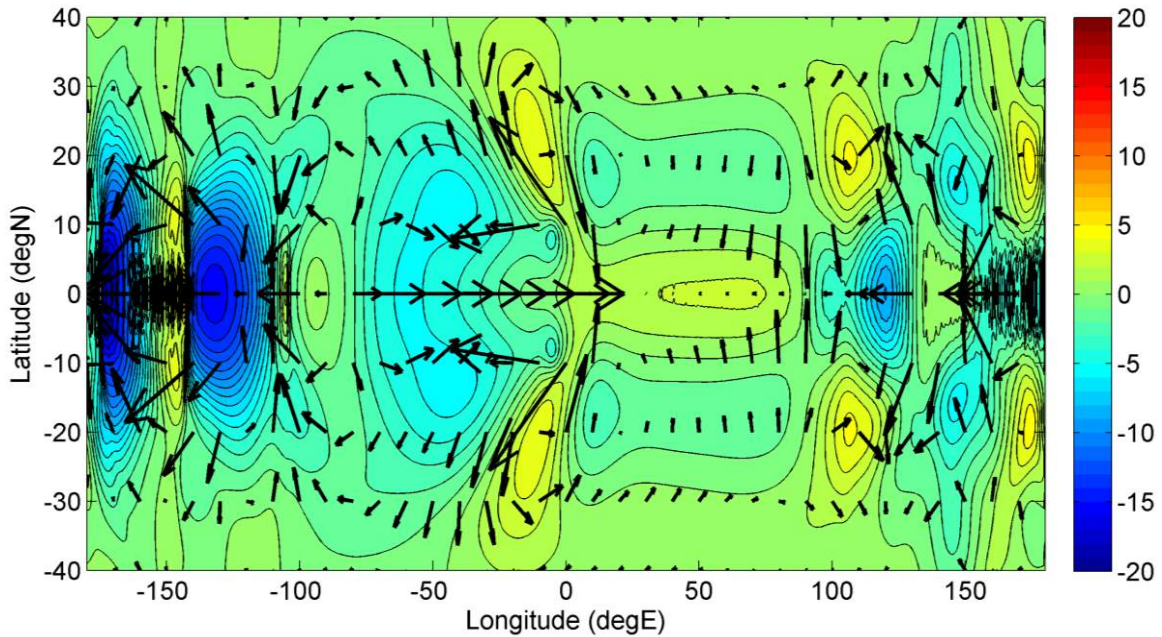

Figure 3.5 Varying heat source with varying amplitude experiment "s3" at 7.75 days. The x-axis is longitude; the y-axis is latitude. Color shading indicates geopotential anomaly in [m] and arrows indicate wind anomaly in [ms$^{-1}$]. The heat source is located at [0N,0E] and has 10 degrees spatial extent. At 20N and 20S are positive geopotential anomalies associated with Rossby wave. Sequence of negative geopotential anomalies west of longitude -25E are related to Kelvin waves which were initiated at the heat source location and circulated around the globe. Strong westerly wind burst can be seen at between [-50E,0E] close to equator and it is related to Kelvin wave and Rossby wave interaction. Westerly bursts can also be seen at longitudes -110E, -150E and 150E, and are related to sequential Kelvin waves interaction. Between longitudes 100E and 150E the structure consisting of two negative and two positive geopotential anomalies is associated with Westerly Inertial Gravity (WIG) wave as in Figure 3.4 but with smaller amplitude because varying source has smaller magnitude in this experiment.

Shallow water model does not have explicit representation of the physical processes responsible for the variability of the source, but it has been established in previous sections that diurnal cycle of the fluxes at the ocean surface may affect diurnal evolution of atmospheric convection. Therefore, the development of the upper ocean diurnal warm layer is one of plausible explanation for diurnal variability in heat source responsible for forcing a sequence of Kelvin waves. Thus, the upper ocean diurnal cycle may impact the structure and variability of Kelvin waves. This will be addressed by observational data analysis in the following section.

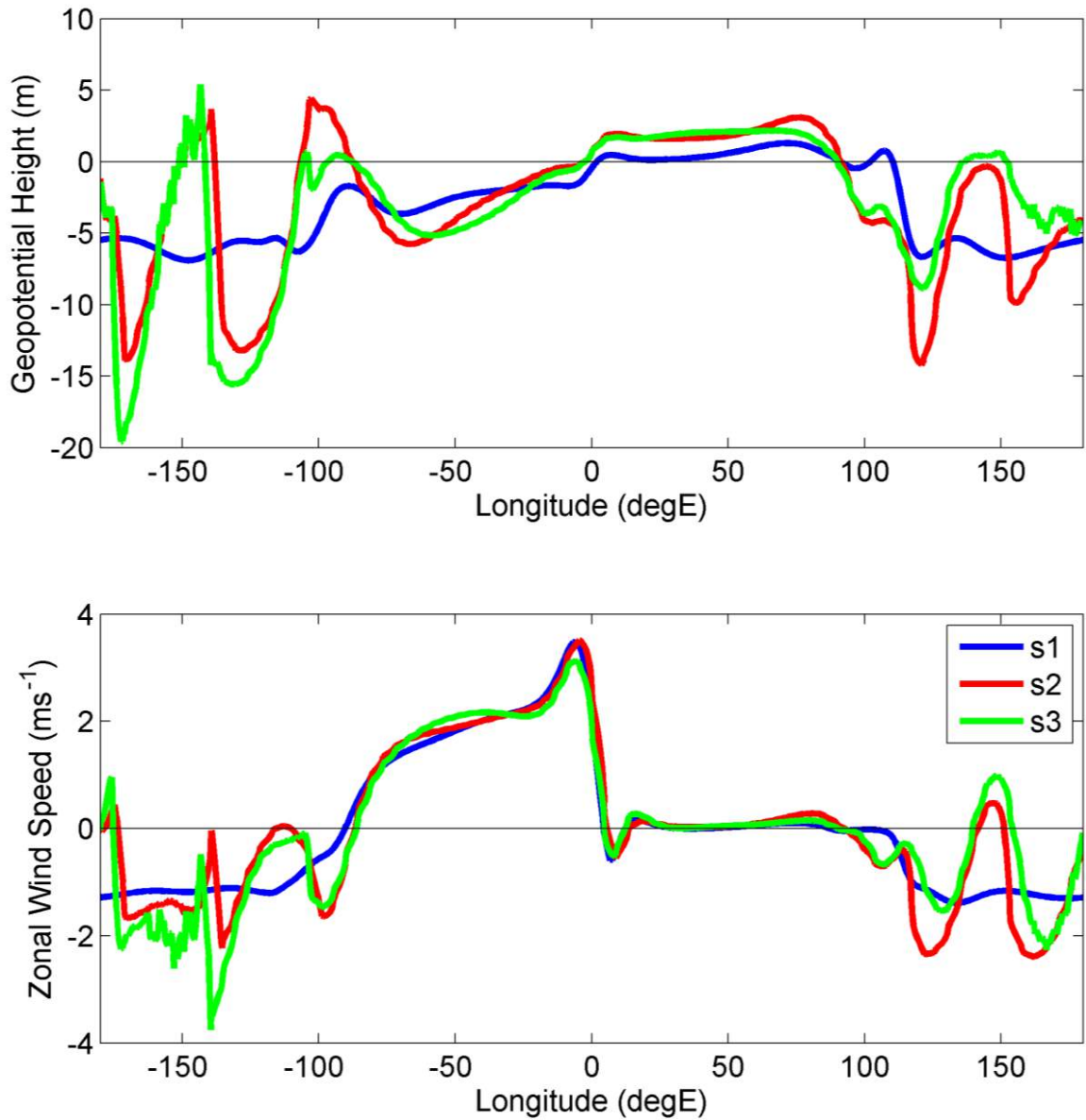

**Figure 3.6 (a)** Zonal section of geopotential height [m] along the equator for control simulation "s1" (blue), varying source simulation "s2" (red) and varying source with varying amplitude simulation "s3" (green). The x-axis is longitude in degree east; the y-axis is geopotential height in [m]. (b) is the same as (a) but for zonal wind speed in [ms$^{-1}$].

## 3.4. Mechanisms of sequential Kelvin waves initiations

We have shown that oscillating and localized source may initiate Kelvin waves from the same region within short period of each other. In this section we will compare these results with the observational data. Several cases of Kelvin wave initiation are selected and used to diagnose atmospheric and oceanic conditions at the ocean surface in order to understand these initiation mechanisms. Variability of the surface wind speed and latent heat flux and magnitude of the upper ocean temperature diurnal cycle and SST in the vicinity of the Kelvin wave initiation are of primary interest. The Kelvin wave trajectories analyzed here are developed using a special databases described in Appendix C. The ocean surface flux data and the diurnal SST variability analyzed in the vicinity to the Kelvin wave have been prepared using methodology described in Appendix D.

The Kelvin wave Eulerian and Lagrangian databases (desribed in Appendix C) allow analysis of individual Kelvin waves and statistics associated with occurrence of convectively coupled Kelvin waves. The primary interest of this Chapter is the Indian Ocean basin, therefore the analysis focuses on the characteristics relevant to Kelvin wave activity in that basin.

The two databases (Appendix C) contain 1948 Kelvin wave trajectories that have occurred between January 1998 and December 2012. Analysis of the areas of origin and end of individual Kelvin wave trajectories reveals that Indian Ocean basin has the highest number of trajectory initiation and that the Maritime Continent is one of the major dissipation areas. Out of the 1948 trajectories worldwide, 810 were active in the Indian Ocean basin defined as the area between 40E (east coast of Africa) and 100E (Sumatra), what makes it the region of the highest Kelvin wave activity observed in our data base. Figure 3.7 shows 3 maxima of Kelvin wave activity: over South America at -60E, Atlantic at 20E and the eastern Indian Ocean at 94E. The number of active Kelvin wave trajectories increases eastward through the Indian Ocean basin from 281 at 44E to 590 at 94E. South America, Atlantic Ocean, Maritime Continent and Western Pacific are the regions of large activity. However, the Maritime Continent and Western Pacific are characterized by gradual, eastward decay in Kelvin wave activity. Thus, many Kelvin waves active there originated in fact over the Indian Ocean. Globally, central and eastern Pacific areas are characterized by smallest number of active Kelvin waves. Figure 3.7a,b present the same data but divided into different categories based on (a) the season of a year and (b) frequency of reoccurrence.

Figure 3.7a shows that in comparison with other regions, Indian Ocean basin has the largest number of summer (JJA) and autumn (SON) Kelvin waves. These two seasons account for more than 50% of all Kelvin waves in that region. Other regions are dominated by winter (DJF) and spring (MAM) Kelvin waves. At the same time, number of Kelvin waves in DJF and MAM is relatively similar in the South America, Atlantic, Indian Ocean and Maritime Continent regions. Hence, most of cross basin differences in number of Kelvin waves is related to JJA and SON seasons.

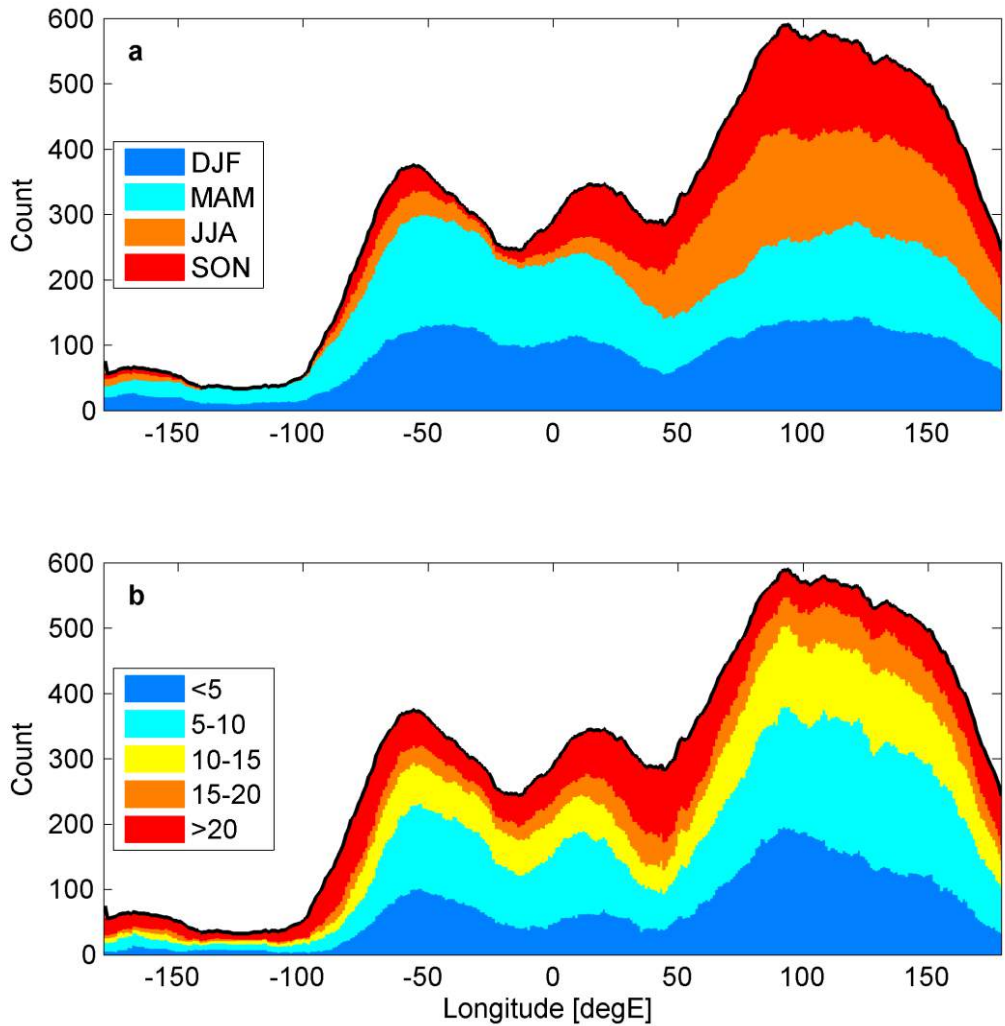

Figure 3.7 Number of active Kelvin waves for each longitude. The x-axis is longitude in degrees East, y-axis is Kelvin wave count. (a) Season of the year; blue is December, January and February; light blue is March, April and May; orange is June, July and August; red is September, October and November. (b) Time between two Kelvin waves: blue is for waves occurring within 5 days of each other; light blue is for waves reoccurring between 5 and 10 days after each other; yellow is for waves reoccurring between 10 and 15 days after each other; orange is for waves reoccurring between 15 and 20 days after each other; red is for waves reoccurring more than 20 days after preceding one.

In Figure 3.7b the longitudinal distribution of Kelvin waves trajectories depending of the frequency of occurrence is shown. Reoccurrence indicates the time which passed from preceding Kelvin wave passage. The blue color in this figure represents number of Kelvin wave trajectories occurring within 5 days after passage of previous Kelvin wave and red color represents number of those trajectories that were not preceded by any other Kelvin wave within 20 days. Light blue, yellow and orange colors represent Kelvin waves which reoccurred within 5 to 10 day, 10 to 15 days and 15 to 20 days, respectively. It can be seen that with increase in overall count of the active Kelvin waves through the Indian Ocean, fraction of the waves with the shortest reoccurrence time (blue and light blue) increases as well. A priori, there is no reason to expect that the areas of high Kelvin wave activity should have different reoccurrence fractions. We will associate this with inhomogeneities in local environment variability.

The fraction of Kelvin waves with reoccurrence time shorter than 5 days is the highest in the eastern Indian Ocean. Kelvin waves reoccurring within less than 10 days (blue and light blue) account in that area for more than 60% of all active Kelvin waves. It can be seen that the fraction of the Kelvin waves reoccurring within 10-20 days is the same globally. There is small number of waves reoccurring with longer periods. Therefore, the majority of the cross basin differences in Kelvin wave activity are associated with Kelvin waves which come quickly after each other and this is exceptionally true in the Indian Ocean basin. The fact that there are many Kelvin waves in that basin and that they come relatively fast after each other indicates that local environment favors these waves and that it may be modified by these waves contributing to the air-sea interaction.

The Indian Ocean is characterized by high Kelvin wave activity and large number of waves propagating, within short time period, between each other. Such waves travel through the basin in several days. The analysis of initiation of such waves is divided into two categories: multiple initiation and spin-off initiation, both of which may depend on interactions between waves and surrounding environment.

### 3.4.1. Multiple initiation

Multiple initiations occur when two or more Kelvin wave trajectories begin in the same area within few days from each other. In this study we define the thresholds to be 5 days and 10 degrees for temporal and spatial proximity respectively. It means that if two Kelvin waves are initiated within 5 days and distance between their areas of origin is less or equal to 10 degrees, they both fit into this category. Sensitivity study has been performed on the threshold definitions and the values were found to be optimal. Although our multiple initiation criteria may seem too relaxed, one should note that even in our numerical experiments the heat source was 10 degrees wide.

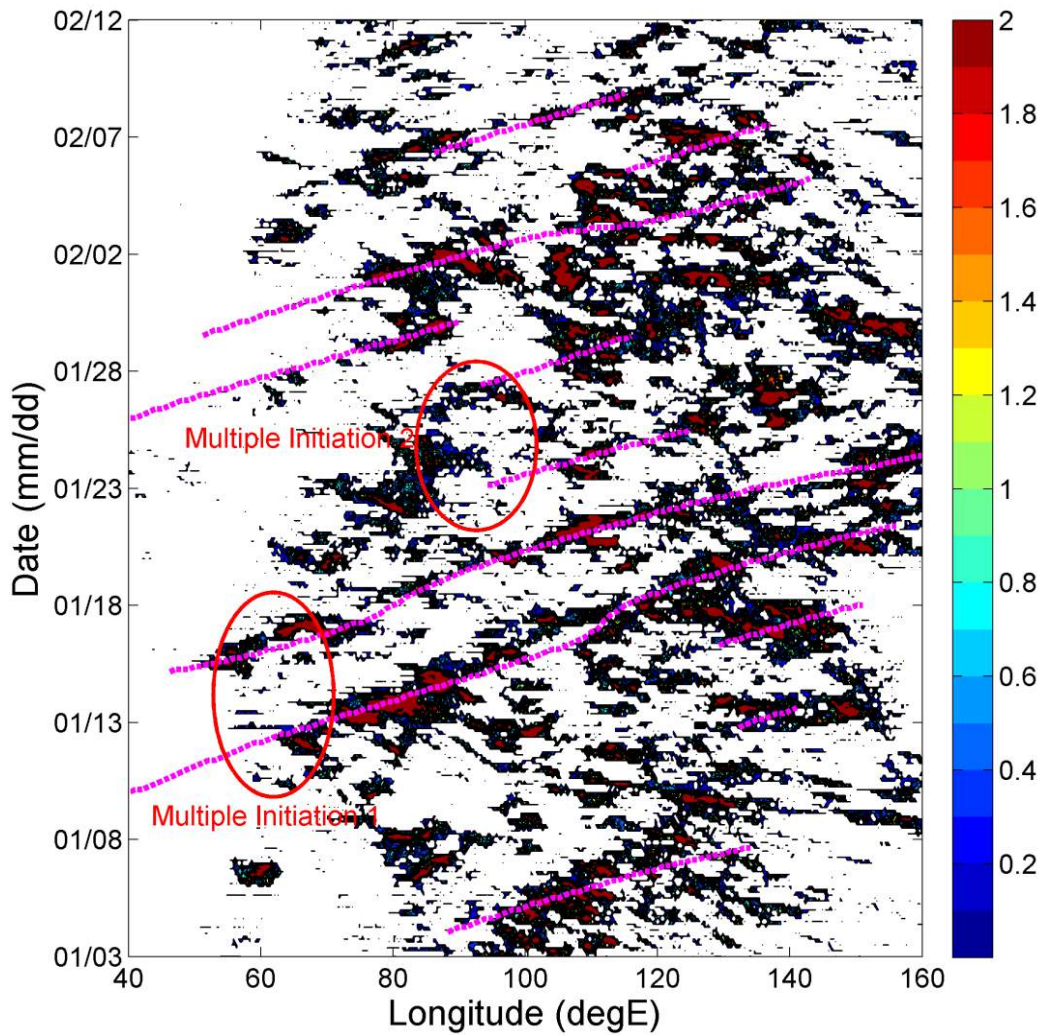

**Figure 3.8 TRMM precipitation in [mmhr$^{-1}$] during the period between January3 and February 12, 2000. The x-axis is longitude in [degE]; the y-axis is time in [mm/dd]. Magenta lines indicate Kelvin wave trajectories.**

An example of multiple initiations is presented in Figures 3.8 – 2.13. In January 2000, a few Kelvin waves were initiated over western Indian Ocean (around 60E). Although some trajectories (marked with magenta lines) began west of that location, the full precipitation associated with those disturbances began over that area. As an example, the precipitation in 2 Kelvin waves was initiated around 60E in mid January. The precipitation associated with the first Kelvin wave was initiated on January 12 at 63E and the precipitation in the second Kelvin wave began on January 15 at 54E. Figure 3.10 shows that the area between 60E and 80E during period between January 5, 2000 and February 5, 2000 was characterized by relatively weak wind speeds in the first half of that period and slightly higher wind speeds in the second half of that period. Variability in wind speeds is reflected in the variability of the latent heat flux at the ocean surface, which was small in the first half of January and slightly higher in the latter one (Figure 3.11). The SST was in the 28.5-29.0 °C range, which is high but smaller than SST over eastern Indian Ocean during the same period (Figure 3.12). This time period is also characterized by high magnitudes of diurnal SST variability with large number of days for which daily mean SST anomaly at the ocean surface exceeded 0.35 °C (Figure 3.9). The location of the initialization of the precipitation associated with both identified Kelvin waves agrees well with time and location of the area of strong upper ocean temperature diurnal cycle and weak latent heat flux. This suggests that diurnally increased surface fluxes due to development of diurnal warm layer at the ocean surface might have contributed to a subsequent development of an atmospheric convection, which formed two individual Kelvin waves. Such mechanism agrees very well with hypothesized mechanism presented in numerical experiments in previous section.

Period selected for this case study shows 2 more examples of interactions between variability of the fluxes at the ocean surface and development of atmospheric convection and subsequent precipitation. Analysis of the trajectories database and full precipitation shows two Kelvin waves (different than summarized above) that initiate within few days around January 23 at 90E. At that time and location wind speed is increased and thus higher values of latent heat flux at the ocean surface are present. The diurnal SST variability is not strong in the area identified as Kelvin waves initiation. However, the analysis of full precipitation shows that area directly west of the beginning of the trajectories is associated with strong precipitation. This suggests that these two Kelvin waves may have been initiated west of calculated beginning of their trajectories. Increased values of the magnitude of the diurnal SST variability are observed around 70E, where convective precipitation began. This would again agree well with hypothesized mechanism presented in numerical experiments.

In summary, we considered two cases of multiple initiations. The results show that one pair of multiple Kelvin waves was initiated in the area of increased SST variability and relatively weak wind speed and latent heat flux forcing. The other pair of multiple Kelvin waves was initiated in the area where the SST variability, wind speed and latent heat flux were increased. These results suggest that both the upper ocean temperature diurnal variability and ocean surface fluxes variability shall be considered as important factors for formation of convection and precipitation, especially when it is organized in multiple Kelvin waves. This is consistent with shallow water results that the diurnally varying heat source may trigger multiple Kelvin waves. Thus, it is possible that the diurnal SST variability is related to the varying heat source in the model experiments. Such link between atmospheric convection and upper ocean diurnal variability was noted previously [*Chen et al.*; *Ruppert and Johnson*]. In the subsequent section the statistics of the multiple Kelvin waves initiation will be further investigated.

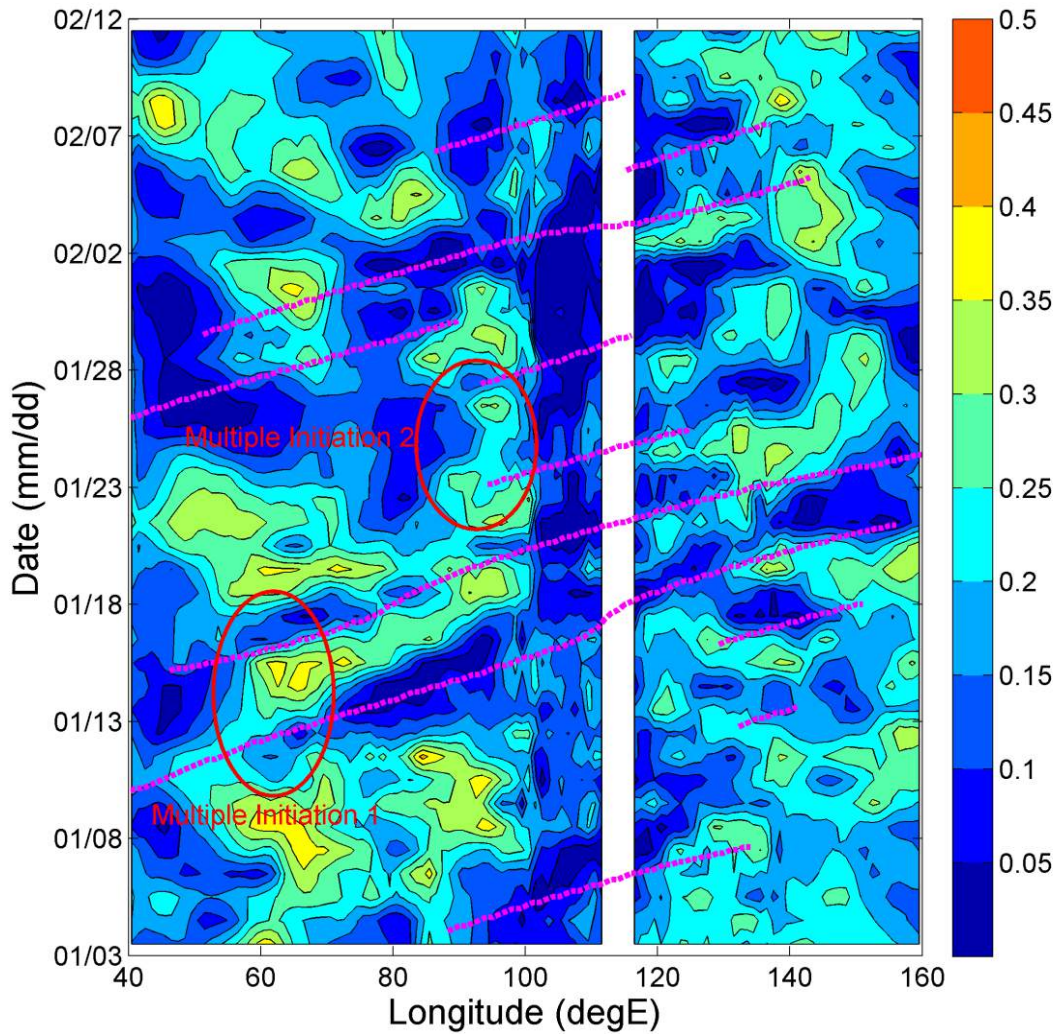

**Figure 3.9 The derived magnitude of the upper ocean temperature diurnal cycle in [°C] during the period between January 3 and February 12, 2000. The x-axis is longitude in [degE]; the y-axis is time in [mm/dd]. Magenta lines indicate Kelvin wave trajectories.**

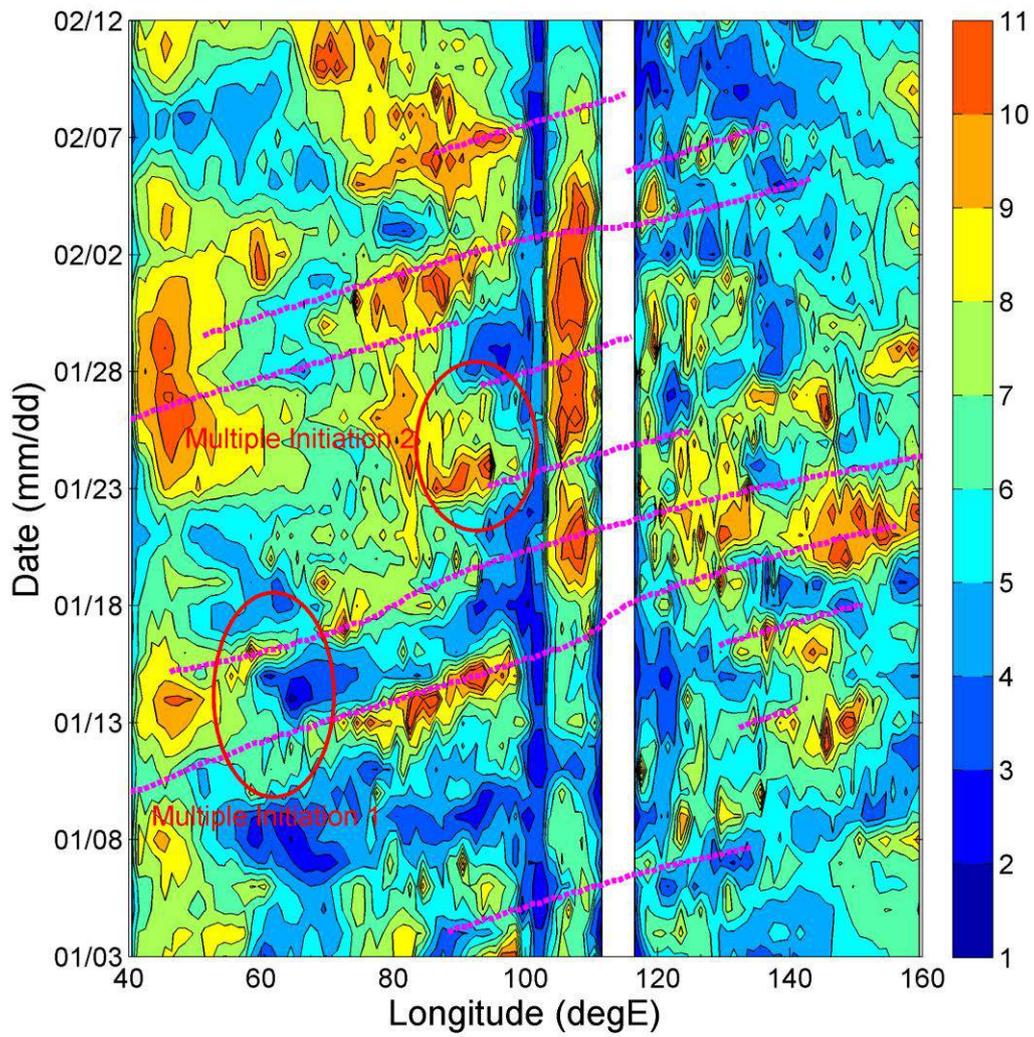

Figure 3.10 The ocean surface wind speed in [ms$^{-1}$] during the period between January 3 and February 12, 2000. The x-axis is longitude in [degE]; the y-axis is time in [mm/dd]. Magenta lines indicate Kelvin wave trajectories.

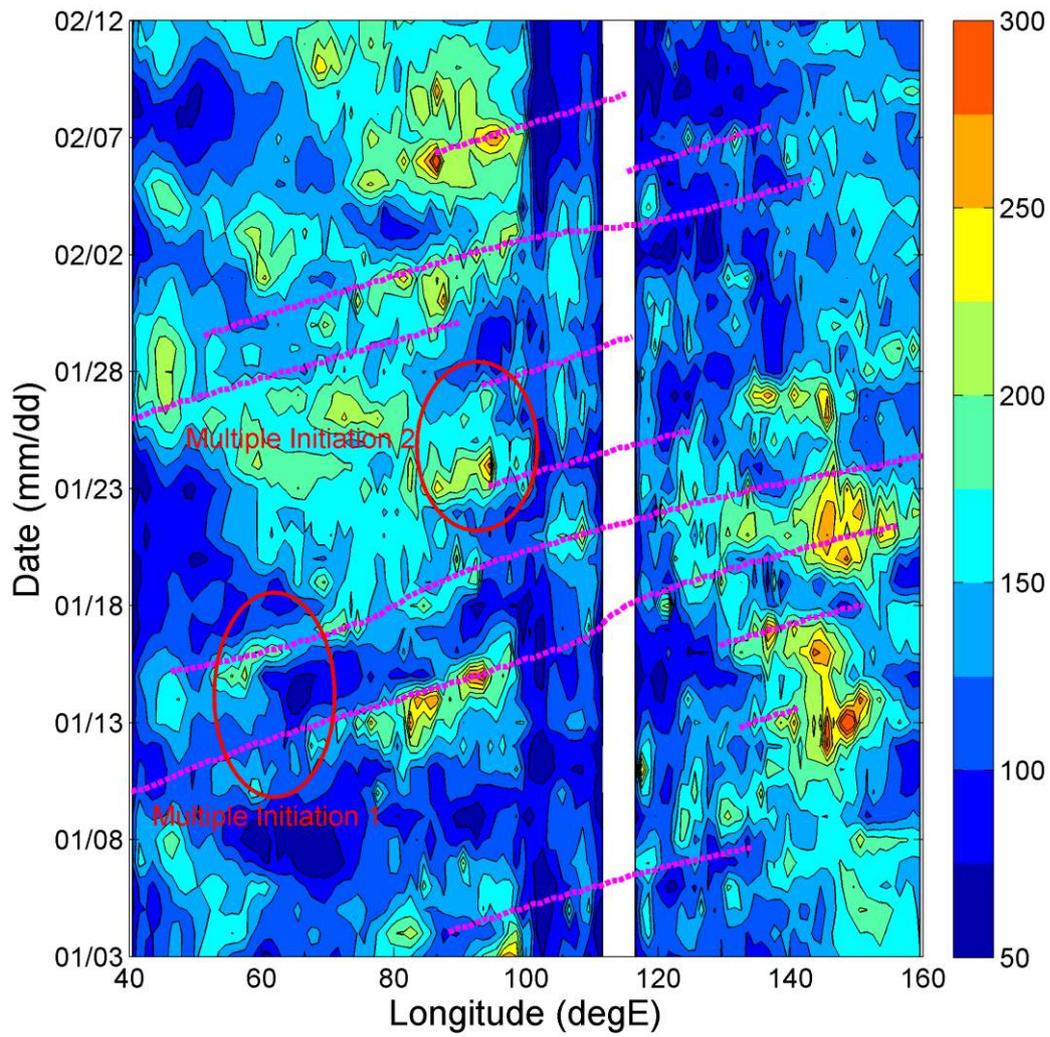

**Figure 3.11 The ocean surface latent heat flux in [Wm$^{-2}$] during the period between January 3 and February 12, 2000. The x-axis is longitude in [degE]; the y-axis is time in [mm/dd]. Magenta lines indicate Kelvin wave trajectories.**

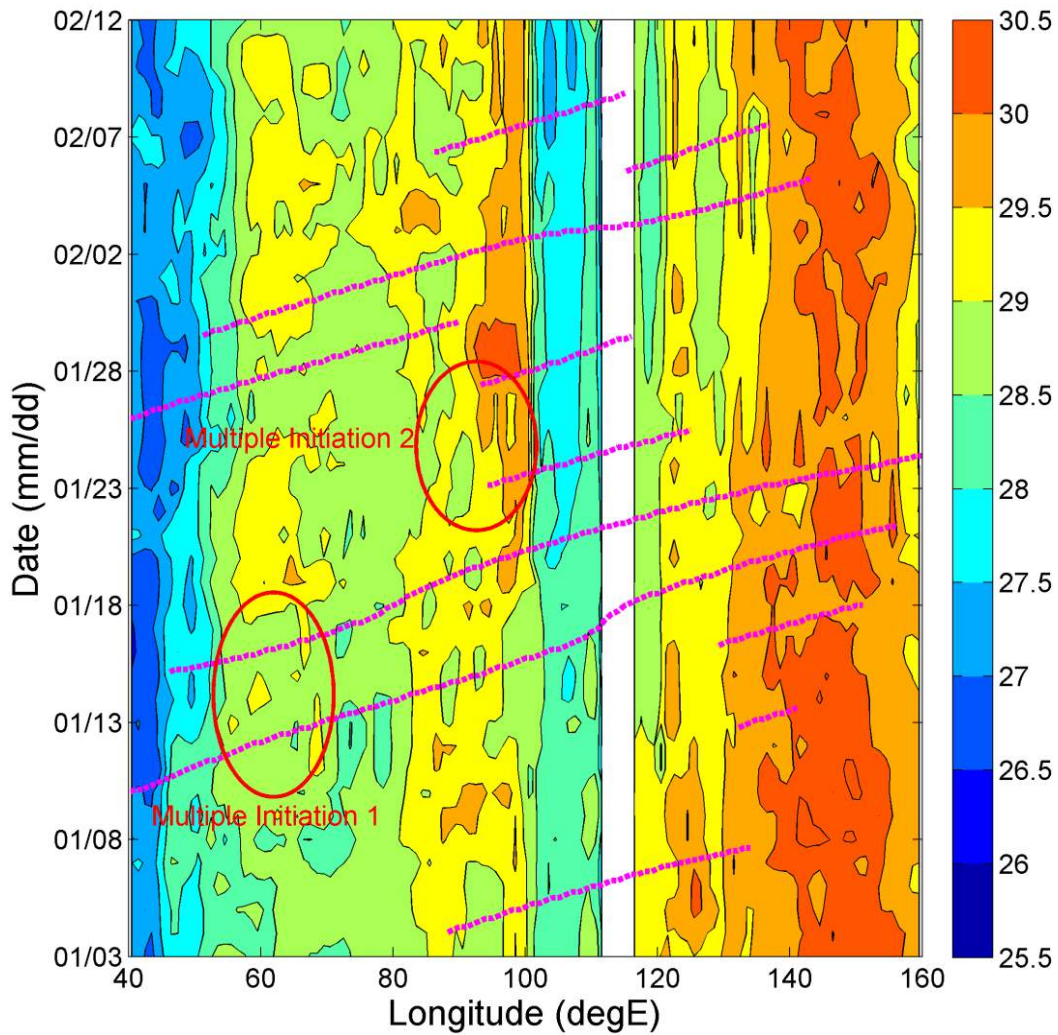

**Figure 3.12** The SST in [°C] during the period between January 3 and February 12, 2000. The x-axis is longitude in [degE]; the y-axis is time in [mm/dd]. Magenta lines indicate Kelvin wave trajectories.

### 3.4.2. Thermodynamic spin off initiation

The thermodynamic spin off initialization takes place when sequential Kelvin waves are forced in temporal proximity to the preceding Kelvin wave trajectory by increased SST caused by strong upper ocean temperature diurnal cycle in the wake of the preceding wave. This ideas is somewhat similar to "diurnal dancing" [*Chen and Houze*, 1997]. Comparing to the previously described category, spatial proximity constrain is relaxed so that the two or more disturbances do not have to initiate in the same area, but the latter one has to begin in location where the first one was active and therefore might have preconditioned the environment or even induced development of the sequential disturbance. Hence the latter Kelvin wave is a spin off from the preceding one. Based on sensitivity studies and to keep constrains comparable between categories, the temporal proximity parameter has been set to 5 days. In order to be able to distinguish between the two categories, initiation of the sequential wave has to take place more than 5 degrees east from the area of origin of the preceding wave. We decided to keep temporal constrain set to 5 days, because beyond that period there is no reason to expect that the diurnal cycle or any other forcing would be induced by the preceding wave. Additionally predominant period of Kelvin wave occurrence is five days, so the clear phase preceding convective phase should be within five day window.

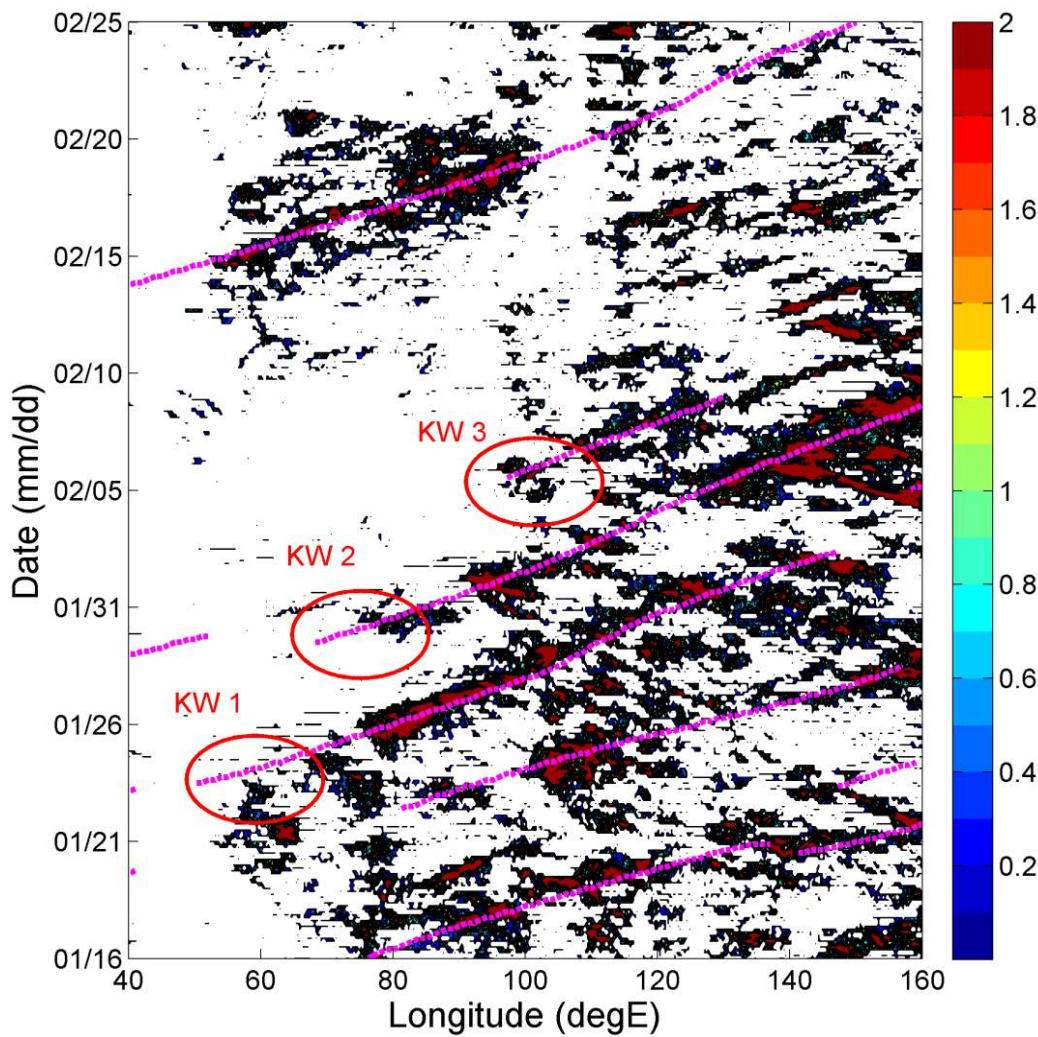

**Figure 3.13 TRMM precipitation in [mmhr$^{-1}$] during the period between January 16 and February 25, 2004. The x-axis is longitude in [degE]; the y-axis is time in [mm/dd]. Magenta lines indicate Kelvin wave trajectories.**

An example of spin off waves is presented in Figures 3.13 – 3.17. Presented are Kelvin waves active in Indian Ocean basin in second half of January and February, 2004. During the period between January 23 and February 4, 3 Kelvin waves initiated between longitudes 51E and 98E. Each of the 3 consecutive Kelvin Waves initiated further east. The first Kelvin wave initiated on January 23 at 51E and passed over location at 70E on January 25. 4.5 days later the second Kelvin wave initiated at 70E. This disturbance passed over location at 98E on February 1. 4 days later the third Kelvin wave originated from this location. Hence, distance between the first and the second or between the second and the third of the consecutive Kelvin waves was bigger than 5 degrees and time between them was less then 5 days which makes latter two of them candidates for a thermodynamic spin off initialization.

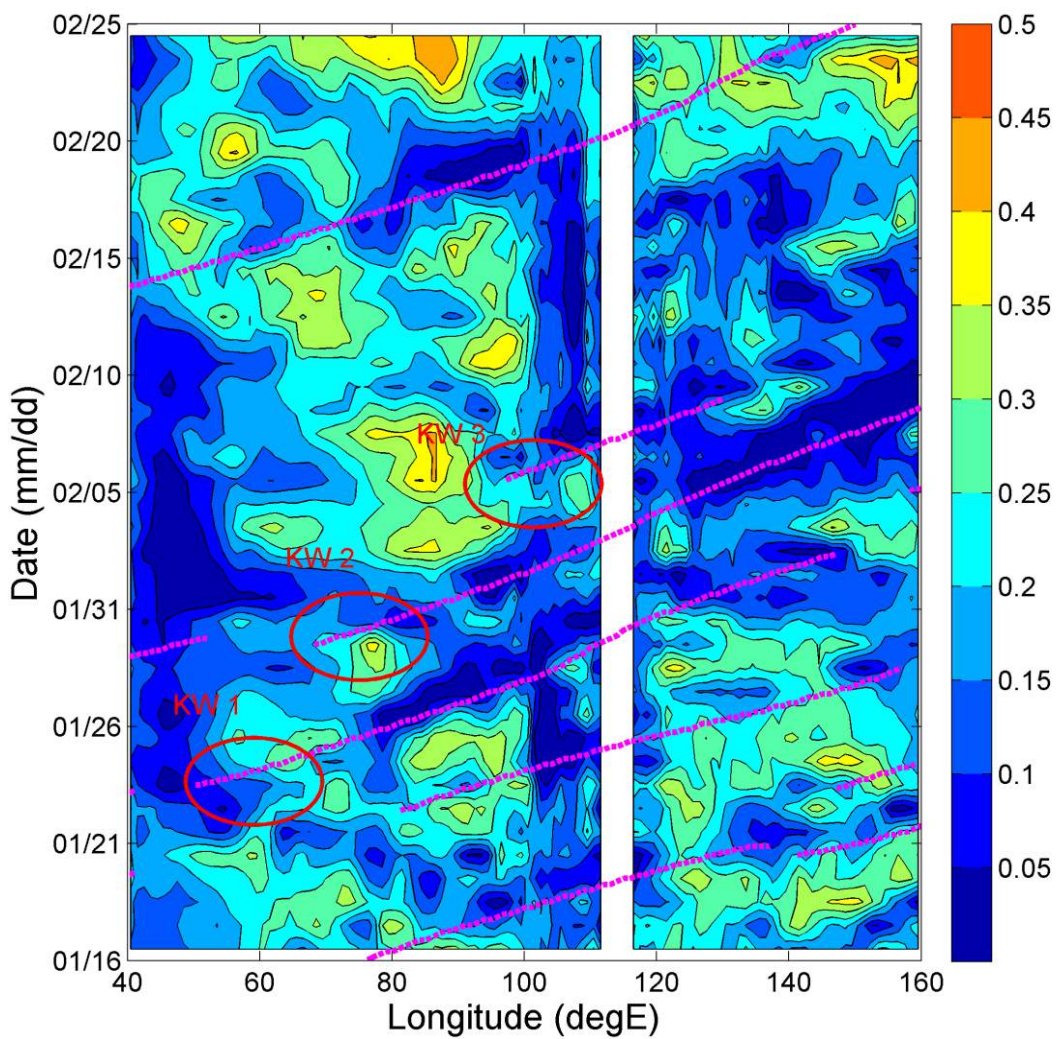

**Figure 3.14 The derived magnitude of the upper ocean temperature diurnal cycle in [°C] during the period between January 16 and February 25, 2004. The x-axis is longitude in [degE]; the y-axis is time in [mm/dd]. Magenta lines indicate Kelvin wave trajectories.**

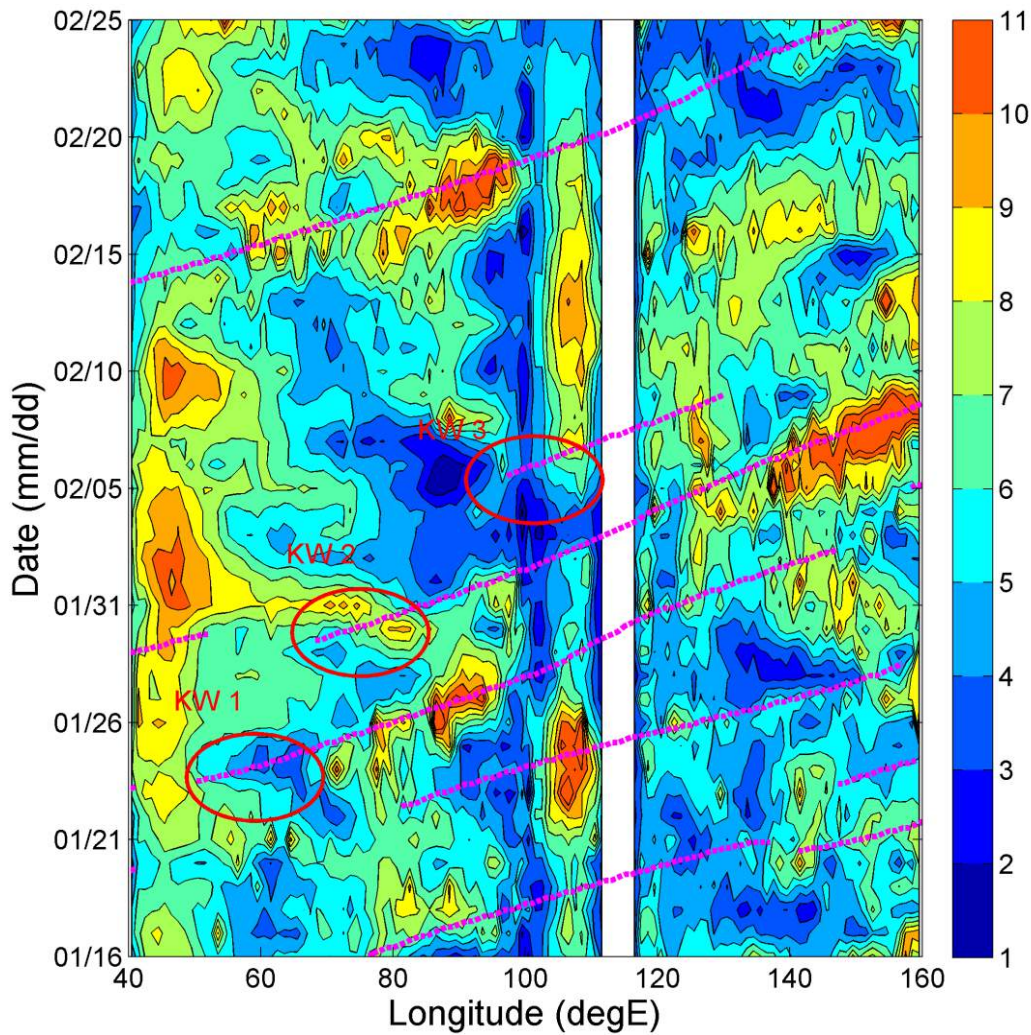

**Figure 3.15 The ocean surface wind speed in [ms$^{-1}$] during the period between January 16 and February 25, 2004. The x-axis is longitude in [degE]; the y-axis is time in [mm/dd]. Magenta lines indicate Kelvin wave trajectories.**

Environmental conditions along the three consecutive Kelvin wave trajectories resemble features linked to typical interactions between Kelvin wave and the surrounding environment which will be discussed in detain in the following Chapter. In Figure 3.14 the areas of interest are marked with red ellipses and individual Kelvin wave trajectories are marked with dashed, magenta line. In the wake of the "KW 1" wave at the longitude 71E, the diurnal SST variability increases; this is represented by green and yellow colors. Similarly, in the wake of KW2 at the longitude 98E the upper ocean temperature diurnal cycle is increased. Location, timing and persistence of the increased values of the diurnal SST variability agrees well with areas and time of the origin of the second ("KW 2") and the third ("KW 3") of the 3 consecutive Kelvin waves. Initiation of the second of the waves, which took place on January 29 at 70E occurred in the area, where the diurnal SST variability, wind speed, latent heat flux at the ocean surface were increased. Moreover, increase in wind speed and latent heat flux at this time and location seems to be in a form of a slow, westward propagating disturbance. The origin of this disturbance may be tracked back to the strong wind speed and latent heat flux anomalies at 95E on January 27, associated with the "KW 1".

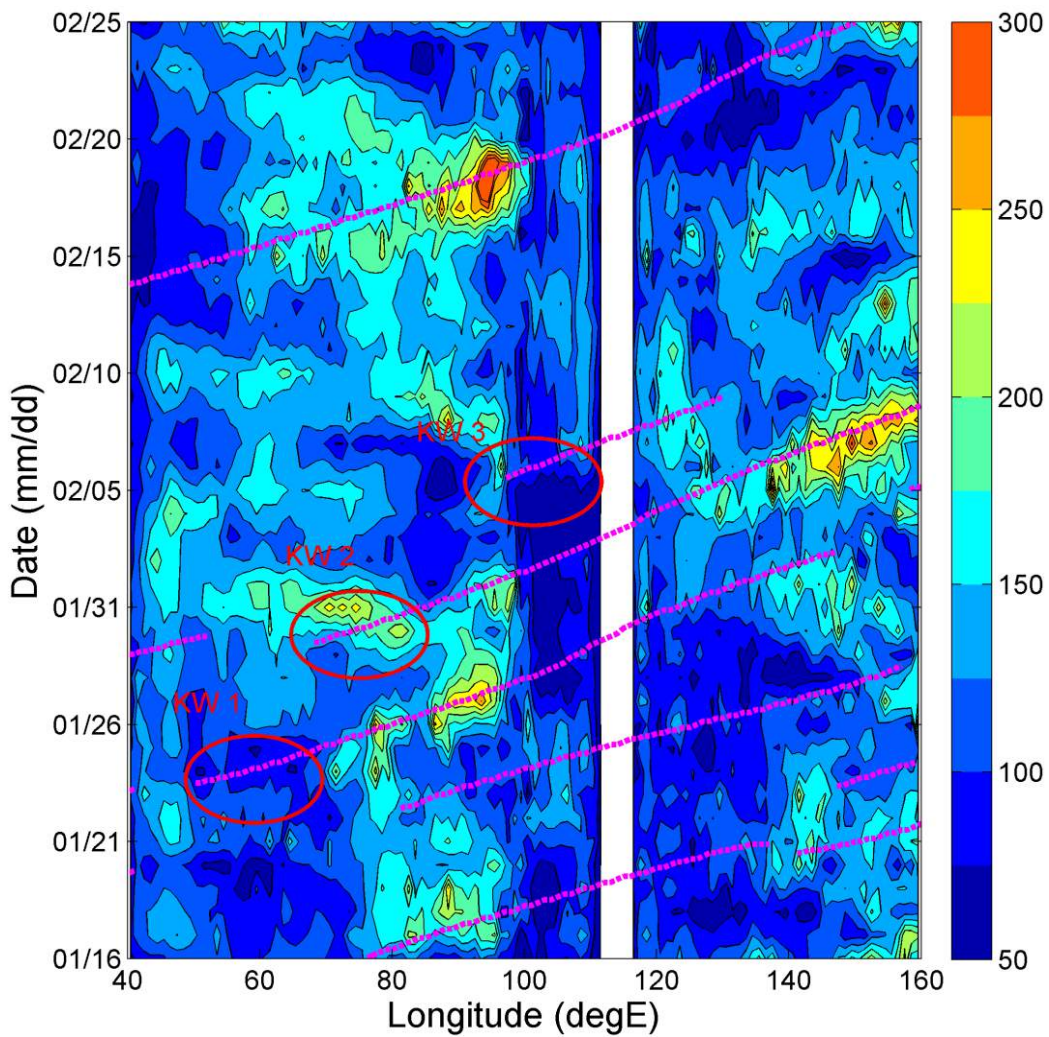

**Figure 3.16 The ocean surface latent heat flux in [Wm$^{-2}$] during the period between January 16 and February 25, 2004. The x-axis is longitude in [degE]; the y-axis is time in [mm/dd]. Magenta lines indicate Kelvin wave trajectories.**

These interactions are visible in Figure 3.15 and Figure 3.16 which present wind speed and latent heat flux at the ocean surface for investigated period. The area of origin of the second ("KW 2") of the consecutive waves was affected by local thermodynamic (increased SST variability) anomaly and propagating dynamic anomaly (increased wind speed and latent heat flux). Both of these anomalies may have been induced by the "KW 1" and support forcing of the "KW 2". Local thermodynamic forcing in the wake of a moving convective disturbance is likely associated with the area of subsidence, surrounding the area of deep convection [*Kiladis et al.*, 2009]. When convective part of a Kelvin Wave leaves the area, it is followed by its non convective part, which favors mild environmental conditions with high insolation and weak wind speeds at the ocean surface. Such environmental conditions favor development of strong diurnal cycle of the temperature in the upper ocean.

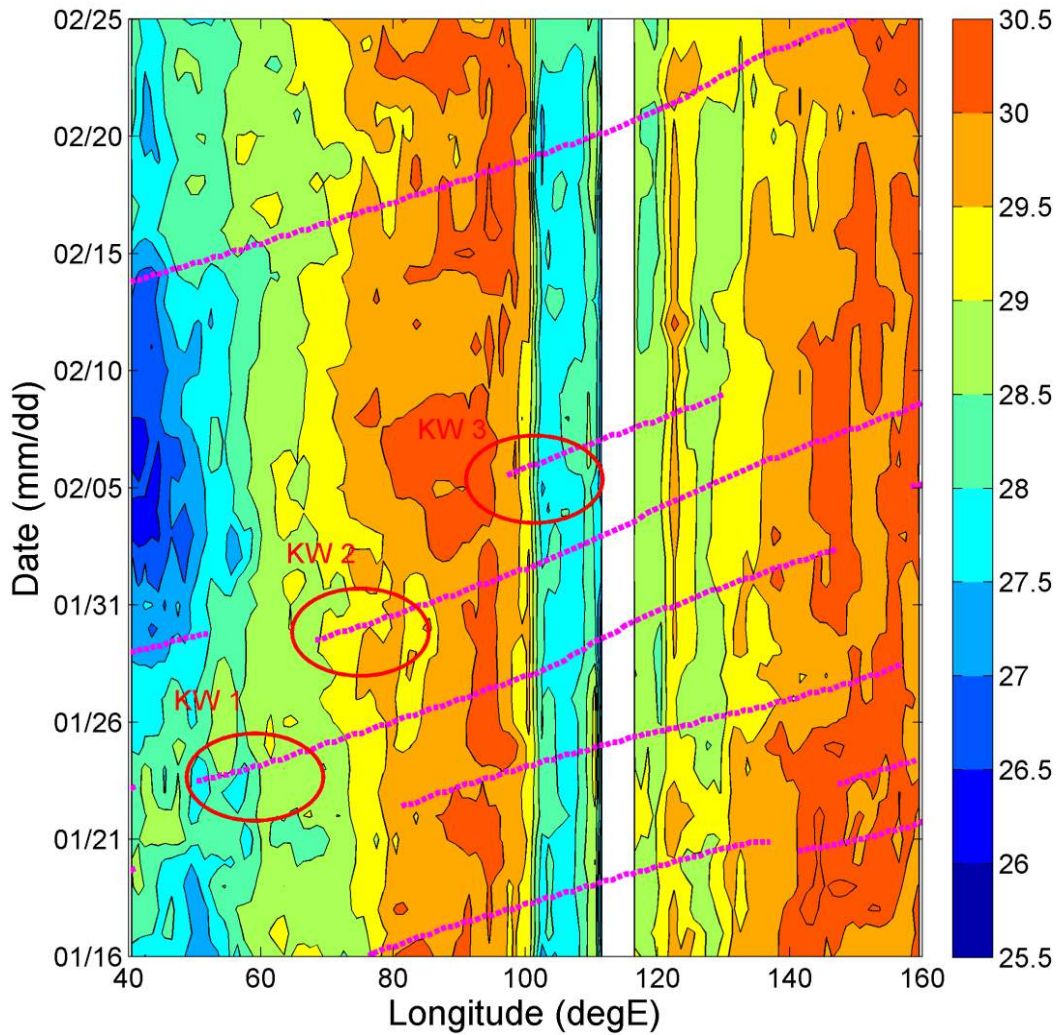

**Figure 3.17** The SST in [°C] during the period between January 16 and February 25, 2004. The x-axis is longitude in [degE]; the y-axis is time in [mm/dd]. Magenta lines indicate Kelvin wave trajectories.

Based on numerical experiments, in the previous section we have shown that a Kelvin wave activity is often associated with formation of an off equatorial circulation in form of a Rossby wave. We have shown that such disturbance propagates westward and is associated with so called westerly wind burst. Although circulation and convection associated with Rossby wave is mostly off equatorial, and therefore cannot be tracked using database used in this study, the signature of its existence can persist in wind speed and latent heat flux signals within equatorial belt. Westward propagation prior to the formation of the "KW 2" of the three consecutive Kelvin waves suggests that this may be the case here. Hence, it is likely that formation of the second the three consecutive Kelvin waves has been supported by both increased diurnal cycle in the upper ocean and increased surface fluxes at the ocean surface, both of which may have been likely induced by the "KW1".

In the case of the "KW 3", the analysis of the fluxes and their variability in the area of the initiation shows that wind speed and latent heat flux were small around 98E at the time of the initiation. However, the magnitude of the diurnal SST variability was large just before the formation of "KW 3". It can be seen in Figure 3.14 that the increased magnitude of the diurnal SST variability at the area of the initiation of the third Kelvin wave agrees with the increased diurnal SST variability propagating eastward in the wake of the "KW 2" between longitudes 80E and 100E. In this case there is no indication of the dynamic interaction with the preceding Kelvin wave, because wind speed and latent heat flux are relatively small. However, there is strong evidence that area of increased diurnal SST variability has been induced by the "KW 2". Hence, local thermodynamic forcing, which may have induced by preceding wave, was likely to contribute to the development of the "KW 3".

During the CINDY2011/DYNAMO campaign, a research version of the fully air-ocean-wave coupled COAMPS provided a real-time forecast support for the field campaign. COAMPS used a 6 hourly data assimilation cycle in the atmosphere and ocean from 17 Aug, 2011 to 15 Jan, 2012 and issued a 4-day forecast once a day at 1200 UTC. The atmospheric data assimilation system in COAMPS is Navy Atmospheric Variational Data Assimilation System (NAVDAS) [*Daley and Barker*, 2001] and the ocean data assimilation system is the NAVY Coupled Ocean Assimilation (NCODA) [*Cummings*, 2005]. The CINDY2011/DYNAMO soundings, Airborne/Air Expendable Bathythermograph and dropsondes deployed from NOAA P3 aircraft were assimilated into COAMPS as well as many other routinely available in-situ and remotely sensed atmospheric observations. In particular, the total precipitable water is assimilated in the atmospheric component. For the ocean data assimilation, satellite sea surface temperature, in-situ profiles, drifters, buoys, ships, AXBT, and Modular Ocean Data Assimilation System (MODAS) synthetic profile are assimilated in NCODA. The initial and boundary conditions are derived from the 0.5° Navy Global Atmospheric Prediction System (NOGAPS) [*Hogan and Rosmond*, 1991] and 1/8° global NCOM forecasts.

Figure 3.18 presents an analysis for the November 27, 2011. The spin off initiation appeared to be responsible for the initiation of the second Kelvin wave observed during the November 2011 MJO. The second Kelvin wave during this episode was triggered by the westerly burst related to the tropical cyclone TC05 in the northern hemisphere and the southern hemisphere cyclonic circulation on November 27. These cyclonic circulations developed in relation to the first Kelvin wave which passed over longitude 75E on November 24, and as seen in the Figure 3.18, by November 27 already approached Maritime Continent. This analysis provides yet another example of spin off initiation consistent with our case study.

This analysis is an alternative to the explanation provided by [*Kerns and Chen*, 2013] who considered both waves to be the part of one system separated by dry air intrusion related to the Rossby gyres.

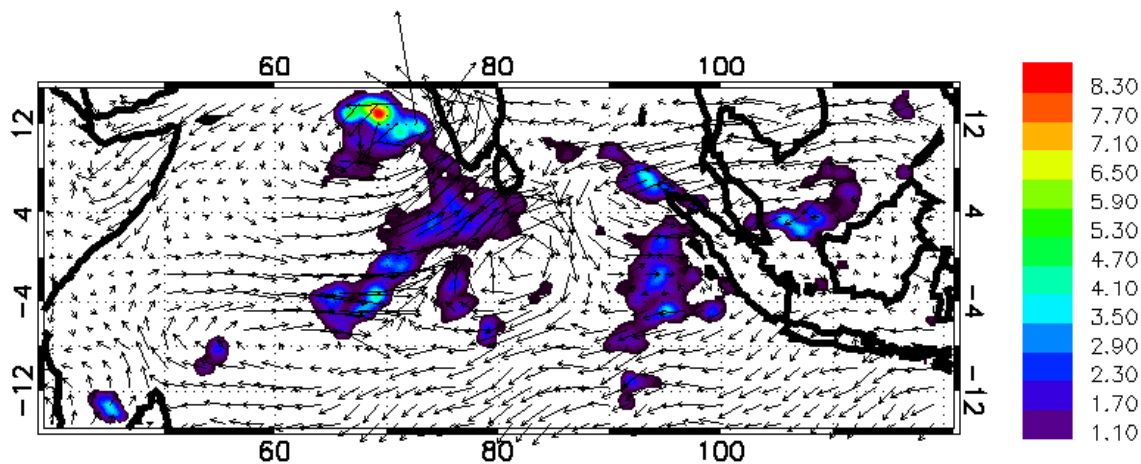

**Figure 3.18** **The example of the spin off initiation as seen in analysis of mesoscale atmosphere –ocean coupled model (COAMPS): COAMPS 850mb winds and TRMM daily precipitation analysis for November 27, 2011. The x-axis is longitude [degE]; the y-axis is latitude [degN]. The enhanced precipitation in the second Kelvin wave develops in the region of strong westerly wind related to cyclonic circulations associated with the first wave.**

In summary, we have shown that the propagating Kelvin wave may trigger another Kelvin wave which we call spin off process. The preceding ("spinning") wave may influence the environment in its proximity and the sequential ("spin off") wave may develop. The analyzed case studies show that the interaction between preceding wave and the environment where spin off wave develops may rely on both the dynamic and local thermodynamic forcings. The dynamic forcing includes propagation of area of the increased wind speed and latent heat fluxes likely related to the Rossby wave. The local thermodynamic forcing is due to increased diurnal SST variability in the wake of the preceding Kelvin wave. We have shown two examples of spin off initiation. The first case indicates that both dynamic and local thermodynamic forcing may have contributed to the spin off Kelvin wave development. The second case suggests that the local thermodynamic forcing was the major trigger of the spin off Kelvin wave. In the subsequent section, based on composite analysis, we will show that such spin off initiations are common and indeed rely on both dynamic and local thermodynamic forcing.

## 3.5. Statistics of Kelvin wave initiations

In previous chapter we have shown that Kelvin wave activity is different between various regions. The zonal variability is strongest for the Kelvin waves which occur within short time period after each other. Such waves have the highest chance to be part of either multiple initiation or spin off initiation category. In this section statistics of the Kelvin wave initiation are investigated. We use Kelvin wave trajectory database, described in Appendix C, to detect all trajectories that satisfy both categories. Our case studies revealed that some trajectories are over extended if compared with full precipitation along them. Therefore here the Kelvin wave initiation is defined as the first precipitation onset along trajectory. The precipitation onset has to satisfy two threshold criteria: 3hourly precipitation has to exceed 0.5 mmhr$^{-1}$ and 24h running mean precipitation along the trajectory has to exceed 0.25 mmhr$^{-1}$. Such Kelvin initiation definition guarantees that initiation is associated with the precipitation. We investigate two initiation categories: multiple initiation and spin off initiation. Multiple initiations occur when two or more Kelvin wave trajectories initiate in the same area within the short period of time. As an example if one Kelvin wave initiated at 60E at day 0 and the other Kelvin wave began at 63E four days later, they both would be considered as multiple initiations. Spin off initiations occur when Kelvin wave initiates in the area over which another Kelvin wave propagated few days earlier. As an example if a Kelvin wave initiates at 87E at day 0, and over that location another Kelvin wave propagated three days earlier, such initiation would be considered spin off initiation. Multiple and spin off initiation categories are disjunctive in a sense that one Kelvin wave initiation can not be considered multiple and spin off at the same time.

Globally there were 362 Kelvin wave trajectories that satisfy multiple initiation criteria and 277 Kelvin waves that matched spin off initiation criteria. This means that about 20% of all Kelvin waves are multiple initiation and about 15% of all Kelvin waves are spin off initiation trajectories. Indian Ocean basin has the largest number of frequently propagating Kelvin waves. The total number of multiple Kelvin wave initiation cases for Indian Ocean basin is 126. That means that about 21% of all Kelvin waves initiated over Indian Ocean matched multiple initiation criteria. The total number of spin off Kelvin wave initiation cases for Indian Ocean basin is 106. That means that about 18% of all Kelvin waves initiated over Indian Ocean matched multiple initiation criteria. Together multiple initiations and spin off initiation account for almost 40% of all Kelvin wave initiations in the Indian Ocean basin.

Figure 3.19 presents zonal distribution of Kelvin wave initiations. All numbers represent total number of Kelvin wave initiations which occurred within 10 degrees wide zonal box. Blue line indicates all Kelvin wave initiations, red line presents multiple initiations and green line represents numbers for spin off initiation. It can be observed that both multiple initiation and spin off initiation categories follow the zonal distribution of all Kelvin waves initiations. However, the maximum number of spin off initiations tends to occur further to the east than maximum number of multiple initiations. For example in the Indian Ocean basin the multiple Kelvin waves have maximum number of initiations between longitudes 50E and 60E and spin off Kelvin waves have maximum number of initiations between longitudes 80E and 90E. The Indian Ocean basin is characterized by the largest fraction of multiple and spin off initializations compared to total number of Kelvin wave initialization. It can be seen that the central and eastern Indian Ocean has large number of both multiple and spin off Kelvin wave initiations, therefore the further analysis is concentrated on that region.

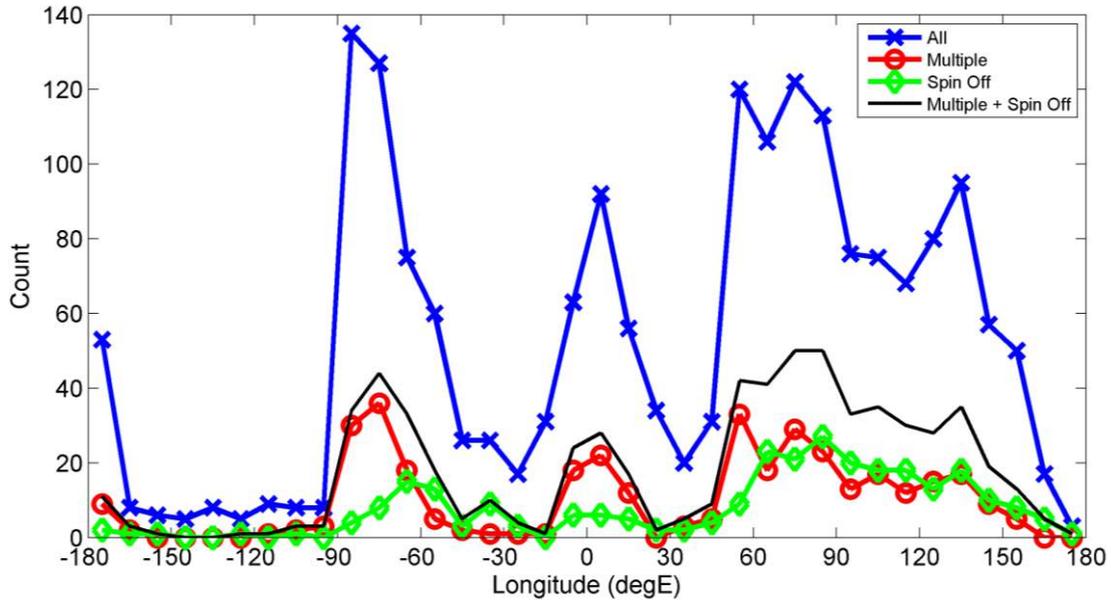

**Figure 3.19 Zonal distribution of number of Kelvin waves initiations. The x-axis is longitude; the y-axis is count of Kelvin wave initiations over each 10 degree wide area. Blue line is number of all Kelvin wave initiation; red line is number of multiple initiation Kelvin waves; green line number of spin off initiation Kelvin waves. Black line indicates sum of multiple and spin off initiations at each longitude.**

Figure 3.20, Figure 3.21 and Figure 3.22 present composite analysis of the magnitude of the diurnal SST variability, wind speed and latent heat flux at the ocean surface during Kelvin wave initiations. Each of the figures consists of 3 panels: (a) is a composite for all Kelvin waves initiation, (b) is a composite for multiple Kelvin waves initiation only and (c) is a composite for spin off Kelvin wave initiations only. All relevant Kelvin waves that initiated between longitudes 60E and 100E were used in composite calculations. The x-axis on each panel represents the longitude relative to the longitude of a Kelvin wave initiation and the y-axis represents the time relative to the date of a Kelvin wave initiation.

Figure 3.20 presents comparison of the magnitude of the upper ocean temperature diurnal cycle in the proximity of the Kelvin wave initiation. The composite for all Kelvin waves (a) shows decreased SST variability between longitudes 0 and 4 in the proximity of the time equal 0. This is manifestation of the initiation and passage of the Kelvin wave that is able to suppress the upper ocean diurnal cycle. The eastward propagation is not clearly visible because we consider area limited to vicinity of the Kelvin wave initiation through which Kelvin wave propagates within few hours. Because temporal resolution of the ocean surface data is 1 day, the propagation of the Kelvin wave appears instantaneous. However, the change in the magnitude of the SST variability is consistent with Kelvin wave phase speed. It can be seen that similar structure is present in panels (b) and (c).

However there are substantial differences between the 3 panels in magnitude of the diurnal SST variability anomaly prior to the wave development. The composite for all Kelvin wave initiations (a) shows increased diurnal SST variability two days prior to wave initiation and two days after the initiation. The maximum of the diurnal SST variability is located 3 degrees to the west in on comparison with location of the wave initiation. In this case, the SST variability is larger after the wave passage. Although, the overall structure is reproduced in multiple initiations composite, there are differences in magnitudes of each diurnal SST variability anomalies. In this case, the maximum prior to wave development is stronger and extends from the location of the initiation to the west. The magnitude of the maximum observed after Kelvin wave passage does not change, but one can see the decrease of the magnitude of the SST variability 3 days after initiation. This decrease is likely due to initiation of sequential Kelvin wave.

In comparison with all Kelvin waves and multiple Kelvin waves, the spins off Kelvin waves display different SST variability. Although the suppressed diurnal cycle due to Kelvin wave development and propagation is still apparent, the time prior to the Kelvin wave development is dominated by low SST variability, likely related to preceding Kelvin wave. The increased upper ocean temperature diurnal structure is present 2 days prior to Kelvin wave initiation, but the magnitude is much lower in comparison with previous composites. The SST variability increases two days after the initiation.

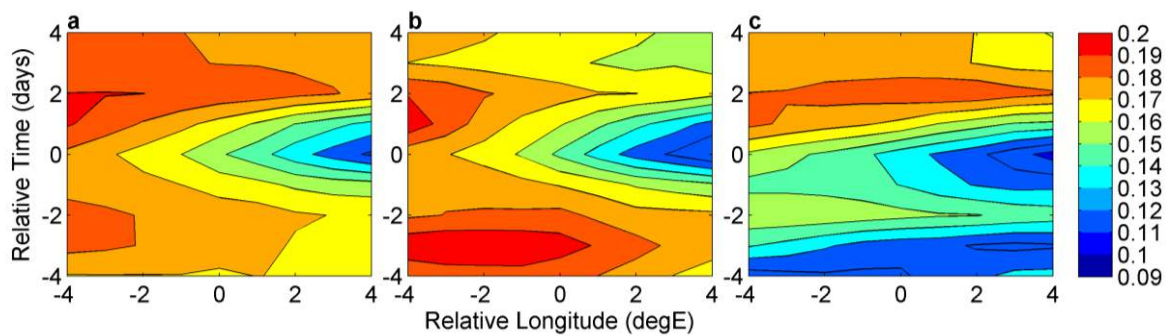

**Figure 3.20** The composite of the upper ocean temperature diurnal cycle in [°C] (color shading) over the area between longitudes 60E and 100E during (a) all Kelvin waves initiations; (b) multiple Kelvin wave initiations; (c) spin off Kelvin wave initiations. In each panel the x-axis in longitude in [degE] relative to the longitude of the Kelvin wave initiation and the y-axis is time in [days] relative to the time of the Kelvin wave initiation.

Figure 3.21 presents composites of the ocean surface wind speed for the initiation of (a) all Kelvin waves, (b) multiple Kelvin wave and (c) spin off Kelvin waves. It can be seen in panel (a) that on the average, the wind speed is increased during Kelvin wave initiation and eastward propagation. Prior and after the initiation, the values are lower. It consistent with suppressed SST variability at the time of Kelvin wave development shown in Figure 3.20. The multiple Kelvin waves initiation composite shows ocean surface wind speed increase at the time on wave development consistent with all Kelvin waves composite. However, prior to the initiation, the wind speed during multiple Kelvin wave initiation is even lower than in all Kelvin waves composite. The increased wind speed 3 days after the development is likely due to passage of the sequential wave. Figure 3.21c shows that wind speed is increased prior to the spin off Kelvin wave initiation. This increase is likely due to passage of the preceding wave. After the spin off Kelvin wave development and passage, the wind speed decreases rapidly.

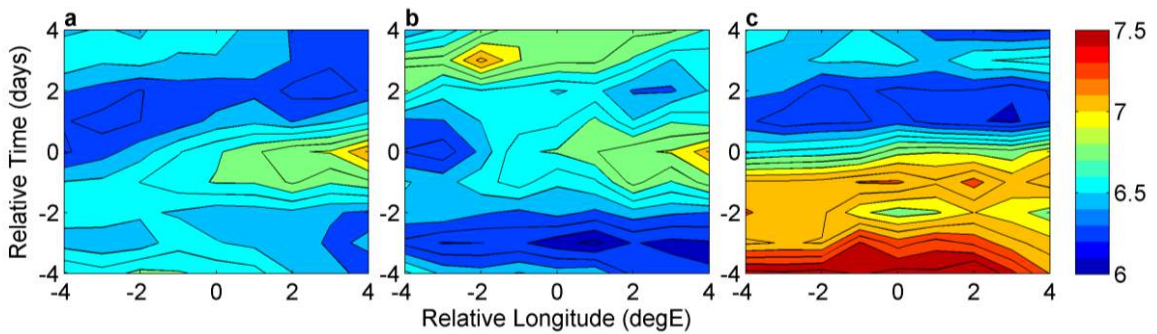

**Figure 3.21 The composite of the ocean surface wind speed in [ms$^{-1}$] (color shading) over the area between longitudes 60E and 100E during (a) all Kelvin waves initiations; (b) multiple Kelvin wave initiations; (c) spin off Kelvin wave initiations. In each panel the x-axis in longitude in [degE] relative to the longitude of the Kelvin wave initiation and the y-axis is time in [days] relative to the time of the Kelvin wave initiation.**

Figure 3.22 presents composites of the ocean surface latent heat flux variability associated with the initiation of (a) all Kelvin waves, (b) multiple Kelvin waves and (c) spin off Kelvin waves. It can be seen that zonal and temporal variability associated with each category is very similar to one shown for the wind speed variability. For (a) all Kelvin waves and (b) multiple Kelvin waves, the increased latent heat flux at the time of the initiation and at the location, and to the east of the location of the wave initiation is apparent. For (a) all Kelvin waves, the latent heat flux is decreased prior to and after wave development. The decrease prior to the wave initiation is stronger for multiple Kelvin waves initiation. The increased latent heat flux values are visible two to four days after wave development. It is likely a manifestation of the sequential wave development and passage. The (c) spin off initiation is associated with large latent heat flux consistent with time and region of the strong winds.

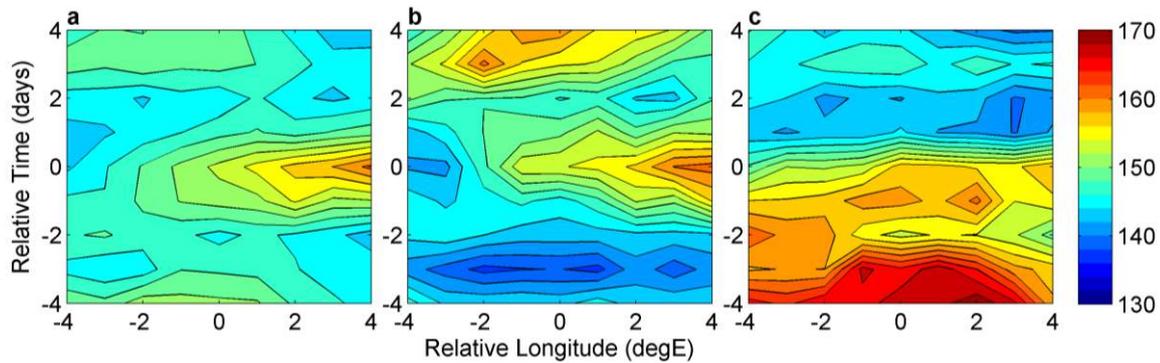

**Figure 3.22** The composite of the ocean surface latent heat flux in [Wm$^{-2}$] (color shading) over the area between longitudes 60E and 100E during (a) all Kelvin waves initiations; (b) multiple Kelvin wave initiations; (c) spin off Kelvin wave initiations. In each panel the x-axis in longitude in [degE] relative to the longitude of the Kelvin wave initiation and the y-axis is time in [days] relative to the time of the Kelvin wave initiation.

In summary, the composite analysis suggests that multiple initiation may be forced by increased diurnal SST variability. Prior to the wave initiation, the upper ocean temperature diurnal structure is increased. At the same time the ocean surface wind speed and latent heat flux are decreased. This suggests that local thermodynamic processes are likely responsible for triggering the Kelvin wave in multiple initiation case. This is consistent with shallow water model results. It has been shown [*Li and Carbone*, 2012] that maxima of the spatial SST gradient are related to development of atmospheric convection. Our results agree with that conclusion and show that the multiple Kelvin waves are initiated when the diurnal SST variability is the largest.

On the other hand, the diurnal SST variability during spin off Kelvin wave initiation is secondary to the strong ocean surface wind speed and latent heat flux variability. Therefore, the triggering of the spin off Kelvin waves is likely to have strong dynamic component with some contribution from local thermodynamic forcing.

## 3.6. Summary

In this Chapter the initiation of the sequential Kelvin waves was investigated on the basis of the global finite element shallow water model [*Giraldo*, 2000], individual case studies and 15-year statistics based on Kelvin wave TRMM precipitation derived trajectories. The upper ocean temperature diurnal variability was assessed using T$^\dagger$ index, which combines wind speed and insolation on the ocean surface. This metric has been introduced in Chapter 2. The environmental conditions, which include ocean surface wind speed and latent heat flux, has been obtained from the OAFlux product [*Yu et al.*, 2008] and used to assess changes in the environment during initiation of the sequential Kelvin waves.

The numerical results show that the characteristics of the heat source may affect Kelvin waves zonal and temporal distribution, but have very limited impact on Rossby waves. The diurnally oscillating heat source forced the sequence of Kelvin waves. The relative wind speed and geopotential height anomalies related to every wave within the sequence were larger than in control simulation, which had constant heat source. Changes in the relative magnitude of the oscillations have impact on the magnitude of the response, but do not change its spatial variability. This suggests that temporal variability of the heat source is an important modulator of the spatial Kelvin wave structure.

Although the shallow water model does not have explicit representation of the physical processes responsible for the variability of the source it has been established that the diurnal cycle of the fluxes at the ocean surface may affect diurnal evolution of atmospheric convection [*Ruppert and Johnson*]. Therefore, the development of the upper ocean diurnal warm layer is one plausible physical process that could account for diurnal variability in heat source. The upper ocean diurnal cycle may impact the structure and variability of the Kelvin waves and this hypothesis was investigated using observational data.

It was identified that sequential Kelvin wave initiation may be divided into two distinct categories: multiple initiations and spin off initiations. A multiple initiation occurs when two or more Kelvin waves develop over the same area within a short time period. The spin off initiation happens when a Kelvin wave develops in the wake of another Kelvin wave.

A pair of multiple Kelvin wave was found to initiate in the presence of the increased SST variability and relatively weak wind speed and weak latent heat flux forcing. Another pair of multiple Kelvin waves was initiated in the area where the SST variability, wind speed and latent heat flux were increased. These results suggest that both the upper ocean temperature diurnal variability and ocean surface fluxes variability should be considered as important factors for formation of convection and precipitation. A composite analysis of the 83 multiple Kelvin wave initiations that occurred over central and eastern Indian Ocean in 15 years shows that the upper ocean temperature diurnal variability is high and that wind speed and latent heat fluxes at the ocean surface are relatively weak prior to the development of such waves. The local thermodynamic processes are likely responsible for triggering Kelvin waves during multiple initiations Therefore it has been shown that the diurnal SST variability may be related to the varying heat source which triggers sequential multiple initiation Kelvin waves.

Interestingly, the case studies of Kelvin waves that initiate as a spin off from another Kelvin wave were found to be triggered by both dynamic forcing and local thermodynamic forcing. The preceding ("spinning") wave may influence the environment in its proximity and the sequential ("spin off") wave may develop. The dynamic forcing results from propagation of the increased wind speed and latent heat fluxes likely related to the Rossby wave. The local thermodynamic forcing is due to increased diurnal SST variability in the wake of the preceding Kelvin wave. We have investigated two case studies of spin off initiation. In the first of them both dynamic and local thermodynamic forcing components were strong. In the second, the local thermodynamic forcing dominated. The composite analysis of 91 spin off initiations that occurred over central and eastern Indian Ocean in 15 years shows that dynamic forcing is dominant component to which local thermodynamic forcing may contribute.

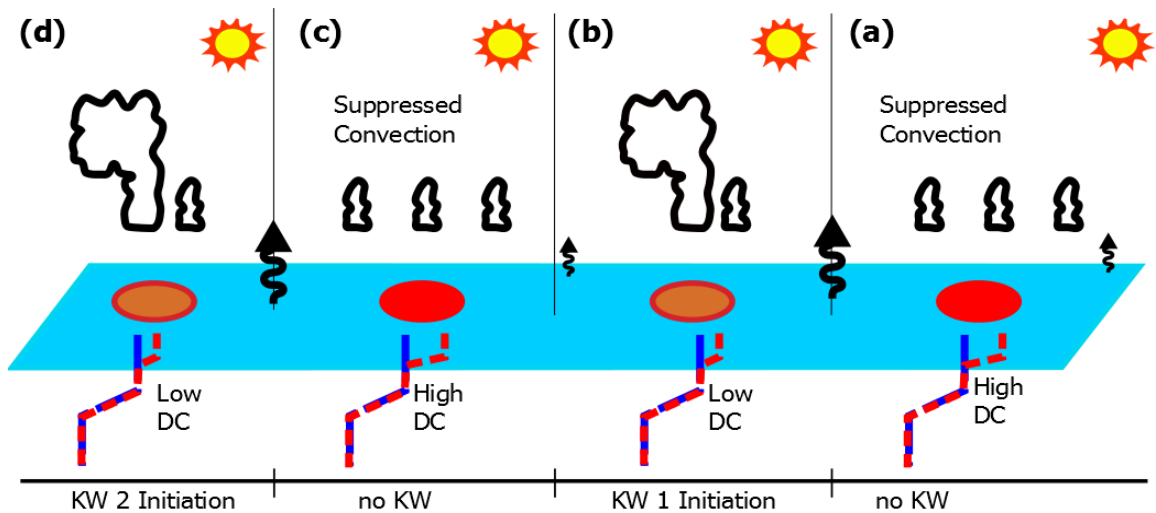

Figure 3.23 Schematic model of multiple Kelvin wave initiation: (a) and (c) represent conditions of suppressed atmospheric convection with high insolation, weak winds and strong diurnal cycle development, (b) and (d) represent days of Kelvin wave initiation preceded by increased air-sea fluxes due to warm layer development.

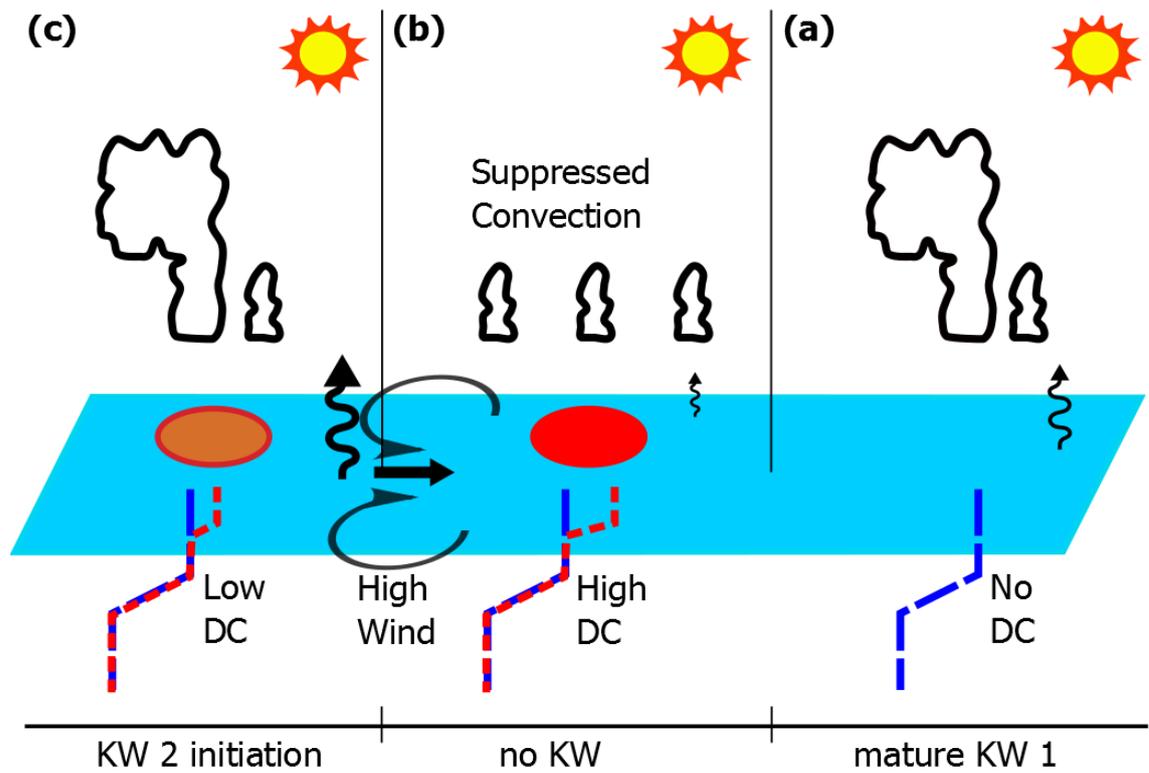

**Figure 3.24 Schematic model of spin off Kelvin wave initiation: (a) mature Kelvin wave which suppresses upper ocean diurnal cycle, (b) conditions of suppressed convection in the wake of the mature Kelvin wave. Such conditions favor high insolation, weak winds and strong warm layer development. The two gyres represent Rossby wave forced by mature Kelvin wave and lead to increased wind speed and air sea fluxes at the ocean surface at the equator, (c) initiation of the spin off Kelvin wave in response to the increased air-sea flux.**

Figure 3.23 summarizes the key characteristics identified in case studies of multiple Kelvin waves initiation. In presence of the suppressed atmospheric convection Figure 3.23 (a) and (c) the upper ocean temperature diurnal cycle is high and as a result air-sea fluxes are high. This leads to development of Kelvin wave on the following day, Figure 3.23 (b) and (d). On days when Kelvin wave initiates the diurnal cycle is weaker than the preceding days, but it is not suppressed completely. The day to day variability of the magnitude of the upper ocean temperature diurnal cycle forces variability of the fluxes from the ocean to the atmosphere. However, on day which exhibit warm layer development, the air-sea fluxes response is delayed in comparison to SST response. This is because there is energy storage in the warm layer, which lasts into the night and enables anomalous flux after the sunset as shown in Chapter 2. This sequence of events shows that the variability of the heating source for the atmospheric convection may be related to the upper ocean temperature diurnal variability. We have used such a diurnally varying heat source in the shallow water model simulations. To the best of our knowledge such link between diurnally varying heat source and Kelvin wave is novel, but the relation between diurnal SST variability and development of atmospheric convection during MJO suppressed phase has been recently noted [*Chen et al.*, 2014; *Ruppert and Johnson*, 2014].

Figure 3.24 presents schematic of spin off Kelvin wave initiation in the certain area, which summarizes the key characteristics identified in the case studies. The preceding ("spinning") Kelvin wave (a) suppresses the diurnal SST variability due to high cloudiness and strong winds. On the following days (b) the convection is suppressed and wind speed is low which results in strong upper ocean diurnal temperature cycle. However, the flux from the ocean to the atmosphere is increased in the presence of strong winds. One possible explanation of the strong wind speed is that the Rossby wave, which was initiated by the preceding Kelvin wave, approaches the area. This interaction enables rapid release of the energy accumulated in the upper ocean for spin off Kelvin wave development (c).

# Chapter 4. Impact of atmospheric equatorial convectively coupled Kelvin waves on the upper ocean variability

## 4.1. Introduction

In this Chapter we study interactions between the upper ocean and ocean surface and Kelvin waves beyond their point of origin. The MJO itself strongly affects the ocean variability and the possibility of air-sea interaction was suggested as one of the mechanisms contributing to MJO development and propagation, during, both, boreal winter and boreal summer [*Flatau et al.*, 1997; *Flatau et al.*, 2003; *Fu and Wang*, 2004; *Shinoda et al.*, 1998]. During the active phase of the MJO increased convective activity leads to reduced shortwave radiation, enhanced surface fluxes, and upper-ocean mixing cool the ocean surface while ahead of the propagating system the large radiative fluxes and weak insolation lead to increased ocean temperature associated with large diurnal SST variability [*Shinoda et al.*, 2013a]. This diurnal SST warming appears to influence the distribution of tropical precipitation contributing to the coupled feedbacks between ocean and atmosphere that maintain the basic state, the timing of the seasonal cycle of SST and trade winds in the tropical Pacific [*Bernie et al.*, 2008]. The impact of the diurnal variability on the modeled tropical convection can be observed even if no other oceanic processes are included [*McLay et al.*, 2012]. As shown by Bernie *et al.* [2007] the inclusion of diurnal SST variability is also important in MJO forecasting. In Woolnough *et al.* [2007] experiments the MJO forecast is improved with the increased resolution of the oceanic mixed layer that leads to the better representation of the diurnal SST variability. Also, resolving diurnal SST variability improves simulations of other tropical phenomena such as the Indian summer monsoon [*Klingaman et al.*, 2011].

Similar air-sea interaction mechanisms may operate during the passage of the Kelvin waves on the scales of several days. On this time scale the upper ocean processes are especially important, and purpose of this paper is to investigate the impact of Kelvin waves on the upper ocean behavior. To this end we use the simple predictive model of the existence and strength of the diurnal warm layer, $T^{\dagger}$ introduced in Chapter 2. Results from Chapter 2 indicate that due to the multi-scale nature of tropical convection the diurnal warm layer developed in both, active and suppressed phase of MJO during the period of low winds and high insolation (for the wind speed smaller than 6 ms$^{-1}$ and daily mean surface solar radiation larger than 80 Wm$^{-2}$). It was more often observed during the suppressed MJO period.

During the DYNAMO period 15 Kelvin waves were identified with exceptionally strong "double barrel" waves related to the November MJO episode [*Johnson and Ciesielski*, 2013]. Nine of these events were associated with the active phase of the MJO while 7 developed during the suppressed phase [*Gottschalck et al.*, 2013]. Radar observations [*DePasquale et al.*, 2014] were used to analyze the DYNAMO period Kelvin waves. These authors noted the difference in structure and intensity of the waves developed during the suppressed and active phases of MJO. In the active phase, the waves were more intense with higher moisture and more clouds, possibly impacting the radiative effects of the wave. In the suppressed MJO phase the waves developed in the drier atmosphere and therefore were less intense. In these waves, unlike in the waves associated with the active MJO, the moisture buildup ahead of the convective activity appeared to play the role in the wave development.

This Chapter investigates the variability of the diurnal cycle of the SST during the passage of Kelvin waves observed in DYNAMO and extends the investigation to 15 years for which diurnal warm layer strength as well as the ocean surface wind speed and latent heat flux were diagnosed using methodology described in Appendix D. The new convectively coupled Kelvin wave trajectories development is described in Appendix C.

In this chapter we discuss observational results based on the DYNAMO field project, in particular the case of sequential Kelvin waves observed during the November 2011 MJO. At the end we extend the results to 15 years Kelvin waves climatology and present simplified conceptual model of our findings.

## 4.2. Results

### 4.2.1. Statistics

Based on zonal distribution of Kelvin wave activity, in Chapter 3 it was shown (Figure 3.7) that the Indian Ocean has the largest number of Kelvin wave trajectories and the largest number of trajectories occurring quickly after each other. In addition, we know that the Indian Ocean has globally the largest intraseasonal variability [*Wheeler and Kiladis*, 1999]. Therefore, it is natural to consider implications of the intraseasonal variability on Kelvin waves activity as defined by our methodology.

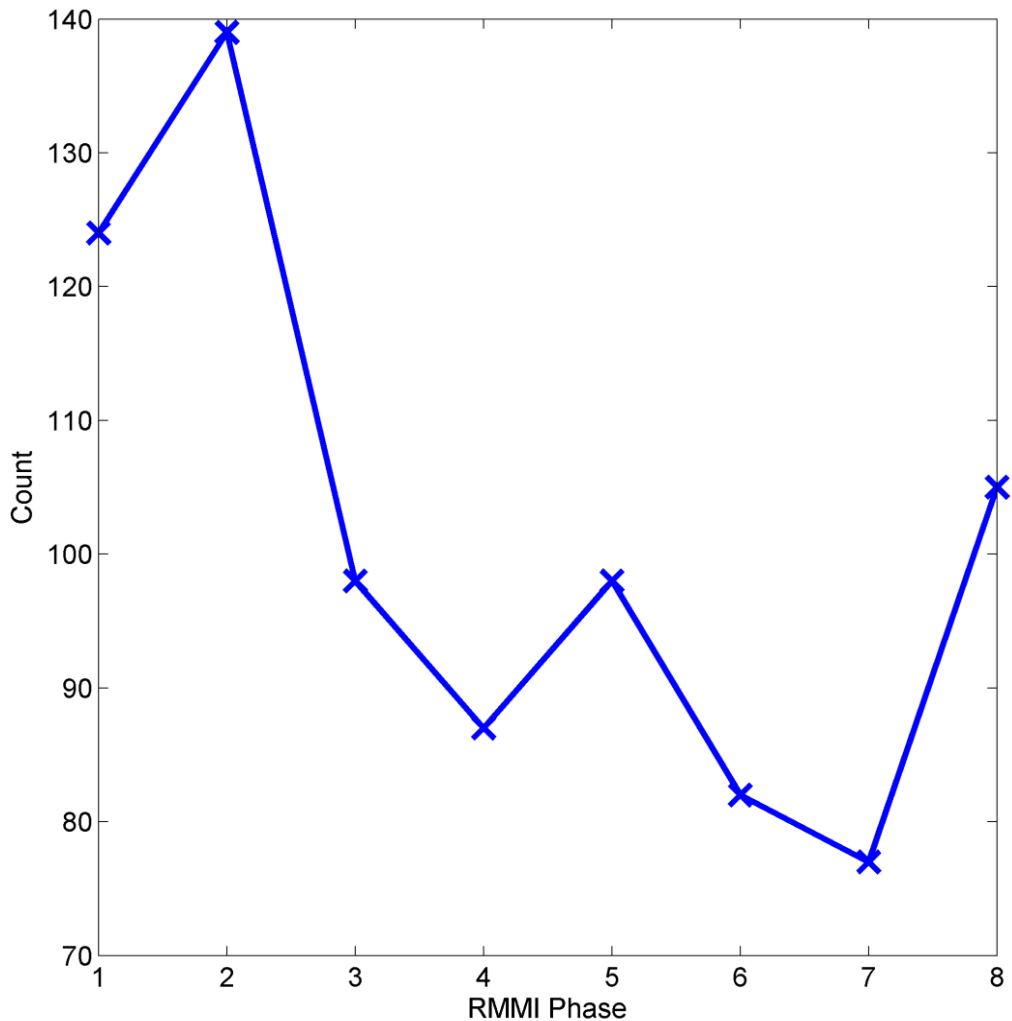

**Figure 4.1 Count of the active Kelvin waves in the Indian Ocean as a function of a RMMI phase of the MJO.**

Figure 4.1 presents the number of Kelvin waves active in Indian Ocean basin (between 40E and 100E) as a function of the RMMI phase. The x-axis represents 8 distinct RMMI phases indicating areas of active MJO phase and y-axis shows number of the active Kelvin waves in the Indian Ocean basin. The RMMI phases 1 to 3 represent convection over Eastern Africa and Indian Ocean, phases 4 to 8 represent Maritime Continent, Western Pacific and Western Hemisphere. It can be seen that the highest number of Kelvin Waves coincide with the MJO phases 1 and 2, which represent intraseasonal convection over eastern Africa and western Indian Ocean. Kelvin Waves count over Indian Ocean drops by 40% for phases which represent intraseasonal convection centered over Maritime Continent and further west (RMMI phases 4 to 7). For the RMMI phase 7, which represents active convection over Atlantic number of Kelvin wave trajectories in the Indian Ocean is the smallest. Beginning with RMMI phase 8, which represents active convection over Africa, the count of the Kelvin waves in Indian Ocean basin starts to increase. The results presented in Figure 4.1 refer to MJO phases with no respect to the RMMI amplitude. Similar analysis for Kelvin Waves coinciding with intraseasonal variability characterized by the RMMI amplitude only bigger than one has been performed with no significant change of the results. These results indicate that intraseasonal variability of tropical convection is strongly linked with sub intraseasonal activity of Kelvin waves in Indian Ocean. There are also zonal differences in reoccurrence periods. Thus, environmental conditions and MJO activity are important modulators of the Kelvin wave activity. On the other hand, since Kelvin wave activity appears to be the largest ahead of MJO, the modification of the ocean conditions by Kelvin wave could impact the MJO development.

In summary, we identified 1948 Kelvin wave trajectories, more that 40% of which were active in the Indian Ocean basin. We have noticed that many Kelvin waves propagate in groups with only short period of time between individual waves in the group. Such waves are most likely to interact with each other and local environment due to air-sea interactions. It is shown that Kelvin waves which propagate shortly after each other account for most of the cross basin differences in the overall Kelvin wave activity. This is particularly true in the Indian Ocean basin. Furthermore, we have shown that intraseasonal variability of tropical convection is strongly linked with sub intraseasonal activity of Kelvin waves in the Indian Ocean. These results implicate that local environmental conditions and MJO activity are important modulators of the Kelvin wave activity.

### 4.2.2. The DYNAMO field project: case of frequent sequential Kelvin waves

During the DYNAMO field campaign 3 MJO events were intensively measured [*Gottschalck et al.*, 2013; *Johnson and Ciesielski*, 2013; *Kerns and Chen*, 2013]. The November 2011 MJO was associated with two sequential atmospheric Kelvin waves (Appendix C). Figure 4.2 presents atmospheric and oceanic conditions during second half of November 2011 over SeaGlider location. It was located in vicinity of the equator, hence enabling recognition and analysis of the equatorial Kelvin waves and oceanic response to their forcing. Figure 4.2a presents Latent Heat Flux (LHF), ShortWave Flux (SWF) and Wind Speed interpolated to the SeaGlider location. Figure 4.2b presents SST, surface salinity SSS, $T^\dagger$ from SeaGlider and TRMM precipitation at the same location. The whole time period presented in Figure 4.2 corresponds with active phase of the MJO. However, two distinguishable anomalies manifested by wind bursts (red line, panel a) on November 24 and November 29 represent two sequential atmospheric Kelvin waves that were part of this MJO event.

It can be seen that precipitation pattern (green curve in Figure 4.2b) is also divided into two smaller precipitation periods. These two periods correlate with variation in fluxes. During each of the convective periods increased magnitude of negative LHF and decreased SWR are observed. Between these convective periods, two days of strong solar radiation, decreasing LHF and decreasing wind speed are apparent.

Black line in Figure 4.2b shows that the SST decreases during the active phase of the MJO. The decrease over the 10 days is of the order of 0.3 °C. At the same time some of the individual days exhibit strong diurnal cycle which has the same order of magnitude as the decreasing trend throughout that period. SeaGlider measurements show also the decrease in SSS correlated with Kelvin wave precipitation events.

On November 24, 25, 28, 29 the diurnal cycle is suppressed. On these days daily maximum $T^\dagger$ does not exceed 0.05 °C and daily mean $T^\dagger$ is less then 0.025 °C. During calm, non-convective period between sequential atmospheric Kelvin waves, the recovery of the strong diurnal signal in upper ocean temperature (red line in Figure 4.2b) is observed. High magnitudes of measured $T^\dagger$ are observed on November 26 and 27, when daily maximum exceeds 0.25 °C and 0.15 °C, respectively. Daily mean $T^\dagger$ for these two days exceeds 0.05 °C. This analysis shows that over investigated period the diurnal cycle provides short term daily pulses of increased SST, with magnitude comparable to the cooling trend.

That confirms that Kelvin Waves interact with the upper ocean but on the shorter time scales and over smaller depths than MJO. Nevertheless, the short term impact on the SST is substantial and $T^\dagger$ is a useful metric of such short term SST variability. Therefore, evaluation of $T^\dagger$ variability in regions and for other periods, where direct upper ocean measurements are not available may provide insight on air-sea interactions in equatorial convectively coupled waves. Black solid and dashed lines in Figure 4.2a represent the comparison between the daily averaged $T^†$ measured in-situ by the SeaGLider and $T^†$ index described previously. The overall comparison shows that derived index reproduces well the measured signal. Thus, $T^†$ provides a metric which enables studying characteristics of upper ocean diurnal variability in the absence of direct in-situ measurements.

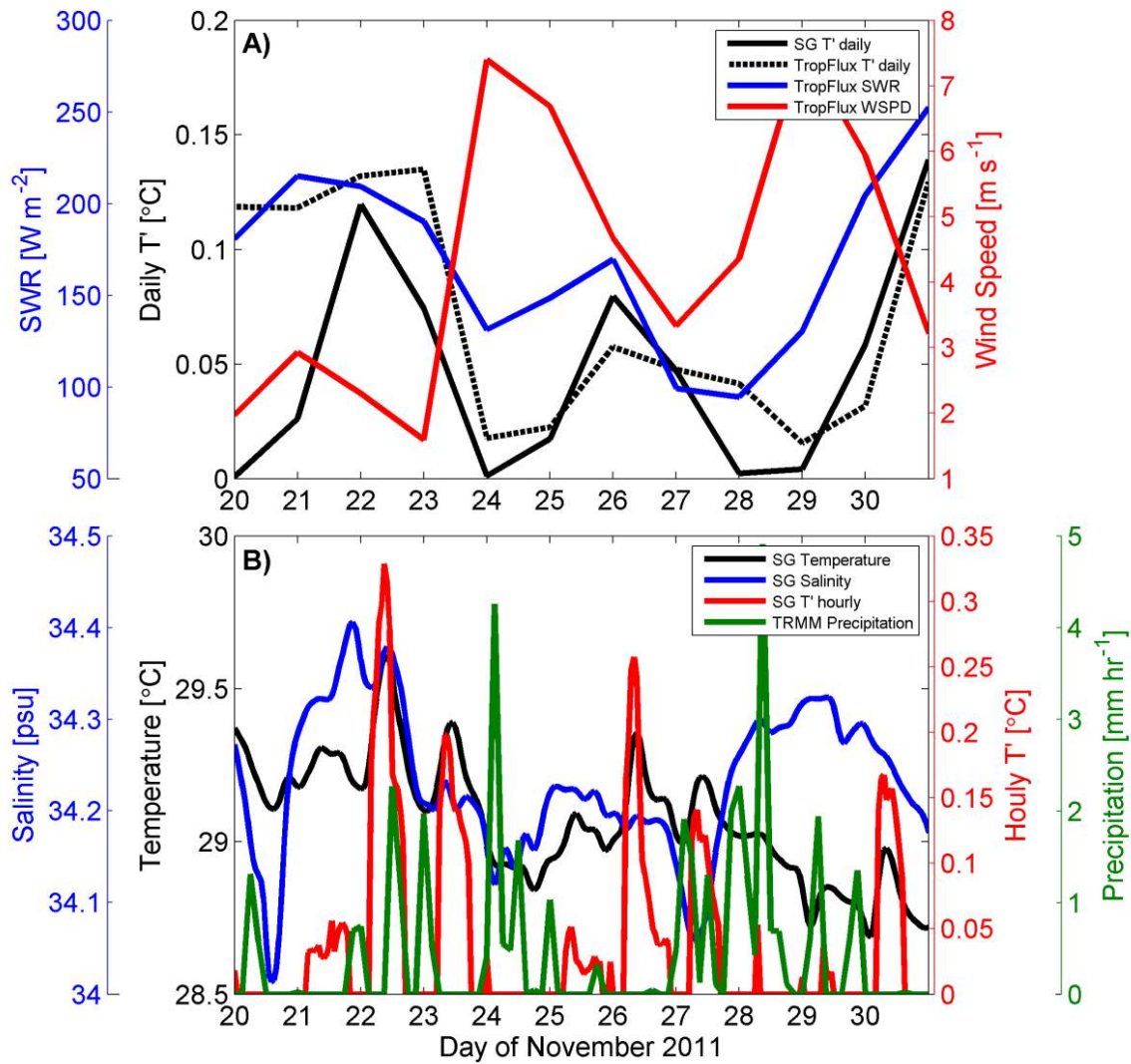

Figure 4.2. (a) Evolution of daily $T^†$ (black solid line), derived $T^†$ (black dashed line), daily SWR (blue line) and daily wind speed (red line) for the period of November 20 – December 1, 2011. (b) SeaGlider measurements of surface (at 0.5m depth) temperature (black line), surface salinity (blue line), $T^†$ and TRMM measurement of precipitation (green line) at the SeaGlider location for the period of November 20 – December 1, 2011.

## 4.2.3. Kelvin waves impact on surface wind speed and latent heat flux

In this section we analyze Kelvin wave impact on the upper ocean for the entire wave trajectory and analyze impact of the convective activity associated with propagating Kelvin wave on ocean surface wind speed and latent heat flux.

Two time periods are considered. For the DYNAMO case the period between November 11 and December, 21 2011 is investigated and analysis covers the equatorial band between Africa and Western Pacific, though our primary interest is Indian Ocean basin. Figure 4.3 and Figure 4.4 present wind speed and latent heat flux at the ocean surface for the investigated area and time. This period is characterized by MJO event passing over Indian Ocean. In contrast to DYNAMO, period between January 3 and February 21, 2000 was characterized by weak intraseasonal activity. The RMMI amplitude stayed below 1 and phase was random, which indicates no apparent areas of neither active nor suppressed convection in the tropics. Figure 4.5 and Figure 4.6 present wind speed and latent heat flux at the ocean surface for this period. Figures 4.3 – 4.6 are presented in form of Hovmöller diagram, for which the x-axis is longitude and the y-axis is time increasing upwards. Surface fluxes data preparation for these analysis is provided in Appendix D.

Such differences in the mean state of the atmospheric circulation over Indian Ocean allow us to compare ocean surface flux response to the Kelvin wave passage during various environmental conditions. Here, we will concentrate mainly on two sequential waves propagation observed during both of these periods. The first case we consider is the passage of the two Kelvin waves observed during the November 2011 DYNAMO MJO episode. These waves are shown in Figure 4.3 and Figure 4.4 with wave trajectories indicated by solid magenta lines. Two Kelvin waves pass over 80E on November 24 and November 28, 2011. Preceding the passage of these waves, during the suppressed phase of MJO, wind speed and latent heat flux is small over the entire Indian Ocean basin. Times and locations of slightly increased values of wind speed and latent heat flux are scattered. Period between November 21 and December 1 is characterized by eastward propagation of the area of increased wind speed (Figure 4.3) and increased LHF (Figure 4.4). These propagating anomalies are related to MJO envelope. Interestingly, during the transition period between the two sequential Kelvin wave trajectories the areas of decreased wind speed and decreased latent heat flux can be observed. In particular, at 80E the wind speed varies around 5-6 ms$^{-1}$ and LHF varies around 125-150 Wm$^{-2}$ prior to the approach of the first Kelvin wave; during the first Kelvin wave passage wind speed increase to more than 11 ms$^{-1}$ and latent heat increase to 225-250 Wm$^{-2}$; during the transition period wind speed decreases to 6-7 ms$^{-1}$ and LHF decreases to 125-150 Wm$^{-2}$; during the second Kelvin wave passage wind speed increases to 9-10 ms$^{-1}$ and LHF increases to 175-200 Wm$^{-2}$. This variability is consistent with the precipitation, SST and salinity variability from in-situ measurements shown in Figure 4.2.

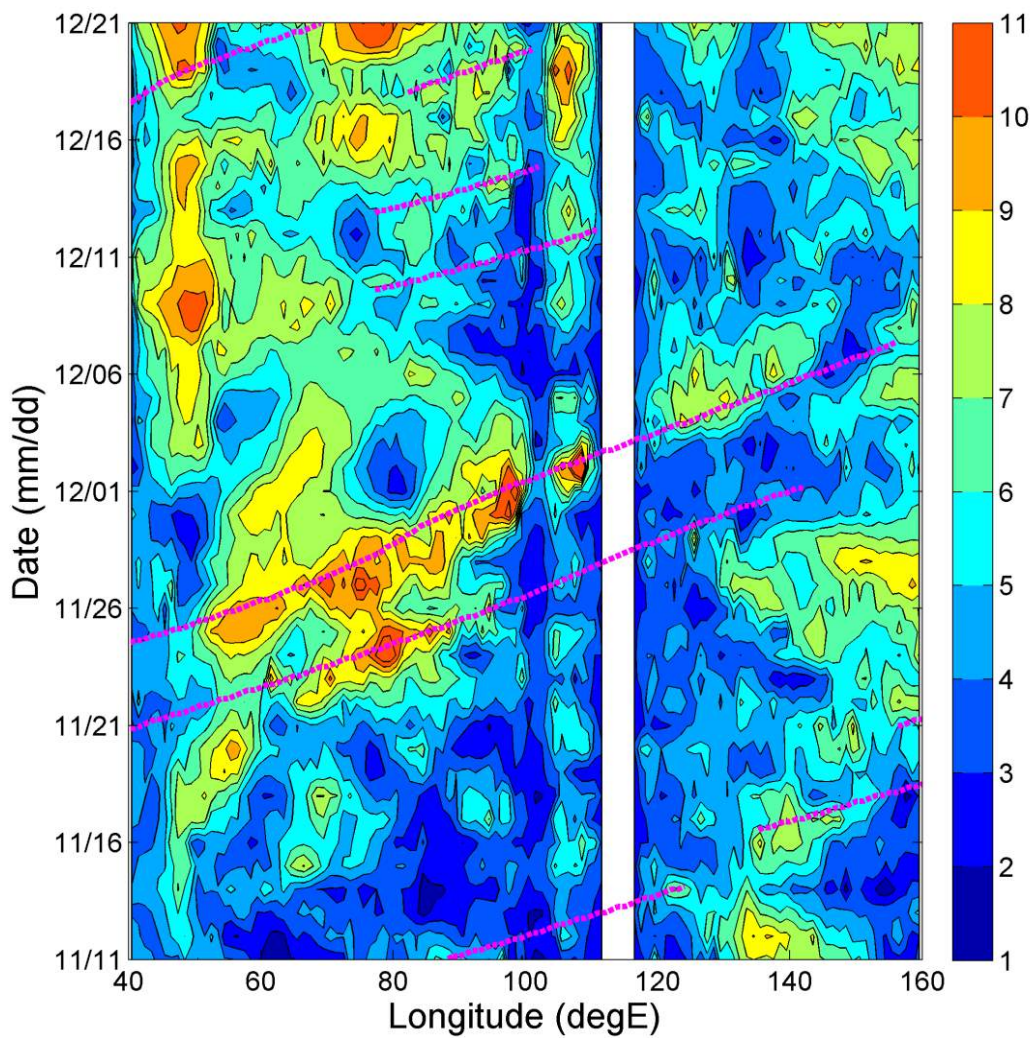

**Figure 4.3 Wind Speed for period 11 Nov – Dec 2011. x axis is longitude, y axis is date. Color scale indicates daily averaged wind speed in [ms$^{-1}$]. Kelvin wave trajectories are marked with solid magenta lines.**

Let us now consider period of the weak intraseasonal activity between January 3 and February 21, 2000 for which it can be observed that although wind speed and LHF show a lot of temporal and zonal variability, there are no clear periods dominated by low values of wind speed and LHF. However, as shown in Figure 4.5 and Figure 4.6 the Kelvin wave activity can be observed in the Indian Ocean over the entire period. Multiple Kelvin wave trajectories can be seen with two Kelvin waves resembling the sequential trajectories observed during November 2011 DYNAMO case. These two sequential Kelvin waves are traversing Indian Ocean during the January 3 – February 21, 2000 period; they are passing over 80E on January 14 and January 18, 2000. Hence, the trajectories are separated by 4 days, which is similar to time between the sequential Kelvin waves observed during the November 2011 event; in both cases they begin west of the 40E and make transition over Maritime Continent.

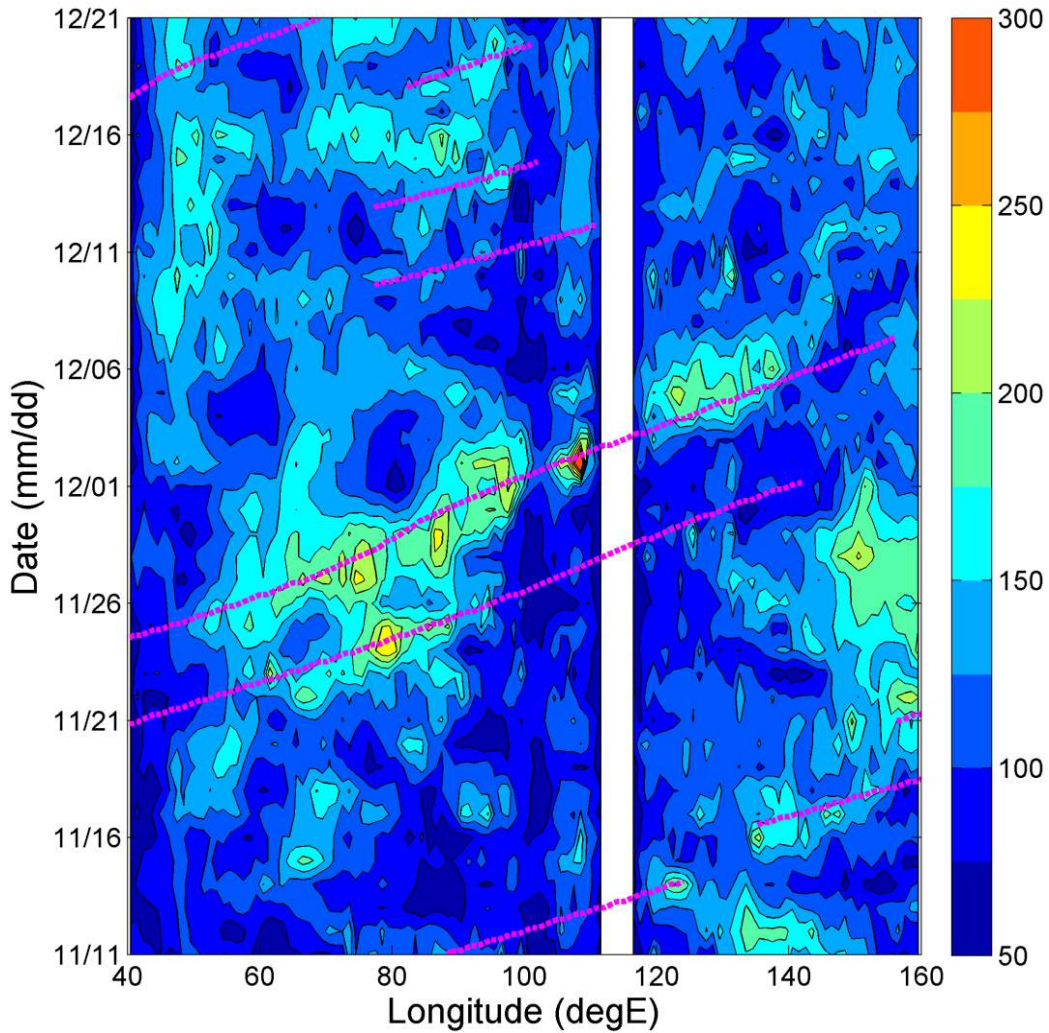

**Figure 4.4 Latent Heat Flux for period 11 Nov – Dec 2011. x axis is longitude, y axis is date. Color scale indicates daily latent heat flux in [Wm$^{-2}$]. Kelvin wave trajectories are marked with solid magenta lines.**

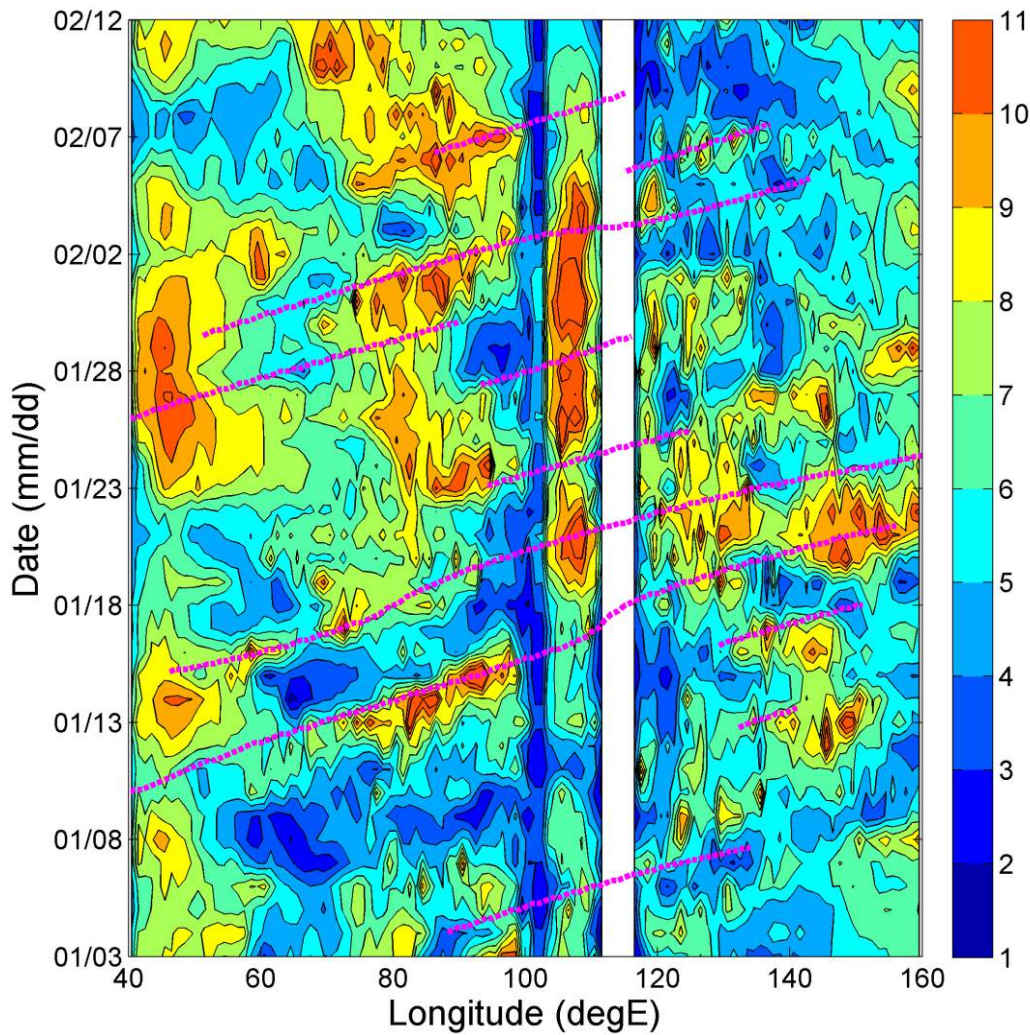

**Figure 4.5** Wind Speed for period 3 Jan – 12 Feb 2000. x axis is longitude, y axis is date. Color scale indicates daily averaged wind speed in [ms$^{-1}$]. Kelvin wave trajectories are marked with solid magenta lines.

Figure 4.5 and Figure 4.6 present wind speed and LHF variability associated with these two Kelvin waves. Wind speed and LHF show increased values directly at longitudes and times defined by each trajectory and decreased values between them. Observed behavior is not limited to the specific area, but accompanies the trajectories through the central and eastern Indian Ocean, between 60E and 100E, which is consistent with the behavior observed during the November 2011 case. Variability at 80E is such that the wind speed varies around 6-7 ms$^{-1}$ and LHF varies around 125-150 Wm$^{-2}$ prior to the approach of the first Kelvin wave; during the first Kelvin wave passage wind speed increases to more than 11 ms$^{-1}$ and latent heat increases to 250-275 Wm$^{-2}$; during the transition period wind speed decreases to 5-6 ms$^{-1}$ and LHF decreases to 100-125 Wm$^{-2}$; during the second Kelvin wave passage wind speed increases to 7-8 ms$^{-1}$ and LHF increases to 125-150 Wm$^{-2}$.

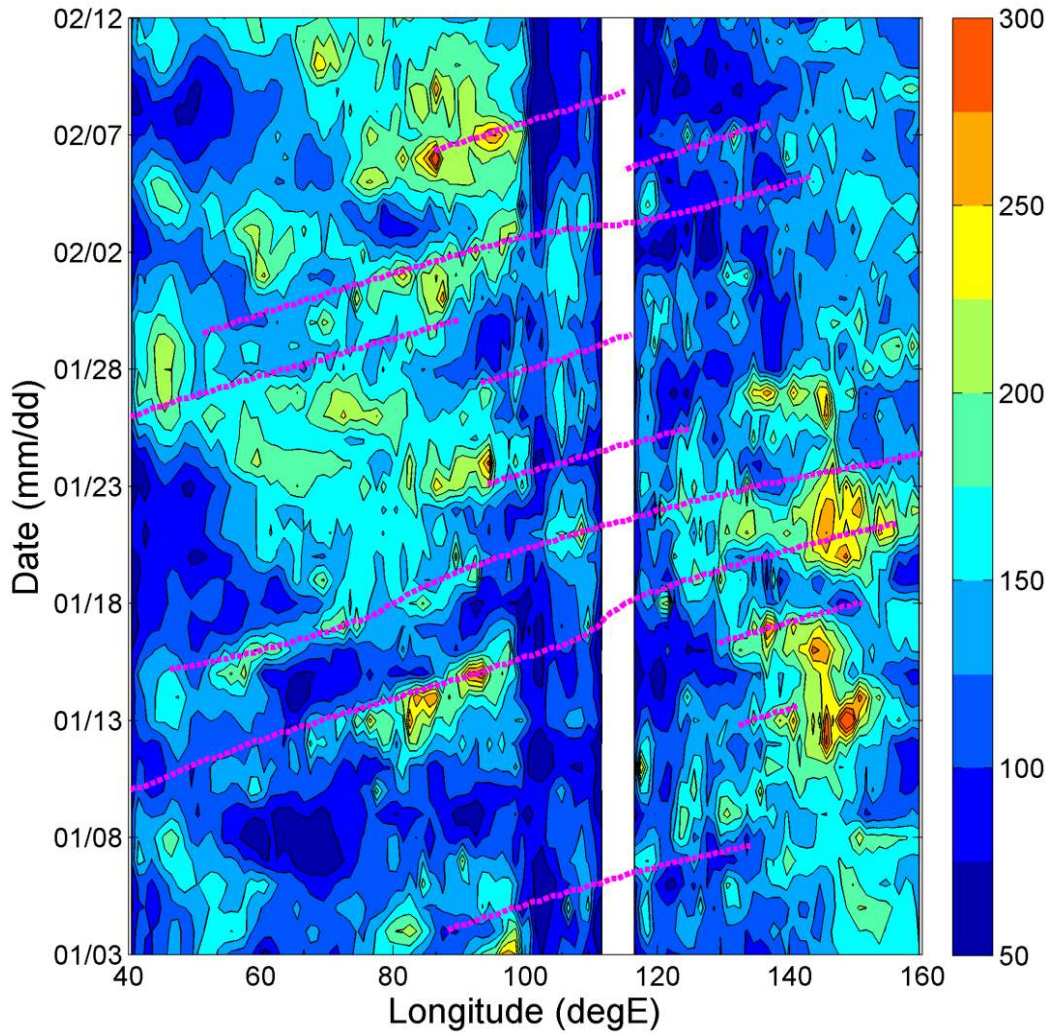

**Figure 4.6 Latent Heat Flux for period 3 Jan – 12 Feb 2000. x axis is longitude, y axis is date. Color scale indicates daily latent heat flux in [Wm$^{-2}$]. Kelvin wave trajectories are marked with solid magenta lines.**

Analysis of these two cases shows that wind speed and latent heat flux exhibit substantial variability associated with the Kelvin wave propagation. Wind speed may double and exceed 11 ms$^{-1}$ at the Kelvin wave trajectory and associated latent heat flux is likely to exceed 200 Wm$^{-2}$, which is increase of more than 50% from the pre-Kelvin wave values.

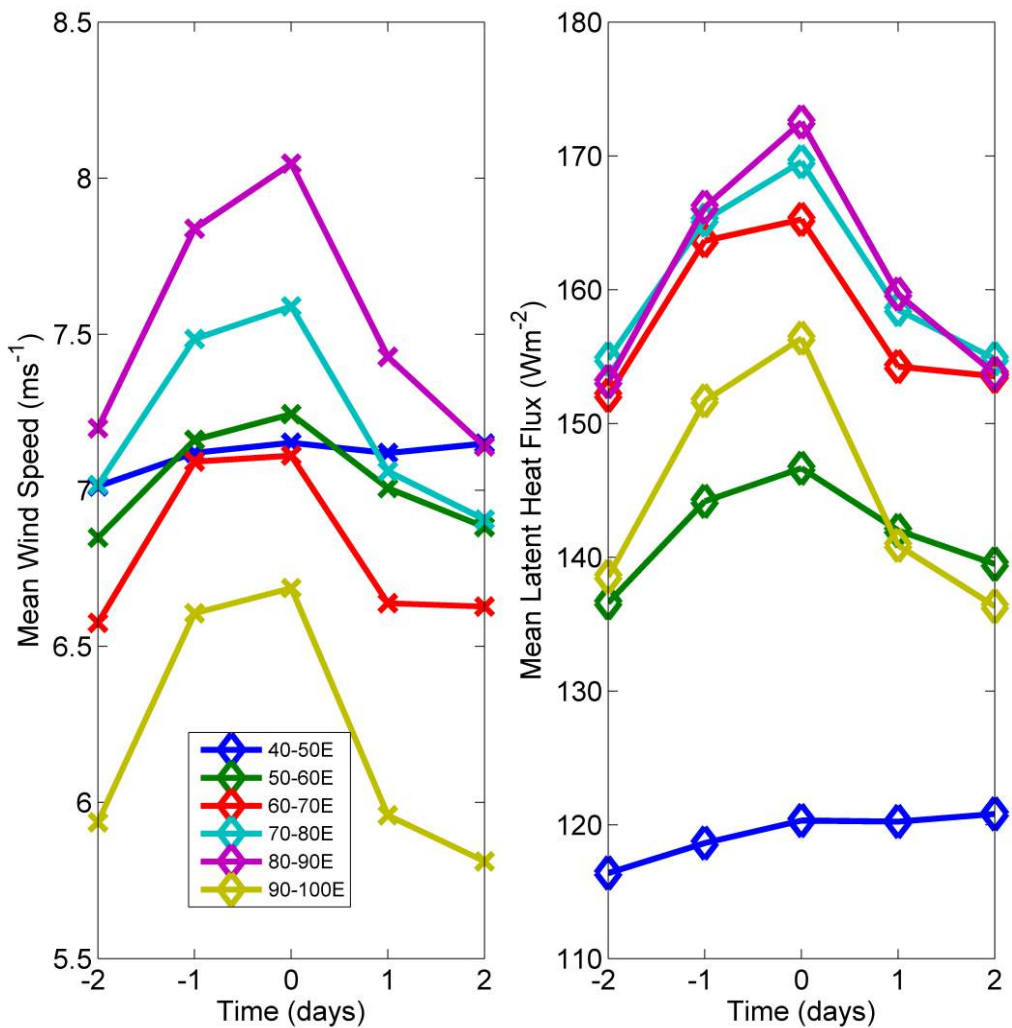

**Figure 4.7 Composite wind speed (left) and latent heat flux (right) for all Kelvin waves active in the Indian Ocean. Time (x axis) is relative to the Kelvin wave trajectory.**

So far we have analyzed two cases. In order to extend this analysis we will derive composite characteristics of Kelvin wave air-sea interactions based on 15 year statistics described previously. Thus, composites are based on all Kelvin wave trajectories. Figure 4.7 shows composite of the ocean surface wind speed (left panel) and latent heat flux (right panel) response to the Kelvin wave activity divided into six 10 degree wide boxes over the entire Indian Ocean. To limit biases associated with FFT filtering, only trajectories active through the entire box were used in composite calculations. Results show that surface fluxes response is weakest in western Indian Ocean. Box 40-50E (blue lines) show no significant variability associated with Kelvin waves passage and box 50-60E show that magnitude of the changes in surface fluxes is smaller in comparison with central and eastern

Indian Ocean. It can be seen that surface fluxes are at the same level 2 days prior and 2 days after Kelvin wave passage, but the actual values vary zonally. Areas east of 60E exhibit consistent increase in wind speed and latent heat flux during the Kelvin wave activity. Composite analysis shows more than 0.5 ms$^{-1}$ increase in surface wind speed and more than 10 Wm$^{-2}$ for all boxes east of 60E. Maximum values are achieved in 80-90E box with increase of 0.85 ms$^{-1}$ in wind speed and nearly 20 Wm$^{-2}$ in latent heat flux.

## 4.2.4. Kelvin wave impact on T$^{\dagger}$

In this section we analyze variability of the ocean surface temperature represented by T$^{\dagger}$. To this end we study the same two time periods as previously.

Figure 4.8 presents T$^{\dagger}$ for November 2011 DYNAMO case. The two studied Kelvin waves are the same as previously described and pass over 80E on November 24 and November 28, 2011. Figure 4.8 shows that derived T$^{\dagger}$ is high in the entire Indian Ocean basin during the November 11 – 21 time period. This is the period of heat gain by the ocean and MJO suppressed phase. Period between November 21 and December 21 is characterized by eastward propagation of the area of decreased upper ocean diurnal variability as measured by T$^{\dagger}$ index (Figure 4.8). This period is dominated by the November MJO event. However, it can be seen that areas of increased T$^{\dagger}$ coincide with transition period between the two sequential Kelvin waves, so the recovery of the diurnal variability is apparent shortly after wave passage. Analysis of derived T$^{\dagger}$ confirms the conclusion from Chapter 2 that within the active MJO phase, the upper ocean diurnal cycle is not entirely obliterated by convective activity but can develop on certain days in certain areas. The variability within the active phase of the November MJO event is correlated with the two sequential Kelvin waves associated with this MJO. The bulk SST changes were investigated too, but the SST variability was dominated by zonal gradient and gradual temporal changes due to intraseasonal and seasonal forcing (not shown).

Variability at 80E for this case is such that T$^{\dagger}$ prior to approach of the first Kelvin wave is of the order of 0.25-0.30 °C, during the first Kelvin wave it decreases to less than 0.05 °C, and increases during the transition period to 0.05-0.10 °C and decreases again during passage of the second Kelvin wave to less than 0.05 °C.

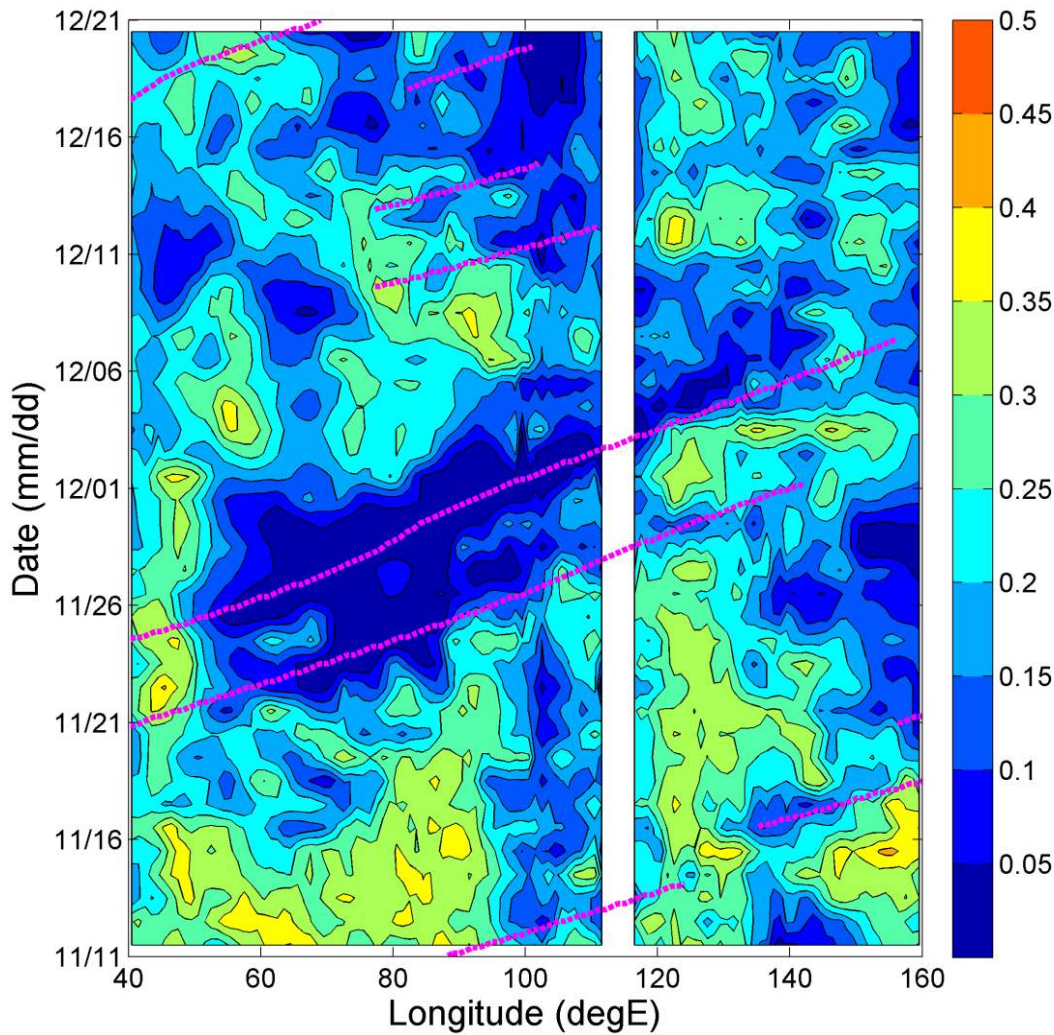

**Figure 4.8** Derived T† index for the November 11 – December 21, 2011 time period. The x-axis is longitude, y-axis is date. Color scale indicates daily averaged $T^{\dagger}$ in [°C]. Kelvin wave trajectories are marked with solid magenta lines.

Figure 4.9 presents analysis of the $T^{\dagger}$ variability in January 2000 for the weak MJO activity case. It can be observed that this period is characterized by a lot of temporal and zonal variability, with no convectively suppressed period preceding the wave development. However, Kelvin waves passage is associated with typical variability of $T^{\dagger}$, similar to that discussed above. In Figure 4.9 typical variability of $T^{\dagger}$ index associated with the two strong sequential Kelvin waves can be seen. The $T^{\dagger}$ index is decreased during passage of each Kelvin wave and increased during transition period between them. This behavior is not limited to the specific area, but accompanies the trajectories through the central and eastern Indian Ocean between 60E and 100E.

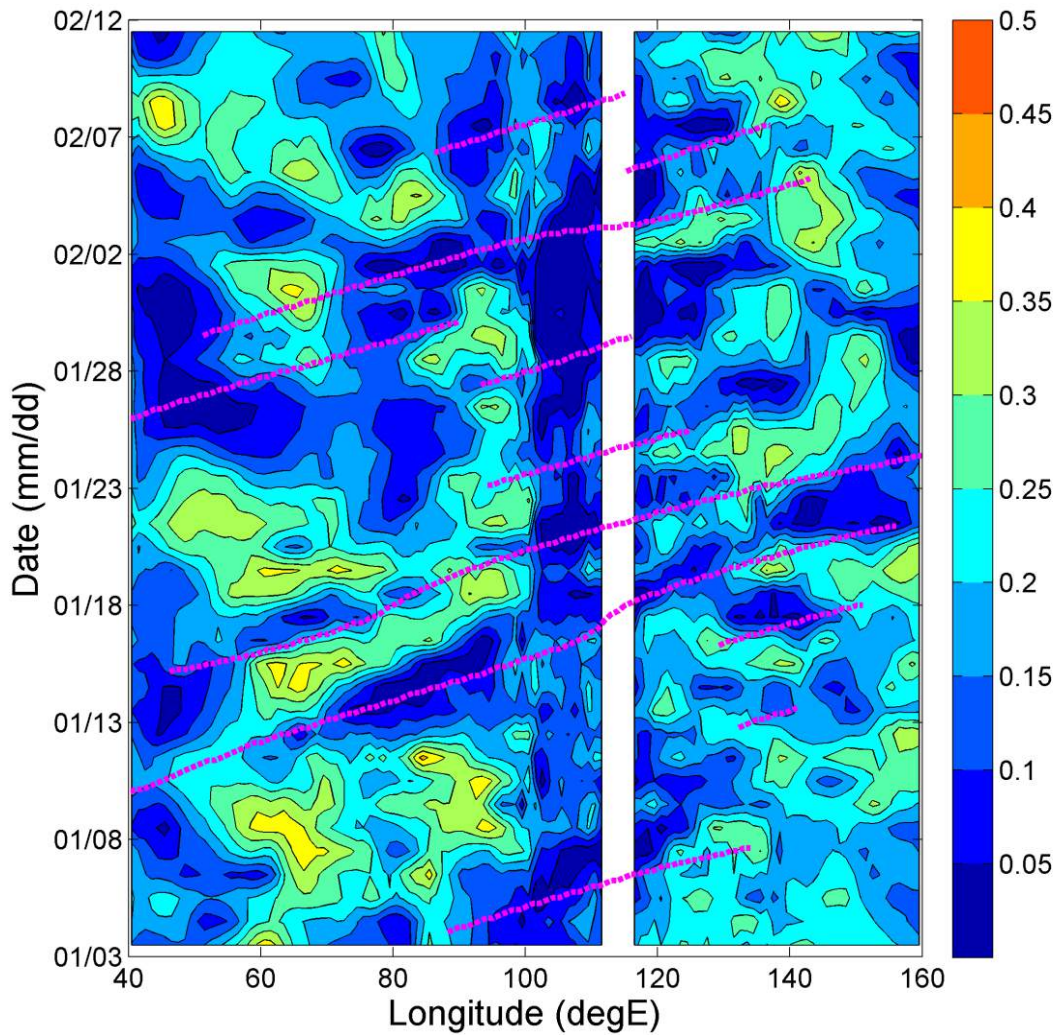

**Figure 4.9** $T^\dagger$ index for the January 3 –February 12, 2000 time period. The x-axis is longitude, y-axis is date. Color scale indicates daily averaged $T^\dagger$ in [°C]. Kelvin wave trajectories are marked with solid magenta lines.

At 80E, prior to the approach of the first Kelvin wave $T^\dagger$ index is about 0.35-0.40 °C, during the first Kelvin wave it decreases to less than 0.05 °C, increases during the transition period two waves to 0.20-0.25 °C and decreases again during passage of the second wave to 0.10-0.15 °C. Although the magnitude and absolute values of the variability associated with the two sequential Kelvin waves in this case is a little different from observed for the DYNAMO case, the overall structure of the variability is exactly the same. For the January 2000 the $T^\dagger$ index values, although clearly smaller during the passage of the second wave, are not suppressed to 0 °C. To summarize – the analysis shows that upper ocean temperature diurnal cycle exhibit substantial variability associated with Kelvin wave propagation. $T^\dagger$ is likely largely suppressed during the Kelvin wave passage and this effect may be limited to the vicinity of its trajectory. On the other hand, the overall the bulk SST structure is dominated by zonal gradient and gradual changes due to seasonal and intraseasonal forcing. The cases shown above do not exhibit large differences between the effect of the wave associated with active MJO phase, in November 2011 and no active MJO associated with January 2000 case. The previous research [*DePasquale et al.*, 2014] shows different behavior of Kelvin waves depending on MJO case. Therefore it could be expected that their effect on the ocean may also differ. To examine this relationship we will extend case studies to the composite

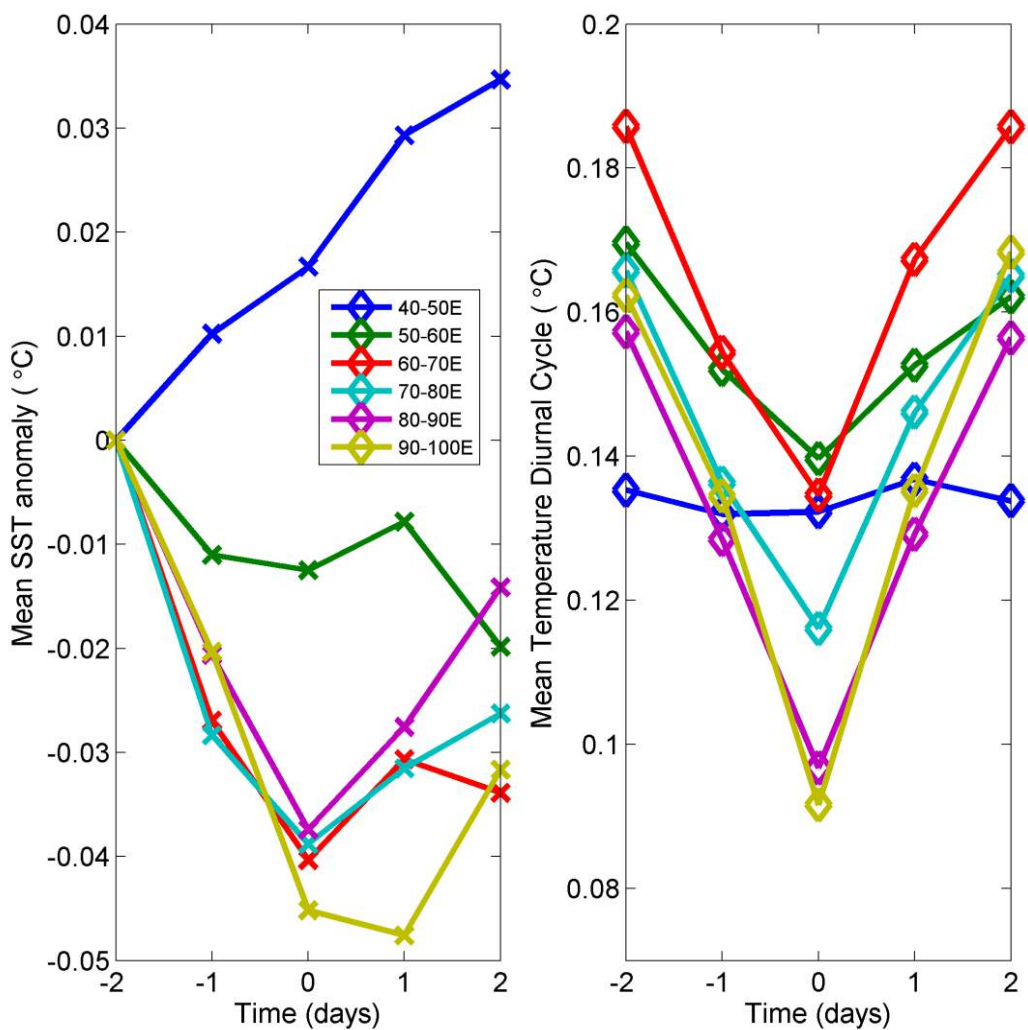

**Figure 4.10 Composite SST anomaly (left) and $T^+$ index (right) for all Kelvin waves active in the Indian Ocean. Time (x-axis) is relative to the Kelvin wave trajectory with 0 indicating time of the wave passage.**

The composite characteristics of Kelvin wave impact on bulk SST and $T^+$ based on 15 year statistics, for all waves as well as active and suppressed MJO periods. Figure 4.10 presents composite of the SST anomaly (left panel) and derived $T^\dagger$ (right panel) response to the Kelvin wave activity divided into 6 boxes, 10 degrees wide, over the entire Indian Ocean. The SST anomaly is calculated using the difference between the SST observed during the Kelvin wave passage and the SST obtained from the multiyear mean for the given longitude. Furthermore the 5 day evolution during Kelvin wave passage is presented with respect to the mean value for day -2 in order to compare the SST variability associated with Kelvin wave passage. Otherwise the variability is dominated by a zonal SST. To reduce biases associated with the FFT filtering, only trajectories active through the entire length of the box are used in composite calculation. Results show that SST and $T^\dagger$ response is weakest in the western Indian Ocean. The 40-50E box (blue lines) shows no significant variability associated with Kelvin waves passage and the 50-60E box shows that magnitude of the typical variability is smaller in comparison to that observed in the central and eastern Indian Ocean. Magnitude of the diurnal cycle and SST anomaly are increasing 2 days after a composite (mean) Kelvin wave passage, but still, they are below the values observed 2 days before the Kelvin wave. The typical values vary zonally. The areas east of 60E show consistent decrease in SST and $T^\dagger$ during the composite Kelvin wave passage. Composite analysis shows decrease in SST anomaly of the order of 0.04 °C and decrease in $T^\dagger$ of more than 0.05 °C for all boxes east of 60E. Decrease in the SST anomaly is similar over the central Indian Ocean but $T^\dagger$ change is not as uniform. The maximum impact of the Kelvin waves on the upper ocean temperature is observed in the 90-100E box where decrease in the order of 0.7 °C in mean diurnal cycle and decrease of nearly 0.05 °C in SST can be seen.

To summarize, during the transition period values of $T^\dagger$ are higher in case of January 2000 than November 2011. Structure of the variability of the $T^\dagger$ index, wind speed and LHF confirms that forcing associated with convection within the Kelvin wave has impact on the upper ocean diurnal cycle development and therefore influences the upper ocean energy budget. Convection, precipitation, strong winds and high latent heat flux associated with Kelvin wave suppress the development of the diurnal cycle. In the wake of the Kelvin wave or during the transition period between them, the convection is weaker and precipitation smaller. Areas of such conditions are associated with weak winds, small LHF and high insolation, which favor diurnal warm layer development. That suggests that although averaged mixed layer temperature is not affected by fast propagating Kelvin waves, its diurnal variability is highly sensitive. It can be suppressed fast but recovers fast, i.e. the diurnal cycle doesn't have the "wake" effect (memory). On the other hand SST recovers slowly (wake effect).

## 4.2.5. Intraseasonal modulation of subsurface response to the Kelvin wave forcing

In this section the analysis of the average responses of the wind speed, latent heat flux and $T^\dagger$ to the Kelvin wave passage is continued to address the question of the impact of the intraseasonal variability on the observed behavior. In previous sections we have shown case studies and the average response of the ocean surface to the Kelvin wave activity. We have also shown how Kelvin wave activity is modulated by intraseasonal variability. Here we will address the question of the changes in ocean surface response to the Kelvin wave passage due to different phases of the intraseasonal variability. To this end, we will analyze composites of wind speed, latent heat flux and $T^\dagger$ response to the Kelvin waves propagating during active and suppressed phases of the MJO. In the Indian Ocean basin, the MJO is active during RMMI phases 2 and 3. For the RMMI phases 6 and 7, Indian Ocean basin is the most suppressed [*Wheeler and Hendon*, 2004]. Our results suggest that, as shown in Figure 4.1, the Kelvin wave activity in the Indian Ocean basin is the highest during RMMI phase 2 and lowest during RMMI phase 7. Therefore, for active phase composites, only Kelvin waves that were propagating over Indian Ocean during RMMI phase 2 were used in calculations. For suppressed phase composites, only Kelvin waves that were propagating over Indian Ocean during RMMI phase 7 were used.

Figure 4.11 presents composite wind speed response to the Kelvin wave passage during (a) active and (b) passive phases of the MJO. It can be seen that zonal differences in response are similar for both phases. During both active and passive MJO phases the response is weakest over the western Indian Ocean, with wind speed values increasing between the 60-70E and 80-90E boxes and dropping significantly in the 90-100E box. Both composites show that wind speed increase during Kelvin wave passage. In addition wind speed 2 days after Kelvin wave passage is below those observed 2 days before its approach. However, during the active phase of MJO composite wind speed change due to the Kelvin wave passage is lager 60-70E box but smaller in 70-80E and 80-90E boxes than during the suppressed phase. Comparison between active and suppressed phase composites shows that, in general, wind speed is larger during the active phase. This is likely due to the fact that active phase of the MJO favors more convective conditions with increased wind speeds [*Wheeler and Hendon*, 2004].

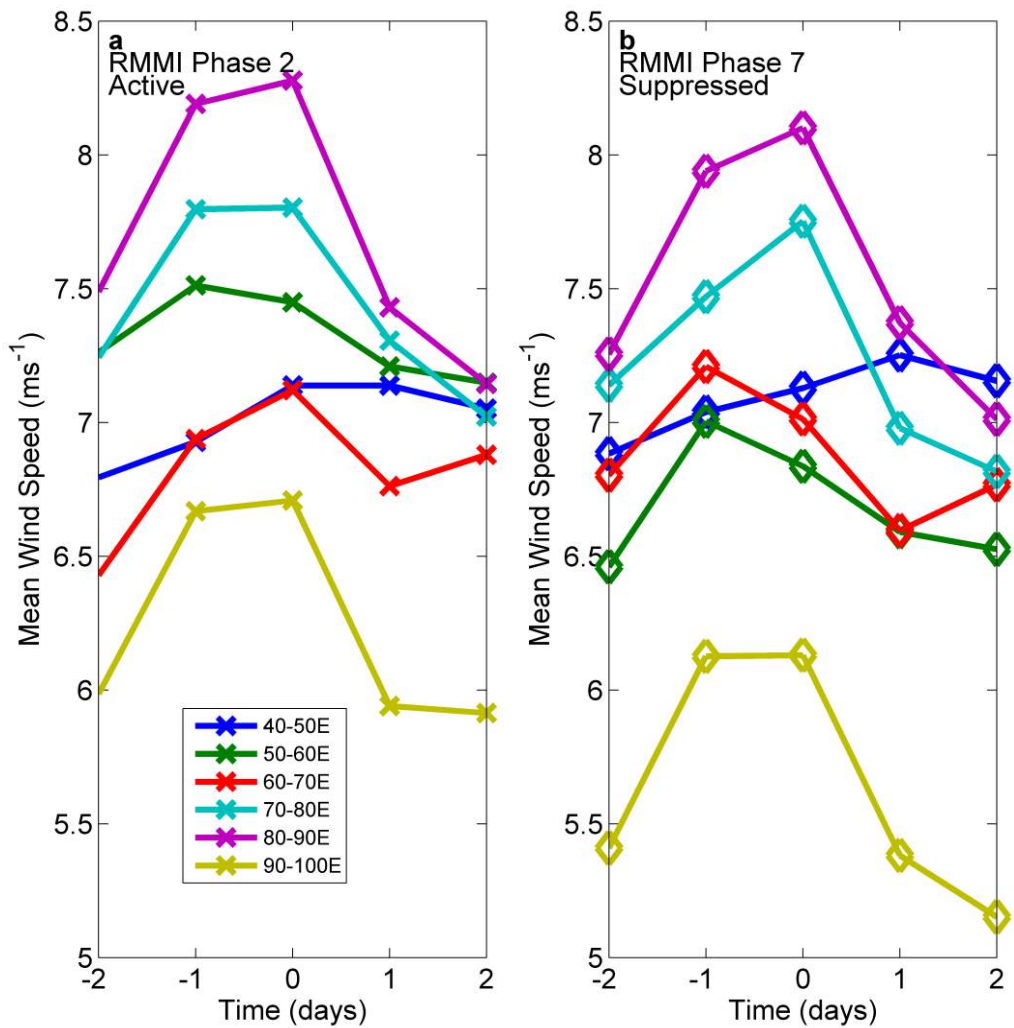

**Figure 4.11 Composite wind speed; (a) Kelvin waves during active MJO in Indian Ocean (RMMI phase 2). (b) Kelvin waves during suppressed MJO in Indian Ocean (RMMI phase 7). Time (x-axis) is relative to the Kelvin wave trajectory with 0 indicating time of the wave passage.**

Figure 4.12 presents composite latent heat flux response to the Kelvin wave passage during (a) active and (b) suppressed MJO phase. The differences are similar to those presented in Figure 4.11 for wind speed. Zonal variability is similar between active and suppressed MJO phases. Western Indian Ocean is characterized by small changes during Kelvin wave activity, the LHF values are largest in the Central Indian Ocean that is between boxes 60-70E and 80-90E However, the latent heat flux decreases significantly in 90-100E area due to the small wind speeds. In the 60-70E box change in LHF during Kelvin wave passage during the active phase is larger by about 10 Wm$^{-2}$ in comparison to the suppressed phase. In boxes 70-80E and 80-90E the change in LHF is larger during the suppressed phase by about 4 Wm$^{-2}$ in comparison to the active phase. In 90-100E box the changes are similar, but overall values are lower during the suppressed phase. This is especially true for longitudes 50-60E and 90-100E. Again, the differences in typical values of the LHF between active and suppressed phases of the MJO may be attributed to the differences in environmental conditions related to the intraseasonal variability because atmospheric convection associated with active MJO phase favors large surface fluxes.

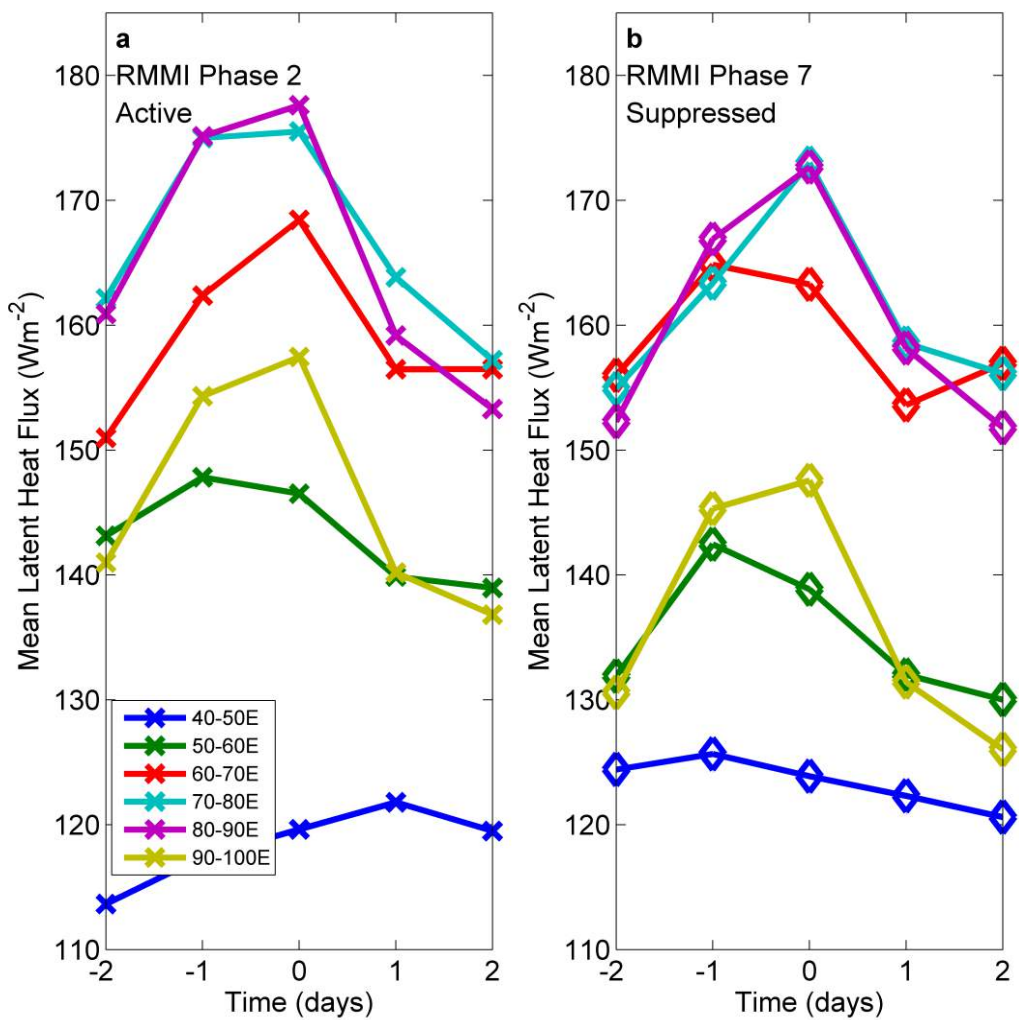

**Figure 4.12 Composite Latent Heat Flux; (a) Kelvin waves during active MJO in Indian Ocean (RMMI phase 2). (b) Kelvin waves during suppressed MJO in Indian Ocean (RMMI phase 7). Time (x-axis) is relative to the Kelvin wave trajectory with 0 indicating time of the wave passage.**

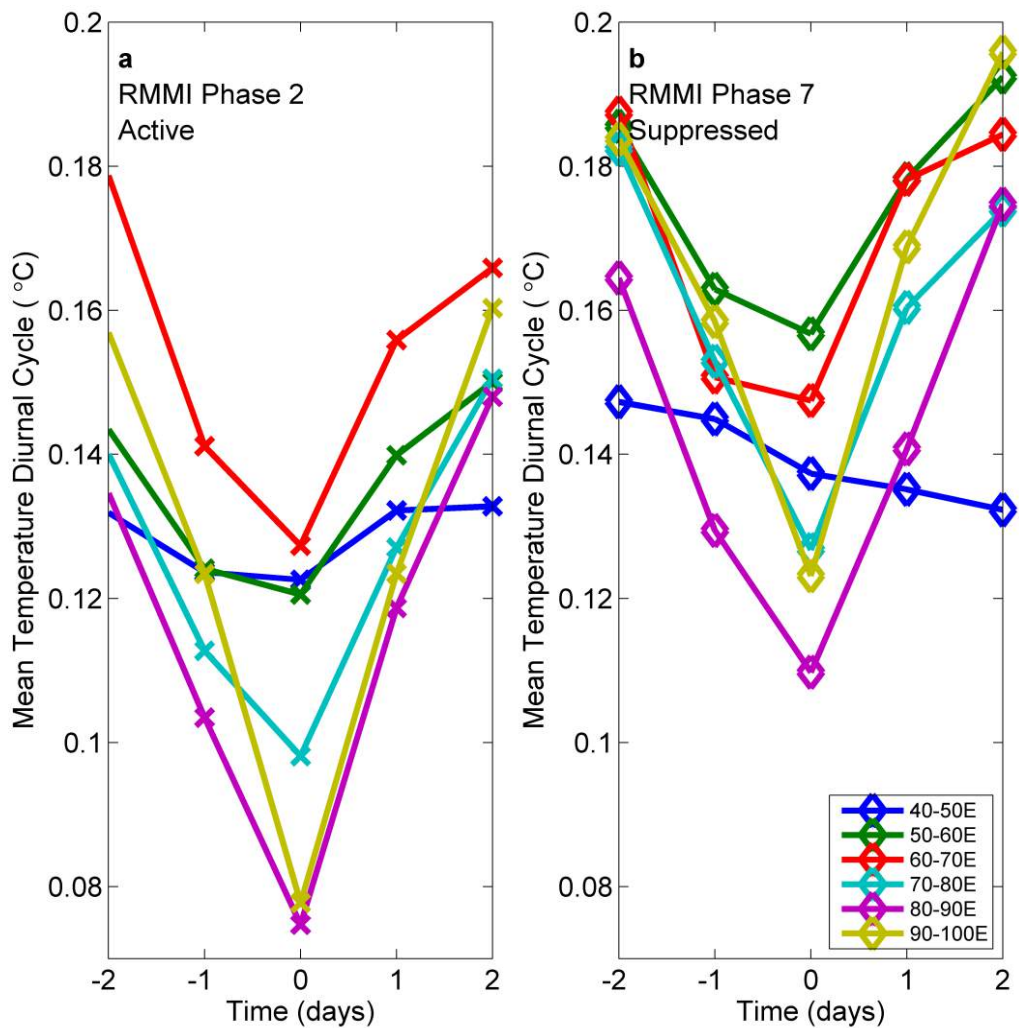

**Figure 4.13 Composite T† index; (a) Kelvin waves during active MJO in Indian Ocean (RMMI phase 2). (b) Kelvin waves during suppressed MJO in Indian Ocean (RMMI phase 7). Time (x-axis) is relative to the Kelvin wave trajectory with 0 indicating time of the wave passage.**

Figure 4.13 presents composite $T^{\dagger}$ index response to the Kelvin wave passage during (a) active and (b) suppressed MJO phases. During the active phase (Figure 4.13a) the mean diurnal cycle is weaker than during the suppressed phase (Figure 4.13b). The difference in response between the western and the central-eastern Indian Ocean is persistent in both RMMI phases. For both, active and suppressed MJO phases, the diurnal cycle response to Kelvin waves is very small for longitudes 40-50E and small for those in the 50-60E range. The strongest response is observed in central and eastern Indian Ocean, where changes in $T^{\dagger}$ exceeds 0.05 °C. During active MJO phase $T^{\dagger}$ index is decreased to below 0.1 °C in the central and eastern Indian Ocean. During suppressed MJO phase minimum value is above 0.1 °C everywhere, but the change in $T^\dagger$ is similar in both cases. This is because the overall values of $T^\dagger$ are higher during passive phase in comparison to active MJO phase. The difference of $T^\dagger$ values between phases may be explained by environmental conditions associated with intraseasonal variability. $T^\dagger$ is the magnitude of the diurnal cycle which is sensitive to the wind speed and solar radiation (Chapter 2). Active MJO phase favor higher wind speed and smaller insolation at the ocean surface. Passive MJO phase favors lower wind speed and higher insolation at the ocean surface. Therefore, active MJO phase favors lower values of the $T^\dagger$ and suppressed phase is associated with higher values of the $T^\dagger$. In spite of the difference between Kelvin waves in active and suppressed phase of MJO, especially in the phase propagation speed [*Roundy*, 2008] their impact on the ocean surface appears to be similar.

In summary, we can state that the differences in upper ocean responses to the Kelvin wave passage between the two case studies considered previously are in agreement with the difference in composite analyses of about 600 Kelvin waves for active and suppressed MJO phases. Although changes in wind speed, LHF and $T^\dagger$ have similar signature in various MJO phases, the intraseasonal variability modulates the typical response; active phase favors higher values of wind speed, LHF and $T^\dagger$ and suppressed phase is more likely associated with lower values of wind speed, LHF and $T^\dagger$. Our analysis shows, that during the active phase the mean diurnal cycle is weaker than during the suppressed phase. Surprisingly the typical variability associated with Kelvin waves is similar in active and inactive phases of the MJO. This is unexpected because Kelvin waves tend to propagate faster during the suppressed phase in comparison to those propagating in the active phase [*Roundy*, 2008]. Therefore, one could expect stronger air-sea interactions during passage of slower Kelvin waves.

## 4.3. Summary and discussion

In this Chapter atmosphere-ocean interaction during propagation of the atmospheric equatorial convectively coupled Kelvin waves is investigated. In particular we are interested in direct impact of the Kelvin wave activity on the ocean surface fluxes and subsurface temperature variability. Our study is based on in-situ measurements of the air-sea interaction during the DYNAMO, a field project in the Indian Ocean, conducted in 2011 and analysis of the atmosphere and ocean variability along Kelvin wave trajectories contained in the database which we developed for this project.

First we have analyzed the November 2011 MJO event which was extensively measured during the DYNAMO field experiment. We note that surface fluxes were disturbed by 2 Kelvin waves that were part of the MJO envelope. The results show that even though during 10-day period SST decreased by about 0.3 $^{\circ}$C, several days exhibited strong diurnal warm layer development. The magnitude of the diurnal warm layer changes in these cases was of the same order as observed synoptic cooling. Therefore, if the diurnal warm layer develops, the effects of intraseasonal cooling trend may be temporarily reversed. Furthermore, we have shown that Kelvin wave activity suppresses the diurnal warm layer development and blocks short term increases in SST. Hence, Kelvin waves are able to disturb redistribution of the energy within the oceanic mixed layer.

Our second goal was to compare two contrasting periods - November 2011 (DYNAMO) representative of the active MJO phase and January 2000 representative of the weak intraseasonal variability. Although the typical variability associated with Kelvin waves investigated in these two case studies was similar, there were differences in mean values of the fluxes. The November 2011 case was characterized by active intraseasonal convection in the Indian Ocean basin and associated with higher values of wind speed and latent heat flux, and lower values of the magnitude of the upper ocean diurnal cycle. Kelvin waves in January 2000 were associated with relatively smaller values of the wind speed and latent heat flux and higher values of the magnitude of the upper ocean diurnal cycle.

Motivated by the results for specific time periods based on the DYNAMO and January 2000 we extend our analysis and derive composite characteristics of Kelvin wave air-sea interactions based on 15 year statistics. The composites are based on all Kelvin wave trajectories identified in our database. Changes observed during the DYNAMO were found to be typical throughout central and eastern Indian Ocean.

Based on analysis of all available Kelvin wave trajectories, typical variability of the surface wind speed, latent heat flux and the magnitude of the upper ocean diurnal cycle in the Indian Ocean have been analyzed. Wind speed and latent heat flux are increased during passage of a Kelvin wave and decreased prior to its approach and after it away. The magnitude of the diurnal cycle is decreased during passage of the Kelvin wave and increased prior to its approach and after it moved away. The whole cycle is quite fast, with the diurnal variability of the upper ocean returning to the initial value after 2 days of wave passage. The results show the zonal inhomogenity in ocean surface response to the Kelvin wave activity. The typical variability is very small in the western Indian Ocean and strongest in the eastern Indian Ocean. Our results show that in the eastern Indian Ocean, during the Kelvin wave passage wind speed increases of about 0.85 ms$^{-1}$, latent heat flux increases of about 20 Wm$^{-2}$ and magnitude of the upper ocean diurnal cycle decreases of about 0.7 °C in comparison to pre-Kelvin wave conditions.

Finally we have investigated impact of the MJO phase on the ocean surface response to the Kelvin wave passage. The Kelvin wave activity in the Indian Ocean basin with respect to the intraseasonal variability showed substantial changes between various RMMI phases. In RMMI phase 7 (suppressed convection in the Indian Ocean) the number of Kelvin wave trajectories in the Indian Ocean decreases by about 45%, compared with RMMI phase 2 (which is associated with largest Kelvin wave activity). Thus, RMMI phase was found to modulate the upper ocean response to the Kelvin wave passage. Although typical variability, including magnitude of the changes forced by Kelvin wave passage, is similar in both active and suppressed phases, the mean state varies between them. Active phase favors stronger convection and larger precipitation. Therefore it is associated with stronger wind speeds, larger latent heat flux, smaller insolation and weaker upper ocean diurnal cycle. Suppressed phase favors calm and clear sky conditions and it is associated with lower wind speeds, lower latent heat flux, higher insolation and stronger upper ocean diurnal cycle.

Our results show that although Kelvin waves are characterized by fast propagation and their effect on daily SST is small in comparison to zonal, intraseasonal and seasonal changes, they have noticeable impact on the ocean surface fluxes and short term upper ocean energy distribution. We have found that Kelvin wave activity is able to completely suppress upper ocean diurnal cycle and limit impact of the warm layer development on the air-sea fluxes. This is independent but consistent with results from Chapter 2. Intraseasonal variability not only modulates number of the Kelvin waves in the Indian Ocean basin, but it also modulates the variability of the ocean response as well.

# Chapter 5. Phase locking of convectively coupled equatorial atmospheric Kelvin waves over Indian Ocean basin

## 5.1. Introduction

The Maritime Continent is located within the tropical warm pool and separates Indian Ocean Basin from the Western Pacific. It includes the archipelagos of Indonesia, New Guinea and Malaysia surrounding shallow seas with some of the warmest ocean temperatures in the world. The importance of the Maritime Continent to the global climate was identified by Ramage [1968]. The strong convection in this region provides the heat source for circulation anomalies and through teleconnections influences the global circulation [*Neale and Slingo*, 2003].

In MJO modeling, the Maritime Continent is recognized as a "predictability barrier" due to inability of dynamic models to properly represent the passage of MJO through combination of shallow seas, islands and large topographic features [*Seo et al.*, 2009]. Since this problem always exists in dynamic forecasts but is not evident in the statistical models is was even proposed that dynamic models should be augmented by the results of statistical forecast to improve the MJO predictions [*Seo et al.*, 2009]. In the recent TIGGE ensemble evaluation progress was shown in the MJO modeling [*Matsueda and Endo*, 2011] but the predictability barrier persisted even in the best of the forecasts. It was hypothesized previously [*Inness and Slingo*, 2006] that propagation of MJO through the Maritime Continent depends on ability CCKW to cross the Maritime Continent barrier. CCKW are important part of the atmospheric equatorial dynamics. In fact, near the equator CCKW are the leading modes of eastward moving convection on time scales between several days to three weeks [*Kiladis et al.*, 2009]. CCKW, together with other equatorial eastward and westward propagating perturbations form "building blocks" of the active phase of Madden Julian Oscillation [*Majda and Khouider*, 2004; *Mapes et al.*, 2006]. The orographic effects of Maritime Continent, especially Sumatra act to block the propagation of MJO by damping the Kelvin wave response, that is surface pressure signal and easterlies to the east of the enhanced convection [*Inness and Slingo*, 2006]. This effect is enhanced in dynamic models contributing to the "predictability barrier". It was also shown that convective [*Ridout and Flatau*, 2011a; b] events that modify the "dynamic equator" can allow CCKW to propagate around Sumatra, contributing to MJO eastward propagation.

One of the important characteristics of Maritime Continent convection is the very strong diurnal variability that could interfere with convective heating of the approaching MJO. It became possible to resolve the diurnal variability over the Maritime Continent with the availability of the TRMM data [*Fujita et al.*, 2011; *Ichikawa and Yasunari*, 2007; *Teo et al.*, 2011]. Recently, it was shown [*Peatman et al.*, 2014] that eastward propagation of MJO envelope breaks down over the Maritime Continent islands, with the enhanced precipitation related to the strong diurnal cycle "jumping" from island to island during the MJO suppressed phase preceding the development of convection.

In this paper we study of interaction between Kelvin waves and diurnal variability of convection in the Indian Ocean basin and over the Indonesia to further shed a light on the problem of contribution of Kelvin waves to MJO propagation through the Maritime Continent. In Chapter 4 climatology of propagating atmospheric CCKW with the emphasis on air-sea interaction and the impact of Kelvin waves on underlying ocean surface has been studied. We created special Kelvin wave database augmented by relationship between the atmospheric surface fluxes and upper ocean behavior (Chapter 2, Appendix C,D). In this Chapter we focus our attention on the diurnal variability of convection in the Kelvin waves as they cross the Indian Ocean and approach Sumatra.

First, we discuss the structure of the high spatial and temporal resolution Kelvin wave database developed using methodology described in Appendix C and used in this project. Next we show climatological features of Kelvin waves that may impact the propagation through the Maritime Continent. In the following section we show how the phase locking between the diurnal variability of convection in Kelvin wave and over the Maritime Continent can impact its propagation.

## 5.2. Precipitation data and AmPm Index

Here we describe an index based on full TRMM precipitation that will be used to quantify effect of Kelvin wave effect on precipitation patterns in the Indian Ocean basin. Development of the full TRMM dataset is explained in Appendix C. Figure 5.1 presents the mean diurnal cycle of the full precipitation. We use full 15-year long record of 3-hourly data from full TRMM dataset. For every available longitude and every 3-hourly interval, the 15-year long record of precipitation is averaged. In form of an equations it can be presented as

$$\overline{P}_{DC}(x_i, h_j) = mean(P(x = x_i, h = h_j)) \qquad (5.1)$$

where $\overline{P}_{DC}$ is diurnal cycle of precipitation, P is full TRMM database, x is full longitude vector and h is full UTC hour vector. $x_i$ and $h_j$ are specific longitude and 3-hourly interval.

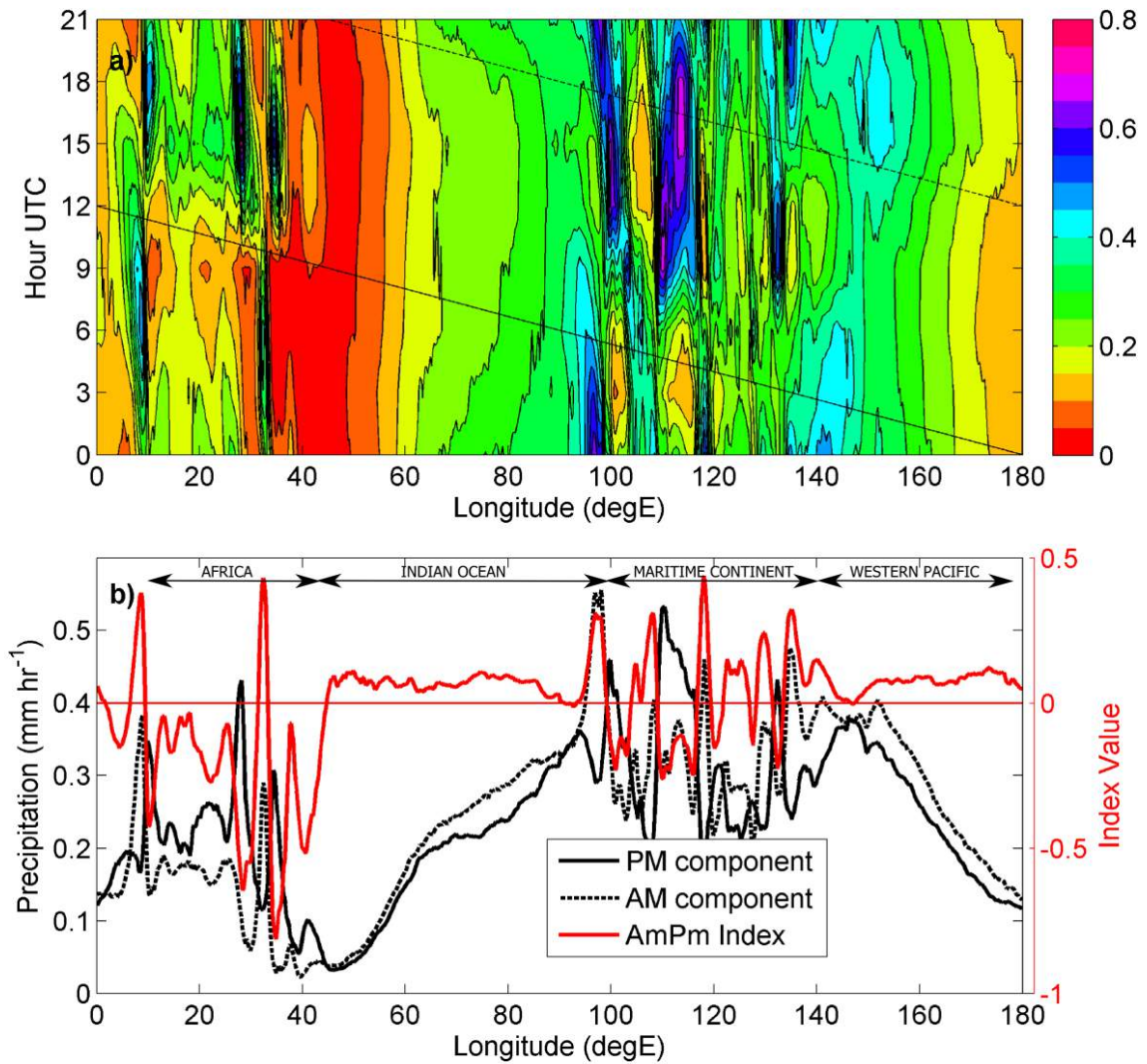

**Figure 5.1 (a) The diurnal cycle of the total precipitation based on full TRMM database. The x-axis is longitude; the y-axis is UTC hour of the day. Local noon is marked with black solid line and local midnight is marked with black dashed line. (b) average local nighttime precipitation (am) and average local daytime precipitation (pm) based on (a). Black dashed line defines "am" precipitation, black solid line defines "pm" precipitation. The x-axis is longitude; y-axis (black, on the left) is magnitude of the signal. The red line defines value of the AmPm index (y-axis on the right).**

Figure 5.1a presents full precipitation diurnal cycle, for which the x-axis is longitude, y-axis is time of the day in hour UTC. Local solar time (LST) noon is marked with black solid line

and LST midnight is marked with black dashed line. For any longitude, anything between the solid and dashed lines is LST afternoon and evening. Hereafter it is referred to as "daytime" or "pm". For any longitude, anything above the black dashed line and below the black solid line is LST night and morning. Hereafter it is referred to as "nighttime" or "am". Figure 5.1a is not a regular Hovmöller diagram, because the y-axis represent an hour of a day so it imposes periodic boundary conditions on the top and bottom of this plot. Different color scale is used to distinguish between this type of longitude – time diagrams and regular Hovmöller diagrams (f.e. Figure 5.2). The analysis presented in Figure 5.1a is simplified by dividing the signal into nighttime ("am") and daytime ("pm") components presented in Figure 5.1b. At any longitude nighttime component is average of LST "am" signal and daytime component is average of LST "pm" full precipitation presented in Figure 5.1a. Figure 5.1b has longitude on the x-axis and magnitude of the nighttime and daytime part of the full precipitation on the left, black y-axis. The nighttime component is presented with solid black line and daytime component is presented with dashed black line. It can be seen that over oceans nighttime precipitation dominates over daytime component. An example is Indian Ocean. Although both diurnal and nocturnal components increase towards east, the nighttime component dominates throughout the entire basin. Interestingly continuous increase in both components and slight domination of the nocturnal precipitation breaks in vicinity of the Maritime Continent. Area directly to the west of the island of Sumatra, which is the first of major land masses within Maritime Continent looking from Indian Ocean in eastward direction, is characterized by strong decrease in daytime precipitation and at the same time strong increase in nighttime component. It results in strong domination of the nighttime precipitation. Although the longitude 97E, which is the area where the difference between the two components is the largest, is still within Indian Ocean, the rapid changes in both precipitation components is likely an implication of the interaction with Maritime Continent.

The area of Maritime Continent is characterized by changes in domination between nighttime and daytime components. The areas 100-104E and 109-116E are dominated by daytime components. These two locations agree with positions of the two major Maritime Continent islands at the Equator: Sumatra and Borneo. The ocean area between them is dominated by nighttime precipitation. Throughout the Maritime Continent the differences between nighttime and daytime precipitation are larger than over ocean. Such strong variability extends to the longitude 140E. Although there are no more major land masses at the Equator east of the island of Borneo, the northern tip of the island of New Guinea is at 1.42S and therefore may impact observed behavior. East of longitude 140E is ocean area of Western Pacific which is characterized by gradual decrease in both nighttime and daytime components of the precipitation and domination of the nighttime component.

The area of Arica, which extends from 9E to 43E, is dominated by daytime precipitation. The only exemption to the domination of the daytime precipitation over Africa is at the 32E. This location agrees with the location of Lake Victoria, which is second largest freshwater lake in the Earth.

Clear differences in diurnal cycle between ocean and land surface areas are visible. In order to improve quantitative analysis, we present an AmPm Index which has a following form

$$AmPm = \frac{Var_{am} - Var_{pm}}{P_{am} + P_{pm}}, \qquad (5.2)$$

where in the numenator $Var_{am}$ and $Var_{pm}$ refer to nighttime and daytime components of the variable which can be either full of filtered precipitation; in the denominator Pam is nighttime component of the full precipitation, $P_{pm}$ is daytime component of the full precipitation; AmPm is value of the index. Because full precipitation is always positive, the denominator in Eq. (5.2) is positive too. Therefore, if the nighttime component dominates over daytime one, the AmPm index is positive. If daytime component is larger than nighttime component, the AmPm index has negative values. The magnitude of the index indicates the relative difference between the two components: the greater the difference, larger the value of the absolute of the AmPm index. In boundary case for full precipitation, if the precipitation is solely daytime or nighttime, the index would have absolute value of one. In boundary case for filtered precipitation, if the precipitation is solely daytime or nighttime and solely due to filtered precipitation component, the index would have absolute value of one.

It can be seen in Figure 5.1b that AmPm Index calculated for the mean diurnal cycle of the precipitation very well represents key features. The AmPm Index has its y-axis (red) on the right side of the figure. The value of the index is above 0 over Indian Ocean which is reflection of the domination of the nighttime component over daytime component. Over Maritime Continent, the AmPm Index changes signs in the multiple locations, which represents zonaly varying excess of either nighttime (positive AmPm Index values) or daytime (negative AmPm Index values) components. In our opinion, the complex structure of zonaly varying diurnal cycle over land and ocean are well reproduced by this measure.

## 5.3. Kelvin wave longitude-diurnal cycle phase locking

### 5.3.1. Individual Kelvin wave case study

Example period of May 2012 is presented in Figure 5.2 and Figure 5.3. This is an example of the comparison between filtered precipitation from TRMM_k14 (Figure 5.2) and full precipitation (Figure 5.3). The full and filtered precipitation data preparation is provided in Appendix C. On both figures the magenta lines indicate calculated Kelvin wave trajectories. Additionally set of 2 dashed black lines indicate 24 hour window around the trajectory at each longitude. By definition, the trajectory follows continuous, eastward propagating ridges (local signal maximums) which exceed threshold of 0.15 mmhr$^{-1}$ rain rate in filtered precipitation. Comparison with full precipitation shows overall good performance of trajectory calculation algorithm. The Kelvin wave of interest begins at 52E on November 26, 2012. The TRMM_k14 value associated with this Kelvin wave increases and achieves maximum value of 0.55 mmhr$^{-1}$ at 120E. The trajectory ends at 147E, which means that this particular Kelvin wave made successful transition into Western Pacific. The full precipitation pattern associated with this event shows that precipitation begins at 58E. Although it varies, the overall structure is strong and coherent along the entire trajectory.

It should be noted that trajectory based on TRMM_k14 filtered precipitation begins few degrees west from the area where full precipitation patterns begins. From the area where full precipitation begins (58E) eastward, the TRMM_k14 filtered precipitation represents very well eastward propagating part of the full precipitation. The difference between filtered and unfiltered signals in the beginning of the trajectory results from the FFT filtering. Thus, the beginnings and ends of trajectories are handled with care in sequential analysis. Also, please note that the slopes of magenta line representing Kelvin wave phase speed in these Hovmöller diagrams are different from slopes on any such diagrams in Chapter 3 or Chapter 4. However, this is because those Hovmöller diagrams presented 40 days each and Figure 5.2 and Figure 5.3 present data for 15 days only.

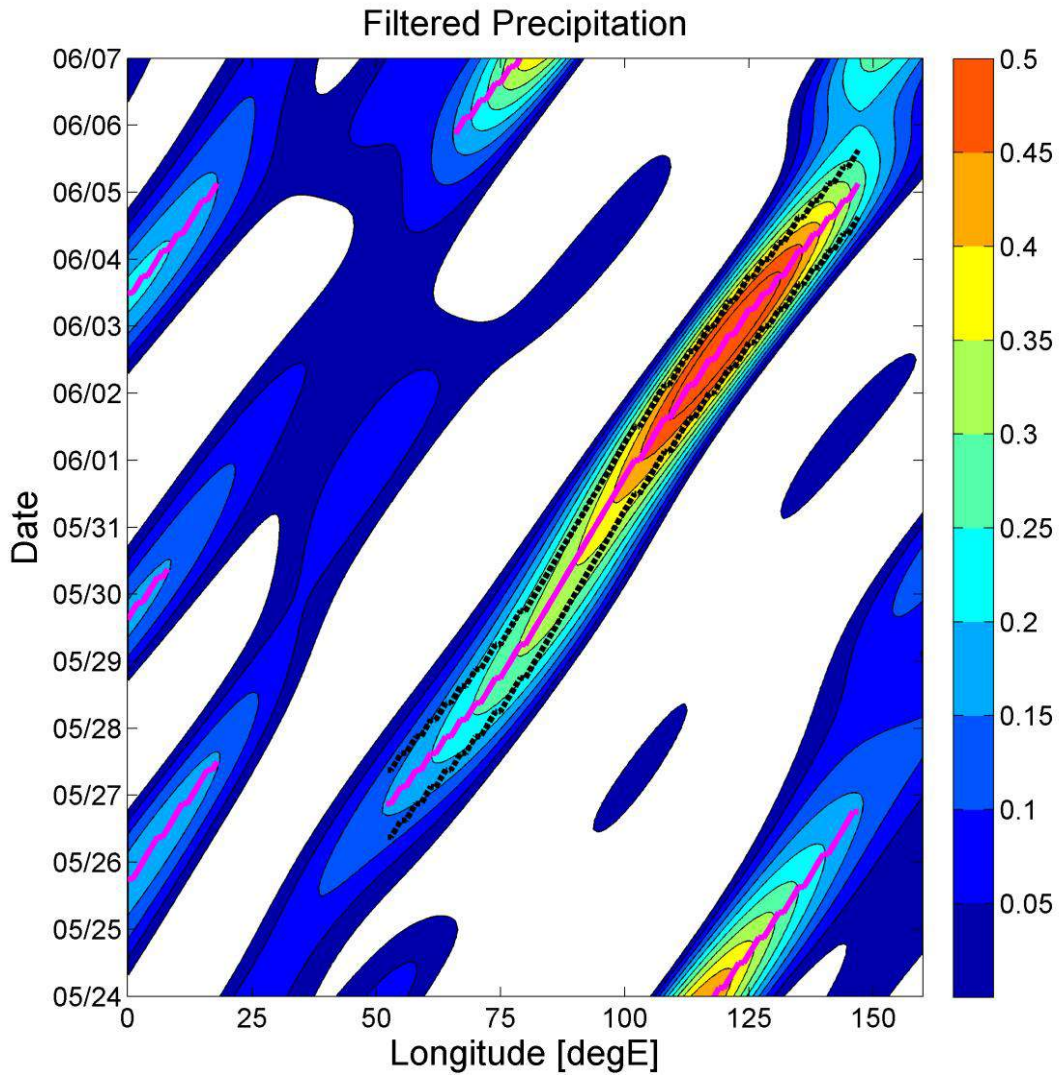

**Figure 5.2 Filtered precipitation during May 2012 case. Negative values are masked. The x-axis is longitude; the y-axis is date. Magenta lines connect maxima and form trajectories. Two black dashed lines indicate ±12h window with respect to selected Kelvin wave trajectory.**

Figure 5.4 and Figure 5.5 present diurnal cycle of TRMM_k40 filtered precipitation and full precipitation respectively for the example period of May 2012. The diurnal cycles were calculated using formula provided by Eq (5.1) applied to single Kelvin wave introduced in Figure 5.2 and Figure 5.3. In order to assess the diurnal cycle associated with propagating Kelvin wave, the 24h window, which has been marked in Figure 5.2 and Figure 5.3, is applied along the trajectory.

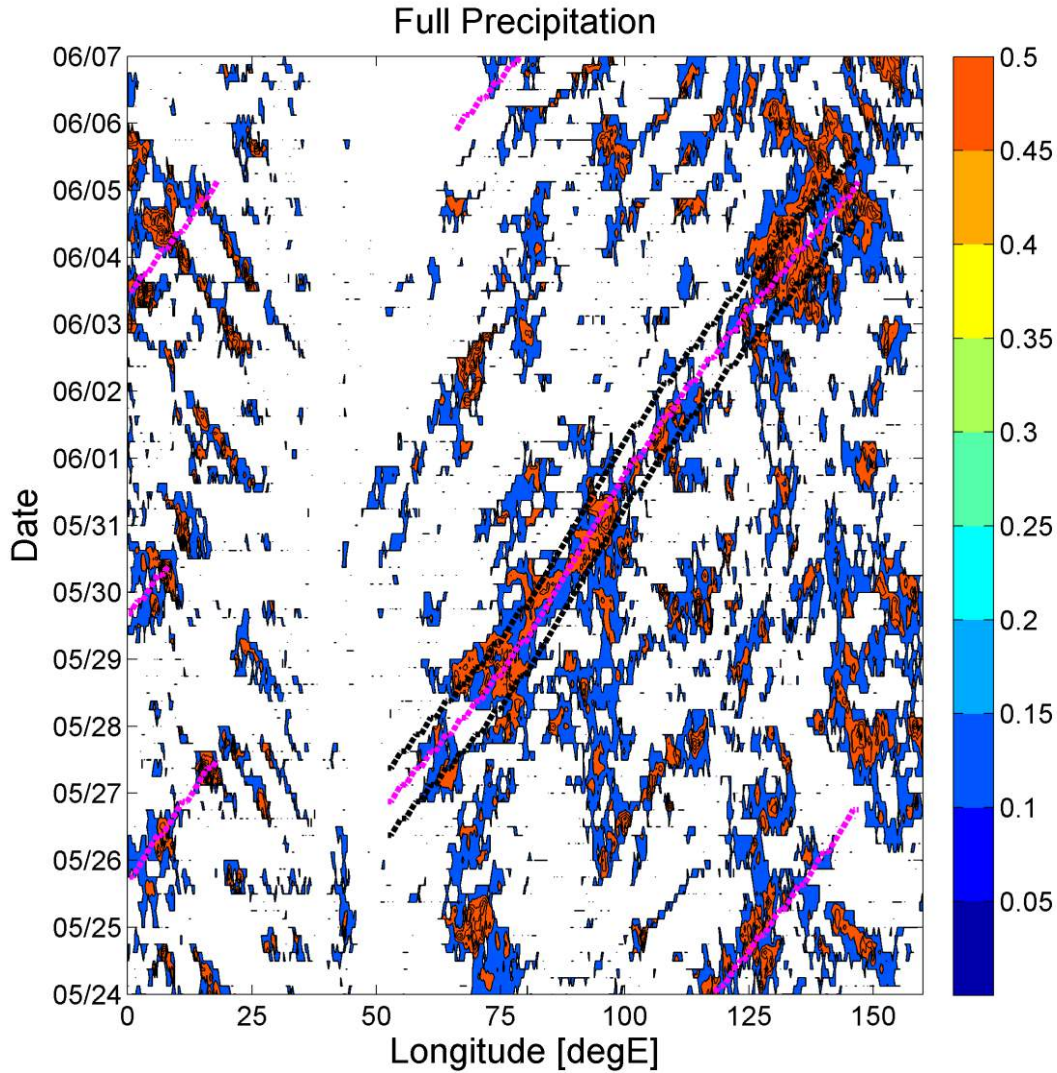

Figure 5.3 Full precipitation during May 2012 case. The x-axis is longitude; the y-axis is date. Magenta lines indicate trajectories calculated from filtered precipitation. Two black dashed lines indicate ±12h window with respect to selected Kelvin wave trajectory.

In Figure 5.4 the passage of the Kelvin wave is clearly seen. Maxima of TRMM_k40 filtered precipitation show eastward propagation and the tilt agrees with the phase speed of the Kelvin wave. Thus, the timing of the filtered precipitation over given longitude depends mostly on the timing of the wave passage over that location. It should be noted that the values of the filtered precipitation presented in Figure 5.4 are different from those presented in Figure 5.2. It is because Figure 5.2 shows TRMM_k14 filtered precipitation which was used to calculate the trajectory. Figure 5.4 presents TRMM_k40 filtered precipitation, which includes finer zonal and temporal scales.

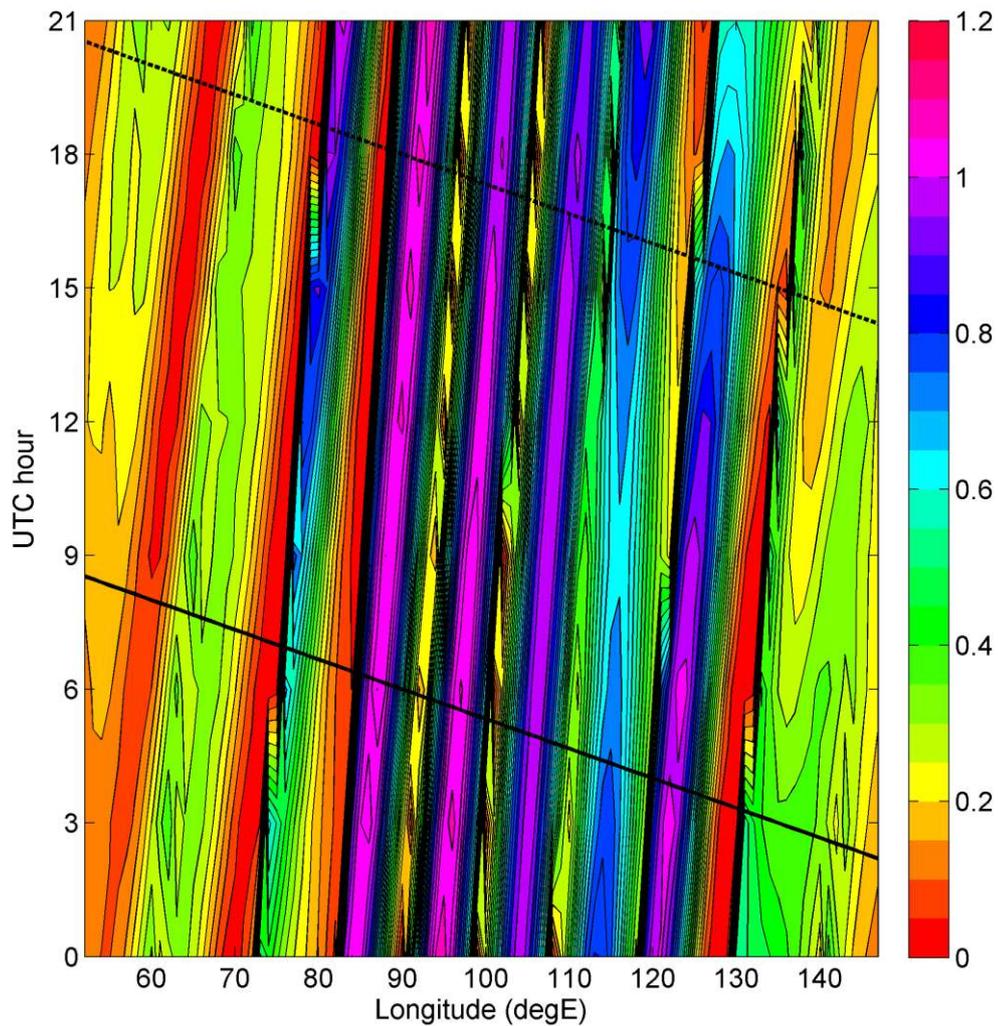

**Figure 5.4 Longitude-diurnal cycle contour plot for TRMM_k40 filtered precipitation for selected trajectory. The x-axis is longitude in degrees east; the y-axis is UTC hour of the day. Black solid line indicates local noon; black dashed line indicates local midnight. Color shading indicates precipitation in [mmhr$^{-1}$].**

Although the full precipitation along the trajectory is less coherent than filtered precipitation, Figure 5.5 shows that the trajectories are still apparent in longitude-diurnal cycle contour plot. However, eastward propagating maxima are visible over Indian Ocean and Maritime Continent, the diurnal distribution of precipitation along the trajectory agrees with the diurnal cycle of convection over land and ocean. This is especially true over the island of Sumatra and just west of it. Over Sumatra the afternoon precipitation dominated and the area just west of the island shows strong nighttime peak, which agree very well with mean diurnal cycle of full precipitation presented in Figure 5.1a.

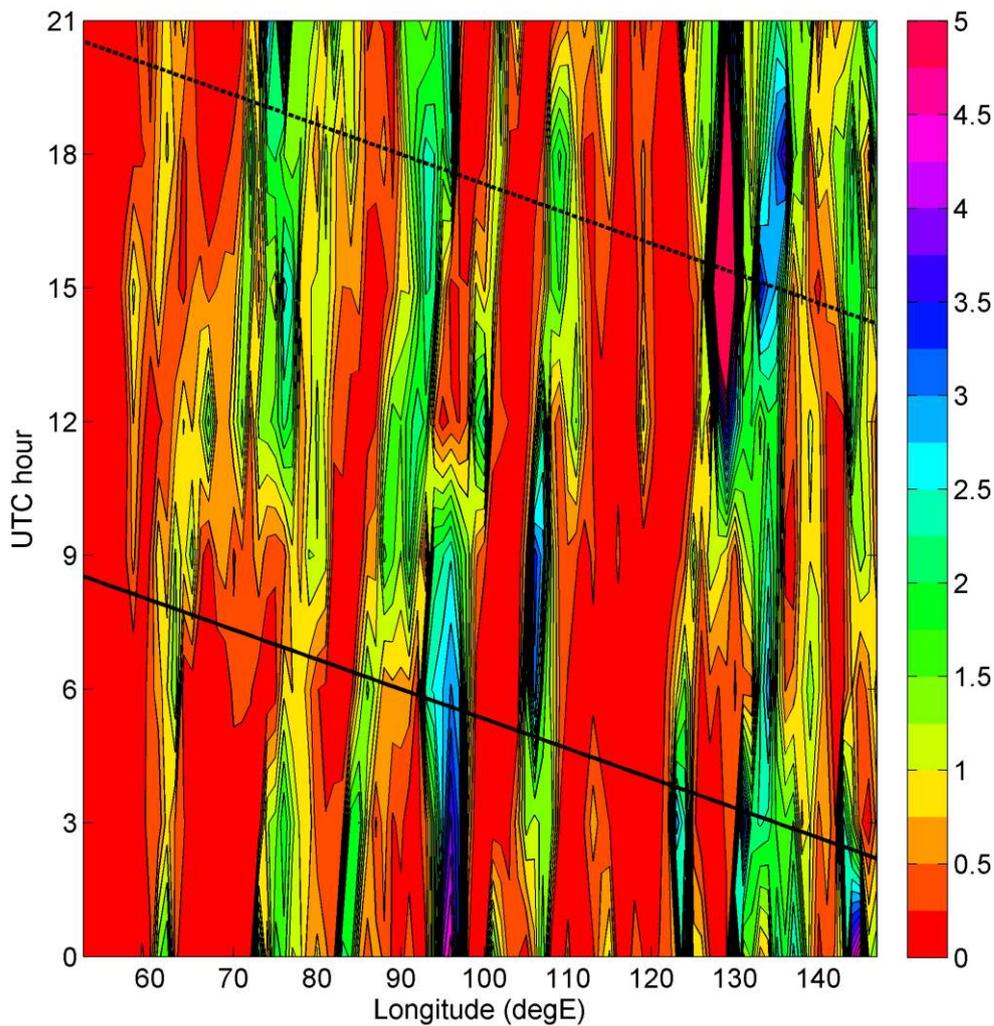

**Figure 5.5** Longitude-diurnal cycle contour plot of full precipitation for selected trajectory. The x-axis is longitude in degrees east; the y-axis is UTC hour of the day. Black solid line indicates local noon; black dashed line indicates local midnight. Color shading indicates precipitation in [mmhr$^{-1}$].

Figure 5.6 presents AmPm indices calculated for TRMM_k40 filtered precipitation (blue) and full precipitation (red). It can be seen that filtered precipitation AmPm index shows sequential maxima and minima, which indicate propagation of the wave. Distance between two sequential maxima, representing nighttime precipitation, or two sequential minima, representing daytime precipitation, is about 10 degrees, which agrees with phase speed of the Kelvin wave. Similar behavior is observed in AmPm index for full precipitation over Indian Ocean basin. The two indices show high correlation, which confirms that applied methodology works and AmPm index is good metric to study diurnal cycle in propagating waves. Over Maritime Continent differences between the two indices are larger, which may be attributed to the fact that full precipitation has larger component not associated with Kelvin wave in that region, even for single Kelvin wave event. In two locations, west of longitude 56E and at longitude 120E, absolute value of AmPm index for filtered precipitation exceeds one. This indicates that Kelvin wave component of the precipitation exceeds the full precipitation which is unrealistic. This behavior is likely artifact of FFT filtering. The area west of 56E is the beginning of the trajectory which is known to introduce some noise as described Appendix C. On the other hand, the area around 120E is where the filtered precipitation achieves maximum. Figure 5.3 shows a gap in full precipitation around that are in otherwise coherent eastward propagating signal. In this case, the due to zonal wavenumber limit on the filtering, this gap is not taken into account in filtered signal. However such behavior of the AmPm index is limited to the very specific areas and vanishes when multiple Kelvin waves are averaged. AmPm index for filtered precipitation reflects very well the behavior observed in contour longitude-diurnal cycle plots. In case of AmPm index for full precipitation the observed behavior is even enhanced, proving this metric useful in such analysis.

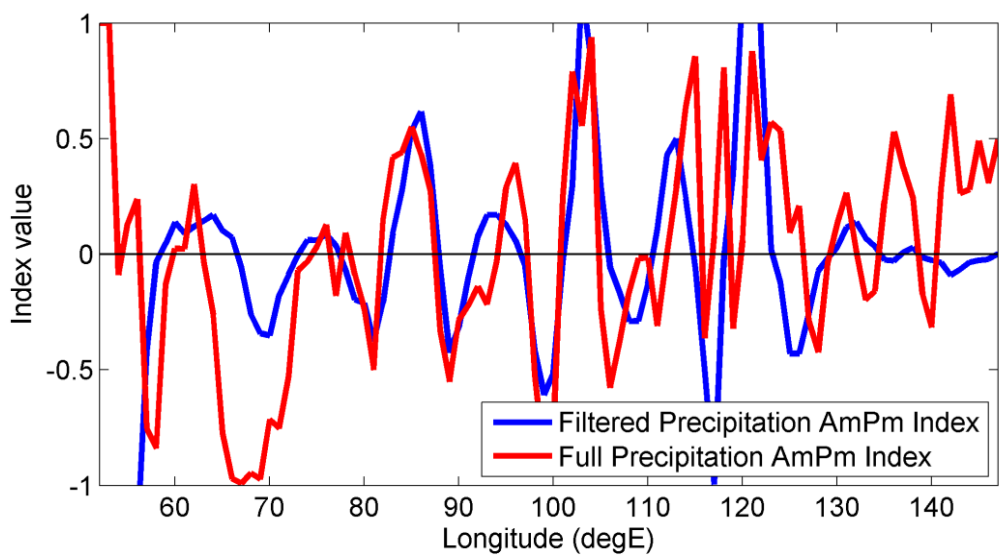

**Figure 5.6 Zonal evolution (blue line) of the AmPm index for TRMM_filt40 filtered precipitation for the selected trajectory. Red line is for zonal evolution of the AmPm index for full precipitation for the selected trajectory. The x-axis is longitude; the y-axis is value of the index. Positive values indicate domination of nighttime (am) precipitation component and negative values indicate domination of daytime (pm) precipitation component.**

To summarize: we track a single Kelvin wave. For given longitude we plot filtered and full precipitation of +/- 12 hours with respect to the trajectory (it resembles Lagrangian description of the propagating wave). For example let us consider wave which passes the longitude 80E on 29 May, 2012 at 6UTC. We use full and filtered precipitation data from 18UTC on 28 May until 18 UTC on 29 May to construct local diurnal cycle at 80E associated with this single wave. This is repeated for every longitude of this trajectory. A propagation of a single Kelvin wave is apparent in such a longitude-diurnal cycle diagram of full and filtered precipitation. One can see in these plots an eastward moving signal and the phase speed can be calculated from the tilt of the consecutive maxima and minima. For the filtered precipitation the diurnal variability at the single longitude is not an effect of the local variability of convection but it is due to propagation of the convectively coupled Kelvin wave. For the full precipitation both effect of moving wave and local variability are included. This difference is due to the fact that filtered precipitation contains signal from this Kelvin wave only. Although the full precipitation is calculated in the vicinity of the Kelvin wave, it includes other components of convective variability. Thus, for full precipitation contributions from local variability and waves other than Kelvin waves are included. The AmPm indices for full and filtered precipitation behave similarly in the case of a single Kelvin wave. Characteristics of wave propagation are reflected in sequential maxima and minima in AmPm index.

### 5.3.2. Mean diurnal cycle of Kelvin wave precipitation

In this section the analysis of the diurnal cycle of the filtered precipitation and full precipitation is extended to the whole database. All available data from 15-year-long databases are used.

Although typical diurnal and zonal variability associated with single Kelvin wave trajectory is not surprising, the existence of such diurnal and zonal variability associated with average of many Kelvin waves would be highly interesting. We have shown that Kelvin waves have certain areas of origin and that one of such areas is Indian Ocean. However, the Kelvin waves can develop over large part of this region. Therefore one would expect the diurnal distribution of the Kelvin wave activity due to their propagation to be random. If it were true, multiple Kelvin waves would be active at certain location at different times of the day. The opposite would mean that Kelvin waves tend to be active at certain longitude at certain time of the day. In such a case, the signal from multiple Kelvin waves would add, recreating the pattern similar to shown for individual Kelvin wave (Figure 5.4). To determine whether the zonal and diurnal distribution of Kelvin waves activity is random or not, we perform composite analysis of all available Kelvin wave data. The goal of this analysis is to examine the existence of the longitude-diurnal cycle phase locking of Kelvin waves.

Analysis of the filtered precipitation from the entire TRMM_filt40 database is presented in Figure 5.7. In the absence of the diurnal cycle in filtered precipitation, the signal averaged independently for every location and UTC time of the day should be more or less uniform. That would represent random diurnal distribution and reflect random passage of Kelvin waves. However, the analysis presented in Figure 5.7 shows coherent diurnal structure that is visible throughout much of the presented region. One can see that the entire presented region, between longitudes 0E and 140E, exhibits the structure of the diurnal and zonal variability similar to one observed for the single Kelvin wave (Figure 5.4). This coherent structure has a form of sequence of the alternating maxima and minima of the filtered precipitation with individual "stripes" tilted rightward. This rightward tilt represents eastward propagation and its angle agrees with the phase speeds of a convectively coupled Kelvin wave. For the specific location or for specific UTC hour, this structure has a form of oscillation in time or longitude respectively. Magnitude of this oscillation is highest between 100E and 120E. The magnitude of the oscillation is also increased between longitudes 0E and 20E. Both of these regions agree with the location of land areas of Maritime Continent and Africa. Although such coherent structure is visible over the entire region, the relative magnitude varies zonally. The magnitude of the signal is smaller over ocean, but the longitude-diurnal cycle locking is still visible.

The amplitude of the diurnal oscillation of the filtered precipitation is bigger than 0.1 $mmhr^{-1}$ over Maritime Continent and about 0.03 $mmhr^{-1}$ over Africa. The amplitude of the diurnal oscillation over ocean is in order of 0.015 $mmhr^{-1}$. The magnitude of the diurnal oscillation over the Indian Ocean may seem small, but one needs to remember that this is 15 year average of the filtered precipitation from the entire TRMM_filt40 database which included nearly 2000 Kelvin waves. Therefore despite the low amplitude, oscillatory character of the diurnal cycle over Indian Ocean is highly surprising, because it indicates that Kelvin waves tend to pass over certain locations at certain times of the day. High values of the diurnal oscillation over Maritime Continent and Africa are likely results from the strong diurnal cycle of precipitation over land (Figure 5.1).

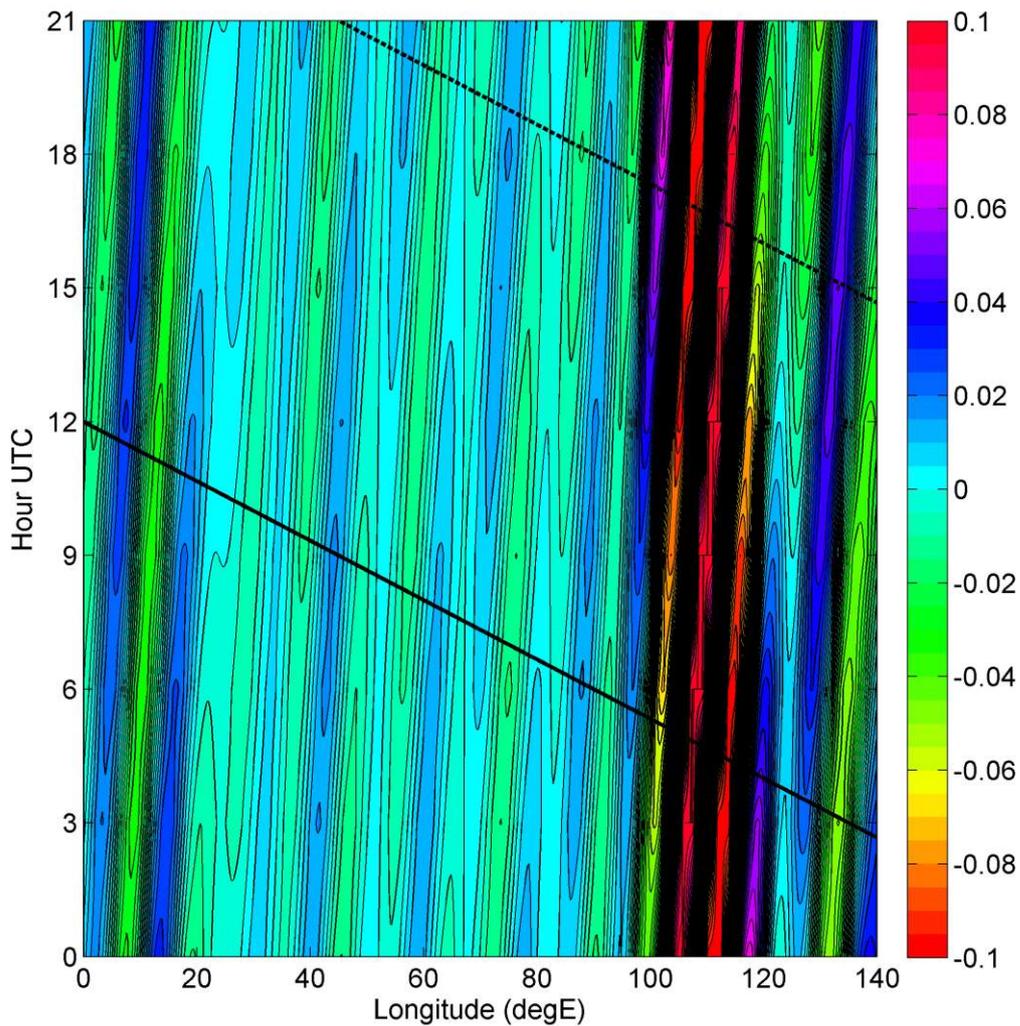

Figure 5.7 The mean diurnal cycle of the filtered precipitation based on TRMM_filt40 database. The whole available record of TRMM_filt40 database is used. The x-axis is longitude; the y-axis is UTC hour of the day. Local noon is marked with black solid line and local midnight is marked with black dashed line. Color shading from -0.1 to 0.1 indicates precipitation anomaly related to Kelvin waves in [mmhr$^{-1}$]. Tilted "stripes" indicate longitude-diurnal cycle phase locking.

Figure 5.8 presents the diurnal cycle of the filtered precipitation from TRMM_filt40 database limited to the vicinity of all Kelvin wave trajectories, which were active between 0E and 140E. Such data selection excludes most of small and negative values from the TRMM_filt40. The data used in this calculation include ± 12 hours window at each longitude to represent the entire diurnal cycle around individual trajectory. Analysis presented in Figure 5.8 shows strong magnitudes of the signal over Maritime Continent and Africa. The signal over the ocean is weaker and zonal gradient of the mean filtered precipitation along Kelvin wave trajectories is visible. Western Indian Ocean (30E-60E) is characterized by the mean values in the order of 0.2 mmhr$^{-1}$ and central and eastern Indian Ocean (60E-100E) exhibit values in the order of 0.3 mmhr$^{-1}$. The zonal gradient of the mean filtered precipitation along Kelvin wave trajectories is likely due to differences in precipitation rates between western and eastern Indian Ocean. Because western Indian Ocean tend to have lower precipitation rates and eastern part of the basin is characterized by higher precipitation rates (Figure 5.1) this precipitation gradient is reflected in filtered precipitation along Kelvin wave trajectories. The tilted "stripes" representing wave propagation are visible over Africa, eastern Indian Ocean and Maritime Continent. The diurnal variability is strong over Maritime Continent and Africa. The diurnal variability of the filtered precipitation along Kelvin wave trajectories over Indian Ocean is weaker, though eastern part of the basin exhibits characteristics of the longitude – diurnal cycle locking.

Figure 5.9 presents the diurnal cycle of the full TRMM precipitation along all Kelvin wave trajectories, active between 0E and 140E. The data selected here include ± 12 hours window around each trajectory to account for the entire diurnal cycle in its vicinity. This figure can be compared to Figure 5.1, which shows the diurnal cycle of the full TRMM precipitation from the entire TRMM database. Data presented in Figure 5.9 are specific subset of the data presented in Figure 5.1, such that they are limited to Kelvin wave trajectories. In comparison to the entire TRMM dataset, the full precipitation associated with Kelvin waves has similar zonal structure of the diurnal cycle. The area of Africa (10E – 40E) is dominated by local afternoon precipitation in the range of 0.6-1.0 mmhr$^{-1}$. The area just west off the coast of Africa is dominated by high nighttime precipitation in the order of 1.6 mmhr$^{-1}$. Nighttime maximum similar in structure and strength is visible in the area off the west coast of Sumatra (95E). The Maritime Continent area shows daytime precipitation maxima at the locations of the islands of Sumatra and Borneo. The daytime maximum at 100E (Sumatra) slightly exceeds 1.0 mmhr$^{-1}$ and the daytime maximum at 114E (Borneo) has precipitation rate of about 1.1 mmhr$^{-1}$. Between the two daytime maximums, the nighttime maximum is visible at the longitude 107E with the precipitation rate of about 0.9 mmhr$^{-1}$. Across the Indian Ocean, the eastward increase in average of the full precipitation along Kelvin waves is visible and diurnal cycle is dominated by nighttime precipitation over that area between longitudes 40E and 100E. Although, the structure of the zonal and diurnal variability of the full precipitation associated with Kelvin wave activity is similar to the structure observed for the mean precipitation (Figure 5.1), the differences in magnitudes are observed. The overall magnitudes of the precipitation rate are higher for the full precipitation associated with Kelvin waves. This is because we limit the analysis to systems which are associated with high precipitation rates.

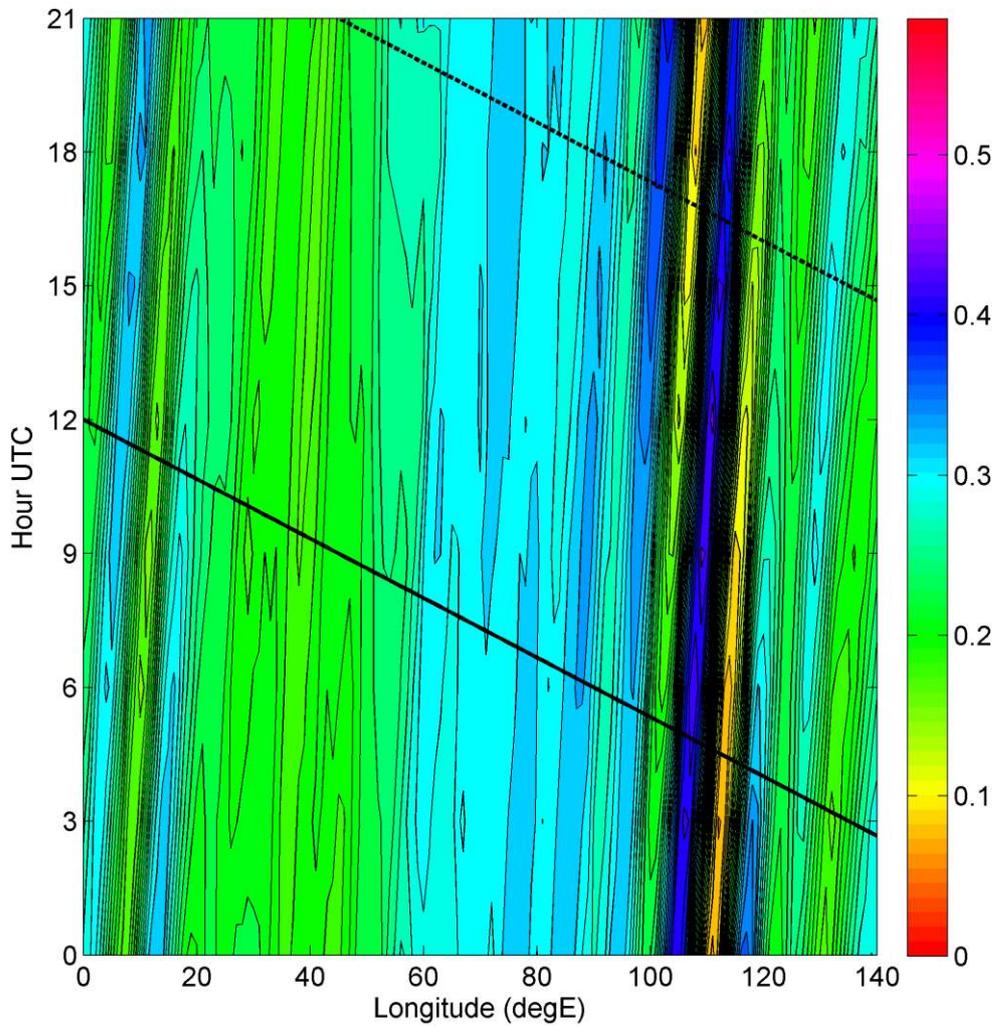

**Figure 5.8** The diurnal cycle of the filtered precipitation along all Kelvin wave trajectories based on TRMM_filt40 database. The 24hour window around the trajectory is used. The x-axis is longitude; y-axis is UTC hour of the day. Local noon is marked with black solid line and local midnight is marked with black dashed line.

Figure 5.10 presents analysis of the AmPm indices calculated for the mean diurnal cycle of the full and filtered precipitation. The blue line represents AmPm index of full TRMM precipitation from the entire TRMM dataset ("TRMM index"). It is the same index as presented in Figure 5.1b and is used here as a reference. The red line represents AmPm index of the filtered precipitation from the entire TRMM_filt40 dataset ("TRMM_filt40 index") and shows the index for the data in and Figure 5.1a. The green line represents AmPm index of full TRMM precipitation along the Kelvin wave trajectories ("TRMM KW index"). It is based solely on data presented in Figure 5.9. The black line presents AmPm index of filtered TRMM precipitation along the Kelvin wave trajectories ("TRMM_filt40 KW index"). It is based on data presented in Figure 5.8 and Figure 5.9.

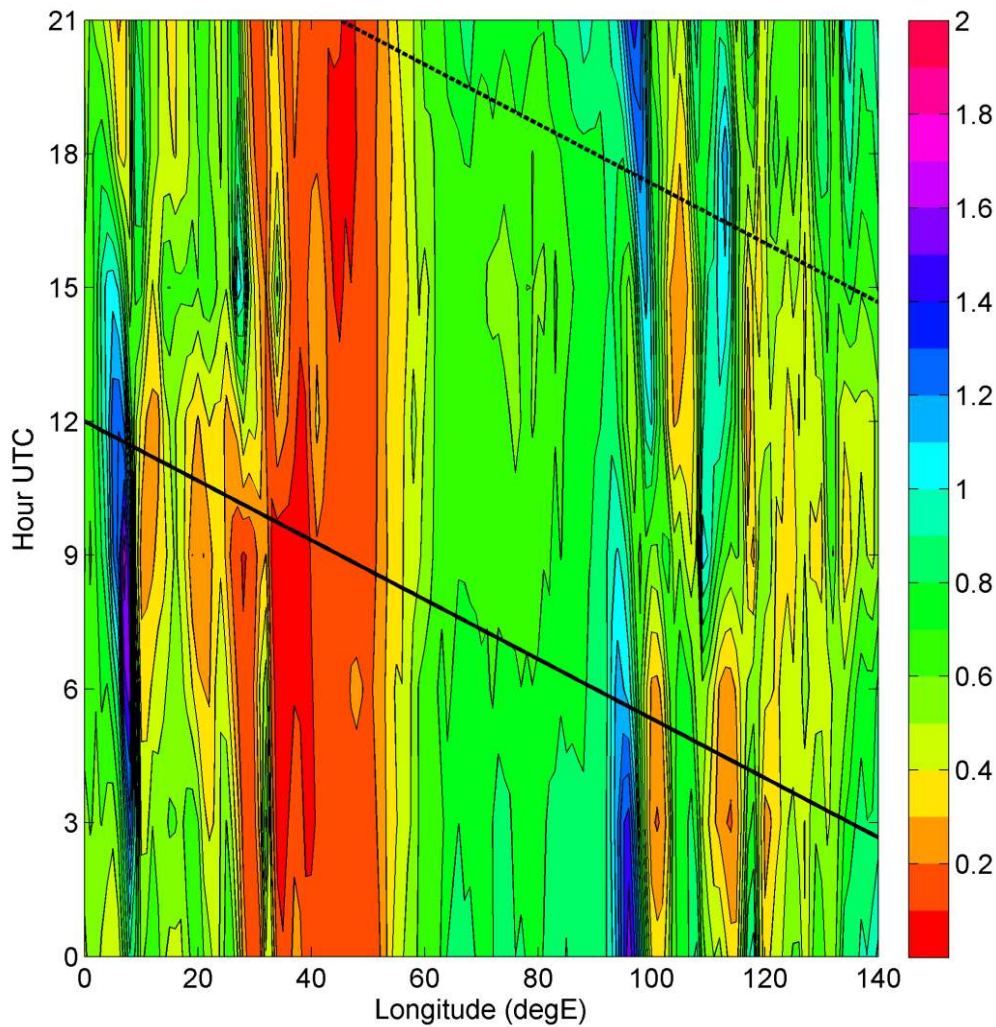

**Figure 5.9 The diurnal cycle of the full precipitation along all Kelvin wave trajectories based on TRMM database. The 24hour window around the trajectory is used. The x-axis is longitude; y-axis is UTC hour of the day. Local noon is marked with black solid line and local midnight is marked with black dashed line.**

To avoid confusion we would like to remind that our definition of the AmPm index is such that its numerator is based on filtered or full TRMM precipitation, but its denominator is always based on full precipitation. Table 5.1 presents description of the four AmPm indices including information on data used to calculate numerator and denominator. Indices calculated for filtered precipitation (TRMM_filt40 and TRMM_filt40 KW) combine in fact both filtered and full precipitation characteristics. However the sign of the AmPm index always depends on its numerator. Hence, the change of the sign of the filtered precipitation and full precipitation indices is always reflection of which component, diurnal or nocturnal, dominates in filtered precipitation or full precipitation diurnal variability.

It can be seen, in Figure 5.10, that AmPm indices reflect the complex characteristics of the diurnal and zonal variability of the filtered and full precipitation. It is apparent that over Africa and Maritime Continent AmPm TRMM index and AmPm TRMM KW index are very similar (blue and green lines). The only difference in sign between these two lines is visible between 12E and 18E, where AmPm TRMM KW index is positive and AmPm TRMM index is negative. That indicates that although that area is dominated by daytime precipitation, the precipitation in Kelvin waves tend to be stronger at night. This is related to Kelvin waves propagation, which is confirmed by the positive values of AmPm TRMM_filt40 index and AmPm TRMM_filt40 index, the two indices calculated for filtered precipitation (red and black lines). Over the Indian Ocean, the AmPm TRMM index and AmPm TRMM KW index are positive, which means that the diurnal cycle of the full precipitation in Kelvin waves is dominated by its nighttime component similarly to the mean diurnal cycle of the precipitation in that region. However, the AmPm TRMM KW index shows larger zonal variability, with the length scale of about 10 degrees, across the Indian Ocean than AmPm TRMM index. Although it stays positive, the modification of the amplitude related to Kelvin wave propagation is apparent.

It can be seen that AmPm TRMM_filt40 index and AmPm TRMM_filt40 KW index (red and black lines) behave similarly. Although they have different magnitude, they usually have the same sign. However, when the two are compared with AmPm TRMM KW index calculated, some differences can be noticed. The AmPm TRMM_filt40 KW index change sign in multiple locations over Africa and Indian Ocean, although the full precipitation is dominated by either daytime or nighttime component and AmPm TRMM KW index tend not to change its sign. This means that filtered precipitation indices better reflect propagation of the Kelvin waves. It can be also seen that over Maritime Continent, the AmPm TRMM_filt40 KW index is a little bit ahead of the AmPm TRMM KW index. This behavior is particularly apparent off the west coast of the island of Sumatra between longitudes 95E and 100E, where the AmPm TRMM_filt40 KW index is negative and the AmPm TRMM KW index is positive. However, the local minimum of the AmPm TRMM_filt40 index value at the longitude 99E is likely related to the local minimum of the AmPm TRMM KW index at 100E. The local maximum of the AmPm TRMM KW index at the longitude 97E represents strong nocturnal precipitation rate maximum just off the coast of Sumatra that can be seen in Figure 5.9. This maximum is likely related to the convection initiated over Sumatra and propagating westward.

| AmPm Index | Numerator data | Denominator Data | Meaning |
| --- | --- | --- | --- |
| TRMM | entire TRMM dataset | entire TRMM dataset | Average diurnal cycle of the full TRMM precipitation |
| TRMM_filt40 | entire TRMM_filt40 dataset | entire TRMM dataset | Average diurnal cycle of the filtered TRMM precipitation |
| TRMM KW | TRMM dataset for Kelvin wave trajectories only | TRMM dataset for Kelvin wave trajectories only | Average diurnal cycle of the full precipitation associated with Kelvin wave trajectories only. |
| TRMM_filt40 KW | TRMM_filt40 dataset for Kelvin wave trajectories only | TRMM dataset for Kelvin wave trajectories only | Average diurnal cycle of the filtered precipitation associated with Kelvin wave trajectories only |

**Table 5.1 Description of AmPm indices presented in Figure 5.10**

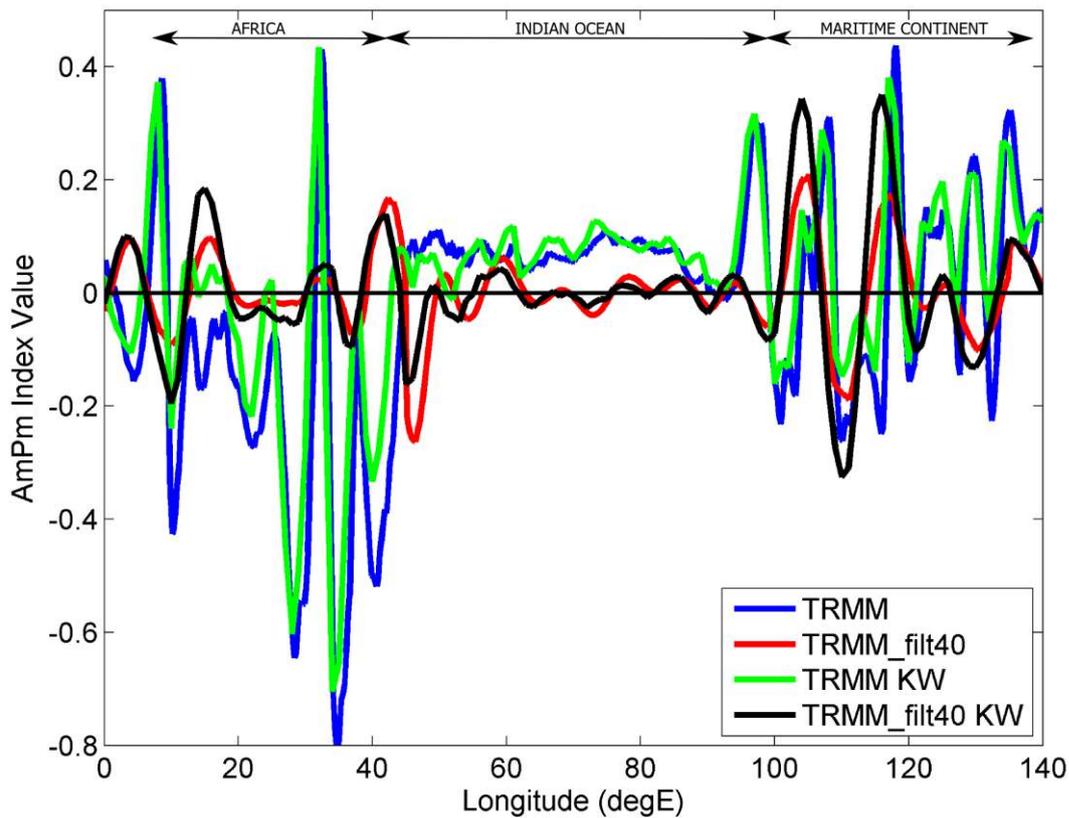

**Figure 5.10 The zonal variability of the AmPm index over Africa, Indian Ocean and Maritime Continent. Blue line is an index based on full precipitation from the whole TRMM database, red line is an index based on filtered precipitation from the whole TRMM_filt40 database, green line is an index based on full precipitation around (+/- 12h) all Kelvin wave trajectories, black line is an index based on filtered precipitation around (+/- 12h) all Kelvin wave trajectories and its magnitude is multiplied by 2 to compare it with other indices. The x-axis is longitude; the y-axis is value of the AmPm index.**

To summarize: in this section similarly to the case of a single Kelvin wave the diurnal cycle of full and filtered precipitation has been calculated for all Kelvin waves identified in the 15 years of TRMM data. The mean diurnal cycle of the full and filtered precipitation associated with Kelvin waves activity was calculated. Interestingly, the diurnal cycle of the precipitation is still apparent for averaged Kelvin waves. Although the diurnal cycle associated with individual Kelvin wave is easy to explain, the fact that signal doesn't vanish when almost 2000 waves are averaged is unexpected, especially over the ocean. The magnitude of the diurnal cycle is the strongest over land areas of Maritime Continent and Africa for both full and filtered precipitation which is likely a reflection of daytime precipitation maximum over the land. The magnitude of the diurnal cycle over the Indian Ocean is smaller, but its structure of alternating nighttime and daytime maxima is coherent. That indicates that Kelvin waves are longitude – diurnal cycle phase locked and tend to be active at certain location during certain time of the day. The full precipitation associated with Kelvin waves differs from full precipitation diurnal cycle climatology over Indian Ocean. Although both signals are dominated by nighttime components, the zonal variability of the length of about 10 degrees is visible in precipitation associated with Kelvin waves. Similar behavior is observed over Africa where the climatology is dominated by daytime precipitation, but full precipitation associated with Kelvin waves exhibits nighttime maximum at 15E. The zonal structure of such differences between full precipitation climatology and full precipitation associated with Kelvin waves is an indication of Kelvin waves propagation.

Our analysis shows that diurnal cycle of filtered precipitation for the entire TRMM_filt40 database and the filtered precipitation calculated along the Kelvin wave trajectories only, provide similar information about the zonal variability of the diurnal cycle. Therefore, in next section we will limit our discussion to the filtered precipitation along the trajectories. In addition, full precipitation along Kelvin wave trajectories will be investigated.

We have demonstrated that the AmPm index is able to capture the key characteristics associated with Kelvin waves propagation and longitude – diurnal cycle phase locking. However, some information provided by four indices described in Table 5.1 is redundant. Therefore further analysis will be limited to only one AmPm index for the full precipitation and one AmPm index for the filtered precipitation. The indices calculated for precipitation associated with Kelvin wave only (TRMM KW and TRMM_filt40 KW) were selected for further analysis because they can be calculated for individual trajectories.

### 5.3.3. Cross Maritime Continent propagation

In this section the impact of longiture – diurnal cycle locking on Kelvin waves propagation over Maritime Continent is investigated. The Kelvin waves active in the Indian Ocean basin are divided into two distinct categories: Kelvin waves that made successful transition over Maritime Continent (hereafter non-blocked Kelvin waves) and Kelvin waves that did not make transition over Maritime Continent (hereafter non-non-blocked Kelvin waves). The non-blocked Kelvin waves are defined as waves which trajectories began west of 90E and ended east of 140E. The blocked Kelvin waves are defined as waves which trajectories began west of 90E and ended between 90E and 140E.

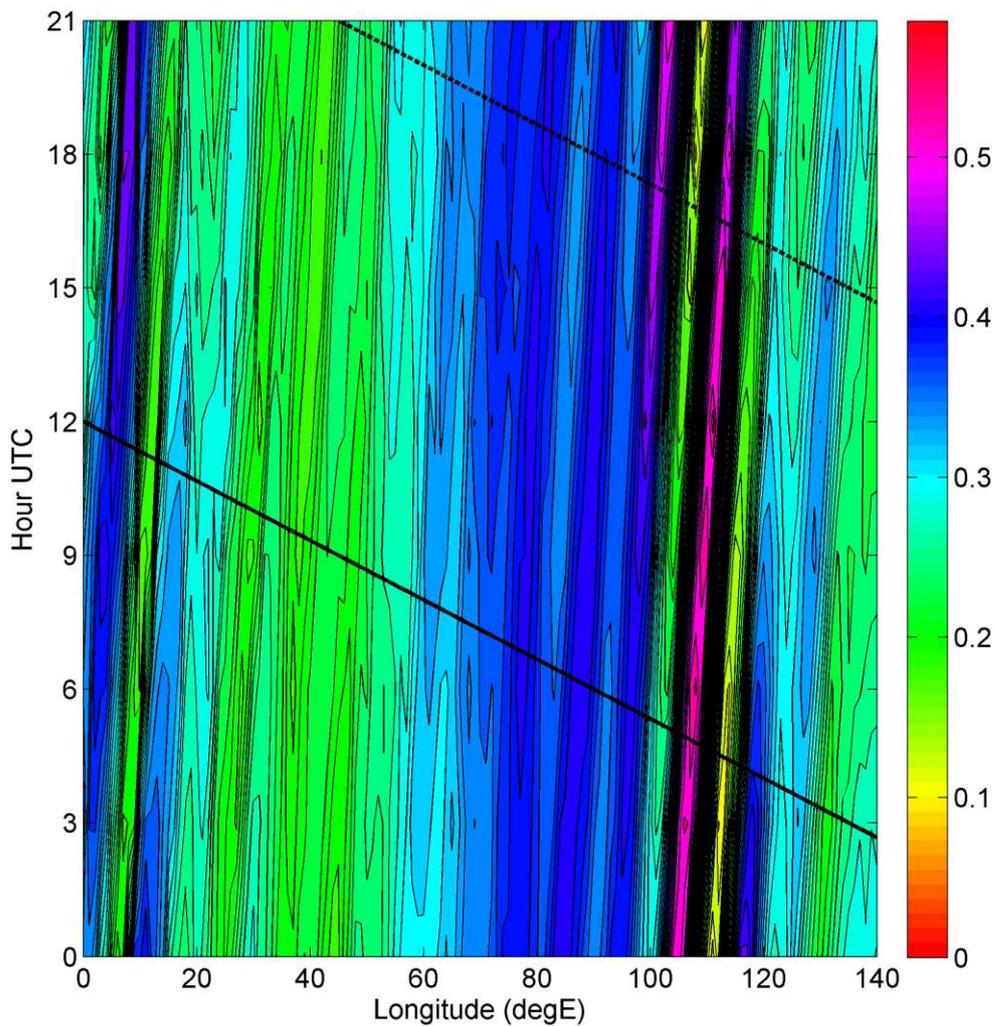

**Figure 5.11 The diurnal cycle of the filtered precipitation along the "non-blocked" Kelvin wave trajectories, which begin west of 90E and end east of 140E based on TRMM_filt40 database. The 24hour window around the trajectory is used. The x-axis is longitude; the y-axis UTC hour of the day. Local noon is marked with black solid line and local midnight is marked with black dashed line.**

The composite analyses of the characteristics of the non-blocked Kelvin waves are presented in Figure 5.11, Figure 5.13 and Figure 5.15. The composite analyses of the characteristics of the blocked Kelvin waves are presented in Figure 5.12, Figure 5.14 and Figure 5.16. Figure 5.11 and Figure 5.12 present mean diurnal cycle of the filtered precipitation along non-blocked and blocked Kelvin wave trajectories, respectively. The differences between non-blocked and blocked Kelvin waves are apparent. The diurnal cycle for non-blocked Kelvin waves is characterized by overall higher filtered precipitation rates than for blocked Kelvin waves. Although in both non-blocked and blocked Kelvin waves strong coupling with diurnal cycle over Maritime Continent and Africa is apparent, values of filtered precipitation rates are about 40% larger over Maritime Continent and about 15% larger over Africa for non-blocked Kelvin waves. Over central and eastern Indian Ocean values of the filtered precipitation are 15 – 40% larger in non-blocked Kelvin waves than in blocked Kelvin waves. In addition non-blocked Kelvin waves exhibit stronger propagation features over that region than blocked Kelvin waves. Although features of Kelvin wave propagation can be seen around 60E for blocked Kelvin waves, the eastern Indian Ocean (70E – 90E) has no such signature. On the other hand, the non-blocked Kelvin waves present coherent propagation signature throughout central and eastern Indian Ocean.

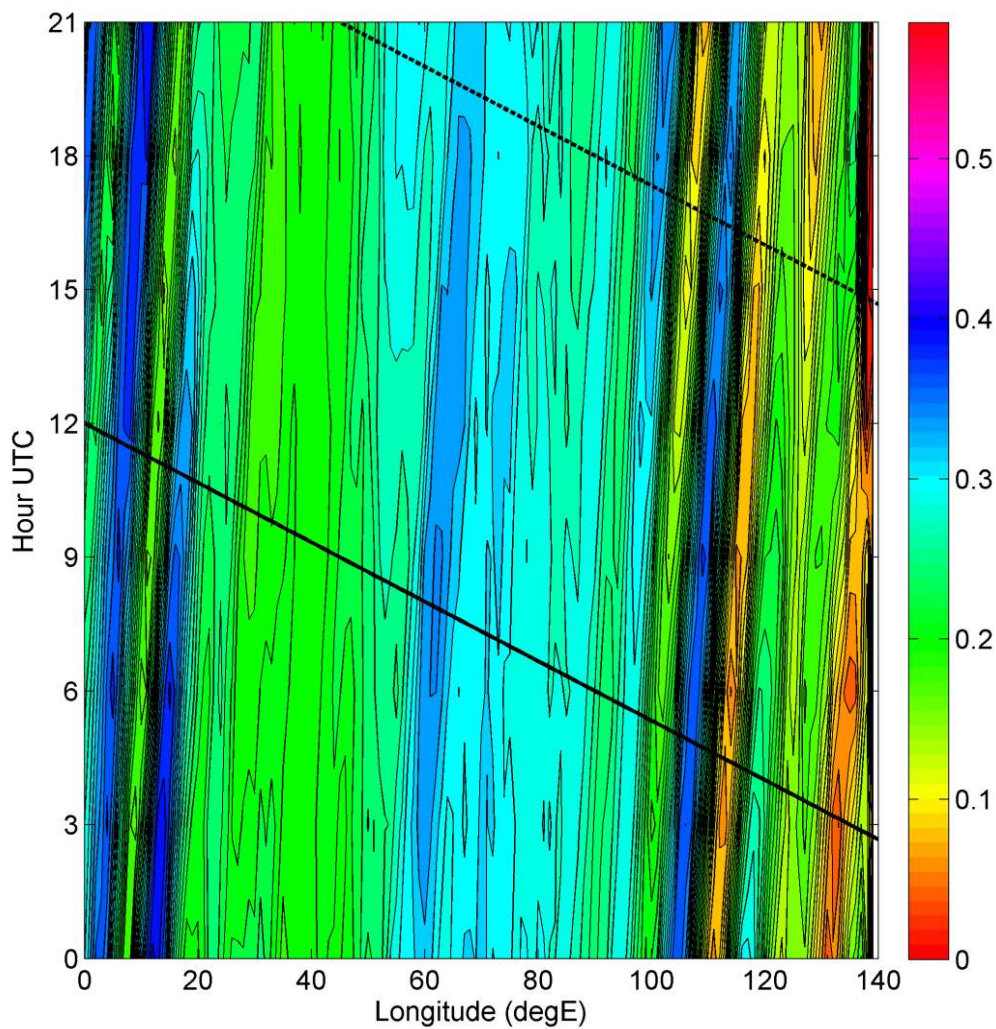

**Figure 5.12 The diurnal cycle of the filtered precipitation along "blocked" Kelvin wave trajectories, which begin west of 90E and end between 90E and 140E, based on TRMM_filt40 database. The 24hour window around the trajectory is used. The x-axis is longitude; the y-axis is UTC hour of the day. Local noon is marked with black solid line and local midnight is marked with black dashed line.**

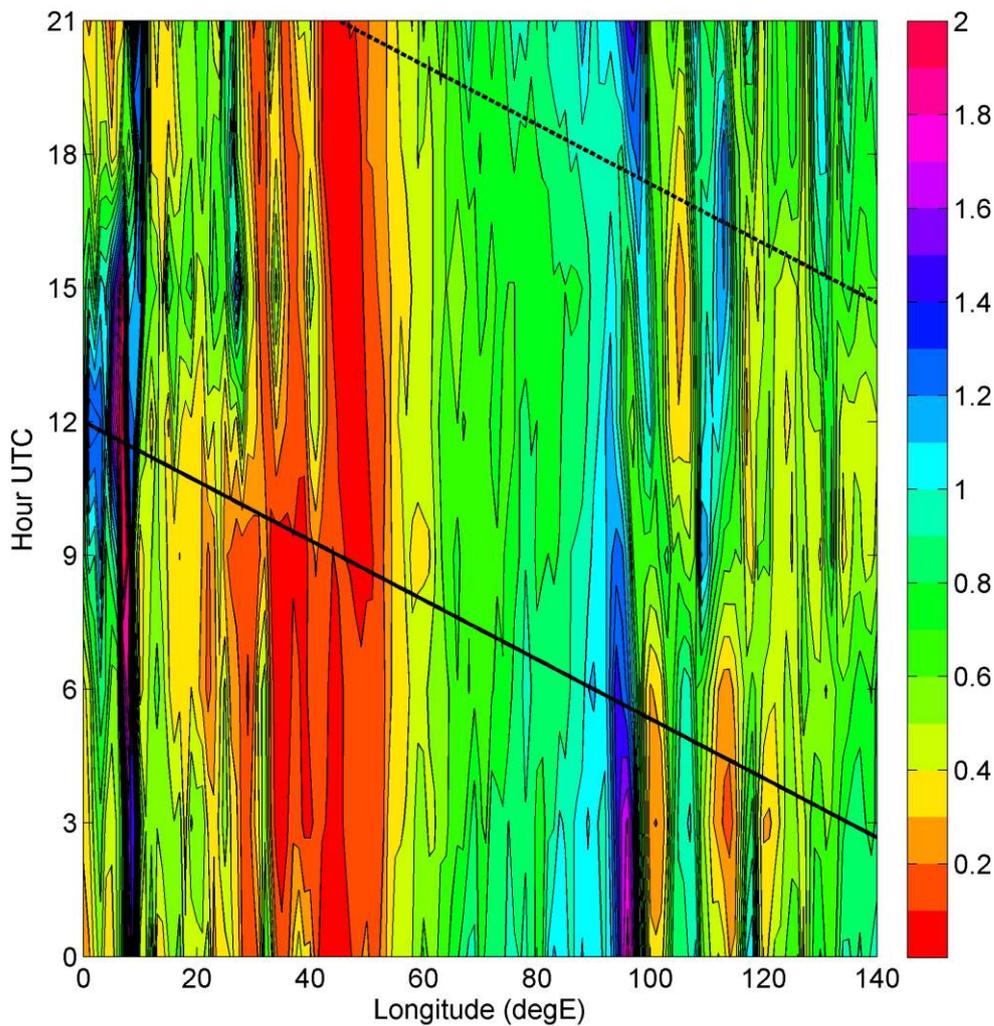

**Figure 5.13 The diurnal cycle of the full precipitation along "non-blocked" Kelvin wave trajectories, which begin west of 90E and end east of 140E, based on TRMM database. The 24hour window around the trajectory is used. The x-axis is longitude; the y-axis is UTC hour of the day. Local noon is marked with black solid line and local midnight is marked with black dashed line.**

Figure 5.13 and Figure 5.14 present mean diurnal cycle of the full precipitation along non-blocked and blocked Kelvin wave trajectories, respectively. It can be seen that similarly to filtered precipitation comparison, the non-blocked Kelvin waves are associated with higher full precipitation rates over Maritime Continent and Africa than non non-blocked Kelvin waves. However, the diurnal and zonal structure is similar in non-blocked and blocked Kelvin wave over these two regions. Over entire Indian Ocean, blocked Kelvin waves are dominated by nocturnal precipitation. The nighttime precipitation rates over central and eastern Indian Ocean are about 25% larger in non-blocked than in blocked Kelvin waves. In the same region, the difference in daytime precipitation rates between non-blocked and blocked Kelvin waves is in the order of 40-60%.

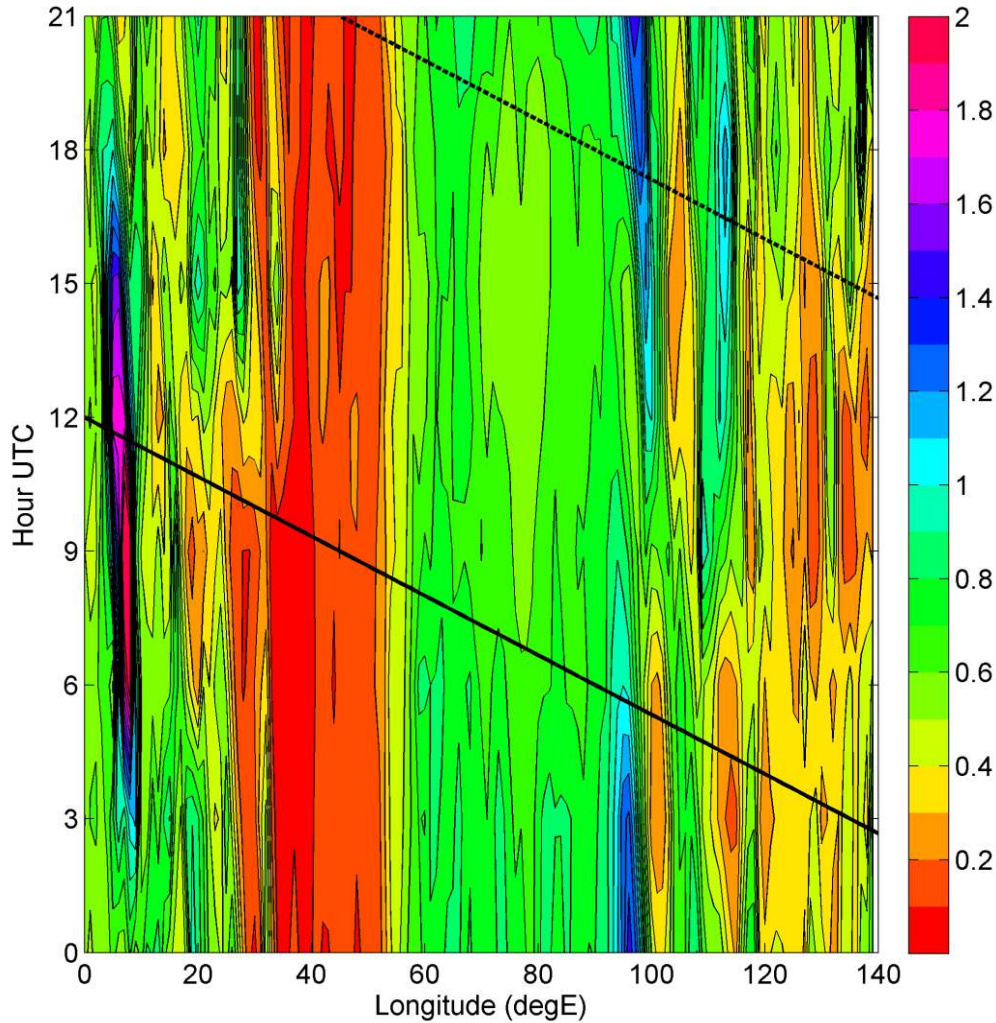

**Figure 5.14 The diurnal cycle of the full precipitation along "blocked" Kelvin wave trajectories, which begin west of 90E and end between 90E and 140E, based on TRMM database. The 24hour window around the trajectory is used. The x-axis is longitude; the y-axis is UTC hour of the day. Local noon is marked with black solid line and local midnight is marked with black dashed line.**

Figure 5.15 and Figure 5.16 present zonal variability of the AmPm difference calculated for the filtered precipitation and full precipitation respectively. Each figure presents differences for the non-blocked (solid blue line) and the blocked (dashed red line) Kelvin wave trajectories. In Figure 5.15 it can be seen that over Maritime Continent the AmPm index differences for non-blocked and blocked Kelvin waves are very similar. Full precipitation differences (Figure 5.16) are also similar which suggests that the diurnal cycle of the precipitation associated with both non-blocked and blocked Kelvin waves is dominated by the diurnal cycle of the convection over Maritime Continent. The AmPm difference for filtered precipitation for non-blocked Kelvin waves presents more regular, periodic structure over Africa, than the AmPm difference for blocked Kelvin waves. That indicates that longitude – diurnal cycle phase locking is stronger in non-blocked Kelvin waves than in blocked Kelvin waves.

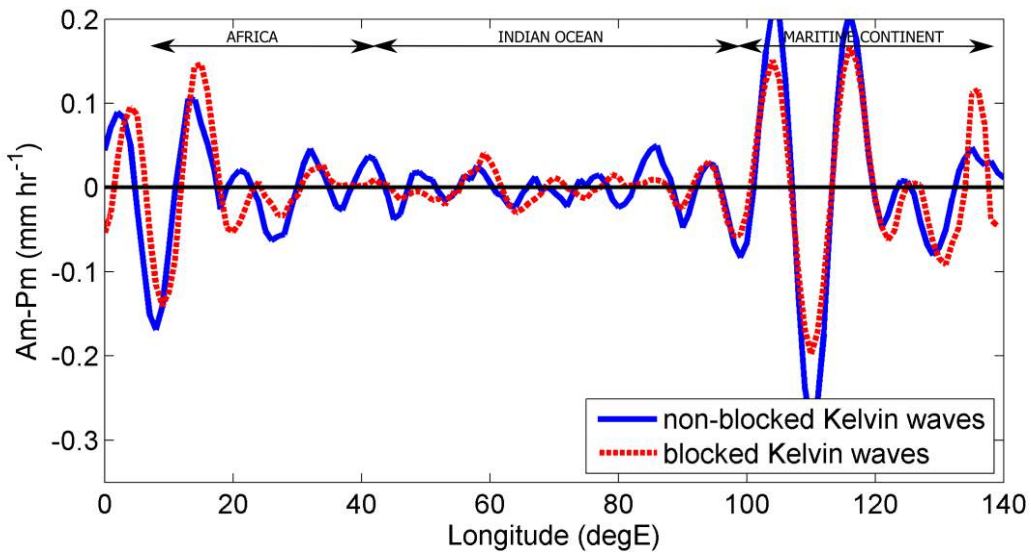

**Figure 5.15** The zonal variability of the difference between nighttime and daytime components of filtered precipitation in [mmhr$^{-1}$] along (blue solid) non-blocked Kelvin wave trajectories and (red dashed) blocked Kelvin wave trajectories. The x-axis is longitude in [degrees east]; the y-axis is difference between "am" and "pm" components.

Over the Indian Ocean there is a contrast between non-blocked and blocked Kelvin waves in AmPm index difference for the filtered precipitation (Figure 5.15) and the full precipitation (Figure 5.16). The AmPm index difference for the filtered precipitation in non-blocked Kelvin waves exhibits large zonal oscillation which suggests strong influence of wave propagation. The distance between the two sequential maxima or two sequential minima is of the order of 8-12 degrees which agrees very well with phase speed of Kelvin waves. For the blocked Kelvin waves the AmPm difference exhibits smaller zonal oscillation and distance between sequential maxima has larger range. That suggests that in comparison with non-blocked Kelvin waves, the blocked Kelvin waves have weaker longitude – diurnal cycle phase locking. The AmPm difference for the full precipitation in non-blocked and blocked Kelvin waves tend to be above 0 over the Indian Ocean. It means that the diurnal cycle of the precipitation is dominated by its nighttime component. At the same time, the AmPm index difference for full precipitation in non-blocked Kelvin waves exhibits strong zonal variability which is consistent with the periodicity in AmPm index difference for filtered precipitation. The AmPm index difference for full precipitation in blocked Kelvin waves exhibits no such zonal variability. The contrast between non-blocked and blocked Kelvin waves in AmPm index differences for full precipitation is particularly apparent at longitudes 64E and 93E, where their signs are opposite. That means that at these two locations, diurnal cycle of the full precipitation along non-blocked Kelvin waves is dominated by its daytime component. This is not the case neither for the diurnal cycle in blocked Kelvin waves nor for the mean diurnal cycle of the precipitation.

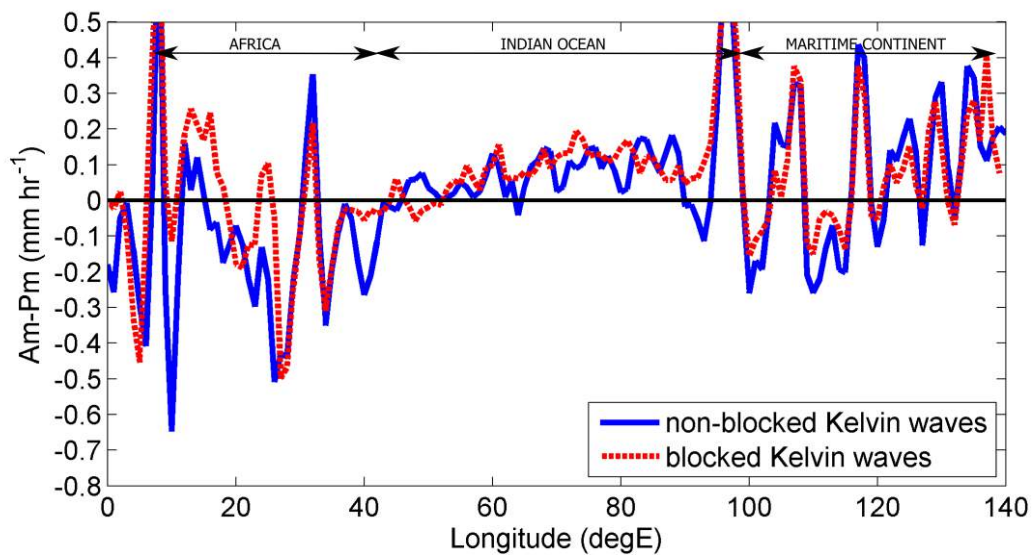

**Figure 5.16 The zonal variability of the difference between nighttime and daytime components of full precipitation along (blue solid) non-blocked Kelvin wave trajectories and (red dashed) blocked Kelvin wave trajectories. The x-axis is longitude in [degrees east]; the y-axis is difference between "am" and "pm" components.**

This analysis of the AmPm differences for non-blocked and blocked Kelvin waves suggests that the relative increase in daytime precipitation for non-blocked Kelvin waves, seen in Figure 5.13, is likely caused by strong longitude – diurnal cycle phase locking of the propagating Kelvin waves.

To summarize: we have shown that diurnal cycle of the filtered and full precipitation along Kelvin wave trajectories has a coherent structure. Over the Maritime Continent the diurnal cycle of precipitation in propagating Kelvin waves is driven by the local diurnal cycle of convection. This coupling has a form of longitude – diurnal cycle phase locking over that region. This is likely because the zonal distribution of land within Maritime Continent, in particular the distance between the islands of Sumatra and Borneo, agrees very well with the distance travelled by a Kelvin wave over one day. Daytime precipitation maxima over the islands and nighttime precipitation maximum between them are strong and similar every day [*Peatman et al.*, 2014]. Therefore, this land-ocean system creates unique "filter" for convection-driven phenomena such as convectively coupled Kelvin waves. The inner diurnal cycle of precipitation in the propagating Kelvin waves can be either in phase or out of phase with the local diurnal cycle of the precipitation over Maritime Continent. Those events in which the diurnal cycle along the trajectory is in phase with the local diurnal cycle over Maritime Continent should have better chance of making successful transition over that region. Kelvin wave in which the inner diurnal cycle along its trajectory is out of phase with the local diurnal cycle between Sumatra and Borneo will propagate in unfavorable environment.

This is confirmed by the analysis of non-blocked (propagating through Maritime Continent) and blocked Kelvin waves. Kelvin waves which make successful transition to the Western Pacific exhibit strong oscillation in the diurnal cycle of the precipitation along their trajectories over Indian Ocean and Africa (Figure 5.11 and Figure 5.15). The zonal and temporal scale of this variability is consistent with both, phase speed of a Kelvin wave and local diurnal cycle of precipitation over Maritime Continent. Such periodicity is not visible over the Indian Ocean for blocked Kelvin waves.

## 5.4. Blocked and non-blocked Kelvin waves statistics

We have used newly developed Kelvin wave trajectories database in order to calculate statistics associated with blocking of Kelvin waves by Maritime Continent. Total number of 748 trajectories active between longitudes 40E and 90E were identified. Out of these Kelvin waves that were active in the Indian Ocean, 198 made successful transition into Western Pacific and 352 trajectories terminated within Maritime Continent. The rest of the trajectories active in the Indian Ocean terminated west of 90E and such Kelvin waves are considered neither blocked nor non-blocked.

Figure 5.17 presents distribution of the origins of trajectories for all, blocked and non-blocked Kelvin waves. The distribution of all Kelvin waves active in the Indian Ocean has two maxima at the longitudes 45E and 75E. The maximum in the western Indian Ocean at 45E contains 53 Kelvin waves which initiated there. This maximum may be related to the dry, non-convective Kelvin waves that were initiated over Africa and became coupled with convection once they entered warm waters of tropical Indian Ocean. Kelvin wave initiation over central Indian Ocean is characterized by a flat distribution with about 45 Kelvin waves initiated over every 5 degrees wide area between 50E and 70E. The second maximum, located at the longitude 75E is associated with 61 Kelvin waves. The zonal distribution for blocked Kelvin waves has similar structure to the distribution for all Kelvin waves. Although it is characterized by the double maximum at the exact same locations as for all Kelvin waves, the central Indian Ocean is characterized by an increasing number of initiations. On the other hand, the distribution for non-blocked Kelvin waves has only one maximum located a little further to the east in comparison with all and blocked Kelvin waves. It is located at 80E and is associated with 25 Kelvin waves initiated there. This is significantly more initiations than over the eastern and central Indian Ocean.

Figure 5.18 presents comparison of the mean precipitation rate over central-eastern Indian Ocean (70-100E) for blocked, non-blocked and all Kelvin waves. The distribution is normalized by the number of all Kelvin waves within each category. It can be seen that the distribution for non-blocked Kelvin waves is characterized by smaller variance and slight shift towards the higher precipitation rates. Difference in distributions between blocked and non-blocked Kelvin waves is apparent, though maxima are not separated. Thus, the precipitation rates over central-eastern Indian Ocean along a Kelvin wave trajectory do not determine its likelihood of being blocked by Maritime Continent.

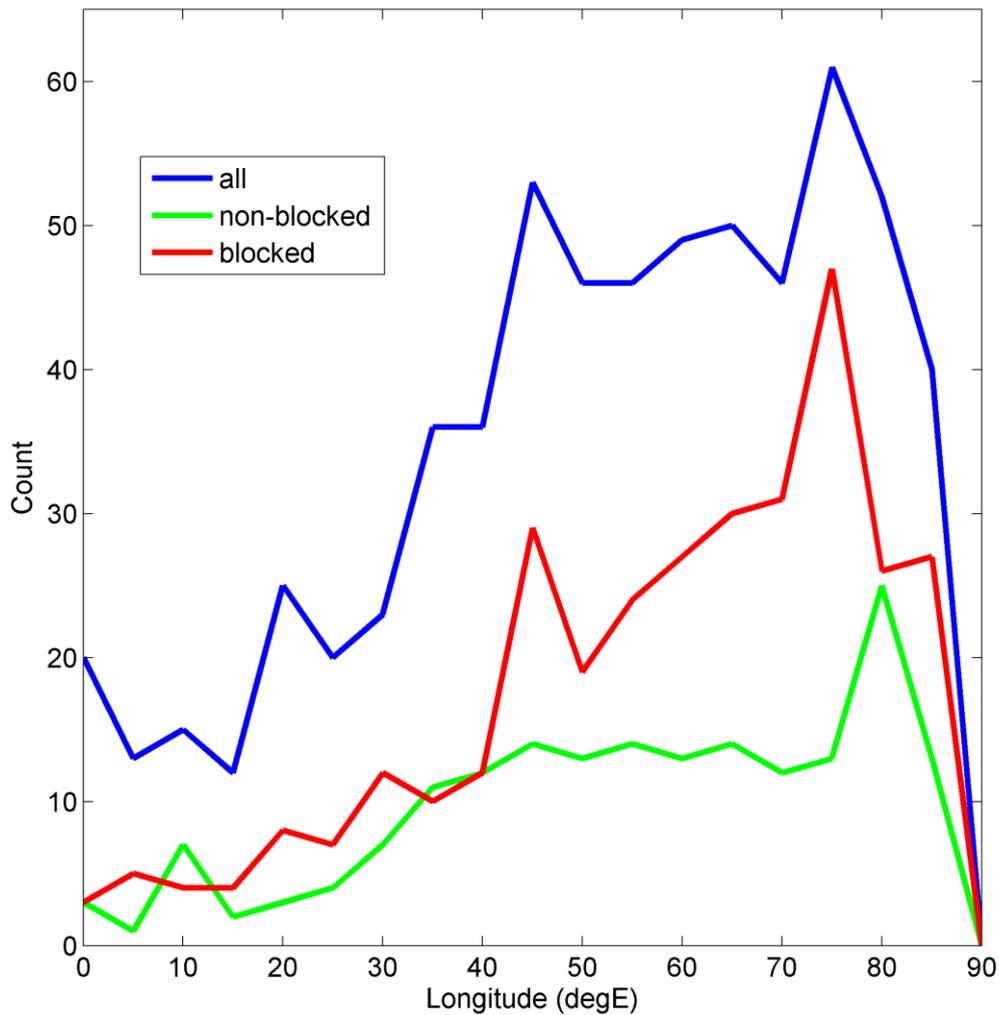

Figure 5.17 Zonal distribution of the longitude of trajectory initiation for (blue) all Kelvin waves active in the Indian Ocean, (green) "non-blocked" Kelvin waves and (red) "blocked" Kelvin waves. The x-axis is longitude in [degrees east]; the y-axis is number of trajectories.

Figure 5.19 presents distribution of the mean Kelvin wave phase speed calculated for all trajectories between longitudes 70E and 100E. Comparison between all, blocked and non-blocked Kelvin waves is presented in the form of a fraction of all Kelvin wave trajectories in a specific category. It can be seen that non-blocked Kelvin waves are characterized by slower phase speed when compared with either blocked or all Kelvin waves. Over 20% of all Kelvin waves that made successful transition over Maritime Continent have mean phase speed of the order of 10.5 degrees per day. On the other hand the largest fraction of blocked Kelvin waves has the phase speed of 11 degrees per day and accounts for less than 16% of all blocked Kelvin wave trajectories. Although the maximum for all Kelvin waves is for trajectories characterized by mean phase speed of 10.5 degrees per day, which is the same as for non-blocked Kelvin waves, the overall structure of the distribution is similar to that of blocked Kelvin waves. It is characterized by a substantial fraction of Kelvin waves propagating with phase speed of 12 degrees per day and more, feature not present for non-blocked Kelvin wave category. Overall comparison shows that the distribution for non-blocked Kelvin waves has smaller variance and is shifted towards smaller phase speeds in comparison with either blocked or all Kelvin waves.

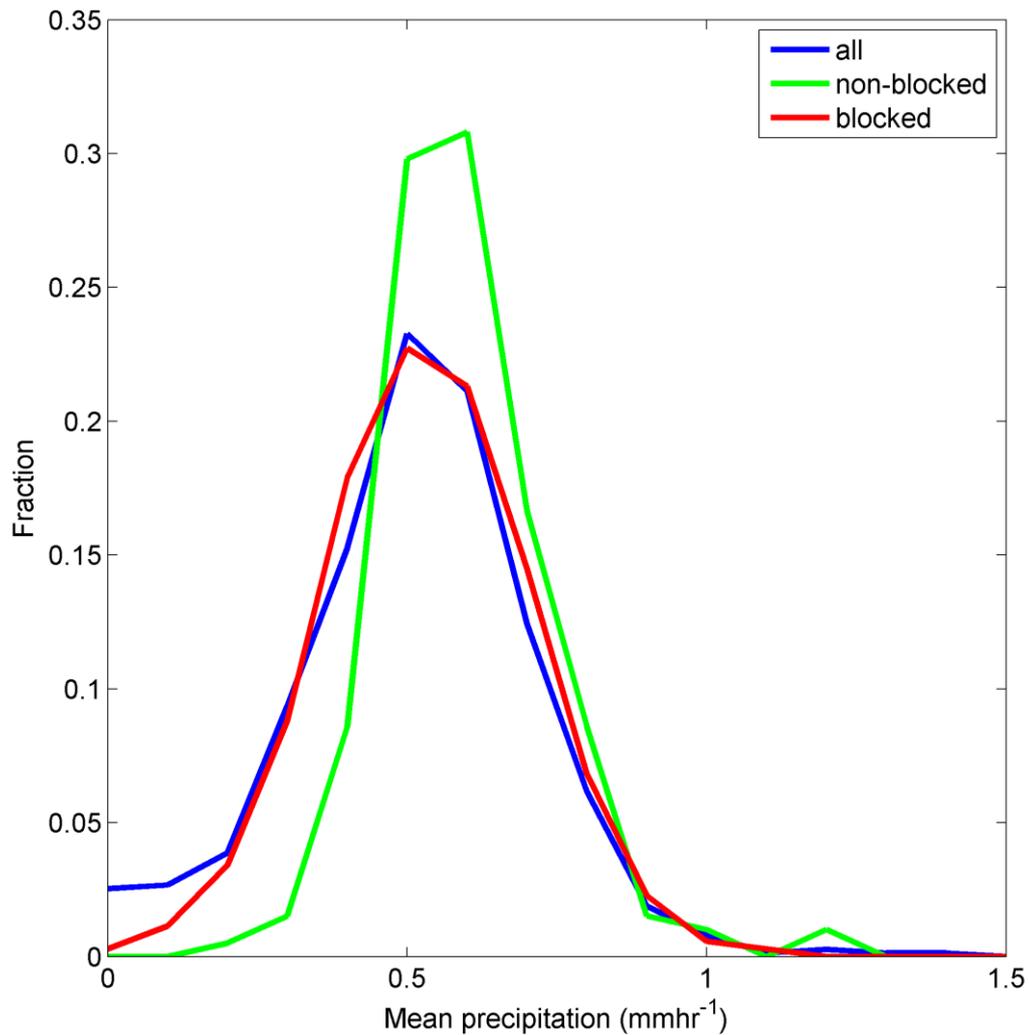

**Figure 5.18 Distribution of the average full precipitation rate along entire Kelvin wave trajectory for (blue) all Kelvin waves active in the Indian Ocean, (green) "non-blocked" Kelvin waves and (red) "blocked" Kelvin waves. The x-axis is mean precipitation rate in [mmhr$^{-1}$]; the y-axis is fraction of trajectories characterized by specific average precipitation rate.**

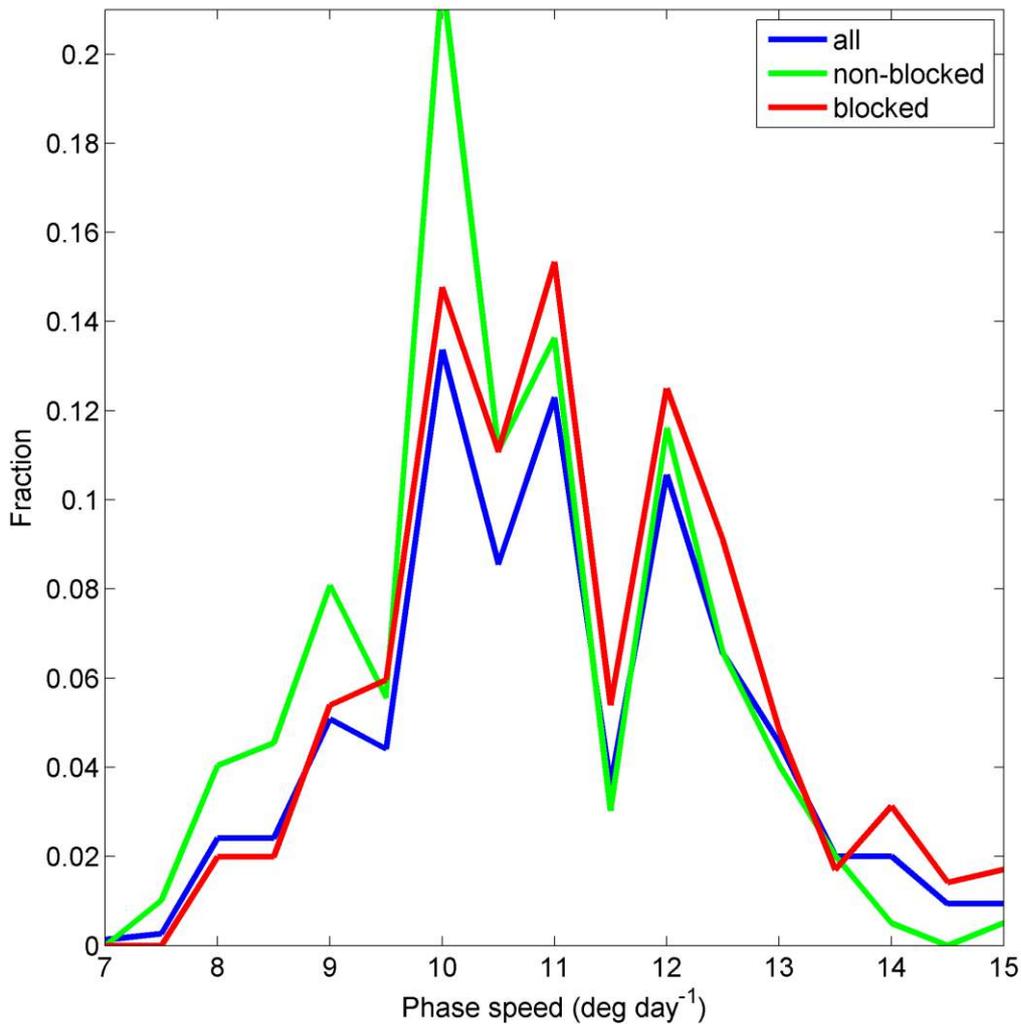

**Figure 5.19** Distribution of the average Kelvin wave phase speed for (blue) all Kelvin waves active in the Indian Ocean, (green) "non-blocked" Kelvin waves and (red) "blocked" Kelvin waves. The x-axis is phase speed in [degrees per day]; the y-axis is fraction of trajectories propagating with specific phase speed.

In order to further investigate the impact of Kelvin wave phase speed on its ability to make transition over Maritime Continent we compare mean phase speed over central-eastern Indian Ocean and over Maritime Continent for waves that made successful transition into Western Pacific. Figure 5.20 presents comparison of the distribution of Kelvin wave phase speeds calculated using: (a) mean phase speed between longitudes 90E and 140E and (b) mean phase speed between longitudes 70E and 100E. Although the distribution was calculated for exactly the same Kelvin waves the two histograms clearly differ. In particular, the distribution of phase speeds over Maritime Continent is shifted towards lower values in comparison to the distribution of phase speeds over central-eastern

Indian Ocean. For the phase speed distribution calculated over Maritime Continent (Figure 5.20a) the peak at 9 degrees per day is apparent. The phase speed calculated over central-eastern Indian Ocean (Figure 5.20b) is dominated by waves propagating with speeds of 10-11 degrees per day. This suggests that Kelvin wave phase speed and its variability is a complicated and important factor influencing its propagation over the Maritime Continent.

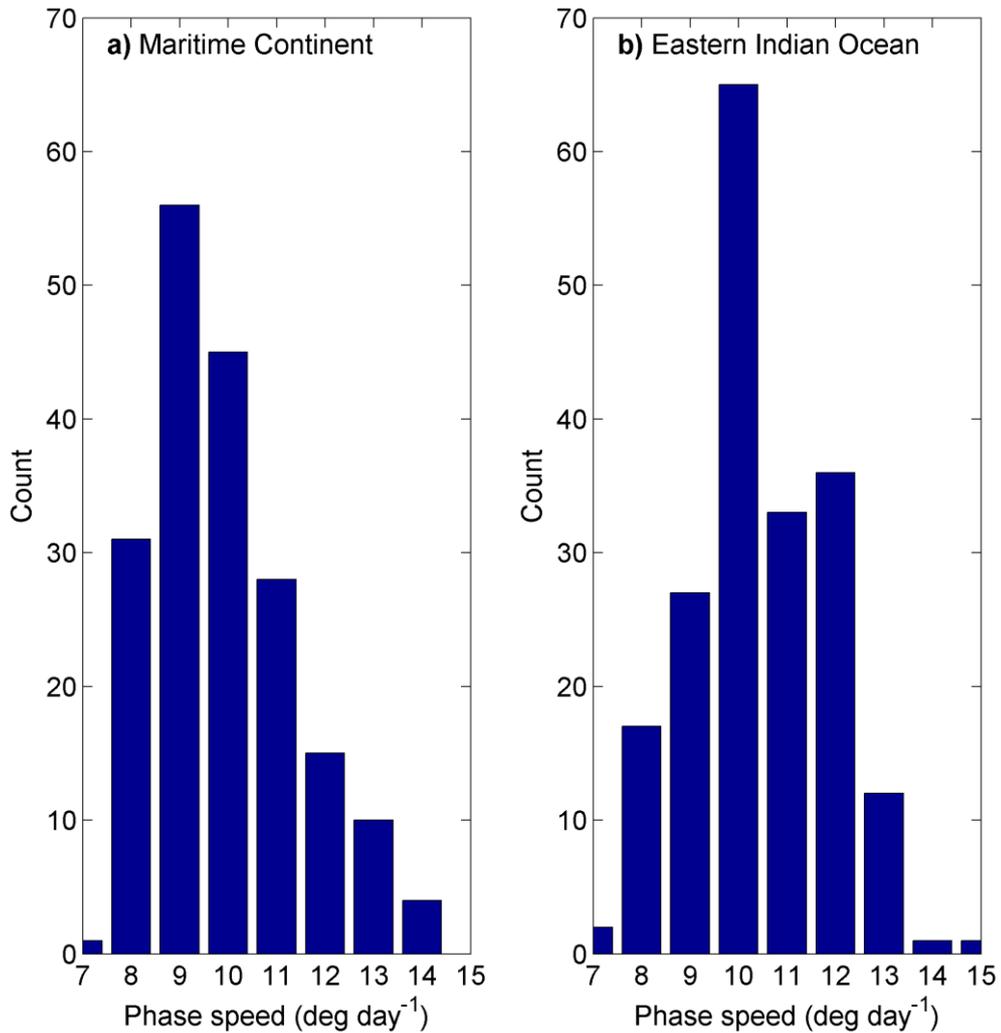

**Figure 5.20** Histogram of average phase speed for "non-blocked" Kelvin waves calculated between longitudes (a) 90E and 140E and (b) 70E and 100E. The x-axis is mean phase speed in [degrees per day]; the y-axis is number of waves propagating with specific phase speed.

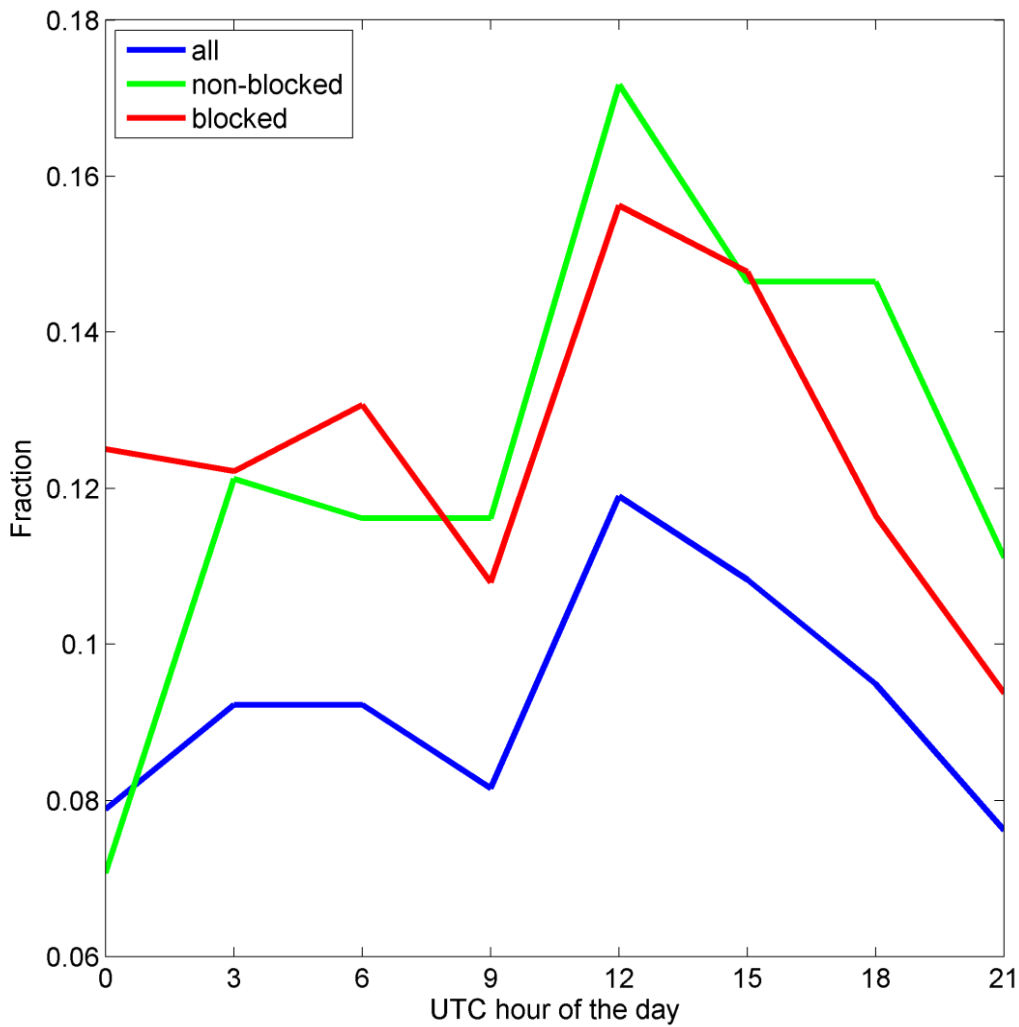

Figure 5.21 Distribution of the UTC hour of the day at which Kelvin wave is active at the longitude 90E for: (blue) all Kelvin waves active in the Indian Ocean; (green) "non-blocked" Kelvin waves; (red) "blocked" Kelvin waves. The x-axis is UTC hour of the day; the y-axis is fraction of trajectories that passed over longitude 90E at the specific UTC hour of the day.

If the Maritime Continent acts as a filter for Kelvin waves, as suggested by the analysis of the longitude – diurnal cycle phase locking, the time of the day when Kelvin waves enter this area may be an important factor. Figure 5.21 presents distribution of the UTC hour at which Kelvin waves were active at the longitude 90E. This distribution is normalized by the number of Kelvin waves in each category. Note however, that fraction of waves active at 90E at a given hour of the day for all Kelvin waves category (blue line) is significantly lower than for either blocked or non-blocked categories. This is because some of Kelvin waves active in the Indian Ocean, which are considered as part of "all" category, terminate west of 90E. This affects neither blocked nor non-blocked categories as they have to be active at 90E by definition. Analysis of the diurnal distribution at 90E shows that Kelvin waves within "blocked" and "all" categories have relatively flat distribution between 0UTC and 18UTC with fraction of waves for each 3-hour interval equal to about 13% for "blocked" category and about 9% for "all" category. On the other hand "non-blocked" Kelvin waves are appearing at 90E more frequently between 9UTC and 18 UTC with maximum visible at 12UTC. This maximum accounts for about 17% of all waves in "non-blocked" category. The smallest fraction of "non-blocked" Kelvin waves is active at the longitude 90E at 0 UTC, when only about 7% of such waves are active. This analysis shows that specific time of the day may favor successful propagation of Kelvin waves over Maritime Continent.

In order to further investigate that matter, the probability of successful transition into Western Pacific is calculated individually for every UTC interval. The probability is calculated as number of Kelvin that were active at 90E and were "non-blocked" by Maritime Continent divided by total number of Kelvin wave trajectories active at that time at 90E. Calculations were performed for all Kelvin wave phase speeds and for those waves that propagated over eastern Indian Ocean with phase speed of 10 to 11 degrees per day. The results are presented in Figure 5.22. Blue solid line presents probability of successful propagation over Maritime Continent for all Kelvin waves that are active at 90E at specific time of the day. Blue dashed line presents similar probability but only for waves that had mean phase speed of 10 to 11 degrees per day over the area between longitudes 70E and 100E. This specific phase speed was selected because the largest fraction of "non-blocked" Kelvin waves had such phase speed (Figure 5.19). It can be seen that for all Kelvin waves, the chance of a wave making successful transition over Maritime Continent increases between 3UTC and 21 UTC. The probability is the largest when waves are active at 90E at 15UTC and the lowest if they are at 0 UTC. The distribution changes when only Kelvin waves with specified phase speed are considered. The probability of successful transition into Western Pacific is about or exceeds 50% if a wave is active at the longitude 90E at 9UTC, 12UTC and 18UTC. The probability is the lowest if a wave arrives at 90E at 0UTC or 15UTC. To further analyze the characteristics of the Kelvin waves that made successful transition over Maritime Continent and have phase speed of 10-11 degrees per day over central-eastern Indian Ocean, the mean phase speed over Maritime Continent is considered. The average phase speed over Maritime Continent calculated for waves that were active at the specific time interval is presented with black solid line and standard deviation of each average is indicated by black vertical bars (Figure 5.22). It can be seen that waves crossing 90E later in the day have higher average phase speed between 90E and 140E. Since we are considering Kelvin waves passing at a specific distance from Maritime Continent, in particular from the islands of Sumatra and Borneo, when a wave arrives at 90E later during the day, it has to move faster in order to get over Maritime Continent in time to interact with its diurnal cycle. This discussion assumes that the Sumatra and Borneo diurnal cycles play an important role in facilitating or dumping Kelvin waves propagation, although other factors may exist as well.

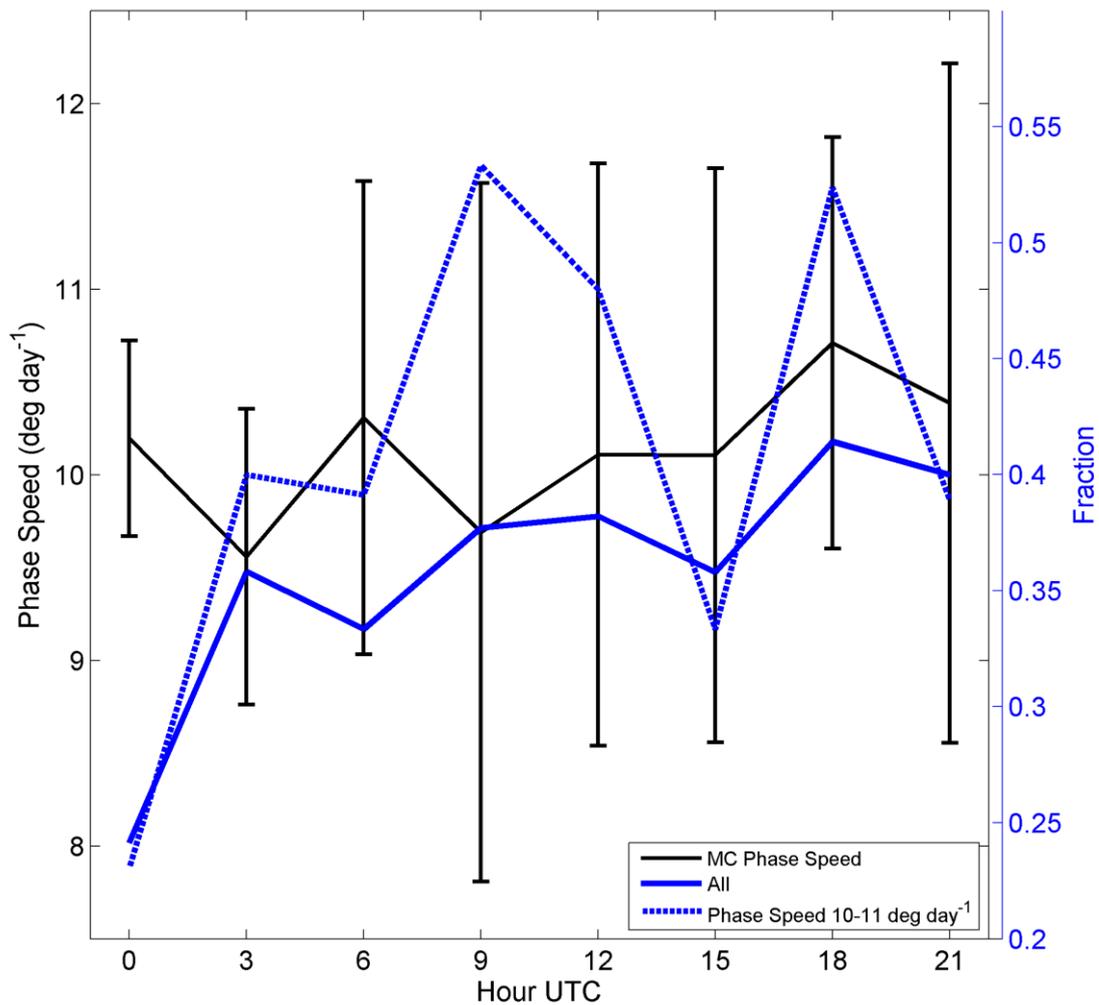

Figure 5.22 Probability of the successful propagation over Maritime Continent for Kelvin waves active at 90E at given UTC hour for (solid blue) all Kelvin waves, (dashed blue) Kelvin waves which have mean phase speed of 10 to 11 degrees per day between 70E and 100E. The x-axis is UTC hour of the day; the y-axis (blue, on the right) is a ration of "non-blocked" Kelvin waves to all Kelvin waves active at 90E at the given UTC hour of the day. The black line indicates the mean phase speed over the Maritime Continent for "non-blocked" Kelvin waves which arrived at 90E at the given time of the day. The vertical black bars indicate the standard deviation of the mean phase speed over Maritime Continent. The y-axis (black, on the left) is mean phase speed over Maritime Continent.

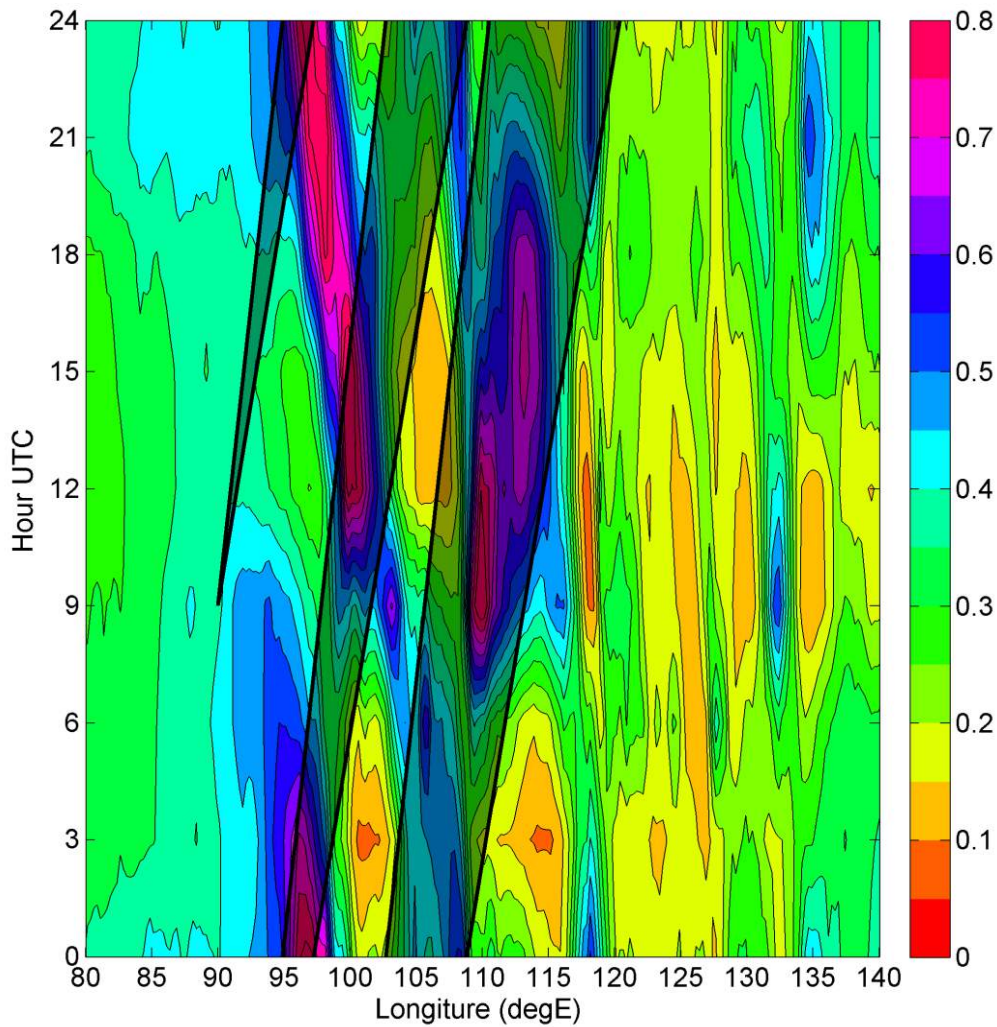

**Figure 5.23** Longitude-diurnal cycle diagram of mean precipitation over Maritime Continent. The x-axis is longitude in [degrees east]; the y-axis is UTC hour of the day. Color shading indicates mean precipitation based on 15 years of TRMM data in [mmhr$^{-1}$]. The two solid black lines and black shading indicate the range of hypothetic trajectories of Kelvin waves active at 90E at 9UTC and propagating with the phase speed over Maritime Continent determined from the Figure 5.22.

Let us now consider the two times of the highest probability of the successful propagation, that at 9UTC and 18UTC. It can be seen that for 9UTC the standard deviation of mean phase speed over Maritime Continent is considerably larger than for 18UTC. This suggests that waves that are active at 90E and 9UTC have larger range of phase speeds over the Maritime Continent for which they can make a successful transition. For waves that arrive at 90E and 18UTC this range of speeds is smaller. We analyze this further on the basis of Figure 5.23 and Figure 5.24. Both figures present the diurnal cycle of the full precipitation between 80E and 140E averaged over 15 years of TRMM data. Superimposed are hypothetic trajectory ranges based on mean and standard deviation for a given hour. Figure 5.23 illustrates hypothetic trajectories that would be active at 90E at 9UTC and Figure 5.24 shows hypothetic trajectories that would be active at 90E at 18UTC. Spread of phase speeds is given by the standard deviation shown in Figure 5.22. Even thought the range of phase speeds for Kelvin waves active at 90E at 9UTC is large, Figure 5.23 shows that such trajectories pass over the maximum daytime convection over land(Sumatra and Borneo) and nighttime convection over ocean (west of Sumatra and between Sumatra and Borneo). Hypothetic trajectories that are active at 90E and 18UTC (Figure 5.24) pass over daytime maximum over Sumatra. Although such trajectories miss nighttime maximum west of Sumatra, they propagate over nighttime maximum over ocean between Sumatra and Borneo and have chance to interact with diurnal maximum over Borneo.

To summarize: about 36% of Kelvin waves active at 90E make transition over Maritime Continent. It is clear that there are differences in characteristics of the "blocked" and "non-blocked" Kelvin waves. The "non-blocked" trajectories tend to begin a little further to the east than the "blocked" ones and are associated with higher precipitation rates over central-eastern Indian Ocean. The "non-blocked" Kelvin waves are also characterized by lower values of the mean phase speed between 70E and 100E when compared with "blocked" Kelvin waves or "all" Kelvin waves active over Indian Ocean. However, examples of "blocked" Kelvin waves show that the mean phase speed of a wave may differ between central-eastern Indian Ocean (70-100E) and Maritime Continent (90-140E). The differences between "blocked" and "non-blocked" waves are more apparent when one considers phase speed of Kelvin wave together with an hour at which it approaches Maritime Continent. We have shown that specific combinations of an hour at which Kelvin wave is active at 90E and its phase speed favor successful propagation. Kelvin waves that occur at 90E at 9UTC, 12UTC and 18UTC and propagate with phase speed of 10-11 degrees per day over the central-eastern Indian Ocean have highest chance (about 50%) to make successful transition into the Western Pacific. Waves that arrive at 90E later during the day, maintain higher mean phase speed over Maritime Continent necessary to encounter favorable conditions over Sumatra and Borneo. Increased likelihood of successful transition into the Western Pacific is linked to interaction with the diurnal cycle of precipitation over Maritime Continent. We have shown that hypothetic trajectories that are active at 90E on 9UTC or 18UTC, given their calculated range of phase speeds, have high chance to interact with the diurnal cycle of the precipitation over Maritime Continent.

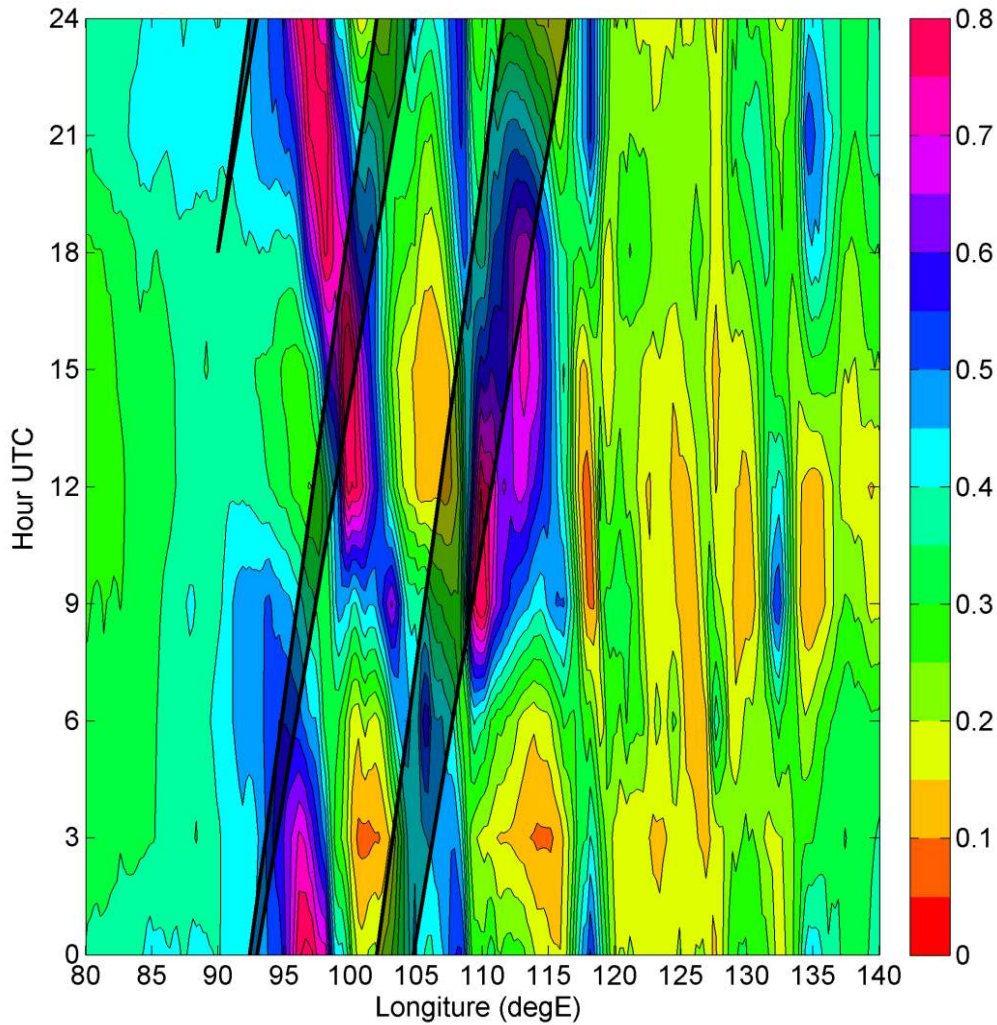

Figure 5.24 Longitude-diurnal cycle diagram of mean precipitation over Maritime Continent. The x-axis is longitude (degrees east); the y-axis is UTC hour of the day. Color shading indicates mean precipitation based on 15 years of TRMM data in [$mmhr^{-1}$]. The two solid black lines and black shading indicate the range of hypothetic trajectories of Kelvin waves active at 90E at 18UTC and propagating with the phase speed determined from the Figure 5.22.

## 5.5. Summary

In this paper the characteristics of atmospheric convectively coupled Kelvin waves have been investigated on the basis of individual case study and composite analysis. An individual Kelvin wave propagation has been detected using longitude–diurnal cycle diagrams for which the diurnal cycle in the vicinity of the trajectory is calculated at every longitude independently. The propagation of a single Kelvin wave is apparent in such diagrams of full and Kelvin wave filtered precipitation. The propagation of a single Kelvin wave is visible as an eastward moving signal for which the phase speed of a wave can be calculated from the tilt of the consecutive maxima and minima.

For the filtered precipitation the diurnal variability at a single longitude is not an effect of the local variability of convection but it is due to propagation of the convectively coupled Kelvin wave. For the full precipitation the effect of moving wave and local variability are included. Additionally, we have derived a simple measure of the diurnal variability- the AmPm index. The AmPm index can be calculated for either full or filtered precipitation and provides information on diurnal variability of the signal at any location along the wave trajectory. The zonal variability of the AmPm index has been shown to reflect Kelvin wave propagation. The index has been tested for averaged full precipitation diurnal cycle based on 15 years of TRMM data and for full and filtered precipitation along an individual Kelvin wave trajectory. In case of averaged diurnal cycle, it reproduces zonal structure of the diurnal cycle of precipitation very well. In particular, the domination of nighttime precipitation over the ocean and domination of the daytime precipitation over the land are apparent. In case of an individual Kelvin wave, the AmPm index for filtered precipitation reproduces features of wave propagation, represented by alternating maxima and minima in the zonal variability of the index. The AmPm index was used to show the characteristics of single Kelvin wave as well as average of multiple Kelvin waves identified in 15 years of TRMM data.

Almost 2000 individual Kelvin waves are identified in the database. Out of these waves, 748 were active in the Indian Ocean basin between 40E and 90E. The longitude-diurnal cycle diagrams are calculated for filtered and full precipitation for each trajectory individually and averaged over all trajectories active over Africa, Indian Ocean and Maritime Continent. Interestingly, the diurnal cycle of the precipitation is still apparent for averaged Kelvin waves. Although the diurnal cycle associated with individual Kelvin wave is easy to explain, the fact that signal doesn't vanish when many waves are averaged is unexpected, especially over the ocean. The magnitude of the diurnal cycle is the strongest over land areas of the Maritime Continent and Africa for both full and filtered precipitation which is likely a reflection of the daytime precipitation maximum over the land. The magnitude of the diurnal cycle over the Indian Ocean is smaller, but its structure of alternating nighttime and daytime maxima is coherent. That indicates that Kelvin waves are longitude–diurnal cycle phase locked and tend to be active at definite location during specific time of the day. The full precipitation diurnal cycle associated with Kelvin waves propagation differs from the full precipitation diurnal cycle climatology over the Indian Ocean. Although both signals are dominated by nighttime components, the zonal variability of the length of about 10 degrees is visible in precipitation associated with Kelvin waves. The zonal structure of such differences between full precipitation climatology and full precipitation associated with Kelvin waves is an indication of wave propagation.

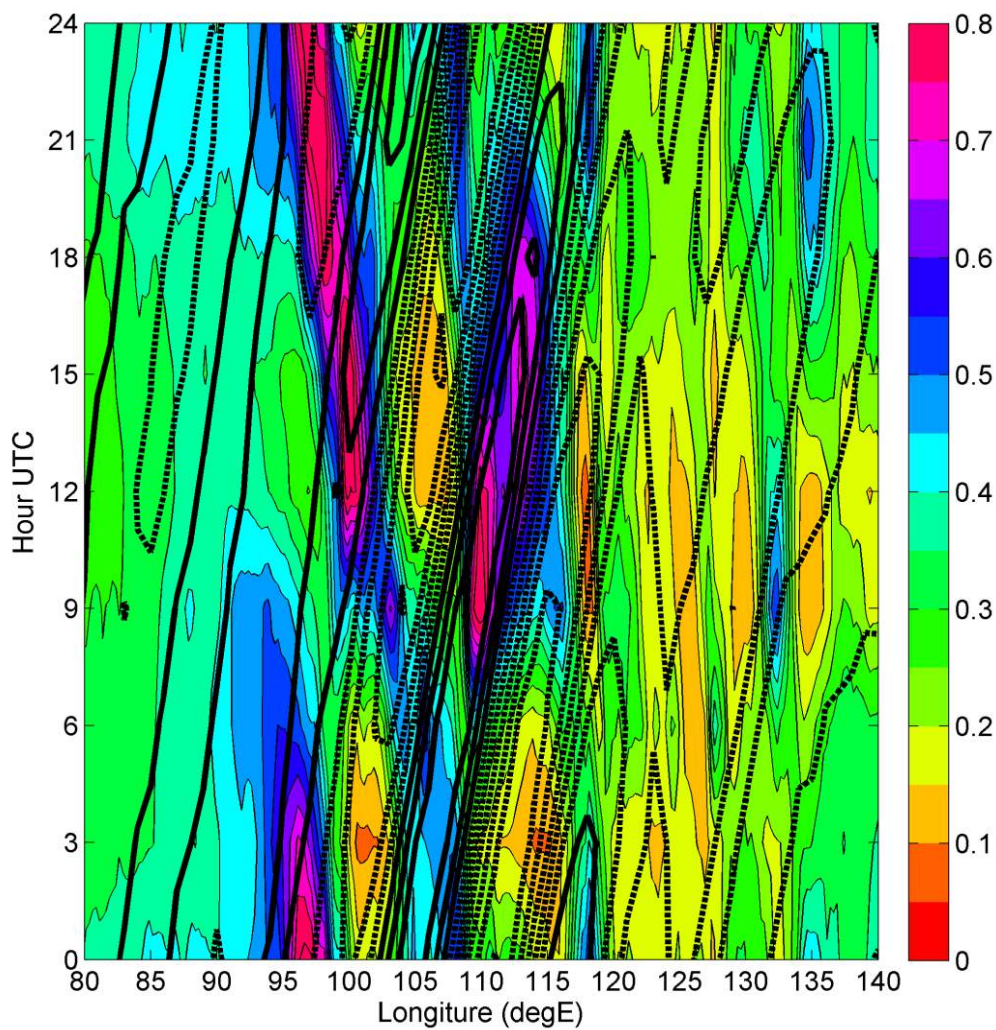

**Figure 5.25** The longitude-diurnal cycle diagram of the average full precipitation based on 15 years of TRMM data. Color shading indicates mean precipitation in [mmhr$^{-1}$]. The black contours indicate mean filtered precipitation for "non-blocked" Kelvin waves. The solid black contours indicate mean precipitation rates above 0.35 mmhr$^{-1}$ and dashed black contours indicate mean precipitation rates below 0.35 mmhr$^{-1}$. The x-axis is longitude in [degrees east]; the y-axis is UTC hour of the day.

The analysis of the AmPm indices calculated from the longitude-diurnal cycle diagrams for multiple Kelvin waves confirms longitude-diurnal cycle phase locking. The amplitude of the zonal oscillation of the AmPm indices, which describes strength of the locking, is the highest over the Maritime Continent and Africa. Although the magnitude of the oscillation is lower over the ocean, the coherent structure of alternating maxima and minima in AmPm index for filtered precipitation over the central-eastern Indian Ocean is reflection of the propagation. The distance between the two sequential maxima or two sequential minima is consistent with the phase speed of a convectively coupled Kelvin wave.

On the basis of the longitude-diurnal cycle diagram we have investigated the impact of the phase locking on the propagation of Kelvin waves through the Maritime Continent. Out of 550 trajectories active at the longitude 90E, 352 were found to be blocked by the Maritime Continent and 198 were found to make successful transition into the Western Pacific. Waves are considered "blocked" if they are active at 90E but terminate west of 140E. A trajectory is considered "non-blocked" if it is active at 90E and terminates at or east of 140E. Based on these definitions the averaged longitude-diurnal cycle diagrams and AmPm indices are calculated for "blocked" and "non-blocked" Kelvin waves individually. Although the longitude-diurnal cycle phase locking is present over the Maritime Continent for both averaged "blocked" and averaged "non-blocked" Kelvin waves, the differences in phase locking over the ocean are apparent. The existence of the longitude–diurnal cycle phase locking over the Maritime Continent in full precipitation for both "blocked" and "non-blocked" Kelvin waves suggests that the diurnal cycle of convection dominates precipitation associated with propagating waves over that region. Thus, the timing and phase speed of a Kelvin wave may impact its ability to successfully propagate over that region. If the wave is in phase with local diurnal cycle of convection between Sumatra and Borneo it might have better chance of making transition into the Western Pacific as it propagates during conditions that favor convection. If, on the other hand, a Kelvin wave propagation is such that convection associated with it is out of phase with local convection over the Maritime Continent, the conditions would be less favorable. Because the distance between Sumatra and Borneo is roughly equal to the distance travelled by a Kelvin wave in one day, the Maritime Continent may act as a filter for Kelvin waves propagating with specific phase speed and approaching it at the specific time of the day. Figure 5.25 presents mean full precipitation diurnal cycle over the Maritime Continent calculated from 15 years of TRMM data (color shading). Superimposed are contours of longitude-diurnal cycle diagram calculated from filtered precipitation for "non-blocked" Kelvin waves only. Solid lines indicate high values (above 0.35 mmhr$^{-1}$) and dashed lines indicate low values (below 0.35 mmhr$^{-1}$) for the mean filtered precipitation. It can be seen that the average "non-blocked" Kelvin wave longitude-diurnal cycle phase locking agrees very well with timing of the maximum daytime precipitation over Sumatra and Borneo and maximum nighttime precipitation between the islands.

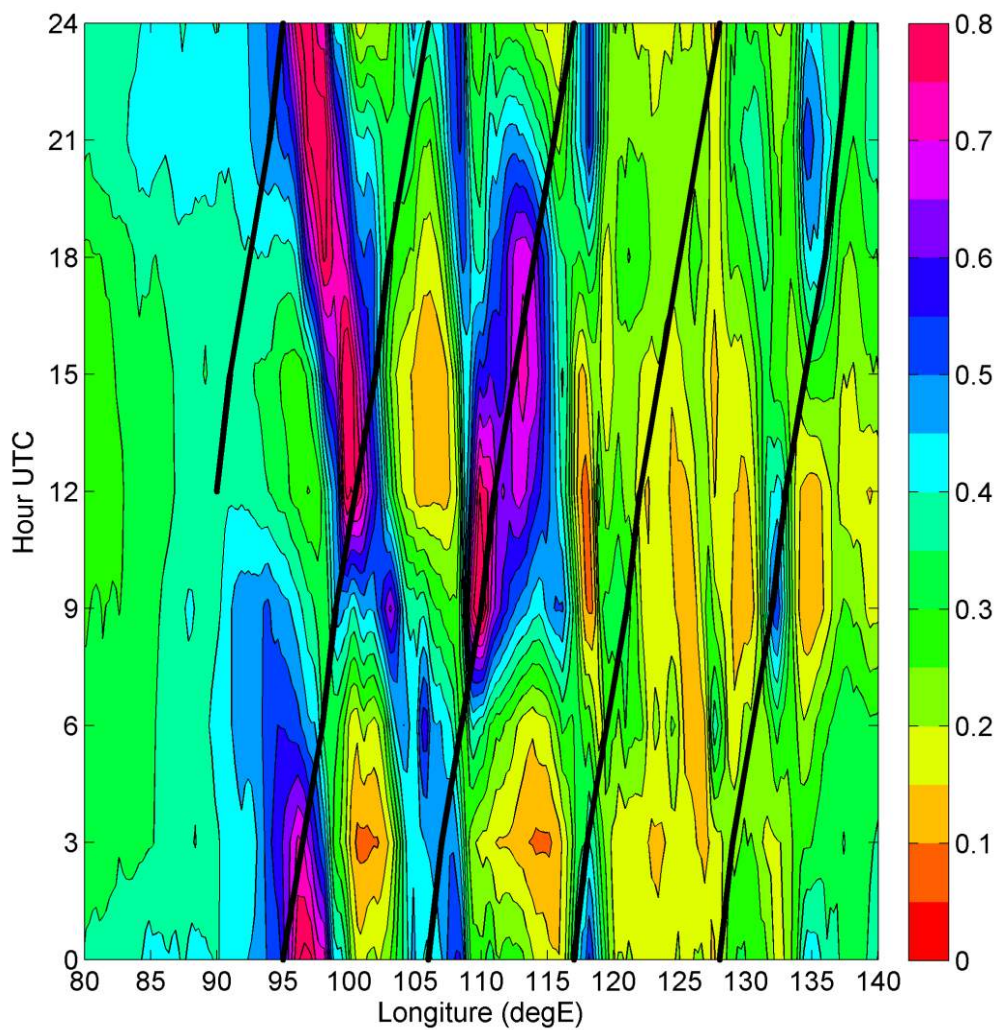

**Figure 5.26** The longitude-diurnal cycle diagram of mean precipitation over the Maritime Continent. The x-axis is longitude in [degrees east]; the y-axis is UTC hour of the day. Color shading indicates mean precipitation based on 15 years of TRMM data in [mmhr$^{-1}$]. The black solid line indicates an example trajectory which was active at 90E on 27 February 2008 at 12UTC. This wave made successful transition into the Western Pacific and encountered strong, local diurnal cycle over Sumatra, Borneo and over the ocean between them.

This mechanism strongly depends on the strength of the local diurnal cycle over the Maritime Continent. Therefore, if for reasons such as intraseasonal variability, monsoon circulation or interactions with ITCZ, the local diurnal cycle over the Maritime Continent is weaker, the proposed mechanism would be weaker too. On the other hand, if diurnal cycle over the Maritime Continent is strong, for example when MJO approaches the region [*Peatman et al.*, 2014], the wave filtering by the Maritime Continent may play an important role in propagation of MJO.

The differences between "blocked" and "non-blocked" Kelvin waves are explored on the basis of the differences in statistics between these two categories. The statistics support composite analysis summarized above. The "non-blocked" Kelvin waves are characterized by lower values of the mean phase speed between 70E and 100E when compared with "blocked" Kelvin waves. However, examples of "blocked" Kelvin waves show that the mean phase speed of a wave may differ between the central-eastern Indian Ocean (70-100E) and the Maritime Continent (90-140E).

The differences between "blocked" and "non-blocked" waves are more apparent when one considers phase speed of Kelvin wave together with an hour at which it approaches the Maritime Continent. We have shown that specific combinations of an hour at which Kelvin wave is active at 90E and its phase speed favor successful propagation. Kelvin waves that occur at 90E at 9UTC, 12UTC and 18UTC and propagate with the phase speed of 10-11 degrees per day over the central-eastern Indian Ocean have the highest chance (about 50%) to make a successful transition into the Western Pacific. Waves that arrive at 90E later during the day maintain higher mean phase speed over the Maritime Continent necessary to encounter favorable conditions over Sumatra and Borneo. Increased likelihood of successful transition into the Western Pacific is linked to interaction with the diurnal cycle of precipitation over the Maritime Continent. We have shown that hypothetic trajectories that are active at 90E at 9UTC or 18UTC, based on their calculated range of phase speeds, have higher chance of interactaction with the diurnal cycle of the precipitation over the Maritime Continent. Figure 5.26 presents an example of the real Kelvin wave trajectory that was active at 90E at 12UTC and maintained the average phase speed of 10 degrees per day over the Maritime Continent. It can be clearly seen that this Kelvin wave encounters positive feedback of the local diurnal cycle of convection over Sumatra and Borneo, as well as over the ocean between the islands.

Our results explain to a certain degree why the Maritime Continent is recognized as a "predictability barrier" for dynamic models. The complex interactions between the local diurnal cycle of convection and Kelvin wave propagation have to be accounted for in order to properly represent Kelvin waves and MJO propagation through that region. However, even modern global climate models have deficiencies in representation of Kelvin wave characteristics [*Hung et al.*, 2013]. For example, the phase speed of a Kelvin wave differs significantly from model to model. In addition the diurnal cycle over the Maritime Continent is difficult to represent in low resolution models [*Love et al.*, 2011]. Therefore, further advancements in direct measurements and modeling efforts are necessary to close the "predictability barrier" over the Maritime Continent. Thus, the consideration of the processes described herein would be advantageous during the Year of the Maritime Continent.

# Chapter 6. Summary

The coupled atmosphere-ocean processes are important factors in understanding atmospheric convection [*Moum et al.*, 2013; *Shinoda et al.*, 1998; *Sui et al.*, 1997; *Weller and Anderson*, 1996]. Development of the atmospheric convection strongly depends on the SST variability [*Cione and Uhlhorn*, 2003; *Fisher*, 1958; *Lau and Sui*, 1997; *Matthews*, 2004]. Therefore, proper representation of atmospheric and oceanic processes which force SST variability is important in tropical meteorology. SST short term variability is driven mostly by changes in solar radiation flux at the ocean surface and mixing between warm oceanic boundary layer and cold thermocline waters [*de Boyer Montégut et al.*, 2004; *Delnore*, 1972; *Price et al.*, 1986; *Price et al.*, 2008]. Additionally, precipitation and evaporation change salinity which impacts the stratification in the upper ocean [*Drushka et al.*, 2011; *Sprintall and Tomczak*, 1992; *Wang et al.*, 2011]. All of the components responsible for short term SST changes (changes in cloudiness, wind speed, fresh water flux and evaporation at the ocean surface) can be linked with convective activity in the tropics, which itself strongly depend on SST [*Gray*, 1968; *Lau and Chan*, 1988; *Webster and Lukas*, 1992]. Therefore, study of these feedbacks is important for proper understanding of processes responsible for convective activity in the tropics, and its variability.

In this study, comprehensive examples of such two-way interactions between atmospheric convection and SST variability have been presented. In Chapter 2 the development and daily evolution of a diurnal warm layer were analyzed on the basis of high temporal resolution in-situ measurements of the upper ocean stratification. We showed that on half of the days in the more than 3 months long study period, a diurnal warm layer developed. The diurnal warm layer was characterized by a temperature structure with a surface maximum that peaked in the mid-afternoon at 1600 LST. The temperature anomaly decayed exponentially with depth over a scale depth of 4–5 m. Further diurnal cycle is such that the diurnal warm layer decays due to nighttime cooling through fluxes of longwave radiation, latent heat and sensible heat and vanishes completely before the sunrise. The energy trapped in the shallow diurnal warm layer is then redistributed throughout the oceanic mixed layer causing its gradual increase. Because the solar heating is effectively trapped in the shallow diurnal warm layer during the day, the daily mean SST is higher than it would have been in the absence of the diurnal warm layer. The mean effective SST anomaly due to existence of the diurnal warm layer is T'≈0.2 °C. The days when a diurnal warm layer developed were characterized by high values of solar radiation flux (above approximately 80 Wm$^{-2}$) and low wind speeds (below approximately 6 ms$^{-1}$). Conversely, the days with no diurnal warm layer had low solar radiation and high wind speed.

Based on in-situ measurements of the diurnal warm layer, predictive model for the existence and strength of the diurnal warm layer was developed, using the gridded solar radiation flux and wind speed data. This predictive model is a useful tool in extrapolating in-situ results into broader spatial and temporal domain. Surface diurnal warm layers are predicted to occur on over 30% of days across the warm pool region, and over the tropical eastern Pacific and eastern Atlantic. However, there is much temporal variability in precipitation and cloud cover due to the multi-scale nature of tropical convection. On days when cloud cover is absent, the wind speed is low enough to allow a surface diurnal warm layer to form. This variability is linked with the MJO with a higher proportion of diurnal warm layer days within the convectively suppressed phase of the MJO compared to the convectively active phase.

This is important because during days on which a diurnal warm layer develops the ocean surface air-sea fluxes change. The daily mean SST anomaly due to the existence of the diurnal warm layer will drive anomalous fluxes of longwave radiation and latent and sensible heat. The anomalous flux is of the order of 4 Wm$^{-2}$ upwards, cooling the ocean and warming the atmosphere. Such anomalous increase of the surface is well established over land, but is novel at the ocean surface.

In Chapter 3, the importance of anomalous flux from the ocean to the atmosphere was further investigated in relation to development of atmospheric convection organized in atmospheric equatorial convectively coupled Kelvin waves. The diurnal warm layer occurrence together with its variability and surface fluxes variability was studied in relation to the initiation of Kelvin waves. This type of tropical disturbances was chosen as an important contributor to the intraseasonal variability and likely suspect to be sensitive to the short term forcing variability. The latter was presented in our idealized numerical model simulations, which showed that diurnally varying forcing changes temporal and spatial characteristics of the Kelvin wave response but is negligible for the Rossby wave response. We hypothesized that the development of the diurnal warm layer and additional anomalous flux from the ocean to the atmosphere is one plausible physical process that could account for diurnal variability of heat source.

The Kelvin wave initiation was investigated on the basis of individual case studies and 15-year statistics based on Kelvin wave trajectories derived from TRMM precipitation data. The upper ocean temperature diurnal variability was assessed using predictive model of the warm layer developed in Chapter 2. The environmental conditions, which include ocean surface wind speed and latent heat flux were used to assess changes in the environment during initiation of the sequential Kelvin waves.

It was found that over Indian Ocean many Kelvin waves pass over the same location shortly after each other. Such waves, which occur within less than 10 days after each other, account for about 40% of all Kelvin wave activity in the eastern Indian Ocean. It was identified that such sequential Kelvin waves initiation is divided into two distinct categories: multiple initiations and spin off initiations. A multiple initiation occurs when two or more Kelvin waves develop over the same area within a short time period. The spin off initiation happens when a Kelvin wave develops in the wake of another Kelvin wave. Analysis of the developed Kelvin wave trajectories database has shown that about 21% of all Kelvin wave initiations in the Indian Ocean are multiple and about 18% of all initiations are of spin off type.

Further, the environmental conditions directly before, during and directly after multiple and spin off initiations were investigated. It was shown, on the basis of 2 case studies, that a pair of multiple Kelvin wave initiates in the presence of the increased diurnal SST variability. A composite analysis of the 83 multiple Kelvin wave initiations that occurred over central and eastern Indian Ocean in 15 years of trajectory data showed that the upper ocean temperature diurnal variability is high and that wind speed and latent heat fluxes at the ocean surface are relatively weak prior to the development of such waves. Hence, the local thermodynamic processes are likely responsible for triggering Kelvin waves during multiple initiations. Thus, it was shown that the diurnal warm layer development and its day to day variability may be related to the varying heat source which triggers sequential multiple initiation of Kelvin waves in model simulations.

Interestingly, analysis of case studies of spin off Kelvin wave initiations showed that contributions from both dynamic forcing and local thermodynamic forcing are important. The preceding ("spinning") wave may influence the environment in its proximity and the sequential ("spin off") wave may develop. The dynamic forcing results from propagation of the increased wind speed and latent heat fluxes likely related to the Rossby wave. The local thermodynamic forcing is due to increased diurnal SST variability in the wake of the preceding Kelvin wave. Composite analysis of 91 spin off initiations that occurred over central and eastern Indian Ocean in 15 years of trajectory data showed that the dynamic forcing is dominant component to which local thermodynamic forcing may contribute.

These results suggest that both the upper ocean temperature diurnal variability and ocean surface fluxes variability should be considered as important factors for formation of atmospheric convection organized in atmospheric equatorial convectively coupled Kelvin waves. Thus, short term variability in the SST forced by changes in wind speed and cloudiness which are manifested as a warm layer at the ocean surface is likely to influence the subsequent development of atmospheric convection.

Eastward propagation of atmospheric equatorial convectively coupled Kelvin waves, away from its area of origin, is associated with typical variability of cloudiness and wind speed at the ocean surface. Organization of the circulation around a Kelvin wave is such that at the location of a Kelvin wave it favors increased cloudiness and increased wind speed at the ocean surface. However, prior and after the passage of a Kelvin wave the circulation is such that it favors subsidence of dry air and therefore calm, clear sky and weak winds conditions. In Chapter 4, typical response of the ocean surface fluxes and daily SST anomaly to the propagation of developed Kelvin waves was investigated on the basis of 2 case studies representing strong and weak MJO activity and composite analysis of 15 years of Kelvin wave trajectories dataset. The DYNAMO period case study, for which MJO was active in the Indian Ocean, showed that typical variability was associated with 2 consecutive Kelvin waves. This typical variability was such that the wind speed and latent heat flux at the ocean surface increased and the diurnal SST variability decreased during passage of each wave. Just before, between and after the two consecutive waves wind speed and latent heat flux at the ocean surface decreased and the diurnal SST variability increased. In the second case study two consecutive Kelvin waves were investigated as well. It was noticed that typical variability in surface fluxes and upper ocean temperature changes associated with them was similar to the DYNAMO case study. However, there were differences in mean values of the fluxes between the two case studies. In the active MJO case study mean values of the wind speed and latent heat flux at the ocean surface were higher and mean value of the diurnal

SST anomaly was lower than in the weak intraseasonal variability case. That suggested that overall ocean surface response to the propagation of a Kelvin wave is modulated by the intraseasonal variability of the tropical convection. Typical variability at the ocean surface is associated with changes in cloudiness and wind speed at the ocean surface during passage of a Kelvin wave. Intraseasonal variability of the tropical convection at its active stage favors convective conditions with increased wind speed and latent heat flux at the ocean surface and increased cloudiness. Weak intraseasonal variability and inactive MJO stage favors clear sky conditions associated with weak wind speed and low latent heat flux at the ocean surface. Therefore, observed differences between case studies for active MJO and weak MJO activity are consistent with changes in environmental conditions on intraseasonal time scales.

Such typical variability at the ocean surface was confirmed by the composite analysis. Additionally, zonal and intraseasonal variability of the ocean surface and upper ocean response to a Kelvin wave passage was studied. It was shown that zonal differences in ocean surface and upper ocean responses to the Kelvin wave were strong. The response is the strongest over eastern Indian Ocean and it is very limited off the coast of Africa. Zonal differences are consistent with zonal gradient of convective activity over the Indian Ocean. Results showed that in the eastern Indian Ocean during the Kelvin wave passage wind speed increases of 0.85 ms$^{-1}$, latent heat flux increases of 20 Wm$^{-2}$ and magnitude of the upper ocean diurnal cycle decreases of 0.7 $^{\circ}$C are expected in comparison to pre-Kelvin wave conditions. Furthermore, we investigated the ocean surface and upper ocean responses to the Kelvin wave in relation to the MJO activity. During the active MJO stage over the Indian Ocean the number of Kelvin wave trajectories in that region decreases by about 45% compared with the suppressed MJO stage of convection. Although typical variability, including magnitude of the changes forced by Kelvin wave passage, is similar in both active and suppressed phases, the mean state varies between them. Active phase favors stronger convection and larger precipitation. Therefore it is associated with stronger wind speeds, larger latent heat flux, smaller insolation and weaker upper ocean diurnal cycle. Suppressed phase favors calm and clear sky conditions and it is associated with lower wind speeds, lower latent heat flux, higher insolation and stronger upper ocean diurnal cycle.

Overall these results show that although Kelvin waves are characterized by fast propagation and their effect on daily SST is small in comparison to zonal, intraseasonal and seasonal changes, they have noticeable impact on the ocean surface fluxes and short term upper ocean energy distribution. We have found that Kelvin wave activity is able to completely suppress upper ocean diurnal cycle and limit impact of the warm layer development on the air-sea fluxes. Intraseasonal variability not only modulates number of the Kelvin waves in the Indian Ocean basin but it also modulates the variability of the ocean response.

In Chapter 5 focus of the research shifted from the Indian Ocean basin to the Maritime Continent. The propagation of atmospheric equatorial convectively coupled Kelvin waves over "propagation barrier" of the Maritime Continent and characteristics of the diurnal cycle of a precipitation along their trajectory was our primary interest. It should be noted that Kelvin waves investigated in Chapter 5 initiated and propagated over Indian Ocean. Therefore, this part is a natural sequel to research presented in Chapters 2 - 4. In Chapter 5, on the basis of the analysis of the longitude-diurnal cycle diagrams for full precipitation and precipitation filtered for Kelvin waves spectrum only, we identified longitude-diurnal cycle phase locking. This specific feature of Kelvin waves is the strongest over land masses of Maritime Continent and Africa. Its magnitude is smaller over Indian Ocean, thought its existence in composite calculations for 15 years of trajectory data is highly surprising. The characteristics of the longitude-diurnal cycle phase locking over the Maritime Continent indicate that Kelvin waves tend to propagate in phase with diurnal cycle of precipitation over that region. The typical diurnal cycle over the Maritime Continent includes local daytime precipitation maxima over the islands of Sumatra and Borneo, and nighttime maxima west of the island of Sumatra and between the two islands. The existence of strong longitude-diurnal cycle phase locking over the Maritime Continent and the fact that phase locking exists over the Indian Ocean suggests that on average Kelvin waves tend to propagate over certain area during certain time of a day. The importance of this phase locking was investigated in relation to the ability of a Kelvin wave to propagate over the Maritime Continent. It was found that Kelvin waves which arrive at the longitude 90E at the particular times of a day (9UTC and 18UTC) and maintain phase speed over the Maritime Continent of around 10 degrees per day have better chance of making successful transition into Western Pacific than waves that arrive there at midnight UTC and maintain different phase speed. Such waves have highest chance to account for local maxima of the diurnal cycle of precipitation. Therefore, Maritime Continent may act as a "filter" for Kelvin waves and increase chance of their successful propagation. We compared characteristics of "blocked" and "non-blocked" (by Maritime Continent) Kelvin waves and showed distinct differences between them. In particular, it was shown that full precipitation over Maritime Continent in both Kelvin wave categories is dominated by diurnal cycle over that region. This confirms that a precipitation in Kelvin wave is dominated by local diurnal cycle over the Maritime Continent. Furthermore, the coherent structure of longitude-diurnal cycle phase locking for "non-blocked" Kelvin waves extends to central-eastern Indian Ocean. The longitude-diurnal cycle phase locking for "blocked" Kelvin waves is limited to the Maritime Continent.

These results explain to a certain degree why the Maritime Continent is "predictability barrier" in dynamic models. The complex interactions between the local diurnal cycle of convection and Kelvin wave propagation need to be accounted for in order to properly represent Kelvin waves and MJO propagation through that region. Proper representation of that interaction is still challenging for modern weather forecasting systems. For example, the phase speed of a Kelvin wave differs significantly from model to model. In addition the diurnal cycle over the Maritime Continent is difficult to represent in low resolution models. Therefore, further advancements in direct measurements and modeling efforts are necessary to close the "predictability barrier" over the Maritime Continent. Consideration of the processes described herein would be advantageous during the incoming field studies such the Year of the Maritime Continent project.

# Appendices

# Appendix A Estimation of surface solar radiation flux from Meteosat7 OLR

The surface solar radiation flux measured at the R/V Roger Revelle was averaged to create an hourly time series. A 10-day sample of this time series (blue solid line in Figure A.1a) shows a clear diurnal cycle. Zero flux at night is typically followed by a smooth curve during the day, peaking at midday at approximately 1000 Wm$^{-2}$. This is indicative of clear sky, cloud free conditions, e.g., on 10 October 2011. However, on some days the solar radiation flux is much reduced, e.g., peaking at only 400 Wm$^{-2}$ on 13 October 2011. This is indicative of cloudy conditions. The green dashed line in Figure 2.10a shows the hourly-mean OLR measured by the Meteosat7 satellite, extracted for the R/V Roger Revelle location. The OLR is constructed from the infrared and water vapor channels, following Roca *et al.* [2002]. The OLR time series are slowly varying (between 260 and 290 Wm$^{-2}$) on clear sky days. However, on cloudy days, such as 13 October 2011, there is a sharp decrease in OLR, down to 120 Wm$^{-2}$, as the infrared emission is from high, cold cloud tops. Hence, during the daytime on cloudy days, low values of OLR correspond to low values of solar radiation flux. This relationship is now exploited to derive a predictive model for solar radiation flux, based on OLR. First, the clear sky diurnal cycle of surface solar radiation flux SWR$_{clear\ sky}$ is calculated for the R/V Roger Revelle location at 0N, 80E (Figure 2.1a). Then a predictand time series y, of the ratio of the measured surface solar radiation flux SWR to the clear sky theoretical flux at the same time of day, is computed

$$y(t) = \frac{SWR(t)}{SWR_{clear\ sky}(t)}. \tag{A.1}$$

For clear sky conditions, y≈1. For cloudy conditions, y<1. A predictor time series x of the OLR difference from its clear sky background value is calculated as

$$x(t) = 1 - \frac{OLR(t)}{OLR_0}, \tag{A.2}$$

where $OLR_0$ is chosen to be a constant 275 Wm$^{-2}$ (Figure A.1a). The analysis is not sensitive to the exact value of this constant. The predictand solar radiation ratio y data are then linearly regressed against the predictor OLR difference x data, as y = ax+b. Data during the night (1800 to 0600 LST) are not used, as SWR = 0 then. Also, shortly after sunrise and before sunset, the solar zenith angle is small so the surface solar fluxes are low and the ratio y is subject to large fluctuations. Hence, only data from 0300 to 1000 UTC (0830 to 1530 LST) are included in the regression calculation, which is performed on a training data set when the R/V Roger Revelle was continuously on station, from 12–19 October 2011. These points are shown by the black crosses in Figure A.1b. The blue dashed line is the best fit regression line. There is a clear negative correlation. Low values of surface solar radiation flux ratio (y) correspond to high values of OLR deficit x, as expected on cloudy days. The large cluster of points near x=0 and y=1 correspond to clear sky days. There are two groups of outliers, corresponding to two physical scenarios where the relationship between surface solar radiation flux and OLR is expected to break down. Shallow (strato-)cumulus clouds will reduce the surface solar radiation flux, but will not have a large effect on OLR as the low cloud tops are still relatively warm. These points form the outliers at low x and low y. High, thin cirrus will not have a large effect on the surface solar radiation flux, but will efficiently absorb upwelling infrared radiation, and re-emit it from the high, cold cloud tops, leading to low OLR. These points form the outliers at high x and high y. Given the physical realism of these outliers, it is reasonable to exclude them from the regression analysis. Hence, the 5% of the data points that lie furthest from the regression lines are labeled with a blue circle in Figure A.1b. The regression was performed on the remaining 95% of the data, and the new regression line is shown by the green dashed line in Figure A.1b. This process was then repeated for four iterations. The final (red) regression line in Figure A.1b is very close to the initial regression line, indicating that these outliers do not have a significant effect on the robust underlying relationship between OLR and surface solar radiation flux. The predictive model for surface solar radiation flux is then

$$SWR_{predicted}(t) = SWR_{clear\,sky}(t)\left\{a\left[1-\frac{OLR(t)}{OLR_0}\right]+b\right\}, \quad (A.3)$$

where a = -1.39, and b = 0.921. This is validated on an independent data set from 11 November to 12 December 2011. The predicted solar radiation flux from a subset of this validation data set (red dashed lines in Figure A.1c) closely matches the measured flux (black solid lines in Figure A.1c), both on hourly and daily mean time scales. Finally, a time series of predicted surface solar radiation flux was constructed for the glider location, using the coefficients from Eq. (A.3). Clear sky and cloudy days can be clearly identified from an example period of this proxy time series (Figure A.1d). There are distinct differences in daily solar radiation values between the location of the R/V Roger Revelle (Figure A.1c) and the glider (Figure A.1d). This confirms that the measured solar radiation flux at the R/V Roger Revelle could not just be used as a proxy for the flux at the glider location. This novel technique could be applied in a range of situations where the surface solar radiation flux is needed at a certain location, and in-situ measurements are taken simultaneously at a different, but nearby, location with similar cloud characteristics.

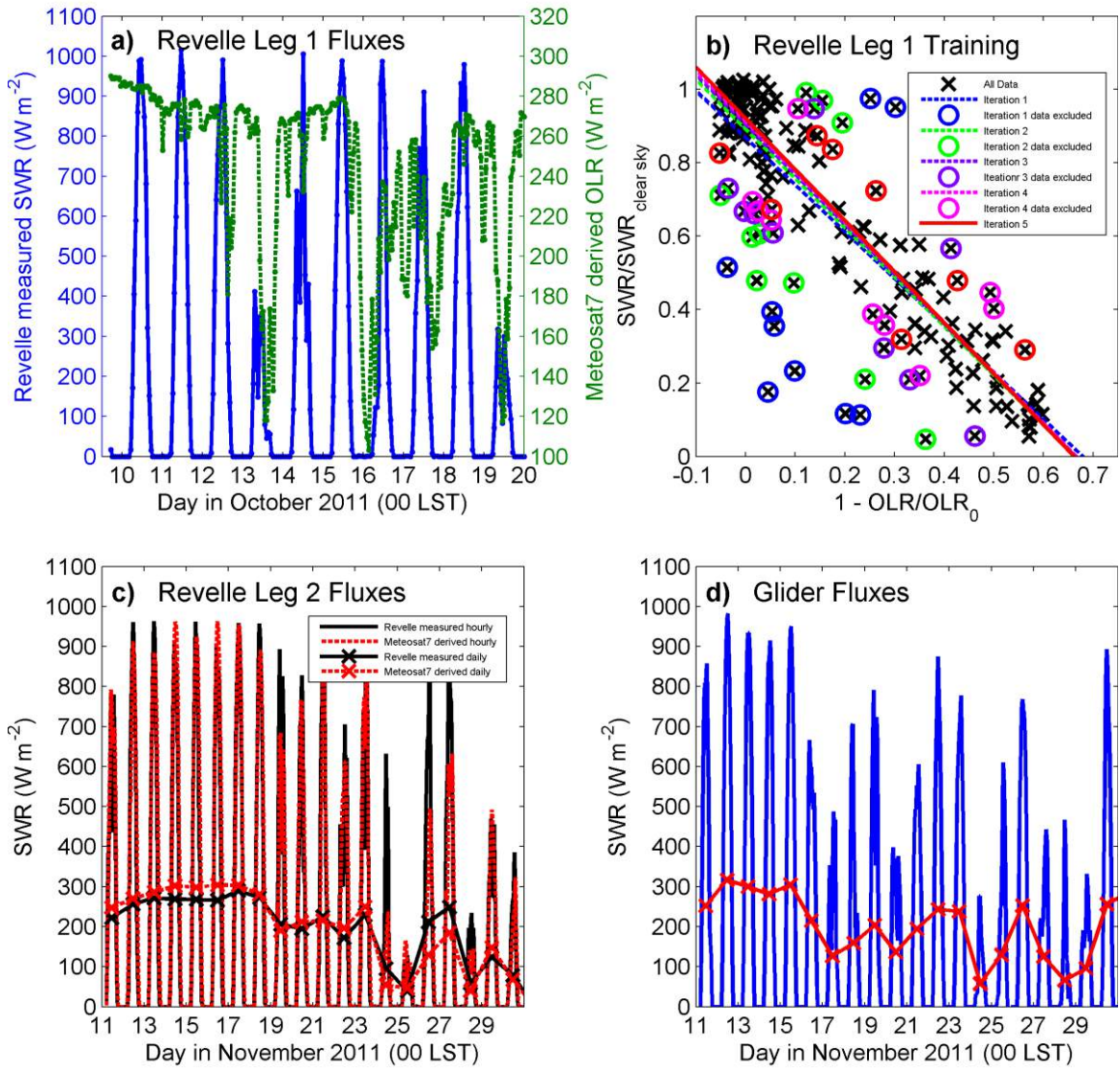

**Figure A.1 (a)** Time series of surface shortwave radiation flux (green dashed line) measured at R/V Roger Revelle, and Meteosat7-derived OLR (blue solid line) at location of R/V Roger Revelle, for the period 10{20 October 2011. Tick marks on the horizontal axis correspond to 0000 LST. **(b)** Scatter plot of Meteosat7-derived OLR difference (OLR–OLR$_0$) at location of R/V Roger Revelle against shortwave radation flux ratio (SWR/SWR$_{clear\,sky}$) measured at R/V Roger Revelle, for the 12{19 October 2011 training period. See main text for details. **(c)** Time series of shortwave radiation flux measured at R/V Roger Revelle (black solid lines), and Meteosat7-derived shortwave radiation flux at location of R/V Roger Revelle (red dashed lines), for part of the validation period. **(d)** Time series of Meteosat7-derived shortwave radiation flux at glider location for November 2011. Hourly data are shown by the blue line, and daily mean data by the red line.

# Appendix B Linearisation of surface fluxes

Here, the surface fluxes are linearised about the mean state to calculate the flux anomalies due to a surface temperature anomaly T'.

## B.1 Surface longwave radiative flux

First, the longwave radiation flux $Q_{LW}$ is given by

$$Q_{LW} = \sigma T_s^4, \tag{B.1}$$

where $\sigma = 5.67 \times 10^{-8}$ W m$^{-2}$ K$^{-1}$ is the Stefan–Boltzmann constant, and $T_s$ is the surface water temperature. This can be linearised about a fixed background temperature $T_b$ = 29.0 °C = 302.15 K (Figure 2.9a), so that $T_s = T_b + T'$ and $T' \ll T_b$. Neglecting terms that are quadratic or higher order in T',

$$Q_{LW} = \overline{Q}_{LW} + Q_{LW}' \approx \sigma T_b^4 + 4\sigma T_b^3 T'. \tag{B.2}$$

Hence, the background longwave radiation flux is $\overline{Q}_{LW} \approx 472$ Wm$^{-2}$, and the perturbation longwave radiation flux is $\overline{Q}_{LW}' \approx 6.3 T'$. Using the value of T'=0.2 °C adopted in section 8, this gives a value of is $\overline{Q}_{LW}' \approx 1.3$ Wm$^{-2}$.

## B.2 Surface latent heat flux

The latent heat flux $Q_L$ is given by the bulk aerodynamic formulae:

$$Q_L = \rho_a L c_E V [q_s(T_s) - q_a], \tag{B.3}$$

where $\rho_a$ is the air density, L=2.5×10$^6$ Jkg$^{-1}$ is the latent heat of vaporization of water, $c_E$=1×10$^{-3}$ is the exchange coefficient, V is the surface wind speed, $q_s$ is the saturation specific humidity, and $q_a$ is the specific humidity. Specific humidity is related to saturation vapour pressure e and air pressure p by

$$q = \frac{\varepsilon e}{p-e} \approx \frac{\varepsilon e}{p}, \tag{B.4}$$

where $\varepsilon$ = 0.622 is the ratio of molecular masses of water vapour and dry air, and e $\ll$ p. The saturation vapour pressure $e_s$ is given by the Clausius–Clapeyron relation

$$e_s = e_{s_0} \exp\left[\frac{L}{R_w}\left(\frac{1}{T_b} - \frac{1}{T_s}\right)\right] \approx e_{s_0}\left(1 + \frac{LT'}{R_w T_b^2}\right), \tag{B.5}$$

where $e_{s0}$ = 4115 Pa is the saturation vapor pressure at $T_b$ = 302.15 K, and $R_w$ =461 Jkg$^{-1}$K$^{-1}$ is the specific gas constant for water vapor. Using the ideal gas law p = $\rho_a R T_b$, where $R = 287$ Jkg$^{-1}$K$^{-1}$ is the specific gas constant for dry air, and $\varepsilon = R/R_w$, the latent heat flux can be decomposed into a background and perturbation term

$$Q_L = \overline{Q}_L + Q_L'$$
$$\approx \rho_a L c_E V [q_s(T_b) - q_a] + \frac{L^2 c_E V \varepsilon e_{s_0}}{R R_w T_b^3} T'. \tag{B.6}$$

Over the tropical ocean, the air temperature is approximately 1 °C below the sea surface temperature and the relative humidity is r = 80%. The specific humidity is $q_a = (r/100) \times q_s(T_a)$, and $V = 3$ ms$^{-1}$ is a typical value on warm layer days when T' = 0.2 °C; Figure 2.10a. Hence, the background latent heat flux is $\overline{Q}_L \approx 54$ Wm$^{-2}$. The perturbation latent heat flux is $Q_L' \approx 13T' \approx 2.6$ Wm$^{-2}$.

## B.3 Surface sensible heat flux

Finally, the sensible heat flux QS is also given by the bulk aerodynamic formulae:

$$Q_S = \rho_a c_p c_E V (T_s - T_a), \tag{B.7}$$

where $c_p = 1004$ J kg$^{-1}$ K$^{-1}$ is the specific heat capacity of dry air. This can also be decomposed into a background and perturbation flux

$$Q_S = \overline{Q}_S + Q_{S'} = \rho_a c_p c_E V (T_b - T_a) + \rho_a c_p c_E V T'. \tag{B.8}$$

The background sensible heat flux is $\overline{Q}_S = 3.5$ Wm$^{-2}$, and the perturbation sensible heat flux is $Q_S' \approx 3.5T' \approx 0.7$ Wm$^{-2}$. This estimate of the anomalous flux is likely to be an overestimate, as it assumes that the air temperature $T_a$ remains constant. In reality, the extra upward flux from the sea surface would warm the atmospheric boundary layer, and reduce the temperature difference $T_S$-$T_a$, reducing the values of the anomalous sensible heat flux. The anomalous latent heat flux would be reduced similarly.

# Appendix C Kelvin wave trajectory detection

## C.1 TRMM precipitation data and filtering

The primary data used to determine Kelvin wave trajectories consists of 15 years of 8 times daily estimates of precipitation from TRMM merged product 3B42 (version 7) [*Huffman et al.*, 2007] extending from January 1998 to December 2013. Such data have often been used to distinguish areas of deep tropical convection and precipitation [*Masunaga*, 2009; *Masunaga et al.*, 2006; *Matthews and Kiladis*, 1999] . The temporal resolution is 3 hours, enabling the analysis of the diurnal cycle. Spatial resolution is 0.25 degree in both longitude and latitude, allowing for representation of high wavenumber features. As Kelvin waves are symmetric around the equator and constrained to the proximity of the equator, the full dataset (latitude, longitude, time) is averaged symmetrically around the equator in the band extending from the latitude 2.5S to the latitude 2.5N. Such methodology had been successfully used in previous studies [*Kiladis et al.*, 2009; *Roundy and Frank*, 2004; *Wheeler and Kiladis*, 1999]. This dataset of time and longitude precipitation is subsequently used for analysis of temporal and zonal variability and is being referred to as the "full TRMM_dataset" - as it contains unfiltered information about the precipitation.

In order to isolate the Kelvin waves from the full TRMM dataset the wavenumber-frequency filtering is performed. The data is transformed into wavenumber-frequency domain using two dimensional FFT; the signal is filtered twice: using methodology described in [*Wheeler and Kiladis*, 1999] and the same methodology but by preserving high frequencies and high zonal numbers related to the diurnal cycle. Thus, two filtered databases - TRMM_filt14 and TRMM_filt40 – are being used, where 14 and 40 refer to the maximum allowed zonal number included in the filtered data. Thus TRMM_filt14 include zonal wavenumbers 1 to 14 and frequencies 1/30 to 0.4cycle per day and TRMM_filt40 include zonal wavenumbers 1 to 40 and frequencies 1/30 to 1 cycle per day. As noted by Roundy [2008] using the symmetric portion of the rainfall, eliminates the signal from the eastward propagation intertia-gravity waves that are present for this wavenumber-frequency range, but have the convective signature asymmetric relative to the equator. Area which is filtered in each of the databases is visible in Figure C.1, which presents Kelvin wave spectrum in wavenumber-frequency domain. Darker shade indicate Kelvin wave spectrum as identified in Wheeler and Kiladis [1999] and further used in dataset TRMM_filt14. Lighter shade indicates Kelvin wave spectrum extended to include high zonal wavenumbers up to 40 and frequencies of up to 1 cycle per day (cpd) which is used in dataset TRMM_filt40.

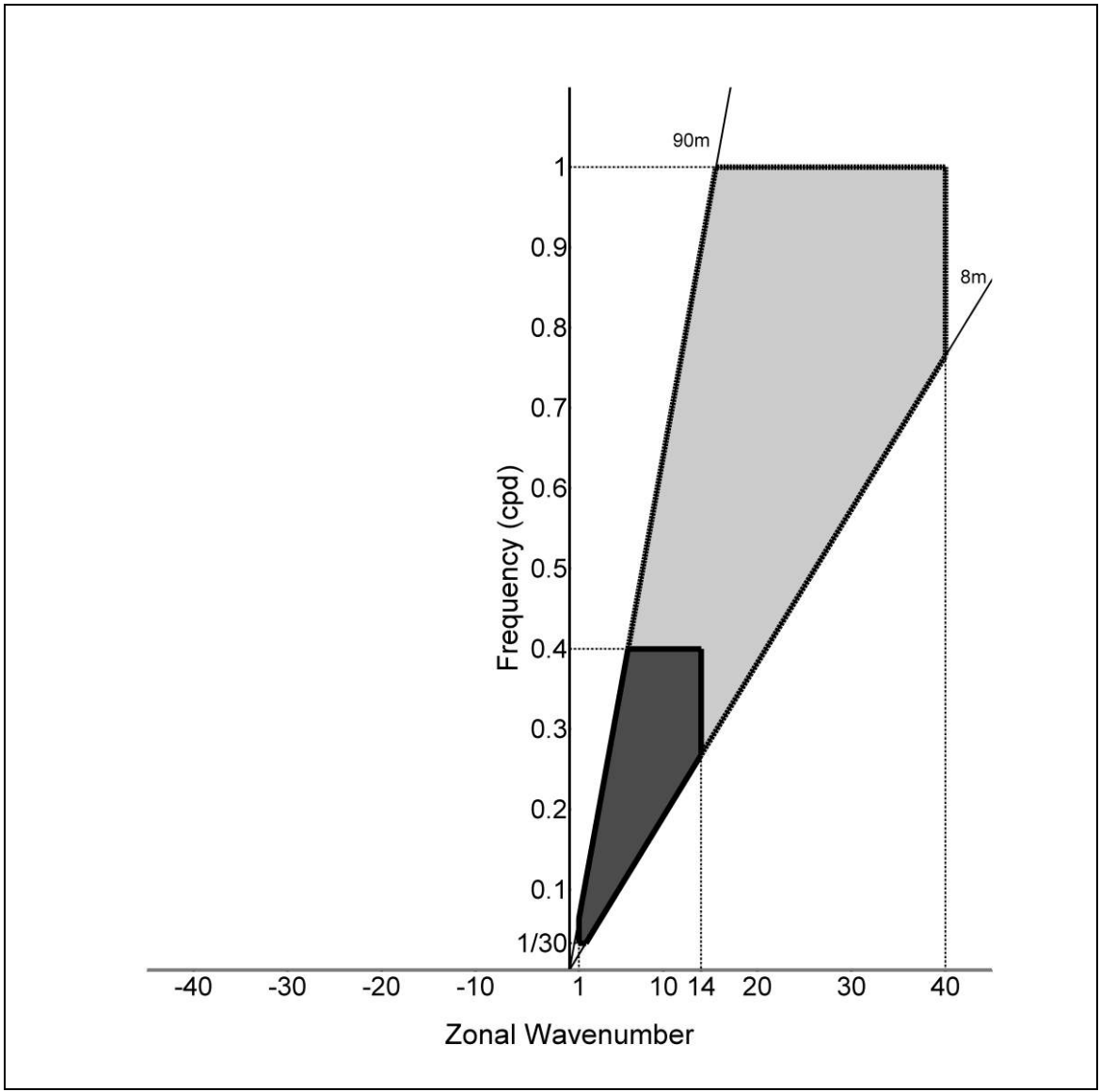

**Figure C.1 Visualization of the Kelvin wave filtering in zonal wavenumber (x axis) and frequency (y axis) domain. Darker shade indicate filtering for TRMM_filt14 limited to zonal wavenumbers less or equal 14 and frequencies less or equal 0.4cpd. Lighter shade indicate extended filtering for TRMM_filt40 which includes zonal wavenumbers up to 40 and frequencies up to 1cpd. Black lines (8m and 90m) indicate equivalent depth derived from shallow water model.**

These datasets represent part of the full precipitation signal associated with the sole Kelvin waves activity. The coarser resolution database (TRMM_filt14) will be used to identify individual Kelvin Wave trajectories, whereas the finer resolution database (TRMM_filt40) includes the diurnal cycle variability of the precipitation. Given the dispersion relationship of Kelvin waves and temporal resolution of the datasets, the minimum effective zonal resolution is approximately 1 degree. Hence, the TRMM_filt14 and TRMM_filt40 data are averaged in the zonal dimension into grid boxes with the resolution of 1 degree.

## C.2 Kelvin waves databases

The TRMM_filt14 dataset is used to detect Kelvin wave activity at every longitude individually. For each longitude local maximum is identified providing the information about the date, hour and amplitude of each individual Kelvin wave. We apply the threshold value of 0.15 mmhr$^{-1}$ to eliminate very small perturbations. This database will be referred to as the "Eulerian database" because the information is given for each longitude-averaged box.

In order to transform the Eulerian database into the Lagrangian coordinates that follow Kelvin wave evolution along its trajectory we consider longitude averaged boxes in close proximity to each other. We have found that in any given location the two sequential Kelvin waves never occur with less then 30 hour time lag. Hence, if the two maxima of the signal are detected in two neighborhood boxes, within the 30 hour timeframe, they are identified as part of a single Kelvin wave trajectory. A trajectory calculation begins with data for which there is no preceding (within the 30 hour threshold) maximum in the westward neighbor of a given location. Such point is defined as the beginning of a trajectory. Calculations are continued until there is no maximum in filtered precipitation to the east of previously defined trajectory point.

Furthermore, trajectories that were less than half day long are excluded from further analysis because they are considered too weak. Once trajectories are known we associate with them TRMM_k14 and TRMM_k40 filtered precipitation and unfiltered precipitation as well as other environmental variables, such as surface fluxes, SST and diurnal SST variability.

An example period of November 2011 during the DYNAMO experiment is presented in Figure C.2 and Figure C.3. Figure C.2 presents filtered precipitation from TRMM_k14 and Figure C.3 presents full precipitation. On both figures the calculated Kelvin wave trajectories are represented by the magenta lines. The trajectory follows continuous, eastward propagating ridges (local signal maxima) which exceed threshold of 0.15 mmhr$^{-1}$ rain rate in unfiltered precipitation. Comparison with the full precipitation signal shows an excellent performance of the trajectory calculation algorithm. However, there are some cases that need to be further discussed. One such category is Kelvin wave trajectories which exhibit small maximum magnitude along its trajectory. For example, the three Kelvin waves visible in filtered precipitation in the eastern Indian Ocean (80-100E) during the first half of December 2011 have maxima below 0.2 mmhr$^{-1}$. Comparison with unfiltered precipitation shows that the two latter trajectories match well with observed precipitation. However, the first trajectory has no precipitation west of 95E and only limited rainfall east of 95E. Such trajectory is biased because its part is not associated with any precipitation and should not be considered a convectively coupled Kelvin wave trajectory. Second category of biases can be discussed using as an example two sequential Kelvin waves which arrive at 80E on 24 November and 28 November. Both trajectories exhibit high maximum values, but comparison with unfiltered precipitation shows that rain in the western Indian Ocean (40-60E) region during their activity is very small. However, both of those trajectories are associated with eastward propagating rain bands west of 40E and are consistent with Kelvin wave filtering criteria (zonal wavenumber and frequency).

Thus, trajectories that have small magnitude and long trajectories for which there is not rainfall in some parts of the trajectory will introduce biases into the database. Such biases are artifacts of the Fourier Transform filtering and are present in all analyses based on that technique. We tested sensitivities of Kelvin wave statistics by changing the threshold criteria and we determined empirically that that 0.15 mmhr$^{-1}$ threshold allows for accounting the Kelvin waves activities without introducing noise that would substantially change the results. Higher value thresholds limit number of detected trajectories and their length and still are suspect to the same biases. Lower value thresholds introduce much more noise to the analysis which can impact and change the results. Thus, although we decided on an empirical threshold value of 0.15 mmhr$^{-1}$ we discuss biases while discussing Kelvin wave characteristics.

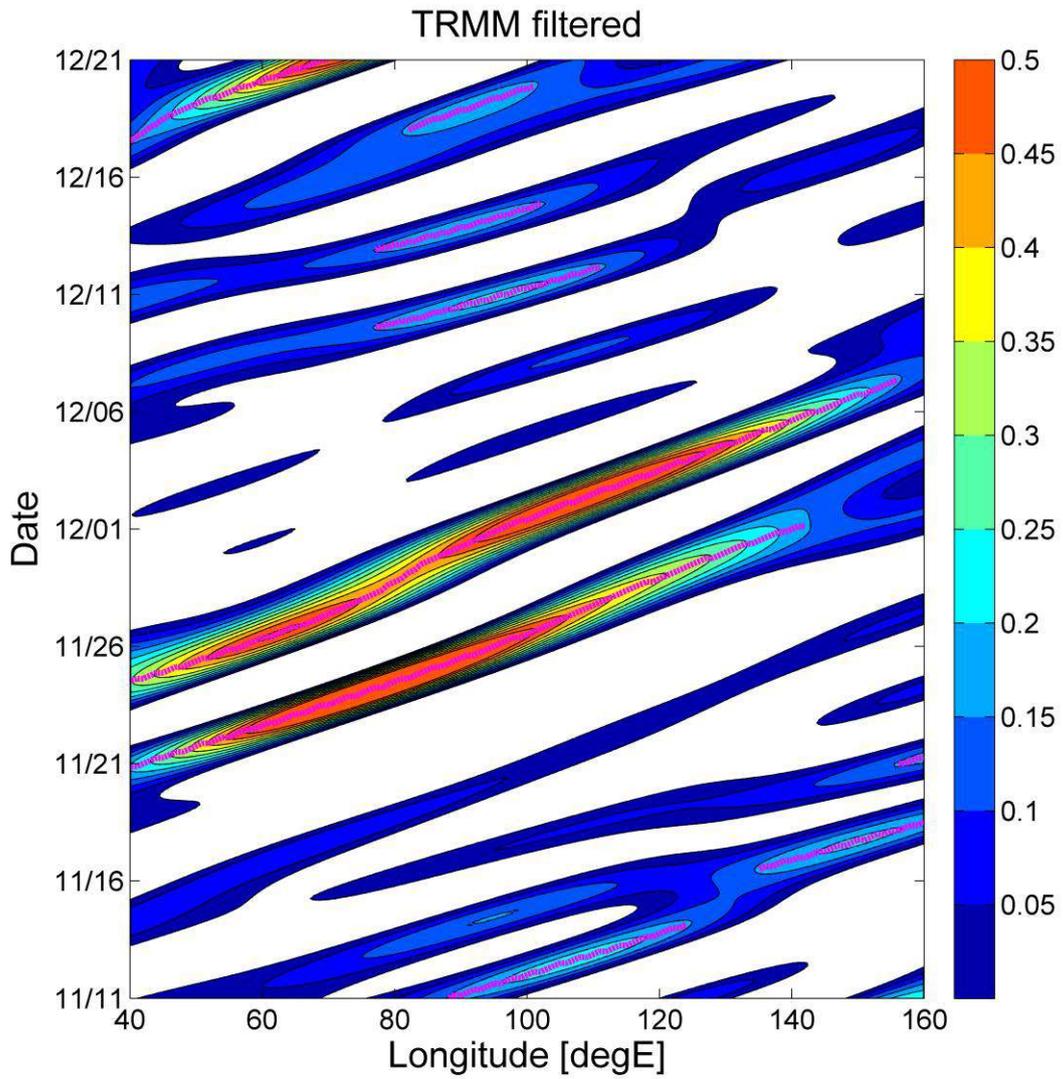

**Figure C.2** Filtered precipitation in November 2011 during the DYNAMO field project. Negative values are masked (white), x-axis is longitude, y-axis indicates the date. Magenta lines connect subsequent maxima and indicate calculated trajectories.

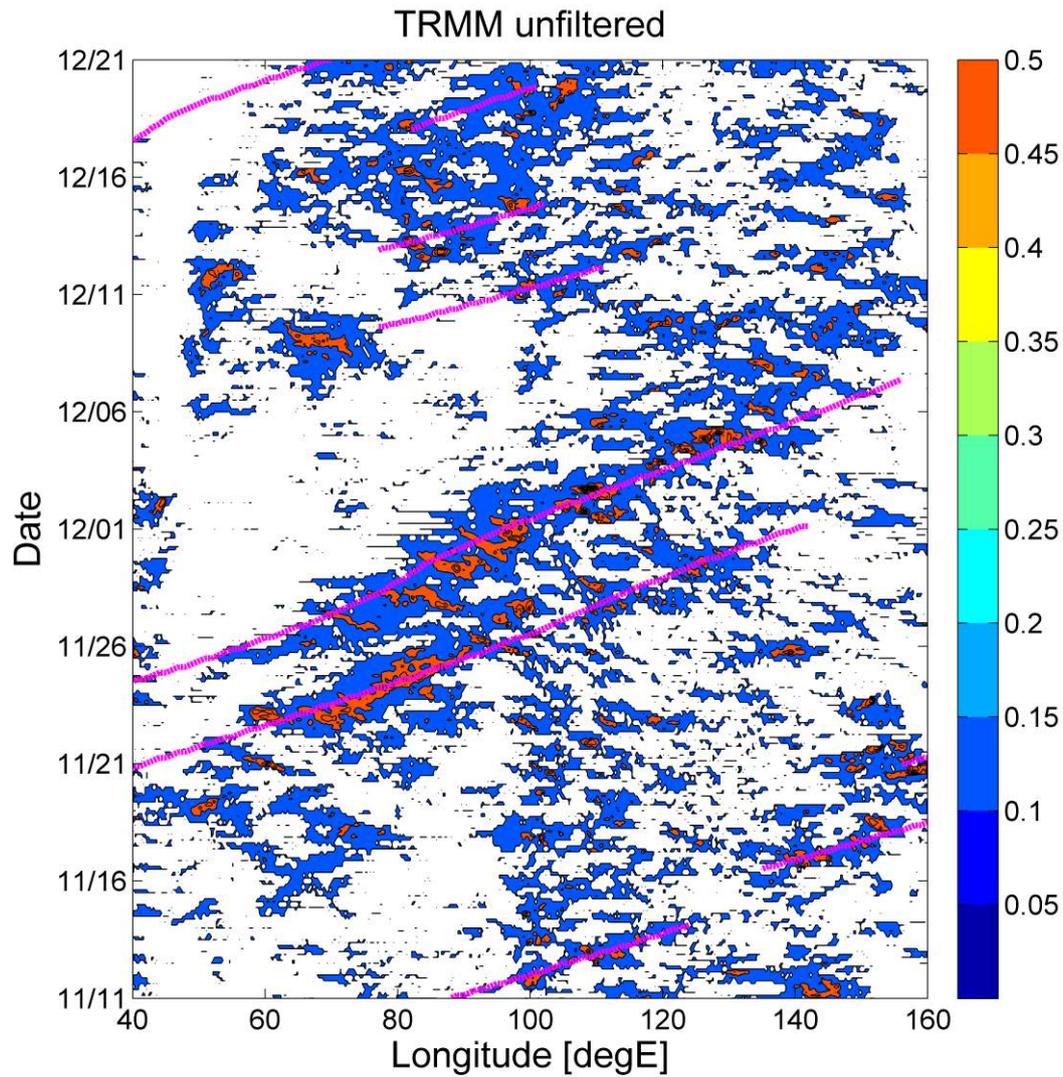

**Figure C.3** Full precipitation for November 2011 during the DYNAMO field project. The x-axis is longitude, y-axis defines the date. Magenta lines represent trajectories calculated from the filtered precipitation.

# Appendix D OAFlux and TropFlux data

OAFlux product [*Yu et al.*, 2008] data provides daily estimates of wind speed, latent heat flux and SST among others data. These data is available on a regular grid with 1×1 degree spatial resolution over oceans. TropFlux product [*Praveen Kumar et al.*, 2013] data provide daily estimates of surface fields including wind speed and shortwave radiation at the ocean surface. These data also have daily temporal resolution and 1×1 degree spatial resolution over tropical oceans extending from 30N to 30S. Our analysis is limited to the proximity of the equator, thus fluxes have to be defined on equator as well. To this end, for every longitude, the maximum value within the equatorial belt of 2.5S – 2.5N is taken. The maximum value is used instead of average over that area, because atmospheric convection is sensitive not only to value of a forcing (i.e. SST) itself but also to the second derivative of the forcing as shown by Li and Carbone [2012] for SST. Therefore, small area within the equatorial belt of high SST gradient may be more important for atmospheric convection development than high average throughout the whole equatorial belt. Wind speed, latent heat flux, SST and magnitude of diurnal SST variability fields were handled similarly. This allowed creating two dimensional time-longitude dataset. These data are used to complement the analysis of the Kelvin wave trajectories.

## D.1 A TropFlux derived $T^{\dagger}$ data

A predictive model of the diurnal SST variability is based on daily mean values of shortwave radiation and wind speed at the ocean surface. This model uses formula derived in Chapter 2 given by Eq (2.18) in order to combine wind speed and shortwave radiation at the ocean surface. Therefore the prescriptive model of the diurnal SST variability ($T^{\dagger}$) provides estimates of daily averaged temperature anomaly near the ocean surface.